\documentclass[a4paper,11pt]{book}
\usepackage{fullpage,makeidx,citesort}
\usepackage{amsmath,amssymb,amsfonts}
\usepackage{epsfig}
\newcommand{\sign}{\operatorname{sign}}
\newcommand{\tr}{\operatorname{tr}}
\newcommand{\str}{\operatorname{str}}
\newcommand{\sdet}{\operatorname{sdet}}
\newcommand{\Tr}{\operatorname{Tr}}
\newcommand{\Arctan}{\operatorname{Arctan}}\newcommand{\Arcth}{\operatorname{Arcth}}
\newcommand{\Li}{\operatorname{Li}}
\newcommand{\diag}{\operatorname{diag}}
\newcommand{\negcdot}{\negmedspace\cdot\negthinspace}
\newcommand{\negtwo}{\negthickspace\negthickspace}
\newcommand{\negthincdot}{\negthinspace\cdot\negthinspace}
\newcommand{\sls}{\negthickspace\not\negmedspace}
\newcommand{\be}{\begin{equation}}
\newcommand{\ee}{\end{equation}}

\makeindex

\begin{document}
\titlepage
$${}$$
\vskip2cm
%
%
%

{\Large
\begin{center}
{\LARGE U}NIVERSITY {\LARGE O}F {\LARGE L}JUBLJANA\\
{\LARGE F}ACULTY {\LARGE O}F {\LARGE M}ATHEMATICS {\LARGE A}ND {\LARGE P}HYSICS
\vskip4cm

\end{center}
}
\begin{center}
{\LARGE
Jure Zupan}
\vskip0.5cm

{\huge {\bf 
 Chiral corrections in electroweak \\
processes with heavy mesons}}\\
\vskip0.5cm
{\LARGE Ph.D. Thesis
\vskip3.7cm
Advisor: Prof. Svjetlana Fajfer\\

\vskip5cm

Ljubljana, 2002
}
\end{center}

\newpage
\thispagestyle{empty}
${}$

\newpage
\thispagestyle{empty}
$${}$$
\vskip10cm
\begin{flushright}
{\LARGE {\bf
TO ANDREJA}}
\end{flushright}
\newpage
\thispagestyle{empty}
$${}$$
\newpage
\thispagestyle{empty}
$${}$$

Many thanks go to the people at the Department of Theoretical Physics at Jo\v zef Stefan Institute. For many enlightening discussions I am especially indebted to Borut Bajc, Damjan Janc, Matja\v z Polj\v sak, Sa\v sa Prelov\v sek Komelj, and of course to my advisor Svjetlana Fajfer, that has managed to guide me through Scylla and Charybdis with many invaluable suggestions. I would also like to thank Damir Becirevic, Jan Olav Eeg, Yuval Grossman, Sourov Roy, and Paul Singer, for widening my horizons with many ideas and insights, as well as for successful collaboration.

It is difficult to thank enough to those, that I hold dear, Andreja, my parents, and friends, who have had enough patience with me over the long hours spent behind the computer screen in preparation of this text.

I would like to acknowledge that this work was supported in part by the Ministry of Education, Science and Sport of the Republic of Slovenia. I am also grateful to the High Energy Physics Group at the Israel Institute of Technology-Technion for the hospitality during winter 2002, where part of this work has been done.

\newpage
\thispagestyle{empty}
$${}$$ 

\chapter*{Abstract}\markboth{ABSTRACT}{}\thispagestyle{empty}
The effective theory based on combined chiral and heavy quark symmetry, the heavy hadron chiral perturbation theory, is applied to $D$ meson decays. In $D^0\to K^0\bar{K}^0$ decay the nonfactorizable contributions are calculated. These arise from chiral loops and products of color-octet currents, while the prediction vanishes in the factorization limit. The approach is confronted with the experimental data. Next, the flavor changing neutral current rare charm decays are considered. The predictions for $c\to u l^+l^-$, $D^0\to \gamma \gamma$, and $D^0\to l^+l^-\gamma$ are given both in the Standard Model as well as for the Minimal Supersymmetric Standard Model with and without $R$ parity conservation. A possible enhancement of order 50 compared to the Standard model prediction is found for the $D^0\to \mu^+\mu^-\gamma$ channel. This makes it an interesting probe of New Physics. 

A modified version of the heavy hadron chiral perturbation theory is used to estimate effects of quenched approximation in the lattice calculations of $B\to \pi, K$ transitions. The relevant form factors, $F_{+,0}$, contain the chiral quenched logarithms that diverge in the chiral limit $m_\pi\to 0$. Behavior of the form factors as functions of $m_\pi^2$ in quenched and full QCD is then found to be substantially different in the region close to the physical pion mass.

In the thesis several technical details are clarified as well. The explicit calculation of three and four-point scalar functions with one heavy-quark propagator is given. Next, existing renormalization group evolutions for $B$ and $K$ meson decays are modified to perform next-to-leading order evolution of Wilson coefficients for charm decays. Also a discussion of gauge invariance in effective theories is given.

\vskip1cm
\noindent
{\ttfamily Key Words:} flavor changing neutral current, weak decays of heavy mesons, heavy meson chiral perturbation theory, rare radiative decays, new physics searches, quenched approximation, lattice quantum chromodynamics
\vskip1cm
\noindent
{\ttfamily PACS:}
13.25.Ft, 
13.20.-v, 
13.60.-r, 
12.60.Jv, 
12.38.Gc 
\newpage
{}\thispagestyle{empty}

\chapter*{Notation}\markboth{NOTATION}{}\thispagestyle{empty}
The characters from the middle of the Greek alphabet $\mu,\nu,\dots$ in general run over space-time indices $0,1,2,3$, while the Latin indices $i,j,k,\dots$ run over spatial indices $1,2,3$.\\
\\
The metric used in the thesis is $\eta^{\mu\nu}=\diag(1,-1,-1,-1)$, where the indices $\mu, \nu$ run over $0,1,2,3$, with $0$ the temporal index.\\
\\
 The Levi-Civita tensor $\epsilon^{\mu\nu\rho\sigma}$ is defined as a totally antisymmetric tensor with $\epsilon^{0123}=1$.\\ 
\\
The Einstein summation over repeated indices is assumed unless stated otherwise. The dot-product $\;p\negcdot k$ denotes $p^\mu k_\mu$.\\
\\
The Dirac matrices are defined so that $\gamma_\mu\gamma_\nu+\gamma_\nu\gamma_\mu=2 \eta_{\mu\nu}$. Also, $\gamma_5=i\gamma_0\gamma_1\gamma_2\gamma_3$. The left and right-chirality projection operators are $P_L=\frac{1}{2}(1-\gamma_5)$ and $P_R=\frac{1}{2}(1+\gamma_5)$. The matrix $\sigma^{\mu\nu}$ is $\sigma^{\mu\nu}=\frac{i}{2}[\gamma^\mu,\gamma^\nu]$. The slash on a character denotes $\;\sls p=p^\mu\gamma_\mu$.\\
\\
The trace $\Tr$ runs over the Dirac indices, while the lower case trace $\tr$ runs over the $SU(3)$ flavor indices.\\
\\
The complex conjugate and Hermitian adjoint of a vector or a matrix $A$ are denoted $A^*$ and $A^\dagger$ respectively. A hermitian adjoint of an operator $O$ is denoted $O^\dagger$. A bar on a Dirac bispinor $u$ denotes $\bar{u}=u^\dagger \gamma_0$.\\
\\
The imaginary and real part of a complex number $z$ are denoted $\Im(z)$ and $\Re(z)$ respectively. \\
\\
The Heaviside function $\Theta(u)$ is defined as $\Theta(u)=1$ for $u>0$ and zero otherwise.\\
\\
Natural units with $\hbar$ and the speed of light taken to be unity are used. The fine structure constant is thus $\alpha_{\text{QED}}=e^2/4\pi\simeq 1/137$.

\newpage
\thispagestyle{empty}
$${}$$

\renewcommand{\figurename}{Figure}
\renewcommand{\tablename}{Table}
\renewcommand{\thesection}{\arabic{chapter}.\arabic{section}}
\chapter{Introduction}
\pagenumbering{arabic}
Particle physics has gone a long way from its beginnings in the first half of the $20^{\text{th}}$ century. From the present perspective it is actually hard to imagine, what the world was like without the ``Standard Model'' of elementary particle physics\footnote{The name was apparently bestowed by Sam B. Treiman \cite{Rosner:2001zy}.}. The gauge-field theoretical description of fundamental electromagnetic, weak, and strong interactions, that emerged in the 1960's, has completely dominated the field ever since. \index{Standard Model} \index{Standard Model! origin of name}

The structure of the Standard Model is as follows. Its building blocks are fermions, leptons and quarks \cite{Gell-Mann:nj}, that come in three families. The Standard Model gauge group is $SU(3)_c\times SU(2)_L\times U(1)_Y$, where the $SU(3)_c$ is the gauge group of Quantum Chromodynamics (QCD) \index{Quantum Chromodynamics} \index{QCD} \cite{Greenberg:1964pe}, $SU(2)_L$ is the gauge group of weak isospin, \index{weak isospin} while $U(1)_Y$ is the gauge group of weak hypercharge \cite{Glashow:tr}. \index{hypercharge} The classification of the leptons and quarks appearing in the Standard Model according to the weak isospin is \index{SU2L@$SU(2)_L$}\index{SU2R@$SU(2)_R$}\index{SU3c@$SU(3)_c$}\index{U1Y@$U(1)_Y$}
\begin{align*}
\begin{matrix}
\text{leptons}\qquad&&&\\
&\dbinom{\nu_e}{e}_L&\dbinom{\nu_\mu}{\mu}_L&\dbinom{\nu_\tau}{\tau}_L\\
&e_R & \mu_R & \tau_R\\
\text{quarks}\qquad&&&\\
&\dbinom{u}{d'}_L&\dbinom{c}{s'}_L&\dbinom{t}{b'}_L\\
&u_R & c_R & t_R\\
&d'_R & s'_R & b'_R
\end{matrix}
\end{align*}
where the binomials with the subscript $L$ denote the weak isospin doublets. Leptons are color singlets, while quarks are in the fundamental representation of $SU(3)_c$. The masses of leptons and quarks are generated via Higgs mechanism \cite{Higgs:ia}. This also gives masses to the $W^\pm$ and $Z$ bosons and breaks the electroweak gauge group $SU(2)_L\times U(1)_Y$ to the electromagnetic $U(1)$. Because there are no right-handed partners of the left-handed neutrinos, these are left massless in the Standard Model (SM), if only renormalizable terms are present. \index{SU2L@$SU(2)_L$}

It is customary to use the mass eigenbasis instead of the weak basis for the quark fields. The rotation to the mass eigenbasis is conventionally conveyed to the down-quark fields
\be
\begin{pmatrix}
d'\\
s'\\
b'
\end{pmatrix}
=
\underbrace{
\begin{pmatrix}
V_{ud}& V_{us}&V_{ub}\\
V_{cd}& V_{cs}&V_{cb}\\
V_{td}& V_{ts}&V_{tb}
\end{pmatrix}
}_{\displaystyle \equiv V_{\text{CKM}}}
\cdot
\begin{pmatrix}
d\\
s\\
b
\end{pmatrix},
\ee
where $V_\text{CKM}$ is a unitary matrix, called the Cabibbo-Kobayashi-Maskawa or CKM matrix \cite{Cabibbo:yz}. \index{Cabibbo-Kobayashi-Maskawa matrix} \index{CKM matrix} It is described by three real mixing angles and a $CP$ \index{CP@$CP$}violating phase. There are several equivalent parameterizations of the CKM matrix, where a very informative one is the so called Wolfenstein parametrization \cite{Wolfenstein:1983yz}, \index{parameterization, Wolfenstein} \index{Wolfenstein parameterization}that takes into account the hierarchical structure of the CKM matrix. Setting $V_{us}=\lambda\sim 0.22$ and then expanding in terms of $\lambda$ to ${\cal O}(\lambda^3)$
\be
V_{\text{CKM}}=
\begin{pmatrix}
1-\lambda^2/2& \lambda & A\lambda^3(\rho-i\eta)\\
-\lambda & 1-\lambda^2/2 &A \lambda^2\\
A\lambda^3(1-\rho-i \eta)&-A\lambda^2 &1
\end{pmatrix},\label{VCKM-Wolf}
\ee
where the parameters $A$, $\rho$, $\eta$ are real and are of order one.

The successes of the Standard Model (SM) description are abundant. To name just the recent few: electroweak precision tests are generally in impressive agreement with the SM predictions \cite{Grunewald:2002,Altarelli:2000ma}, the $CP$ \index{CP@$CP$}violation experiments in $B$ and $K$ meson systems support the CKM description of the Standard Model with one universal phase \cite{Nir:2002,Higuchi:2002gz,Aubert:2002ic}, the discovery of $t$-quark in the mass range predicted from the electroweak precision data was a triumph of the SM \cite{Abe:1994xt,Abe:1994st}. All in all, there is just one missing building block, the discovery of Higgs boson, \index{Higgs boson} that would make the picture complete. The direct searches at LEP \index{LEP} give the current lower limit $m_H>114.4$ GeV at the $95\%$ confidence level \cite{Mariotti:2002,unknown:2001xw}. The indirect experimental constraints are obtained from the precision measurements of the electroweak parameters, which depend logarithmically on the Higgs boson mass through radiative corrections. Currently these measurements constrain the Standard Model Higgs boson mass to $m_H=81\genfrac{}{}{0pt}{}{+51}{-33}$ GeV or to values smaller than $193$ GeV at the $95\%$ confidence level \cite{Electroweak}.

 From a theoretical point of view, the Standard Model has also quite a few very attractive features. First of all, it is a renormalizable theory. This means that it is very predictive. Using a relatively small set of parameters, masses of quarks and leptons, masses of gauge bosons and the values of coupling constants, all in all of order $20$ \footnote{More precisely 3 lepton masses, 6 quark masses, 4 CKM parameters, 3 gauge coupling constants, mass of the Higgs boson and the quartic coupling $\lambda$ give altogether 18 parameters. Counting in also the strong $CP$\index{CP@$CP$} parameter $\theta$ this amounts to 19 parameters of the renormalizable Standard Model.}, one is able, at least in principle, to predict a myriad of processes. Because of renormalizability no additional infinite terms are generated by quantum effects, so that in principle the validity of the SM can be extended to arbitrary high scales. 

\index{beyond the Standard Model|(}
However, we know from the observations, that the SM cannot be the end of story. First of all, gravity is not included in the Standard Model. The quantized description of gravity has proved to be a very challenging subject, that has kept theorists busy for the past two decades with an especially extensive work done in the field of string theories \cite{Polchinski:rq}. No experimental insight is available in this area, though. Next, recent data from Superkamiokande \cite{Fukuda:1998mi} \index{Superkamiokande} and SNO \index{SNO}\cite{Ahmad:2001an,Ahmad:2002jz} have provided a solid experimental evidence for neutrino oscillations. \index{neutrino oscillations} These imply nonzero neutrino masses, contrary to the SM description. Another phenomenological indication of non-Standard Model physics is the unification of strong, electromagnetic and weak couplings in the context of supersymmetric grand unified theories (SUSY GUTs) \index{GUT} at the scales of ${\cal O}(10^{16}$ GeV) \cite{Dimopoulos:1981yj,Murayama:2000dw}. Very solid experimental data suggesting non-SM physics are coming from astrophysics and cosmology. \index{astrophysical constraints, on SM} The astrophysical observations suggest that most of the matter in the Universe is not luminous, but dark \cite{Salati:1999}. Most of the dark matter also is not baryonic. The nonbaryonic dark matter can either be cold or hot, but the general consensus is that most of it must be cold. The Standard Model does not provide a candidate for nonbaryonic cold dark matter, while for instance a very appealing candidate is provided by the lowest supersymmetric candidate, the neutralino. Another evidence pointing toward SM extensions is the generation of baryon-antibaryon asymmetry in the early Universe. To generate it, the interactions between particles should be $CP$ \index{CP@$CP$}violating. The $CP$ violation is present in the SM, but is not strong enough to account for the observed asymmetry \cite{Altarelli:2000ma,Cohen:1993nk}.

There are also some conceptual problems with the structure of the Standard Model. The running of coupling constants suggests the unification scale at $10^{14}-10^{16}$ GeV.\footnote{Note that precise unification of couplings does not occur with running of coupling constants given only by the Standard Model fields \cite{Murayama:2000dw}.} In view of this large scale, the weak scale $1/\sqrt{G_F}\sim 250$ GeV suddenly appears to be very small. The large difference between the two scales cannot be explained ``naturally'' in the context of the SM. This `` hierarchy problem'' is connected to the fact that the theory contains a fundamental scalar field, which receives quadratically divergent loop contributions to the mass parameter. Taking the cutoff in regularization prescription to represent the scale of new physics, the values of the bare mass and the loop corrections to it have to be fine-tuned to give the small physical mass. This is the case, unless the scale of new physics is close to the weak scale. The hierarchy problem can be solved in several different ways. If the fundamental Higsses exist, the theory can be stabilized by TeV scale supersymmetry \cite{Altarelli:2000ma,Murayama:2000dw}. The other option is that Higgs is a composite object, that is either a bound state of fermions or a condensate. Technicolor theories represent a class of proposals along the latter lines \cite{Lane:2000pa}. Another solution to the hierarchy problem has been suggested recently \cite{Arkani-Hamed:1998rs,Antoniadis:1998ig}. If additionally to the usual 3+1 space-time dimensions, ``large'' compact dimensions are assumed, the scale of gravity is much lower than the Planck scale. For two sub-mm extra dimensions the scale of gravity is in the TeV range. \index{signatures, new physics}\index{supersymmetry}

\index{cosmological constraints on SM}

Another challenging conceptual problem is coming from the cosmological observations of distant supernovae type Ia explosions, that suggest a nonzero cosmological constant \cite{Perlmutter:1998np,Riess:1998cb}. The corresponding energy density is of the order of present critical density of the Universe $\rho_c \sim 10^{-26} \text{kg}\; \text{m}^{-3} \sim 10^{-14}\;\text{eV}^4$. If this is to be explained by the vacuum expectation values and the chiral condensates of the SM fields that correspond to energy scales from a few $100$ MeV to a few $ 100$ GeV, one would need an incredibly fine-tuned cancellation between various contributions to arrive at the correct value of the cosmological constant. Note, that a number of alternative explanations for the dimming of supernovae have also been proposed \cite{dust,extradim,evolve,axion-photon}, some of them requiring new physics beyond the SM.

\index{beyond the Standard Model|)}

Given the discussion above, the modern point of view is to consider the Standard Model ``merely'' as an effective field theory. In the effective field theory description one usually has two scales with a very distinct hierarchy and the intermediate scale $\mu$, that separates the two. The physics at the lower scale can then be described by means of a Wilsonian expansion ${\cal L} \sim C_i(\mu) Q_i(\mu)$, where the higher scale physics is hidden in the coefficients $C_i(\mu)$, while operators $Q_{i}(\mu)$ incorporate the lower scale physics. In the Standard Model only the renormalizable operators appear. Operators of higher dimensions are suppressed by the high scale, e.g., by the GUT scale, \index{GUT} and can break the conservation laws of the SM only weakly. 

In the effective field theories, one can distinguish between two approaches, the ``bottom-up'' or the ``top-down'' approach. \index{effective theories!''bottom-up'' approach} \index{effective theories!''top-down'' approach} In the ``top-down'' approach, the high-scale physics is well understood and the coefficients $C_i$ are calculable, for instance in the perturbative framework. The prominent example of this approach is the operator product expansion applied to the weak decays \cite{Buchalla:1995vs}. In this case the ``high scale theory'', the electroweak theory, is well understood. For processes at energies $E\ll\mu_W\sim 90$ GeV, the $W$ and $Z$ bosons can be integrated out. In this way one effectively gets the old Fermi theory of weak interactions, but with calculable corrections to $(V-A)(V-A)$ contact interaction. \index{V-A} Viewing the Standard Model as an effective theory, on the other hand, represents the ``bottom-up'' approach, where little is known about the high energy physics.

An example of the ``bottom-up'' approach is also the application of the effective theory concepts to strong interactions. QCD is a well understood theory, however, the low energy processes are in the nonperturbative region, \index{nonperturbative effects} where an expansion in the coupling constant is no longer applicable. Calculations ab initio, i.e., by starting with the QCD Lagrangian and finishing up with the predictions for physical observables, are still possible through the use of lattice QCD techniques, but are computationally very challenging \cite{Gupta:1997nd}. Lattice methods also have their own limitations. To get meaningful results, calculations have to be done in Euclidean space-time, which makes the calculations of decay processes with more than one hadron in the final state very hard. Also, in order to make the numerical difficulties tractable, a number of approximations have to be made, e.g., by neglecting sea-quark effects, \index{sea-quark} or by working at relatively high pion masses. Another option, that has been commonly used in the past, is to use the symmetries of the QCD Lagrangian to construct effective theories. Unknown couplings in the effective theory are then fixed from experiment. If such an effective theory contains a small expansion parameter, it can be predictable, with more experimental processes predicted than there are parameters to be fixed from experiments. The small expansion parameter for the chiral perturbation theory ($\chi$PT) is provided by the small momenta of interacting Goldstone bosons and by the small masses of $u,d,s$ quarks, with $m_s\sim 100 \; \text{MeV}$ still significantly smaller than the chiral scale $\sim 1$ GeV \cite{Weinberg:1978kz,Gasser:1984gg,Gasser:1983yg,Colangelo:2000zw}. A different approximate symmetry is used to construct the heavy quark effective theory (HQET) \cite{Isgur:vq,Neubert:1993mb,Manohar:dt}. This is obtained when masses of $b$ and $c$ quarks are taken to be very large. Both chiral and heavy-quark symmetries can be combined in processes involving single heavy hadron, resulting in a heavy hadron chiral perturbation theory (HH$\chi$PT) \cite{Casalbuoni:1996pg}. \index{HQET}

\index{approach, used in thesis|(}\index{difficulties!of approach}

In this thesis several applications of the effective theory concepts will be made \cite{Eeg:2001un,Eeg:2001ip,Fajfer:2001ad,Fajfer:2001jq,Fajfer:2002gp,Fajfer:2002bu,Becirevic:2002sc}. The main focus will be on the application of heavy hadron chiral perturbation theory to $D$ meson decays. We use the leading terms in the $1/m_c$ expansion and the expansion of momenta. Note, that in principle also higher order terms in the expansion could be important. Keeping only the leading order terms in the expansion has, however, several important advantages as (i) the set of unknown parameters is relatively small, (ii) there are enough experimental data to fix all of them, (iii) gauge invariance at 1-loop in HH$\chi$PT is obvious. Neglecting higher order terms can then be also viewed as a part of our model. The idea has been tested on the example of $D^0 \to K^0\overline{K^0}$, where good agreement with experiment has been found \cite{Eeg:2001un}. To appreciate this fact, one has to keep in mind that the commonly used factorization approximation predicts vanishing branching ratio for this decay mode, in disagreement with experimental data.
\index{approach, used in thesis|)}

\index{rare decays}
The same theoretical framework is then applied to the rare $D^0\to \gamma\gamma$, $D^0\to l^+l^-\gamma$ decays \cite{Fajfer:2001ad,Fajfer:2001jq,Fajfer:2002gp,Fajfer:2002bu}. The area of rare heavy meson decays has received a boost with the onset of B-factories. \index{B-factories} One of the goals of Belle \index{Belle} and BaBar \index{BaBar} has been to pinpoint the $CP$ \index{CP@$CP$}violating mechanism and to further constrain the CKM matrix elements. But a considerable part of experimental efforts constitute the searches for rare decays \cite{Isidori:2001nd}. Rare decays are especially interesting, if they are connected to a conservation law. Several such selection rules are present in the SM, with $\mu\to e \gamma$ and proton decay for instance completely forbidden at perturbative level in the renormalizable SM, while they can occur in scenarios beyond the SM. But also processes that are not completely forbidden in the Standard Model can be extremely useful as probes of new physics. For instance, the flavor changing neutral currents (FCNCs), i.e., transitions of type $q_i\to q_j +\gamma(\nu\bar{\nu}, l^+l^-)$, do not occur in the SM at tree level. They do, however, occur at the loop level, but are suppressed because of the Glashow-Iliopoulos-Maiani (GIM) mechanism and because of the hierarchical structure of the CKM matrix elements. The FCNCs can be significantly affected by the possible new physics effects, that either contribute at tree level or in the loops. Note that new physics could affect the FCNCs in a substantially different manner than the $\Delta F=2$ and the charged currents, from which the constraints on CKM matrix elements are obtained at present. \index{signatures, new physics}

\index{nonperturbative effects}
Phenomenologically very exciting are the so-called golden-plated modes. For the decay mode to be golden-plated, it has to fulfill several requirements: (i) the SM amplitude has to be either very small or completely forbidden, (ii) it has to be theoretically clean, with small or virtually no uncertainties due to the nonperturbative strong interaction physics, (iii) it has to receive potentially large contributions from new physics scenarios. Examples of such golden modes are $K\to \pi \nu \bar{\nu}$ and $B\to X_s \nu \bar{\nu}$, where theoretical uncertainties due to the nonperturbative physics are very small \cite{Isidori:2001nd}. No such golden modes are present in the charm physics. In rare $D$ decays the nonperturbative physics of light quarks is expected to dominate the decay rates. Consider for instance the case of $c\to u\gamma$ \index{ctougamma@$c\to u\gamma$} transition that occurs only at the one loop level in the Standard Model. The contributions coming from $b,s,d$ quarks running in the loop are
\be
{\cal M}(c\to u\gamma)=\sum_{q=d,s,b}V_{uq}^* V_{cq} {\cal M}_q \sim \left\{
\begin{aligned}
{\cal O}&(\lambda^5 m_b^2)&: b-\text{quark},\\
{\cal O}&(\lambda m_s^2)&: s-\text{quark},\\
{\cal O}&(\lambda \Lambda_{\text{QCD}}^2)&: d-\text{quark},
\end{aligned}\right.\label{ctougamma}
\ee
where we have tentatively set $\Lambda_{\text{QCD}}\sim 300$ MeV instead of $m_u$ for the $u-$quark contribution, anticipating the size of nonperturbative effects. The expansion parameter $\lambda=\sin \theta_c=0.22$ is coming from Wolfenstein parametrization of CKM matrix \eqref{VCKM-Wolf}. The situation is quite different compared to the $s\to d$ or $b\to s$ FCNCs.\index{FCNC} For instance in $s\to d\nu \bar{\nu}$ the same CKM hierarchy is present as in \eqref{ctougamma}, but with $b,s,d$ in \eqref{ctougamma} replaced by $t,c,u$ respectively. Since top quark is very heavy, this outweighs the $\lambda^4$ suppression. The $s\to d\nu \bar{\nu}$ transition is thus dominated by the perturbatively calculable short distance contribution of intermediate top quark. In $c\to u\gamma$ on the other hand, the short distance (SD) contribution comes from the intermediate $b$ quark. Since $b$ quark is much lighter than the top quark, it cannot surpass the $\lambda^4$ suppression. Thus the contributions from the heaviest, $b$-quark, are expected to be the least important. One can then expect that in rare $D$ decays the nonperturbative long distance (LD) effects coming from the lighter two down quarks, $d,s$, will give the dominant contributions. \index{long-distance contributions}

Because LD effects dominate in $D$ decays, no extraction or tests of CKM matrix are possible in these decays. Also, in order to be able to probe new physics, its effects, if present, have to be large. However, there is an important sidepoint to the whole story. Namely, $D$ physics probes the flavor structure of up-quark sector, in contrast to $K$ and $B$ decays. The non-SM extensions of up and down-quark sectors can be very different. In this sense rare $D$ meson decays can prove as a valuable probe of new physics effects. In this thesis possible effects of supersymmetric extensions of the SM to the $c\to u l^+l^-$, $D^0\to \gamma\gamma$, $D^0\to l^+l^-\gamma$ decays will be considered.

Finally, the use of effective theories can be made also in the lattice QCD ab-initio calculations \cite{Kronfeld:2002pi}. This is not surprising, given that there are many different scales present in the problem, the masses of quarks $m_Q,m_q$, the nonperturbative scale $\Lambda$, as well as the UV and IR cutoff scales set by the lattice spacing $a$ and the size of the lattice $L$. In the ideal case they exhibit a hierarchy
\be
L^{-1}\ll m_q\ll \Lambda\ll m_Q\ll a^{-1}.
\ee
In a real numerical simulation, the parameters $L, a, m_q,m_Q$ are varied. If these are not too far from the real world values, the numerical data can be extrapolated to continuum values. Theoretical guidelines for this extrapolation are provided by the effective field theories, that also allow for the estimate of errors. In this thesis we will focus on a particular approximation made in the lattice calculations, the omission of sea-quark effects. This is called the quenched approximation and is widely used in the calculations. Quenched chiral perturbation theory together with heavy quark symmetry \cite{Booth:1994hx,Sharpe:1995qp} will be used to point toward the consequences of quenching in $B \to \pi$ weak transition calculations \cite{Becirevic:2002sc}.

The outline of the thesis is as follows. In the first three chapters we introduce the prerequisites for the phenomenological studies in the subsequent chapters. In chapter \ref{HQET} we introduce the concept of effective field theories with a focus on $\chi$PT and HQET. Gauge invariance is discussed as well. In chapter \ref{scalar-loops} we work out the technical details connected with the integration of two-, three- and four-point functions in HQET. Chapter \ref{weak-int} is devoted to operator product expansion and the application to weak interactions. The factorization approximation is discussed in the same chapter. In chapter \ref{D0K0K0bar} the theoretical framework is applied to $D^0\to K^0\overline{K^0}$ decay, while in chapter \ref{rareD} rare $D$ decays $D^0\to \gamma \gamma$ and $D^0\to l^+l^-\gamma$ are estimated both in the SM and in the supersymmetric extensions. Finally, in chapter \ref{quenching-errors} quenching errors in $B\to \pi$ transitions are discussed. Further technicalities and explicit results of the calculations are relegated to the appendices.

\chapter{Heavy quark effective theory and chiral expansion}\label{HQET}
\section{Effective theories}\label{Generalsection} \index{effective Lagrangian, definition of|(} \index{effective theory approach}\index{heavy hadron chiral perturbation theory|(}\index{nonperturbative effects}
The persisting problem of the phenomenological calculations in hadronic physics is the nonperturbative nature of strong interactions. In the past three decades the approach of effective theories has proved to be an extremely important tool in these considerations. As is usual in contemporary physics, the hard problems are simplified or avoided entirely by the use of approximate and/or exact symmetries. As will be shown below, the use of symmetries is also the common feature of the effective theories.

\begin{figure}
\begin{center}
\epsfig{file=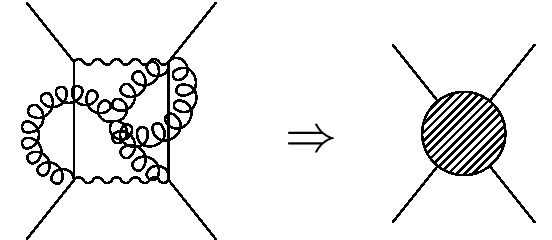, height=2.5cm}
\caption{\footnotesize{One particle irreducible parts of the diagram are replaced by effective vertices- the loops are replaced by the blob, i.e., the tree level effective vertex.}}\label{blob-eff} \index{effective vertices}
\end{center}
\end{figure}

First let us introduce the notion of the effective quantum action. The nonperturbative definition can be found e.g in chapter 16. of \cite{Weinberg:kr}, while we will discuss only the perturbative definition of the \index{effective action} effective action\footnote{The effective action has been first introduced perturbatively in \cite{Goldstone:es}, while a nonperturbative definition was first given in \cite{Jona-Lasinio:1964cw}.}. Let us consider a general quantum field theory with a set of fields $\phi^r$. The observables of the theory are deduced from the appropriate Green's functions. Perturbative calculations of these consist of tree as well as of loop diagrams. The {\it effective quantum action} $\Gamma[\phi]$ is such an action, that reproduces the Green's functions of the original theory exactly, but that is used only at tree level in the perturbative expansion. In terms of path integrals
\begin{equation}
\int [d \phi]\phi^{r_1}\cdots \phi^{r_n} e^{i \int d^4 x {\cal L}(\phi)}=\int_{\text{TREE}} [d\phi]\phi^{r_1}\cdots \phi^{r_n}e^{i\Gamma[\phi]},
\end{equation}
where $[d \phi]$ denotes the path integral measure and the integration on the right-hand-side is only over the tree diagrams. Pictorially, using the effective action $\Gamma[\phi]$ instead of the original action $\int d^4 x {\cal L}(\phi)$ means that the parts of diagrams containing loops can be replaced with blobs representing effective vertices (see Fig.~\ref{blob-eff}). \index{effective vertices} Only tree level diagrams are then left in the calculation.

There is a very useful theorem connected with the effective action.\index{effective action} It states that if the original action $I[\phi]=\int d^4 x {\cal L}[\phi]$ and the integration measure are invariant under the {\it linear} infinitesimal transformations
\begin{equation}
\begin{split}
\phi^r(x)\to \phi^r(x)+&\epsilon {\cal F}(\phi^r;x),\\
& {\cal F}(\phi^r;x)=s^r(x)+\int t^r_{\bar{r}}(x,y) \phi^{\bar r}(y) d^4 y, \label{lintransf}
\end{split}
\end{equation}
where $s^r(x)$ and $t^r_{\bar r}(x,y)$ are c-number functions, then the effective action $\Gamma[\phi]$ is also invariant under the same transformations. Note that this is not necessarily true for nonlinear infinitesimal transformations, where effective action in general will not be invariant under the same transformations as the original action. For proof and further details see \cite{Weinberg:kr}.

\begin{figure}
\begin{center}
\epsfig{file=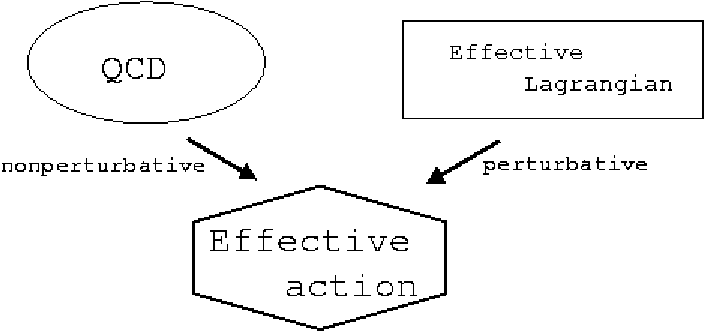, height=4cm}
\caption{\footnotesize{Both QCD and the effective Lagrangian corresponding to QCD lead to the same effective action.\index{effective action} The connection between QCD and the effective action is nonperturbative, between the effective Lagrangian and the effective action on the other hand it is perturbative}}\label{grafikon}\index{nonperturbative effects}
\end{center}
\end{figure}

Now we turn to the notion of effective Lagrangian, with the main idea depicted on Fig.~\ref{grafikon} for the case of QCD. \index{Quantum Chromodynamics} \index{QCD} Let us suppose that the initial Lagrangian ${\cal L}$ consists of a set of fields that transform linearly under a group $G$, and that the Lagrangian itself is invariant under $G$. For the case of QCD the fields will be the light quarks transforming under $SU(3)_L\times SU(3)_R$.\index{SU3L@$SU(3)_L\times SU(3)_R$} Since the realization of $G$ is linear, also the effective action $\Gamma[\phi]$ is invariant under $G$. The calculation leading from the initial Lagrangian ${\cal L}$ to the effective action $\Gamma[\phi]$ can be highly nontrivial, with a possible nonperturbative regime as is the case for the low-energy QCD. It is then useful to construct from the relevant fields $\phi$ (for the case of low-energy QCD, these are the pseudoscalar fields or any other fields relevant for the processes considered) the most general Lagrangian ${\cal L}_{\text{eff}}(\phi)$ invariant under $G$. In general this will consist of an infinite number of terms with unknown couplings. The procedure will be useful if we find a rationale to keep just a finite number of terms. In the case of chiral perturbation theory this is provided by an expansion in momenta, while in heavy quark effective theory the expansion is in the inverse of the heavy-quark mass.

 If the fields in the effective Lagrangian ${\cal L}_{\text{eff}}$ transform linearly under $G$, also the effective action $\Gamma[\phi]$ following from the effective Lagrangian ${\cal L}_{\text{eff}}$ will be invariant under $G$. We can then perform the perturbative calculation using the effective Lagrangian ${\cal L}_{\text{eff}}$ to some fixed order and predict the effective action $\Gamma[\phi]$. At each order a number of unknown couplings have to be determined from the experiment. The number of couplings grows with the higher order contributions, so that the effective Lagrangian approach becomes less and less predictive when going to higher orders. \index{effective Lagrangian, definition of|)}

The method of effective Lagrangians has been very successfully applied to the case of spontaneously broken global symmetries.\index{spontaneous symmetry breaking} In particular it is very successful in the effective description of strong interactions in the low energy regime. We thus illustrate the procedure for this case. We start with the underlying ``fundamental'' Lagrangian ${\cal L}$. Let us suppose that the fields $\phi^r$ transform linearly under some continuous transformation group $G$
\begin{equation}
\phi'=g \phi,
\end{equation}
where $g\in G$, while we have introduced the column $\phi$ containing all the fields. If the initial Lagrangian ${\cal L}(\phi)$ is invariant under $G$, so is the effective action $\Gamma[\phi]$. This is the case also if $G$ is spontaneously broken down to some subgroup $H$. The important artifact of the spontaneously broken global symmetries are the massless Goldstone boson \index{Goldstone bosons} fields that parametrize the $G/H$ right coset space. Since they are massless, it is useful to factor them out of the other fields appearing in the problem. We introduce \index{spontaneous symmetry breaking}\index{xi@$\xi$}
\begin{equation}
\phi(x)=\xi(\varphi(x)) \tilde{\phi}(x),
\end{equation}
where $\varphi$ are the Goldstone boson fields. It is always possible to choose the function $\xi(\varphi(x))$ such that $\tilde{\phi}(x)$ do not contain Goldstone boson fields (for details see chapter 19 of \cite{Weinberg:kr}, or \cite{Coleman:sm,Callan:sn}). Since the subgroup $H$ is unbroken, it is always possible to redefine $\xi(\varphi(x))\to \xi(\varphi(x)) h$, $h\in H$, without changing the effective action. As already stated, the $\xi(\varphi(x))$ is then a representative of the right coset, with $\xi(\varphi(x)) h$ being equivalent for all $h\in H$. A very common parametrization of the right coset space $G/H$ is
\begin{equation}
\xi(\varphi(x))=e^{i \varphi^a(x) x_a},
\end{equation}
with $x_a$ the generators of broken symmetries (the independent vectors in the Lie algebra of $G$ that do not belong to the Lie algebra of $H$). The fields $\tilde{\phi}$, $\varphi$ transform under $G$ as
\begin{equation}
\phi(x)'=g\phi(x)=g\xi(\varphi(x))\tilde{\phi}(x).
\end{equation}
Since $g \xi(\varphi(x))$ is also an element of $G$, it must be in some right coset with the representative $\xi(\varphi'(x))$
\begin{equation}
g\xi(\varphi(x))=\xi(\varphi'(x)) h(\varphi(x),g), \label{transf-gamma}
\end{equation}
where $h(\varphi(x),g)$ is some element of $H$ that depends both on $\varphi$ and $g$, with $h(\varphi(x),h)=h$. From \eqref{transf-gamma} the transformation properties of $\tilde{\phi}(x)$ follow trivially
\begin{equation}
\tilde{\phi}(x)'=h(\varphi(x),g)\tilde{\phi}(x).
\end{equation}
The transformation properties of $\varphi(x)$ and $\tilde{\phi}(x)$ under $G$ are in general very complicated and far from linear. It is thus of great use if one is able to construct functions of Goldstone boson fields and/or other fields that do transform linearly under $G$. These fields will be constructed explicitly for the case of $SU(3)_R\times SU(3)_L$ in the next section. 

Let us now discuss the construction of the effective Lagrangian ${\cal L}_{\text{eff}}(\tilde{\phi},\varphi)$ from fields $\tilde{\phi}(x)$ and $\varphi(x)$. The effective Lagrangian is assumed to be invariant under $G$. Since in general neither $\tilde{\phi}(x)$ nor $\xi(\varphi(x))$ transform linearly under $G$, one could expect that the symmetry properties of the effective Lagrangian ${\cal L}_{\text{eff}}(\tilde{\phi},\varphi)$ could be somewhat distorted in the transition to the description with the corresponding effective action $\Gamma[\tilde{\phi},\varphi]$. This is not the case as can be shown by the following argument. We start with the element of the Lie algebra of $G$
\begin{equation}
\xi^{-1}(\varphi(x)) \partial_\mu \xi(\varphi(x))=i \sum_a x_a D_{a\mu}(x) +i \sum_i t_i E_{i\mu}(x), \label{EandD}
\end{equation}
with
\begin{subequations}
\begin{align}
D_{a\mu}(x)&=\sum_b D_{ab}(\varphi(x))\partial_\mu \varphi_b(x),\\
E_{i\mu}(x)&=\sum_b E_{ib}(\varphi(x))\partial_\mu \varphi_b(x),
\end{align}
\end{subequations}
where $t_i$ denote the generators of Lie algebra of $H$, while $x_a$ denote the other generators in Lie algebra of $G$, as before. It is then fairly easy to show (see chapter 19 of \cite{Weinberg:kr}) that $D_{a\mu}(x)$ and ${\cal D}_\mu \tilde{\phi}(x)=\partial_\mu\tilde{\phi}(x)+i \sum_j t_j E_{j\mu}(x) \tilde{\phi}(x)$ transform as
\begin{equation}
D_{a\mu}(x)'=\sum_b T_{ab}(h(\varphi(x),g)) D_{b\mu}(x), \label{Dtrans}
\end{equation}
\begin{equation}
\big({\cal D}_\mu \tilde{\phi}(x)\big)'=h(\varphi(x),g){\cal D}_\mu \tilde{\phi}(x), \label{calDtrans}
\end{equation}
with $T_{ab}(h)$ the adjoint representation of $H$. Note that $h(\varphi(x),h)=h$, so that $D_{a\mu}(x)$, ${\cal D}_\mu \tilde{\phi}(x)$ in \eqref{Dtrans}, \eqref{calDtrans} transform linearly under $H$. Note also, that $\xi(\varphi(x))$ cannot appear explicitly in the effective Lagrangian ${\cal L}_{\text{eff}}$, because this is assumed to be invariant under the global $G$ transformations and the $\xi(\varphi(x))$ factors can be rotated away. Consider for instance the derivative
\begin{equation}
\partial_\mu \big(\xi(\varphi(x)) \tilde{\phi}(x)\big)=\xi(\varphi(x))\left(\partial_\mu \tilde{\phi}(x)+\xi^{-1}(x)\partial\xi(x)\tilde{\phi}(x)\right). \label{no-deriv}
\end{equation}
The factor $\xi(\varphi(x))$ on the right-hand side can then be rotated away. In the effective Lagrangian ${\cal L}_{\text{eff}}$ only terms with at least one derivative on the Goldstone boson fields or with no Goldstone boson fields will remain. The most general effective Lagrangian ${\cal L}_{\text{eff}}$ invariant under $G$ can then be constructed from $\tilde{\phi}$, $D_{a\mu}$, ${\cal D}_\mu \tilde{\phi}$, ${\cal D}_\mu {\cal D}_\nu\tilde{\phi}$,.... by constructing the most general expression invariant under $H$. Such Lagrangian will then {\it be invariant under the full group G} also. Since these combinations of fields transform linearly under $H$, the effective action $\Gamma$ will be invariant under $H$ as well. Chiral counting\footnote{This is explained at the end of section \ref{heavylightExp}.} prohibits the terms with no derivative on Goldstone fields to appear in the effective action $\Gamma$. The effective action $\Gamma$ will then be an expression constructed from $\tilde{\phi}$, $D_{a\mu}$, ${\cal D}_\mu \tilde{\phi}$, ${\cal D}_\mu {\cal D}_\nu\tilde{\phi}$,...., invariant under $H$ and thus under $G$.

Incidentally, the argument presented above insures also the renormalizability (as understood in the general sense of the word) of the effective Lagrangian approach. No terms that are not already present in the effective Lagrangian ${\cal L}_{\text{eff}}$ can appear in the effective action $\Gamma$. All divergences that appear in the course of the calculation can be reabsorbed into the definitions of the couplings appearing in the effective Lagrangian ${\cal L}_{\text{eff}}$.

\section{Chiral perturbation theory}\label{CHPTsection} \index{chiral!perturbation theory|(} \index{approach, used in thesis|(}\index{CHPT@$\chi$PT|(}\index{chiral!expansion|(}
 One of the earliest and also one of the most successful examples of effective theories is the chiral perturbation theory ($\chi$PT) which we will briefly review in this section. The expansion parameter in the chiral perturbation theory is the momentum exchange in the process, $p^2$. Argument for the validity of this expansion will be given at the end of the next section. 

To start with, let us write down the QCD Lagrangian (we neglect the weak interactions in the following) \index{Quantum Chromodynamics} \index{QCD}\index{flavor}
\begin{equation}
{\cal L}_{\text{QCD}}=-\frac{1}{4} F_{\mu\nu}^aF_a^{\mu\nu}+\sum_n \bar{\Psi}^{(n)}\big(i\gamma^\mu D_\mu-m^{(n)}\big) \Psi^{(n)}, \label{QCD}
\end{equation}
where the summation is over different quark flavors, $\Psi^{(n)}$ are quark fields and the covariant derivative is
\begin{equation}
D_\mu\Psi^{(n)}=\big(\partial_\mu+ig_s G_\mu^a T_a\big)\Psi^{(n)},
\end{equation}
with $G_\mu^a$ the gluon field, $g_s$ the strong coupling \index{gs@$g_s$} and $T_a=\lambda^a/2$ with $\lambda_a$ the Gell-Mann $SU(3)$ matrices, for which $\tr(\lambda^a\lambda^b)=2 \delta^{ab}$. The gauge field strength tensor (curvature tensor) is 
\begin{equation}
T_a F^a_{\mu\nu}\Psi=[D_\mu,D_\nu]\Psi.
\end{equation}
Let us now focus only on the three light quark flavors $u,d, s$. As these quarks are relatively light (with quark masses small compared to the chiral scale as we will see below) the Lagrangian \eqref{QCD} is approximately invariant under the left and right chiral rotations $U(3)_R\times U(3)_L$
\begin{equation}
\Psi^{'n}=\Big[\exp\big({iT_a \theta_R^a \tfrac{1}{2}(1+\gamma_5)}\big)\Big]^{nm}\Big[\exp\big({iT_a \theta_L^a \tfrac{1}{2}(1-\gamma_5)}\big)\Big]^{ml}\Psi^l.
\end{equation}
The axial $U(1)_A$ is anomalous \index{axial anomaly} and is broken by nonperturbative effects. The vector $U(1)_V$ is the global symmetry group of the baryon number and is not needed for further discussion. In the following we then focus on the $SU(3)_R\times SU(3)_L$ global transformation group that is assumed to be spontaneously broken down to the vector subgroup $SU(3)_V$.\index{spontaneous symmetry breaking} Following the general procedure outlined in the previous section we define the quark fields with the Goldstone bosons factored out \index{Goldstone bosons} \index{SU3L@$SU(3)_L\times SU(3)_R$}
\begin{equation}
\Psi=e^{i\gamma_5 \varphi(x)} \tilde{\Psi}. \label{factor-out}
\end{equation}
The Goldstone boson fields $\varphi(x)$ then transform under the $SU(3)_R\times SU(3)_L$ according to (cf. \eqref{transf-gamma})
\begin{equation}
e^{{iT_a \theta_R^a \tfrac{1}{2}(1+\gamma_5)}}e^{{iT_a \theta_L^a \tfrac{1}{2}(1-\gamma_5)}}e^{i \gamma_5 \varphi(x)}=e^{i \gamma_5 \varphi'(x)}e^{i T_a \theta^a_V(\varphi(x))}, \label{transf-all}
\end{equation}
where $\varphi'(x)$ is a $3\times 3$ matrix of transformed Goldstone boson fields, while the dependence of $\theta_V^a(\varphi(x))$ on the parameters of global $SU(3)\times SU(3)$ transformations, $\theta_{L,R}^a$, has not been denoted explicitly in \eqref{transf-all}.
Multiplying with projectors $\frac{1}{2} (1\pm \gamma_5)$ one arrives at
\begin{align}
e^{iT_a \theta_R^a}e^{i\varphi(x)}&=e^{i \varphi'(x)}e^{i T_a \theta_V^a(\varphi(x))}, \label{transf1}\\
e^{iT_a \theta_L^a } e^{-i\varphi(x)}&=e^{-i\varphi'(x)} e^{i T_a \theta_V^a(\varphi(x))}. \label{transf2}
\end{align}
Following \cite{Gasser:1984gg} we introduce $R=\exp\big(i T_a \theta_R^a\big)$, $L=\exp\big(i T_a\theta_L^a\big)$, $U(x)=\exp\big(iT_a\theta_V^a(\varphi(x))\big)$, $\Sigma=\exp(2 i \varphi(x))$. The transformation properties of $\Sigma$ are then
\begin{equation}
\Sigma'=R\Sigma L^\dagger.
\end{equation}
The $\Sigma$ field thus transforms linearly under the global transformations. The quark fields with Goldstone fields factored out transform as 
\begin{equation}
\tilde{\Psi}'=\exp\big(i T_a \theta_V^a(\varphi(x))\big)\tilde{\Psi}=U(x) \tilde{\Psi}. \label{Psitilde}
\end{equation}
To make contact with the previous section we introduce also $\xi(x)=\exp\big(i \varphi(x)\big)$ and vector and axial vector fields \index{xi@$\xi$}
${\cal V}_{\mu}$ and
${\cal A}_\mu$ given by:
\begin{equation}
{\cal V}_{\mu} = \frac{1}{2}(\xi\partial_\mu\xi^\dagger
+\xi^\dagger\partial_\mu\xi) \qquad \qquad
{\cal A}_\mu = \frac{i}{2}
(\xi^\dagger \partial_\mu\xi -\xi\partial_\mu\xi^\dagger),
\label{defVA}
\end{equation}
They transform according to \eqref{transf1}, \eqref{transf2} as
\begin{equation}
{\cal V}'_{\mu} = U{\cal V}_\mu U^\dagger +U\partial_\mu U^\dagger\qquad \qquad
{\cal A}'_\mu = U {\cal A}_\mu U^\dagger.
\end{equation}
The axial vector ${\cal A}_\mu$ and vector ${\cal V}_\mu$ currents are (apart from the constant factor) exactly the $ \sum_a x_a D_{a\mu}(x)$ and $\sum_i t_i E_{i\mu}(x)$ parts of \eqref{EandD} respectively. In other words, the axial vector current ${\cal A}_\mu$ is the $D_{a\mu}$ of \eqref{Dtrans}, while the covariant derivative of \eqref{calDtrans} is $(\partial_\mu +{\cal V}_\mu)\tilde{\Psi}$. \index{axial current}

Up to this point we have been assuming that the $SU(3)_R\times SU(3)_L$ is an exact symmetry of the QCD Lagrangian. However, this symmetry is broken by the mass term in \eqref{QCD}. To introduce the breaking in the effective Lagrangian it is useful to introduce an external field $\chi$ that in the end is set equal to the value of the mass matrix $m=m^{(n)}\delta_{n n'}$. We make the replacement \cite{Gasser:1984gg} \index{SU3L@$SU(3)_L\times SU(3)_R$}
\begin{equation}
\bar{\Psi}m\Psi=\bar{\Psi}_R m\Psi_L+\bar{\Psi}_L m\Psi_R\to \bar{\Psi}_R \chi\Psi_L+\bar{\Psi}_L \chi^\dagger\Psi_R.
\end{equation}
If the external field $\chi$ is assumed to transform according to $\chi\to R \chi L^\dagger$, the ``corrected'' QCD Lagrangian is then invariant under the $SU(3)_R\times SU(3)_L$. This will then be true also for the effective Lagrangian.

Before we write down the final expression for the leading order Lagrangian in the chiral expansion, we rescale the Goldstone boson fields $\varphi=\Pi/f$, where $\Pi$ is a $3\times3$ traceless matrix of light pseudoscalar fields
\begin{equation}
\Pi=\begin{pmatrix}
\frac{\pi^0}{\sqrt 2}+\frac{\eta_8}{\sqrt 6} & \pi^+ & 
K^+\\
\pi^- &-\frac{\pi^0}{\sqrt 2}+\frac{\eta_8}{\sqrt 6} & 
K^0\\
K^- & \bar{K}^0 & -\frac{2}{\sqrt 6}\eta_8 
\end{pmatrix},\label{pimatrix}
\end{equation}
and $f$ is a dimensionful parameter that is determined from experiment. At the leading order it is equal to the pion decay constant $f_\pi=132$ MeV. Later on, in section \ref{coupling}, we will also use the notation $\Pi=\sum_i P^i t^i$, with $t^i=\lambda^i/\sqrt{2}$, where $\lambda^i$ are the $SU(3)$ Gell-Mann matrices. \index{ta@$t^a$}

Since $\Pi/f$ are Goldstone boson fields, the effective Lagrangian does not contain terms without derivatives on $\Pi$, as explained at the end of the previous section (see the discussion below Eq.~\eqref{no-deriv}). For the low energy processes the momentum exchange $p^2$ can then be used as an expansion parameter. In the leading order chiral Lagrangian only the terms with the smallest number of derivatives are kept. Using the counting $p^2\sim m_q\sim m_\pi^2$, one arrives at the usual ${\cal O}(p^2)$ chiral Lagrangian for the light \index{expansion! chiral} \index{Lagrangian! chiral leading}
pseudoscalar mesons \cite{Gasser:1984gg} \index{chiral!Lagrangian}\index{coupling constants! chiral, leading order}\index{mu0@$\mu_0$}
\begin{equation}
{\cal L}_{\text{str}}^{(2)}=\frac{f^2}{8} \tr (\partial^\mu \Sigma
\partial_\mu \Sigma^\dagger)+\frac{f^2 \mu_0}{2}\tr(\chi^\dagger \Sigma
+\Sigma^\dagger \chi),\label{chirallagr}
\end{equation}
where $\Sigma = \exp{(2 i \Pi/f)}$ with
$\Pi $ given in \eqref{pimatrix}, while the trace $\tr$ runs over flavor indices. The external field $\chi$ is set in the calculation equal to the current quark mass matrix ${\cal M}=\diag(m_u,m_d,m_s)$. The coefficient of the first term in \eqref{chirallagr} is fixed by the requirement that the kinetic term of pseudoscalar mesons is properly normalized. The second term on the other hand contains an additional unknown constant $\mu_0$. This terms leads at the leading order to the Gell-Mann--Oakes--Renner relations \cite{Gell-Mann:rz} $m_\pi^2\sim 2\mu_0(m_u+m_d)$, $m_K^2\sim 2\mu_0(m_{u,d}+m_s)$, $m_\eta^2\sim\frac{2}{3} \mu_0 (m_u+m_d+4m_s)$.

The order ${\cal O}(p^4)$ Lagrangian contains ten additional terms \cite{Gasser:1984gg}, which we refer to as counterterms. \index{counterterms! definition of} We will write down explicitly only the terms that contribute to the $\pi$ and $K$ wave function renormalization factors and to the $f_\pi$, $f_K$ decays constants (cf. section \ref{coupling}). Other counterterms will not enter our analysis. The relevant terms are \index{chiral!Lagrangian}\index{coupling constants! chiral, NLO} \index{Lagrangian! chiral subleading}
\begin{equation} 
{\cal L}_{4}=L_4 4 \mu_0\tr(\partial_\mu \Sigma \partial^\mu \Sigma^\dagger)\tr({\cal M}\Sigma^\dagger+\Sigma 
{\cal M}^\dagger)+
L_5 4 \mu_0\tr(\partial_\mu \Sigma^\dagger \partial^\mu \Sigma[{\cal M}\Sigma^\dagger+\Sigma {\cal M}^\dagger])+\dots\label{eqQ:1}
\end{equation}
while the complete ${\cal O}(p^4)$ Lagrangian can be found in \cite{Gasser:1984gg}.

From the chiral Lagrangian
 one can also deduce the form of the 
 light weak current (see e.g. \cite{Donoghue:dd}). At the order ${\cal O}(p)$ this is \index{coupling constants! light current}\index{current, light}\index{light current} \index{hadronization! currents}
\begin{equation}
j_\mu^a \, = \, -i\frac{f^2}{4}\tr(\Sigma \partial_\mu
\Sigma^\dagger\lambda^a),
\label{jX}
\end{equation}
corresponding to the quark current
$j_\mu^a=\bar{q}_{L}\gamma_\mu\lambda^aq_{L}$, with $\lambda^a$ an SU(3)
flavor matrix.\index{chiral!perturbation theory|)} \index{CHPT@$\chi$PT|)}\index{chiral!expansion|)}

\section{Heavy Quark Effective Theory and Chiral Expansion}\label{heavylightExp}\index{expansion! in $1/m_Q$|(}
Since its early applications \cite{Isgur:vq} the heavy quark symmetry has been one of the key ingredients in the theoretical investigations of hadrons containing a heavy quark. It has been successfully applied to the heavy hadron spectroscopy, to the inclusive as well as to a number of exclusive decays (for reviews of the heavy quark effective theory and related issues see \cite{Neubert:1993mb} or \cite{Manohar:dt}). To describe interactions with not too energetic light mesons, the heavy quark symmetry has been combined with chiral symmetries leading to the heavy hadron chiral perturbation theory (HH$\chi$PT) \cite{Casalbuoni:1996pg}. \index{HQET}

The important observation in the heavy quark expansion is that the mesons containing an infinitely heavy quark $Q$ exhibit a set of simple properties. Since a heavy quark is very massive its Compton wavelength is much smaller than the size of the meson. The latter is determined by the wave function corresponding to the light degrees of freedom, the light quarks and the soft gluons. In the limit of an infinitely heavy quark, the wave function of the light degrees of freedom is the solution of QCD field equations for a static triplet color source. It is thus independent of the spin of the heavy quark as well as of its flavor. That is, the solution for the light degrees of freedom does not change if we replace $Q(v,s)$ with $Q'(v,s')$, where $v$ and $s$ denote velocity and spin of the heavy quark respectively.

To get more quantitative, let us consider a hadron with a heavy quark. The major part of the momentum is carried by the heavy quark. This propagates almost unperturbed and interacts with light degrees of freedom only through small exchanges of momenta. In words of Neubert: ``The heavy quark flies like a rock!''\cite{Nubert:Lecture}. It is thus useful to separate the heavy quark momentum $P_Q$ into the momentum due to the movement of the meson $m_Q v$ and the perturbations
\begin{equation}
P^\mu_Q=m_Q v^\mu+k^\mu,
\end{equation}
where $v^\mu$ is the four-velocity of the hadron. The heavy quark propagator is then \index{heavy quark propagator}
\begin{equation}
\frac{i}{{\sls P}_Q-m_Q+i \epsilon}=\frac{i}{v\cdot k+i\epsilon}\frac{1+\sls v}{2}+{\cal O}(k/m_Q)\to \frac{i}{v\cdot k+i\epsilon} P_+,\label{heavy-prop}
\end{equation}
where in the last step the limit $m_q\to \infty$ has been taken and the projectors $P_\pm=(1\pm \sls v)/2$ have been introduced. Since $P_+ \gamma ^\mu P_+=v^\mu$ also the couplings of the heavy quark to gluons $g_s T_a \gamma^\mu$ \eqref{QCD} can be simplified to $g_s T_a v^\mu$ at the leading order in $1/m_Q$. The Lagrangian corresponding to these Feynman rules is
\begin{equation}
{\cal L}=\bar{h}_v (i v^\mu\partial_\mu- g_s v^\mu G^a_\mu T_a)h_v, \label{heavyquark}
\end{equation}
where $h_v$ satisfies $P_+h_v=h_v$, $P_- h_v=0$. This Lagrangian can be obtained from the QCD Lagrangian \eqref{QCD} by projecting to the ``large Dirac components'' and factoring out the trivial phase change due to the hadron movement
\begin{equation}
h_v\sim P_+ e^{im_Q v\cdot x} Q.
\end{equation}
Neglecting terms suppressed by additional powers of $1/m_Q$ this replacement leads to the Lagrangian \eqref{heavyquark}. The heavy quark Lagrangian exhibits the heavy quark spin symmetry. Intuitively this can be expected from the fact that no Dirac gamma matrices appear in \eqref{heavyquark}. Formally, it is easy to show that the Lagrangian \eqref{heavyquark} is invariant under the generators of $SU(2)$ transformations $S^i=\frac{1}{2} \gamma_5 \sls v \sls e^i$, where $i=1,2,3$, while $e^i$ are three vectors orthogonal to the heavy quark velocity $v$, $v\cdot e=0$. For $S^i$ then\footnote{To prove these relations it is best to go to the heavy hadron rest frame \cite{Neubert:1993mb}.}
\begin{equation}
[S^i,S^j]=i \epsilon^{ijk} S^k, \qquad\qquad [\;\sls v,S^i]=0,
\end{equation}
and $S_i^\dagger=\gamma_0S_i\gamma_0$, from which it trivially follows that the Lagrangian \eqref{heavyquark} does not change under the transformation
\begin{equation}
h_v'= (1+i \theta^i S^i)h_v, \label{heavyspintr}
\end{equation}
with both $h_v'$ and $h_v$ satisfying $P_-h_v'=P_-h_v=0$.

To be able to construct the effective Lagrangian on the meson level, we have to consider the transformation properties of the heavy mesons under Lorentz, heavy quark spin and flavor symmetries. We will follow the elegant tensor representation formalism \cite{Falk:1990yz,Falk:1991nq}, but constrain ourselves only to the case of $J^P=0^-,1^-$ mesons.

The mesons consist of the heavy quark Q and a light antiquark $\bar{q}_a$. These are described by the Dirac spinor field $h_v$ for which $\sls v h_v=h_v$ and the light quark field $\bar{q}_a$ for which $\bar{q}_a \sls v=-\bar{q}_a$. The minus sign in the last equation is necessary in order to project out predominantly the antiquark degrees of freedom. The meson field will then be represented by a Dirac spinor-antispinor field $H_a \sim h_v \bar{q}_a$ that in general has 16 components. However the requirements
\begin{equation}
\frac{1-\sls v}{2}H_a=0, \qquad H_a\frac{1+\sls v}{2}=0, \label{projreq}
\end{equation}
reduce the 16 components to only 4 independent components (each of the projectors reduces a Dirac bispinor to a two-component spinor). The four components will then describe the heavy pseudoscalar meson with one entry and the vector meson with three independent degrees of freedom.

The most general field satisfying requirements \eqref{projreq} is then
\begin{equation}
H_a=\frac{1+\sls v}{2} \big(P_a^\mu \gamma_\mu-P_a \gamma_5\big), \label{heavyfield}
\end{equation}
with
\begin{equation}
v^\mu P_\mu=0.
\end{equation}
As expected the $P_a$ is the pseudoscalar field and $P_a^\mu$ the vector meson field. The transformation of the heavy meson field \eqref{heavyfield} under the heavy quark spin transformations \eqref{heavyspintr} is then
\begin{equation}
H_a\to (1+i \theta^i S^i)H_a. \label{mesonheavyspintr}
\end{equation}
To take into considerations also the interactions with the Goldstone bosons\index{Goldstone bosons!interactions of} these are factored out from $H_a$ as outlined in sections \ref{Generalsection}, \ref{CHPTsection} (i.e. the $\bar{q}_a$ is replaced by $\bar{\tilde{q}}_a$, that transforms according to \eqref{Psitilde}). Under the chiral $SU(3)_R\times SU(3)_L$ transformations thus 
\begin{equation}\index{SU3L@$SU(3)_L\times SU(3)_R$}
H_a\to H_b U_{ba}^\dagger, \label{heavychiraltr}
\end{equation}
where we did not write the tilde on the $H_a$ field. Finally, under Lorentz transformations $\Lambda$, the $H_a$ field transforms as $h_v\bar{q}_a$ 
\begin{equation}
H_a\to D(\Lambda)H_a D^{-1}(\Lambda),
\end{equation}
with $D(\Lambda)$ the $(1/2,0)\oplus(0,1/2)$ representation of the Lorentz group $D(\Lambda)=\exp[i \frac{1}{2} \omega^{\mu\nu} \sigma_{\mu\nu}]$.

The most general effective Lagrangian to order ${\cal O}(p)$ in the chiral expansion, that is invariant under the transformations \eqref{mesonheavyspintr}, \eqref{heavychiraltr}, taking into account the restrictions \eqref{projreq}, and is a Lorentz scalar, is \cite{Casalbuoni:1996pg} \index{chiral!Lagrangian}\index{coupling constants! heavy meson sector}\index{Lagrangian! heavy meson}
\begin{equation}
{\cal L}_{\text{str}}^{(1)}=-\Tr(\bar{H}_{a}iv\negcdot D_{ab}H_{b})+g
\Tr(\bar{H}_{a}H_{b} \gamma_\mu {\cal A}_{ba}^\mu \, \gamma_5),
\label{eq-8}
\end{equation}
where $ D_{ab}^\mu H_b=\partial^\mu H_a - H_b{\cal V}_{ba}^\mu$,
 while the trace $\Tr$ runs over Dirac indices.
Note that in (\ref{eq-8}) and the rest of this section
$a$ and $b$ are {\em flavor} indices.

The vector and axial vector fields
${\cal V}_{\mu}$ and
${\cal A}_\mu$ in (\ref{eq-8}) are the same as in \eqref{defVA} and are given by:
\begin{equation}
{\cal V}_{\mu} = \frac{1}{2}(\xi\partial_\mu\xi^\dagger
+\xi^\dagger\partial_\mu\xi) \qquad \qquad
{\cal A}_\mu = \frac{i}{2}
(\xi^\dagger \partial_\mu\xi -\xi\partial_\mu\xi^\dagger),
\end{equation}
where $\xi = \exp{(i \Pi/f)}$, with $\Pi$ defined in \eqref{pimatrix}. The $\overline{H}_{a}$ field is $\overline{H}_{a}=\gamma_0 H_a^\dagger \gamma_0$, with $H_a$ defined in \eqref{heavyfield}.

 The higher order terms (referred to as counterterms) \index{counterterms! definition of} in the expansion in $v\cdot p\sim{\cal O}(p) $ and $m_q\sim {\cal O}(p^2)$ are then up to the order ${\cal O}(p^3)$ \index{chiral!Lagrangian}\index{g@$g$}\index{k12@$k_{1,2}$}\index{Lagrangian! heavy meson}
\begin{equation}
\begin{split}
{\cal L}_3^{\text{heavy}}=&2\lambda_1 \Tr[\bar H_aH_b]({\cal M_+})_{ba}+k_1 \Tr[\bar{H}_a i v \negcdot D_{bc}H_b]
({\cal M_+})_{ca}\\
&+k_2 \Tr[\bar{H}_a i v \negcdot D_{ba}H_b]\tr({\cal M_+})+\delta{\cal L}_g\dots
\end{split}\label{eqQ:3}
\end{equation}
with ${\cal M_+}=\frac{1}{2}
(\xi^\dagger M\xi^\dagger+\xi M\xi)$ and $\delta{\cal L}_q$ given in \eqref{deltaLg}. Dots denote terms that were not written out, as they do not contribute to the $Z_D$, $Z_{D_s}$ wave function renormalization factors and the $f_D$, $f_{D_s}$ decay constants that will be discussed in section \ref{coupling} nor to the decays $D^0\to K^0\bar{K}^0$, $D^0\to \gamma\gamma$, $D^0\to l^+l^-\gamma$ considered in chapters \ref{D0K0K0bar} and \ref{rareD}. The effect of $\lambda_1$ term is to change the heavy meson propagator. In the case of $s$ quark the shift is $v\cdot p \to v \cdot p -\Delta$, where $\Delta=m_{H_s}-m_H$. The $k_1,k_2$ terms will contribute to the wave function renormalization of the heavy mesons. Note that we did not include in the analysis the terms suppressed by $1/m_H$. These are considered to be of higher order in the expansion. 

The $\delta{\cal L}_g$ part of the Lagrangian \eqref{eqQ:3} will contribute a correction to the $D^* D\pi$ coupling from which the value of $g$ will be obtained. Neglecting $1/m_H$ terms, one gets \cite{Boyd:1994pa,Stewart:1998ke}
\be
\begin{split}
\delta{\cal L}_g=&\frac{g \tilde{\varkappa}_1 \mu_0 }{16 \pi^2 f^2} \Tr[\bar{H}_a H_b \gamma_\mu \gamma_5]{\cal A}_{bc}^\mu ({\cal M}_+)_{ca}+\frac{g \tilde{\varkappa}'_1 \mu_0 }{16 \pi^2 f^2} \Tr[\bar{H}_a H_b \gamma_\mu \gamma_5]({\cal M}_+)_{bc}{\cal A}_{ca}^\mu \\
+&\frac{g \tilde{\varkappa}_3 \mu_0 }{16 \pi^2 f^2} \Tr[\bar{H}_a H_b \gamma_\mu \gamma_5]{\cal A}_{ba}^\mu ({\cal M_+})_{cc}+\frac{g \tilde{\varkappa}_5 \mu_0 }{16 \pi^2 f^2} \Tr[\bar{H}_a H_a \gamma_\mu \gamma_5]{\cal A}_{bc}^\mu({\cal M_+})_{cb} \\
+&\frac{\delta_2 }{4 \pi f} \Tr[\bar{H}_a H_b \gamma_\mu \gamma_5]i v\cdot D_{bc}{\cal A}_{ca}^\mu +\frac{\delta_3 }{4 \pi f} \Tr[\bar{H}_a H_b \gamma_\mu \gamma_5]i D_{bc}^\mu v \cdot {\cal A}_{ca}. \label{deltaLg}
\end{split}
\ee
Note that for $D^* \to D\pi$ transition only the $\tilde{\varkappa}_3$ term will be proportional to $m_K^2$, while the others will be proportional to $m_\pi^2$ and thus negligible. On the other hand $\tilde{\varkappa}_3$ is $1/N_c$ suppressed in the large $N_c$ expansion, where $N_c$ denotes the number of colors (for more details see \cite{Gasser:1984gg,'tHooft:1973jz}). In the numerical evaluation the counterterms in \eqref{deltaLg} will be thus set to zero. The error due to this procedure will be estimated by using two renormalization prescriptions as explained in detail in section \ref{coupling}. \index{counterterms! values of}

Next we consider bosonization of currents \index{bosonization} that appear in weak decays. At the leading order in $1/m_H$ and at the next-to-leading order in chiral expansion this is \cite{Wise:hn,Boyd:1994pa} \index{alpha@$\alpha$}\index{current heavy-to-light}\index{heavy-to-light current}\index{kappa12@$\varkappa_{1,2}$}\index{hadronization! currents}\index{bosonization}
\begin{equation}
\begin{split}
\label{current}
\bar{q}_a\gamma^\mu(1-\gamma_5)Q&\to \frac{i \alpha}{2}\Tr[\gamma^\mu (1-\gamma_5) H_b] \xi_{ba}^\dagger \\
&+\frac{i \alpha}{2}\varkappa_1\Tr[\gamma^\mu (1-\gamma_5) H_c ]\xi_{ba}^\dagger ({\cal M_+})_{cb}\\
&+\frac{i 
\alpha}{2}\varkappa_2\Tr[\gamma^\mu (1-\gamma_5) H_b] \xi_{ba}^\dagger \tr({\cal M_+})+\dots 
\end{split}
\end{equation}
Beside the leading order ${\cal O}(p^0)$ current in the chiral counting, given in the first line of \eqref{current}, we display also two ${\cal O}(p^2)$ terms. These will be relevant for the discussion of the $f_D$, $f_{D_s}$ decay constants given later on in section \ref{coupling}. Other ${\cal O}(p^2)$ terms are not written out explicitly. They can be found in \cite{Boyd:1994pa}.

In the same way as the heavy-light current (\ref{current}), operators of more general structure $ (\bar{q}_a\Gamma Q)$, with $\Gamma$ an arbitrary product of Dirac matrices, can be translated into
an operator containing meson fields only \cite{Falk:1993fr}. At the leading order \index{alpha@$\alpha$}\index{hadronization! general forms}
\begin{equation}
\big(\bar{q}_a \Gamma Q\big)\to \frac{i
\alpha}{2}\Tr[P_R \Gamma H_b\xi_{ba}^\dagger] + \frac{i
\alpha}{2}\Tr[P_L \Gamma H_b\xi_{ba}].\label{eq-105}
\end{equation}
For instance the operator $ (\bar{q}_a \sigma_{\mu\nu}
\tfrac{1}{2} (1+\gamma_5)Q)$ proportional to $Q_7$ operator (cf. section \eqref{RG-OPE}) is then translated into \index{Q7@$Q_7$}
\begin{equation}
\big(\bar{q}_a \sigma_{\mu\nu} \tfrac{1}{2}(1+
\gamma_5)Q\big)\to \frac{i
\alpha}{2}\Tr\left[\sigma_{\mu\nu}\tfrac{1}{2}
(1+\gamma_5)H_b\xi_{ba}^\dagger\right]. \label{Q7hadr}
\end{equation}

\index{expansion! in $1/m_Q$|)}
\index{contributions!chiral loop|(} \index{expansion! chiral|(}
Finally, let us give the Weinberg's counting rule \cite{Weinberg:1978kz} for the case of heavy hadron chiral perturbation theory. This counting rule establishes the relative importance of loop contributions. Consider a general diagram with $L$ loops, for which we do the counting in terms of momenta $p\ll 4\pi f$ that flow in the internal lines. The Goldstone boson propagators are of the form $1/(p^2-m_\pi^2)$ and contribute a factor of $p^{-2}$ for each propagator. Similarly the heavy meson propagators $\sim 1/v\cdot p$ contribute a factor of $p^{-1}$. Each loop contributes an integration factor $d^4 p$, so that finally the dimension of the diagram in terms of $p$ is
\be
D=-2 I_b -I_{\text{h.f.}}+4L+\sum_i d_i V_i,\label{begin-D}
\ee
where $I_b$ and $I_{\text{h.f.}}$ are the numbers of internal bosonic and heavy-field lines respectively, while $V_i$ is the number of vertices in the diagram with the dimension $d_i$. The dimension of the vertex $d_i$ is obtained by counting the number of derivatives and quark masses in the corresponding interaction term, where each derivative is counted as $p$ and each $m_q\sim m_\pi^2$ as $p^2$. The dimension of the diagram \eqref{begin-D}, can be rewritten in a more convenient form. To do so, notice that the number of loops is connected to the number of internal lines $I=I_b+I_{\text{h.f.}}$ and the number of vertices $V=\sum_iV_i$. Each internal line contributes an integration over momenta, while each vertex contributes a delta function in the momenta. At the end one overall delta function due to translation invariance is left, connecting incoming and outgoing momenta, so that
\be
L=I-(V-1),
\ee
or
\be
I=L+V-1. \label{rel1}
\ee
Also, the sum over the numbers $n_i$ of incoming heavy field legs attached to a vertex of type $i$ is connected to the number $E_{\text{h.f.}}$ of external heavy field legs in the diagram and to the number of internal lines $I_{\text{h.f.}}$
\be
2 I_{\text{h.f.}}+E_{\text{h.f.}}=\sum_i V_{i} n_{i}^{\text{h.f.}}.\label{rel2}
\ee
Using the relations \eqref{rel1}, \eqref{rel2} in \eqref{begin-D} leads to
\be
\begin{split}
D=&-2I+I_{\text{h.f}}+4 L+\sum_i d_i V_i\\
=&2L+2+\sum_iV_i\left(d_i+\frac{1}{2}n_i^{\text{h.f.}}-2\right)-\frac{1}{2}E_{\text{h.f.}}.
\end{split}
\ee
The number of external heavy legs $E_{\text{h.f.}}$ in a particular process is constant. Also the reduced dimension of interaction vertex $\Delta_i=(d_i+\frac{1}{2}n_i^{\text{h.f.}}-2)$ is always nonnegative as can be easily verified from Lagrangians \eqref{chirallagr}, \eqref{eqQ:1}, \eqref{eq-8}, \eqref{eqQ:3}. The reason is that the pure Goldstone boson Lagrangian contains at least two derivatives or one quark mass. Once the heavy fields are introduced, they always come in pairs and there is also at least one derivative in the interaction. Since reduced dimension is nonnegative, $\Delta_i\ge 0$, adding more vertices will only increase the dimension of the diagram, and thus reducing its importance. The same is true of the loops. Adding one more loop to the diagram makes it of a $p^2$ higher order. Leading order diagrams are thus the tree level ones, as usual in the perturbation theory. \index{chiral!counting}
\index{contributions!chiral loop|)} \index{expansion! chiral|)}

\section{Photon couplings and gauge invariance}\label{gauge}\index{coupling constants!photon coupling} \index{electromagnetic, interactions in approach}\index{gauge interactions in effective theories|(}
The photon couplings are obtained by gauging the Lagrangians
\eqref{chirallagr}, \eqref{eq-8} and the light current \eqref{jX}
with the $U(1)$ photon field $B_\mu$.
The covariant derivatives are
then ${\cal D}_{ab}^\mu H_b=
\partial^\mu H_a +i e B^\mu (Q'H-H Q)_a-H_b {\cal V}_{ba}^\mu$
and
${\cal D}_\mu\xi =\partial_\mu \xi +
i e B_\mu [Q,\xi]$ with $Q=\diag (\frac{2}{3},-\frac{1}{3},
-\frac{1}{3})$
and $Q'$ the heavy quark charge ($Q'=\frac{2}{3}, -\frac{1}{3}$ for the case of $c$ and $b$ quarks respectively). The vector and
axial vector
fields \eqref{defVA} change after gauging and are $ {\cal
V}_{\mu}
=\frac{1}{2}(\xi{\cal D}_\mu\xi^\dagger
+\xi^\dagger{\cal D}_\mu\xi) $ and
$ {\cal A}_\mu = \frac{i}{2}
(\xi^\dagger {\cal D}_\mu\xi -\xi{\cal D}_\mu\xi^\dagger)$.
Similarly, the light weak current \eqref{jX} contains after gauging the covariant
derivative ${\cal D}_\mu$ instead
of $\partial_\mu$. However, the gauging procedure alone does not
introduce
a coupling of the form $D^* D \gamma$ without emission of additional Goldstone bosons. To
describe
this
electromagnetic interaction
we follow \cite{Stewart:1998ke} introducing an additional gauge invariant
contact
term with
an unknown coupling $\beta$ \index{beta@$\beta$}
of dimension -1.
\begin{equation}
{\cal L}_\beta=-\frac{\beta e}{4} \Tr \bar{H}_a H_b \sigma^{\mu
\nu}F_{\mu
\nu} Q_{ba}^\xi -\frac{e}{4 m_Q} Q' \Tr \bar{H}_a
\sigma^{\mu \nu} H_a
F_{\mu\nu},\label{eq-100}
\end{equation}
where $Q^\xi=\frac{1}{2}(\xi^\dagger Q \xi+\xi Q \xi^\dagger)$ and
$F_{\mu\nu}=\partial_\mu B_\nu- \partial_\nu B_\mu$.
The first term concerns the contribution
of the light quarks in
the heavy meson and the second term
describes emission of
a photon
from the heavy quark. Its coefficient is fixed by the heavy quark
symmetry. From \eqref{eq-100} both $H^* H \gamma$ and
$H^*H^*\gamma$ interaction terms arise. Even though the
Lagrangian \eqref{eq-100} is formally of higher order in $1/m_Q$ or chiral expansion,
we
do not neglect it, as it has been found that it gives a sizable
contribution to $D^*(B^*)\to D(B) \gamma\gamma$ decays \cite{Guetta:1999vb}. In chapter \ref{rareD} we will find, that in the
 $D^0\to \gamma\gamma$ decay the Lagrangian terms \eqref{eq-100} give the largest contribution to the
parity conserving part of the
amplitude. However, they do not contribute to the decay rate
 by more than $10\%$. The Lagrangian \eqref{eq-100}
in principle
receives a number of other contributions at the order ${\cal O}(1/m_Q)$,
but these can be absorbed in the definition of $\beta$ for the processes $D^0\to \gamma\gamma$, $D^0\to l^+l^-\gamma$, that will be 
considered in chapter \ref{rareD} \cite{Stewart:1998ke}. \index{approach, used in thesis|)}

\index{gauge invariance!general proof|(}\index{theorem, regarding gauge invariance}
In the following we present two proofs that such gauging procedure of the effective Lagrangian does indeed lead to a gauge invariant effective action and thus to a gauge invariant amplitude. \index{amplitude!gauge invariant|(}\index{diagrammatic, proof of gauge invariance|(} The general proof is just a special case of the proof given in chapter 16 of \cite{Weinberg:kr}, that has already been cited in section \ref{Generalsection} (cf. Eq.~\eqref{lintransf}). The electromagnetic $U(1)$ transformations of fields appearing in the effective Lagrangians \eqref{chirallagr}, \eqref{eq-8} are linear
\begin{align}
H_a&\to e^{ie Q'\alpha(x)}H_a e^{-ie Q_a \alpha(x)}=H_a+ie(Q' H-HQ)_a \alpha(x)+\cdots \label{gaugeone}\\
\xi&\to e^{ieQ\alpha(x)}\xi e^{-i e Q\alpha(x)}=\xi+ie [Q,\xi]\alpha(x)+\cdots\\
B_\mu&\to B_\mu-\partial_\mu \alpha,\label{gaugethree}
\end{align}
with $Q_a$ the $Q_{aa}$ component of $Q=\diag(\frac{2}{3},-\frac{1}{3},-\frac{1}{3})$ and no summation over $a$. The cited proof then states that as long as the effective Lagrangian is gauge invariant under the linear transformations \eqref{gaugeone}-\eqref{gaugethree}, so is the effective action, which is what we wanted to show. 
\index{gauge invariance!general proof|)}

\index{gauge invariance!useful theorem|(}

The general proof does not help us in the calculation, where one wishes to find {\it finite} sets of diagrams, that are already gauge invariant. Here a very useful tool is a diagrammatic proof of gauge invariance, which we state next. Consider an arbitrary off-shell initial Feynman diagram with arbitrary number of loops, heavy lines and photon lines. The sum of the diagrams obtained by inserting an additional photon line everywhere in the initial diagram, where this is permitted by the gauged Lagrangians \eqref{chirallagr}, \eqref{eq-8} and \eqref{eq-100}, is gauge invariant. Finding gauge invariant sets of diagrams in the actual calculation is then straightforward. One starts from an appropriate initial diagram, inserts photon vertices everywhere and ends up with a gauge invariant set.
\begin{figure}
\begin{center}
\epsfig{file=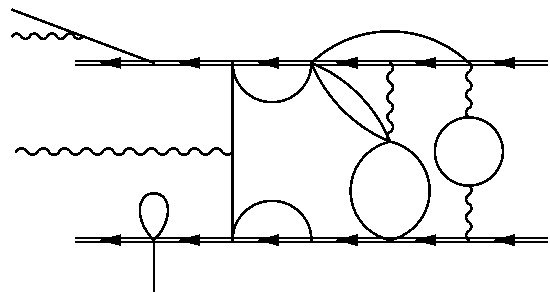, height=4cm}
\caption{\footnotesize{Example of an initial Feynman diagram. The gauge invariant set is obtained by adding a photon leg wherever this is possible. The heavy mesons are depicted by the double lines, while light pseudoscalars are represented by the solid lines.}}
\end{center}
\end{figure}
\index{gauge invariance!useful theorem|)}

\index{gauge invariance!diagrammatic proof|(}
The proof of the above statement follows closely the proof of gauge invariance of QED amplitudes as presented in the textbook of Peskin and Schroeder \cite{Peskin:ev}. The complication is, that we have to deal with two sorts of charged particles, the heavy mesons and light-pseudoscalars, and with an in principle infinite number of couplings between them. We shall prove the statement about gauge invariance only for the vertices with up to three pseudoscalar and/or heavy meson fields, as this will be needed further on in the calculations done in the thesis. At the end we shall present also a discussion concerning more general vertices. 

The expressions for the vertices follow from the effective Lagrangians \eqref{chirallagr}, \eqref{eq-8}. For the coupling of photon(s) to the light pseudoscalars, the coupling is of the form
\be
i e B_\mu(\pi^+\partial^\mu \pi^- -\pi^-\partial^\mu \pi^+)+e^2B_\mu B^\mu (\pi^+\pi^-),\label{lightphoton}
\ee
where $\pi^\pm$ is the pion field. For the kaons the expression is the same as in \eqref{lightphoton}, but with $\pi$ replaced with $K$ field. The coupling of the photon to the heavy meson is of the form
\be
e v\cdot B (Q'-Q_a)(P_{a\nu}^\dagger P_a^\nu-P_a^\dagger P_a),
\ee
while the coupling of the photon to heavy mesons with one pseudoscalar emitted is 
\be
i \frac{egB_\mu}{f}2\big[-iP_{a\alpha}^\dagger P_{b\beta}\epsilon^{\alpha\nu\beta\mu}v_\nu +\big(P_a^{\dagger\mu}P_b+P_a^\dagger P_b^\mu\big)\big]\big([\Pi,Q]\big)_{ba}, \label{heavylightphoton}
\ee
with $\Pi$ given in \eqref{pimatrix} and $Q=\diag(\frac{2}{3},-\frac{1}{3},-\frac{1}{3})$. To the set of couplings \eqref{lightphoton}-\eqref{heavylightphoton} one should add the couplings \eqref{eq-100}. However, these are manifestly gauge invariant due to the presence of $F_{\mu\nu}$ term and need not be considered in the following. 

Let us now consider an arbitrary Feynman diagram with incoming and outgoing legs off-shell, where we limit ourselves to the case of couplings \eqref{lightphoton}-\eqref{heavylightphoton}. Since there are only up to three mesons per each vertex, only two of the meson fields can be charged. To each vertex we can thus associate a charged line with one ingoing and one outgoing charged meson leg and thus with a well defined direction of charge flow. The initial diagram is interlaced with such charged lines. Since to each vertex only one charged line is associated, the charged lines never cross. In other words, to a given charged line only neutral lines attach. Because charged lines are connected only by neutral lines, each charged line can be considered separately. 

Charged line can either form a loop or connect to two external charged legs. To begin with we consider the charged line that begins and ends on the external off-shell charged legs. Such a line of charge flow has a general structure \\
\begin{center}
\epsfig{file=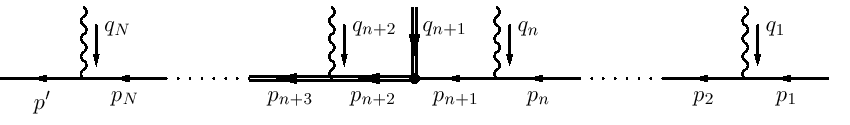}
\end{center}
with double lines representing the heavy mesons, and the solid lines denoting the light pseudoscalars. Lines are arranged so that the charged lines are horizontal with neutral lines attached to it (i.e. the heavy meson carrying momentum $q_{n+1}$ is neutral).

To simplify the problem even further, we shall first consider the charged line of only light pseudoscalar mesons, with coupling to photons given by \eqref{lightphoton}. For simplicity we also assume, that to this initial charged light pseudoscalar line only single photon lines attach 
\\
\begin{center}
\epsfig{file=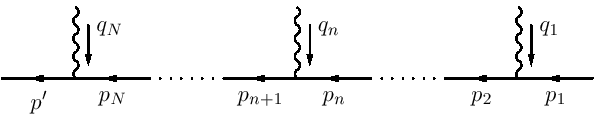}
\end{center}
The part of the amplitude corresponding to this initial charged line is then of the form
\be
\begin{split}
M_0\propto&\frac{i}{p'^2-m^2}(-e)\big(i p_N+i(p_N+q_N)\big)^{\mu_N} \frac{i}{p_N^2-m^2}\dots\frac{i}{p_1^2-m^2}\\
=& \frac{i}{p'^2-m^2}\frac{e(2p_N+q_N)^{\mu_N}}{p_N^2-m^2}\dots\frac{e (2 p_1+q_1)^{\mu_1}}{p_1^2-m^2},\label{M_0}
\end{split}
\ee 
where $\mu_1,\dots,\mu_N$ denote the uncontracted Lorentz indices corresponding to photon lines carrying incoming momenta $q_1,\dots,q_N$. To obtain the complete expression for the off-shell initial diagram, the Lorentz indices $\mu_1,\dots,\mu_N$ would be contracted by photon propagators. 

In the next step we attach an additional external photon line to the initial charged line, wherever this is possible. The Lorentz index $\mu$ of the additional external photon line carrying incoming momentum $k$ is contracted by the polarization vector $\epsilon^\mu(k)$, when the photon line is put on-shell. A gauge invariant amplitude has to be invariant under the change $\epsilon^\mu(k)+c k^\mu$. To test gauge invariance we thus contract the Lorentz index $\mu$ of the additional external photon line with $k^\mu$. This should then give vanishing result for the corresponding amplitude, when external legs are put on-shell.

The additional photon line can be either attached to the pseudoscalar propagator or to the vertex already containing one photon leg \eqref{lightphoton}. When we attach the photon to the $n-$th propagator, all the momenta in the propagators after it get shifted by $k$ 
\\
\begin{center}
\epsfig{figure=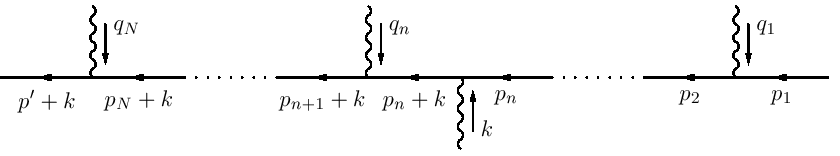}
\end{center}
The invariant amplitude corresponding to the above diagram is then
\be
\begin{split}
M\propto&\frac{i}{(p'+k)^2-m^2}\frac{e(2p_N+2k+q_N)^{\mu_N}}{(p_N+k)^2-m^2}\dots \frac{e(2 p_n+2k+q_n)^{\mu_n}}{(p_n+k)^2-m^2}\frac{e(2p_n+k)\cdot k}{p_n^2-m^2}\dots\frac{e(2 p_1+q_1)^{\mu_1}}{p_1^2-m^2}\\
=&e \frac{i}{(p'+k)^2-m^2}\frac{e(2p_N+2k+q_N)^{\mu_N}}{(p_N+k)^2-m^2}\dots \frac{e(2 p_n+2k+q_n)^{\mu_n}}{p_n^2-m^2}\dots\frac{e(2 p_1+q_1)^{\mu_1}}{p_1^2-m^2}\\
&-e \frac{i}{(p'+k)^2-m^2}\frac{e(2p_N+2k+q_N)^{\mu_N}}{(p_N+k)^2-m^2}\dots \frac{e(2 p_n+2k+q_n)^{\mu_n}}{(p_n+k)^2-m^2}\dots\frac{e(2 p_1+q_1)^{\mu_1}}{p_1^2-m^2}, \label{propphoton}
\end{split}
\ee
where in the last equality we have used $k^2=0$ and $2p_n\cdot k=(p_n+k)^2-m^2-(p_n^2-m^2)$. We then add a photon line also to the $n-$th photon vertex
\\
\begin{center}
\epsfig{figure=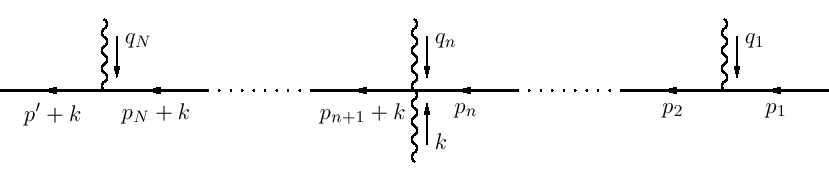}
\end{center}
The amplitude corresponding to this diagram is
\be
M\propto\frac{i}{(p'+k)^2-m^2}\frac{e(2p_N+2k+q_N)^{\mu_N}}{(p_N+k)^2-m^2}\dots \frac{-e^22k^{\mu_n}}{p_n^2-m^2}\dots\frac{e(2 p_1+q_1)^{\mu_1}}{p_1^2-m^2}. \label{photonvert}
\ee
The sum of the two insertions, Eq.~\eqref{photonvert} and Eq.~\eqref{propphoton}, gives
\be
\begin{split}
M\propto&\;e \frac{i}{(p'+k)^2-m^2}\frac{e(2p_N+2k+q_N)^{\mu_N}}{(p_N+k)^2-m^2}\dots \frac{e(2 p_n+q_n)^{\mu_n}}{p_n^2-m^2}\dots\frac{e(2 p_1+q_1)^{\mu_1}}{p_1^2-m^2}\\
&-e \frac{i}{(p'+k)^2-m^2}\frac{e(2p_N+2k+q_N)^{\mu_N}}{(p_N+k)^2-m^2}\dots \frac{e(2 p_n+2k+q_n)^{\mu_n}}{(p_n+k)^2-m^2}\dots\frac{e(2 p_1+q_1)^{\mu_1}}{p_1^2-m^2}.
\end{split}
\ee
After we sum over all such insertions we get
\be
M=eM_0(p',\dots,p_1)-eM_0(p'+k,\dots,p_1+k),\label{sumM}
\ee
with $M_0$ defined in \eqref{M_0}. Now let us consider the case, where the initial and the final leg of the charged line are external legs. The on-shell amplitude is then obtained by subtracting the ingoing and outgoing propagators and multiplying with appropriate wave-function coefficients, according to the Lehmann-Symanzik-Zymmermann (LSZ) reduction formula (see, e.g.,\cite{Peskin:ev}). \index{LSZ} However, the sum of the amplitudes $M$ \eqref{sumM} does not have the double-poles of the form 
\be
\frac{1}{p_1^2-m^2}\frac{1}{(p'+k)^2-m^2},
\ee
that are required to obtain nonzero on-shell amplitudes after the application of LSZ formula. In other words, the on-shell amplitude corresponding to the sum of charged line diagrams with an additional photon attached where possible (and contracted with $k^\mu$), is zero. The on-shell amplitude is thus gauge invariant.

Similar reasoning applies if the line is closed, i.e., if the charged line forms a loop. Then the final amplitude involves also the integration over loop variable, so that 
\be
M\propto \int d^4 q\; \left(eM_0(p'+q,\dots,p_1+q)-eM_0(p'+q+k,\dots,p_1+q+k)\right).
\ee
After the shift of the integration variable in the second term, $q+k\to q$, the amplitude is seen to be equal to zero. If this shift is justified (as is the case in nonanomalous diagrams) the gauge invariance is guaranteed once again.

In the reasoning outlined above there were two crucial steps. First, the propagator identity\\
\begin{center}
\epsfig{file=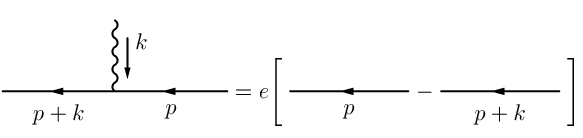}
\end{center}
has to hold for any charged particle. And second, for each amputated vertex multiplied only by the propagator next to it to the right, the following identity has to be true
\\
\begin{center}
\epsfig{file=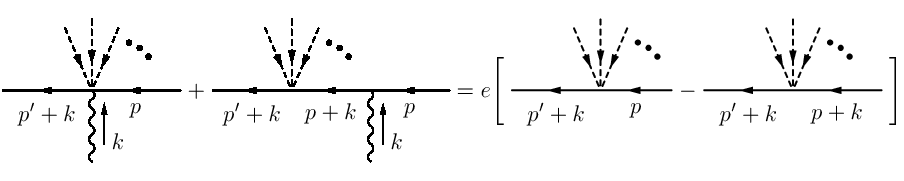}
\end{center}
where an arbitrary number of neutral lines (shown as dashed lines) attach to the vertex. The first identity is needed so that also the photon attached to the last leg in the charged line can be represented as a difference of two propagators.

Let us first prove that the propagator identity holds for the mesons involved in the problem. For the light pseudoscalars we have
\be
\frac{i}{(p+k)^2-m^2} (-i e) 2 p\cdot k\frac{i}{p^2-m^2}=e \Big(\frac{i}{p^2-m^2}-\frac{i}{(p+k)^2-m^2}\Big), \label{prop-ident1}
\ee
which is the result needed. To obtain \eqref{prop-ident1} we have used the identity $2 p\cdot k=(p+k)^2 -m^2-(p^2-m^2)$. For the heavy pseudoscalar mesons
\be
\frac{i}{v\cdot(p+k)-\Delta} (-ie) v\cdot k\frac{i}{v\cdot p-\Delta}=e\Big(\frac{i}{v\cdot p-\Delta}-\frac{i}{v \cdot(p+k)-\Delta}\Big),
\ee
where we have used the identity $v\negcdot k=v\negcdot (p+k)-\Delta-(v\negcdot p-\Delta)$, while for the heavy vector mesons
\be
\frac{-i(\eta_{\mu\mu'}-v_\mu v_{\mu'})}{v\cdot (p+k)-\Delta} i e v\cdot k \frac{-i(\eta_{\mu'\nu}-v_{\mu'} v_\nu)}{v\cdot p-\Delta}= e\Big(\frac{-i(\eta_{\mu\nu}-v_\mu v_\nu)}{v\cdot p-\Delta}-\frac{-i(\eta_{\mu\nu}-v_\mu v_\nu)}{v\cdot (p+k)-\Delta}\Big).\label{prop-ident3}
\ee
As for the vertex identities, we shall consider them all at once. Consider the two diagrams
\\
\begin{center}
\epsfig{file=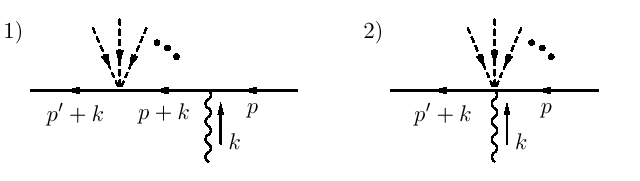}
\end{center}
where again an arbitrary number of neutral lines, shown as dashed lines, can attach to the vertex. The first diagram can be written as
\be
V_3(p'+k,q,p+k)S(p+k)V_\gamma(k)S(p),\label{one}
\ee
where $S(p)$ is the charged meson propagator (either light or heavy), while $V_\gamma$ is the photon-emission vertex, but with the replacement $\epsilon_\mu(k) \to k_\mu$ for gauge invariance considerations. The vertex with neutral lines in the diagram 1) is denoted $V_3(p'+k,q,p+k)$, with $q$ the sum of ingoing neutral line momenta. The neutral lines can be either only mesonic or contain additional photons beside the one carrying momentum $k$. If additional photons are present, the $V_3$ vertex is assumed to represent only one of the possible permutations of photon legs. 

The diagram 2) is 
\be
V_3'(p'+k,q,p)S(p),\label{two}
\ee
where $V_3'$ denotes the photon coupling to the neutral lines vertex. This is obtained through gauging, i.e the derivative $\partial_\mu$ in the relevant term in the Lagrangian is replaced by $ie B_\mu$. For the gauge invariance considerations, we further replace $\epsilon_\mu(k)\to k_\mu$. Effectively this means, that $V_3'$ is obtained from $V_3$ by replacing in $V_3$ the momentum of charged incoming meson with $-ip\to i e k$, and for the charged outgoing mesons $i(p'+k)\to-ie k$. In other words, if $V_3(p',q,p)$ is of the form $V_3(p',q,p)\propto p'_{\mu_1}\dots p_{\mu_n}\dots q_{\mu_N}$, the $V_3'$ is $V_3'(p',q,p)\propto -e\sum_{j=1}^n p'_{\mu_1}\dots k_{\mu_j}\dots p_{\mu_n}\dots q_{\mu_N}$, where the replacement is never done on $q$'s, the momenta of neutral lines. 

We would like to show, that the sum of \eqref{one} and \eqref{two} is
\be
e[V_3(p',q,p)S(p)-V_3(p'+k,q,p+k)S(p+k)]. \label{finalres}
\ee
We have already shown that $S(p+k)V_\gamma(k)S(p)=e[S(p)-S(p+k)]$ in Eqs.~\eqref{prop-ident1}-\eqref{prop-ident3}, so that in order to show \eqref{finalres}, it suffices to show that $e V_3(p'+k,q,p+k)+V'_3(p'+k,q,p)$ equals $e V_3(p',q,p)$. This is straightforward for the case of vertices listed in Eqs.~\eqref{lightphoton}-\eqref{heavylightphoton}. The important thing to note is, that these terms come from Lagrangians \eqref{chirallagr}, \eqref{eq-8} that have only one derivative per each field. The vertex $V_3$ can thus contain not more than one of each charged meson momenta. In general they will be of the form
\begin{align}
e V_3(p'+k,q,p+k)&\sim (p'+k)_\mu(p+k)_\nu,\label{firstV3}\\
V'_3(p'+k,q,p)&\sim -k_\mu p_\nu-(p'+k)_\mu k_\nu,\label{secondV3}
\end{align}
where we have written out only the relevant (i.e. charged) momenta, and \eqref{secondV3} has been obtained according to the replacements described in text below Eq.~\eqref{two}. The sum of \eqref{firstV3} and \eqref{secondV3} then gives $p'_\mu p_\nu\sim V_3(p',q,p)$ as was to be shown. These completes the proof of the statement about gauge invariance for the effective theory couplings with up to two charged fields and up to one derivate acting on any of these fields.

For the vertices with more than two charged lines and more than one derivative on some of the charged fields, additional complications arise. First of all the charge flow lines can now cross each other. Because of the crossing, charged lines are not uniquely defined. In fact, to prove gauge invariance, one has to consider all possible ways of defining charge flow lines. To deal with this, one focuses on one charged line only, defining also which fields in the Lagrangian destroy/create legs of this line and regard other charged lines attached to the chosen charge flow as we did the neutral legs before. If there are not more than one derivative acting on each meson field, everything proceeds as it did above.

Extension of the arguments given above to the case of more than one derivative acting on the charged fields in the Lagrangian is not straightforward. In this case also the propagator identities (counterparts of the Eqs.~\eqref{prop-ident1}-\eqref{prop-ident3}) change and become more complex. The theorem is then easiest to prove on a case by case basis.\index{amplitude!gauge invariant|)} \index{diagrammatic, proof of gauge invariance|)}\index{gauge interactions in effective theories|)}\index{gauge invariance!diagrammatic proof|)}

\section{Determination of the parameters}\label{coupling}
The unknown couplings appearing in the Lagrangians \eqref{chirallagr}, \eqref{eq-8} are obtained from experiment. In the following we shall present the determination of the couplings first at tree level and then also at the one loop level. 

\subsubsection{Tree level}
At tree level $\alpha$ \index{alpha@$\alpha$} \eqref{current} is trivially related to the heavy meson decay constant $f_{B,D}=\alpha/\sqrt{m_{B,D}}$, where the decay constant is defined through an axial current \index{axial current} matrix element. For e.g $f_D$ this is
\begin{equation}
\langle 0| \overline{u}\gamma^\mu \gamma_5 c|D^0 \rangle
=ip_D^\mu f_D,
\label{fD}
\end{equation}
From \cite{Hagiwara:pw,Heister:2002fp} one deduces $f_{D_s}=268\pm 25\;\text{GeV}$, from which $\alpha^{\text{Tree}}=0.38\pm0.04\; \text{GeV}^{3/2}$. 
Note that this value has been extracted from a system with a valence $s$-quark and one expects a sizable 1-loop correction. 

From the CLEO measurement \index{CLEO} of the $D^{*+}\to D^0 \pi^+$ partial decay width \cite{Ahmed:2001xc,Anastassov:2001cw}, the value of $g^{\text{Tree}}=0.59\pm 0.08$ can be deduced, with $g^{\text{Tree}}$ being the $D^* D \pi$ coupling constant (cf. Eq.~\eqref{eq-8}). The pion decay constant is taken to be $f^{\text{Tree}}=f_\pi=130.7\pm0.4 \text{ MeV}$ \cite{Hagiwara:pw}. \index{beta@$\beta$}

In order
to obtain the value of $\beta$ we use the available experimental
data from $D^{*+}\to D^+ \gamma$ and $D^{*0}\to D^{0}\gamma$ decays.
For instance, one can use the recently determined $D^{*+}$ decay
width $\Gamma (D^{*+})=96\pm 4\pm 22\; {\rm keV}$ \cite{Ahmed:2001xc,Anastassov:2001cw}
together with the branching ratio
$\text{Br}(D^{*+}\to D^+ \gamma)=(1.6 \pm 0.4)\%$ \cite{Hagiwara:pw}.
At tree level one has
\begin{equation}
\Gamma(D^{*+}\to D^+ \gamma)=\frac{e^2}{12 \pi} \left(
\tfrac{2}{3} \tfrac{1}{m_c}-\tfrac{1}{3}
\beta\right)^2 k_\gamma^3,\label{eq-102}
\end{equation}
with $k_\gamma=\frac{m_{D^*}}{2}(1-\frac{m_D^2}{m_{D^*}^2})$ the
momentum of the outgoing photon. Using $m_c=1.4 \;
{\rm GeV}$
one
arrives at $\beta=2.9\pm 0.4 \; {\rm GeV}^{-1}$
\renewcommand{\thefootnote}{\fnsymbol{footnote}}
\footnote{There is also
a solution of \eqref{eq-102} $\beta=0.09 \pm 04 \;
{\rm GeV}^{-1}$ that,
however,
does not agree with the determination of $\beta$
from the $D^{*0}$ decay.},
where
the errors reflect the experimental errors.

On the other hand one can also use the ratio of partial decay width in
$D^{*0}$
system $\Gamma(D^{*0}\to D^0\gamma):\Gamma(D^{*0}\to D^0 \pi^0)=(38.1\pm
2.9):(61.9\pm 2.9)$, where the experimental errors are considerably
smaller
than in the previous case. At tree level one has
\begin{equation}
\frac{ \Gamma(D^{*0}\to D^0\gamma)}{\Gamma(D^{*0}\to D^0
\pi^0)}=\frac{e^2}{12 \pi} \frac{ k_\gamma^3}{k_\pi^3}\frac{12 \pi
f^2}{g^2}
\left(\tfrac{2}{3}\beta +\tfrac{2}{3} \tfrac{1}{m_c}\right)^2,
\label{eq-108}
\end{equation}
with $k_\gamma$,$k_\pi$ the momenta of the outgoing photon and pion
respectively.
Using $m_c=1.4 \; {\rm GeV}$, $g=0.59$, $f=f_\pi=130.7 \;
{\rm MeV}$ one
arrives at $\beta=2.3\pm 0.2 \; {\rm GeV}^{-1}$,
\footnote{The other solution is $\beta= -3.6 \pm 0.2 \; {\rm GeV}^{-1}$
that does not agree with $D^{*+}$ data.}
 where the quoted errors again reflect only the errors on the input parameters coming from experiment.
 The $\beta$ coupling coming from
from $D^{*+}$ \eqref{eq-102} and $D^{*0}$ \eqref{eq-108} are in
fair agreement, but not equal. This signals that other
contributions
coming from chiral loops and higher order terms that would alter our
determination of $\beta$ might be important. Since the contribution of
chiral
loops to $\Gamma(D^{*+}\to D^+ \gamma)$ are approximately $50\%$, while
for
$D^{*0}\to
D^0\gamma$ they are about $20\%$ \cite{Stewart:1998ke}, we use
in our numerical calculations the value of
$\beta = 2.3\; {\rm GeV}^{-1}$
obtained from $\Gamma(D^{*0}\to D^0\gamma)$.

\subsubsection{Wave function renormalization}\index{corrections, chiral to couplings|(} \index{counterterms!contributions of|(}\index{renormalization, wave function}\index{wave function renormalization}
The values of couplings at the 1-loop depend on the regularization and renormalization prescriptions. Values for two renormalization prescriptions will be given, for the $\overline{\text{MS}}$ scheme and for the renormalization prescription $\bar{\Delta}=2/\epsilon-\gamma+\ln(4 \pi) +1\to0$ as used by Gasser and Leutwyler in their analysis \cite{Gasser:1984gg}. We will first discuss the calculation of wave function renormalization factors and then move on to the values of couplings $\alpha, g, f$ at one loop.

The wave function renormalization factors are defined as follows. We discuss first the case of light pseudoscalars. Let us define the sum of all one particle irreducible diagrams (1PI)\footnote{1PI diagram is a diagram that does not become disconnected, if any of the internal lines is cut.} contributing to the light pseudoscalar propagator \index{amputated diagram}\index{one particle irreducible diagrams}\index{1PI}
\\
\begin{center}
\epsfig{file=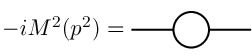}
\end{center}
where the amputated 1PI on the right-hand side is understood, while $p$ is the momentum running through the diagram. An infinite sum of 1PI diagrams represents the full light pseudoscalar propagator 
\\
\begin{center}
\epsfig{file=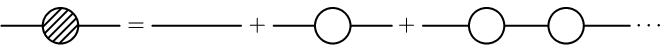}
\end{center}
The full light pseudoscalar propagator is thus a geometric sum of 1PI amputated diagrams and the intermediate bare propagators. After resummation this gives for the full propagator
\be
\frac{1}{p^2-m_0^2-M^2(p^2)},
\ee
where $m_0$ is the bare light pseudoscalar mass that is renormalized by the 1PI $M(p^2)$ term to give the physical mass
\be
m^2=m_0^2+M^2(m^2).
\ee
Close to the physical mass pole, the full pseudoscalar propagator can be approximated by
\be
\frac{1}{p^2-m_0^2-M^2(p^2)}\sim \frac{Z_P}{p^2-m^2}+\dots
\ee
where dots represent regular terms, while $Z_P$ is the pseudoscalar wave function renormalization factor (see, e.g., \cite{Peskin:ev})
\be
Z_P=1+\left. \frac{\partial M^2(p^2)}{\partial p^2} \right|_{p^2=m^2}.
\ee
In a very similar way also the heavy meson wave function renormalization is defined, the only difference being that the heavy meson propagator has only one power of momenta in the denominator \eqref{heavy-prop}. If we denote the sum of amputated 1PI diagrams contributing to the heavy meson propagator by $-i 2\Delta(v\cdot k)$, where $k$ is the propagator momentum, then the heavy meson wave function renormalization factor is
\be
Z_H=1+\left. \frac{\partial \Delta(v\cdot k)}{\partial v\cdot k}\right|_{v\cdot k=0}.
\ee
The wave function renormalization factors $Z_{P,H}$ enter the LSZ formula for the scattering matrix. \index{LSZ} The scattering matrix is calculated using amputated diagrams, that are multiplied by $\sqrt{Z_{P,H}}$ for each external leg (see, e.g.,\cite{Peskin:ev}).

\begin{figure}
\begin{center}
\epsfig{file=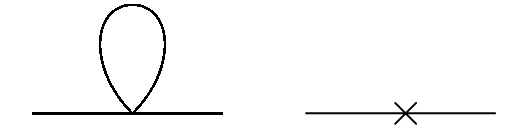, height=1.2cm}
\caption{\footnotesize{The ${\cal O}(p^4)$ 1PI amputated diagrams, that contribute to the light pseudoscalar wave function renormalization. The cross denotes counterterm contributions.}}\label{lightwave} \index{counterterms!contributions of}
\end{center}
\end{figure}

Contributions to the wave function renormalizations for the light pseudoscalars $K$, $\pi$ at the ${\cal O}(p^4)$ order in the chiral counting are shown on Figure \ref{lightwave}. Explicitly they are
\begin{align}
\begin{split}\label{ZK}
Z_{K}&=1-\frac{1}{16\pi^2f^2}\bigg[A_0(m_K^2)+\frac{1}{2}A_0(m_\pi)+\frac{1}{2}A_0(m_\eta^2)\bigg]\\
&\qquad\qquad\qquad\qquad
-8L_5\frac{4\mu_0}{f^2}
(\hat m+m_s)-16L_4\frac{4\mu_0}{f^2}(m_u+m_d+m_s),
\end{split}\\
Z_{\pi}&=1-\frac{1}{16\pi^2f^2}\bigg[\frac{2}{3}A_0(m_K^2)+\frac{4}{3}A_0(m_\pi^2)\bigg]-8L_5\frac{4\mu_0}{f^2}~2\hat m-
16L_4\frac{4\mu_0}{f^2}(m_u+m_d+m_s),\label{Zpi}
\end{align}
with $A_0(m^2)$ function defined in appendix \ref{app-loops}, Eq.~\eqref{A_0}, while $m_{u,d,s}$ are quark masses with $\hat m=\frac{1}{2}(m_u=m_d)$. The counterterm contributions $L_{4,5}$ come from the insertions of terms given in Eq.~\eqref{eqQ:1}. \index{counterterms!contributions of}

\begin{figure}
\begin{center}
\epsfig{file=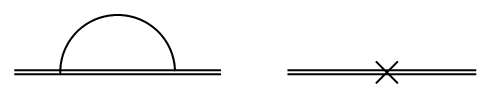, height=1.2cm}
\caption{\footnotesize{The ${\cal O}(p^3)$ contributions to the heavy meson wave function renormalization. The cross denotes counterterm contributions.}}\label{heavywave}
\end{center}
\end{figure}

The one chiral loop contributions to the heavy meson wave function renormalizations are shown on Figure \ref{heavywave}. Explicit expressions for the heavy pseudoscalar and vector mesons $D_a,D_a^*$, containing $\bar{q}_a$ light valence antiquark, are
\begin{align}
Z_{D_a}&=1-\frac{3 g^2}{16 \pi^2 f^2}\sum_i (t^i)_{ab}(t^{i\dagger})_{ba}B'(\Delta_{D_b^*D_a},m_{P_i})+k_1 m_a +k_2(m_u+m_d+m_s),\label{ZDa}\\
\begin{split}
Z_{D_a^*}&=1-\frac{2 g^2}{16 \pi^2 f^2}\sum_i (t^i)_{ab}(t^{i\dagger})_{ba}\big[ \tfrac{1}{2} B'({\Delta}_{D_bD_a^*},m_{P_i}) +B'({\Delta}_{D_b^* D_a^*},m_{P_i})\big]\\
&\hskip9mm+k_1 m_a +k_2(m_u+m_d+m_s),
\end{split}\label{ZDSta}
\end{align}
where the summation over $a$ is suspended, while $\Delta_{H_1H_2}=m_{H_1}-m_{H_2}$, $m_{u,d,s}$ are the quark masses, and $B'(\Delta,m)= \frac{\partial}{\partial
\Delta}\bar{B}_{00}(m,\Delta)$, where $\bar{B}_{00}(m,\Delta)$ can be found
in appendix \ref{app-loops}, Eq.~\eqref{B_00}. The $SU(3)$ matrices $t^i$are defined through $\Pi=P^i t^i$, Eq.~\eqref{pimatrix}. In the heavy quark limit $m_Q\to \infty$ we have $\Delta\to 0$ and the two renormalizations are equal.

\subsubsection{The decay constants}
\begin{figure}[h]
\begin{center}
\epsfig{file=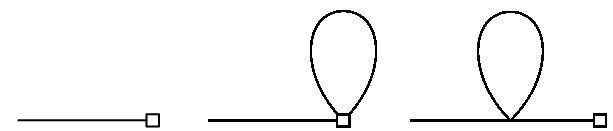, height=1.7cm}
\caption{\footnotesize{Leading and one loop chiral corrections to $f_{\pi}, f_{K}$ decay constants. The light pseudoscalars are represented by solid line, square denotes weak current insertion.}}\label{fpidiagr}
\end{center}
\end{figure}

The light pseudoscalar decay constants receive contributions at one chiral loop level from diagrams on Fig.~\ref{fpidiagr}. For \index{fK, chiral corrections to@$f_K$, chiral corrections to}\index{fpi, chiral corrections to@$f_\pi$, chiral corrections to}
 $\pi, K$ they are
\begin{align}
\begin{split}
f_K&=f\Big(1+\frac{1}{16\pi^2 f^2}[A_0(m_\eta^2)+2A_0(m_K^2)+A_0(m_\pi^2)]+8L_5\frac{4\mu_0}{f^2}(m_s+\hat m)\\
&\qquad\qquad\qquad\qquad\qquad\qquad\qquad\qquad\qquad+
16L_4\frac{4\mu_0}{f^2}(m_u+m_d+m_s)\Big)\sqrt{Z_K},
\end{split}
\\
f_\pi&=f\Big(1+\frac{1}{16\pi^2 f^2}\frac{4}{3}[A_0(m_K^2)+2A_0(m_\pi^2)]+8L_5\frac{4\mu_0}{f^2}2\hat m+
16L_4\frac{4\mu_0}{f^2}(m_u+m_d+m_s)\Big)\sqrt{Z_\pi},\label{eq:q8}
\end{align}
where the wave-function renormalizations are given in \eqref{ZK}, \eqref{Zpi}. We use the expression for the pion decay constant, together with the experimental value $f_\pi=130.7\pm 0.4$ MeV \cite{Hagiwara:pw}, from which at 
one loop $f=0.12\pm0.01 \;\text{GeV}$ \cite{Gasser:1984gg} both in $\overline{\text{MS}}$ and Gasser-Leutwyler prescriptions. The error is due to the poorly known $L_4$ counterterm, that will be discussed latter on in this section in somewhat more detail (cf. Eqs.~\eqref{Li}, \eqref{Litwo}).

\begin{figure}[h]
\begin{center}
\epsfig{file=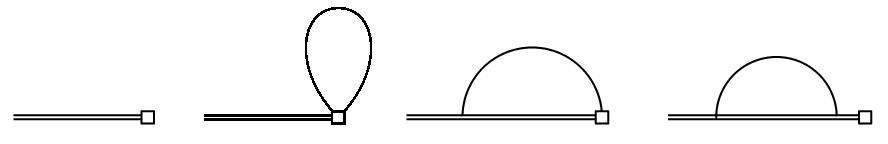, height=1.7cm}
\caption{\footnotesize{The leading contribution and the one loop chiral corrections to $f_{D}, f_{D_s}$ decay constants. The double line represents heavy meson, solid line the light pseudoscalar mesons, while the square denotes the weak current insertion. The diagram with one light pseudoscalar attached to the weak current (second from right) vanishes.}}\label{fDdiagr}
\end{center}
\end{figure}

The decay constants $f_D$, $f_{D_s}$ \index{alpha@$\alpha$|(} receive contributions from diagrams depicted on Fig.~\ref{fDdiagr}. For $\bar{\Delta}\to 0$ these have been calculated in \cite{Casalbuoni:1996pg,Boyd:1994pa}, while the leading logs
have been obtained already in \cite{Grinstein:1992qt,Goity:1992tp}. Taking into account the counterterms, the 1-loop expressions are
\begin{subequations}\label{both-fDs}
\begin{align}
\begin{split}
f_D=\frac{\alpha}{\sqrt{m_D}}\Big[& 1- \frac{3
g^2}{32\pi^2f^2}\Big(\frac{3}{2}
B'(\Delta_{D^*D},m_\pi)+B'(\Delta_{D_s^*D},m_K)+\frac{1}{6}B'(\Delta_{D^*D},m_\eta)\Big)\\
&+\frac{1}{32\pi^2f^2}\Big(\frac{3}{2}A_0(m_\pi^2)+A_0(m_K^2)+\frac{1}{6}A_0(m_\eta^2)\Big)\\
&+\left(\tfrac{1}{2}k_1+\varkappa_1\right)\hat{m}+\big(\tfrac{1}{2}k_2+\varkappa_2\big)(2 \hat{m}+m_s) \Big], \label{fD1loop}
\end{split}
\\
\begin{split}
f_{D_s}=\frac{\alpha}{\sqrt{m_{D_s}}}\Big[& 1- \frac{3
g^2}{32\pi^2f^2}\Big(2
B'(\Delta_{D^*D_s},m_K)+\frac{2}{3}B'(\Delta_{D_s^*D_s},m_\eta)\Big)\\
&+\frac{1}{32\pi^2f^2}\Big(2A_0(m_K^2)+\frac{2}{3}A_0(m_\eta^2)\Big)+\left(\tfrac{1}{2}k_1+\varkappa_1\right)m_s+\big(\tfrac{1}{2}k_2+\varkappa_2\big)(2 \hat{m}+m_s) \Big],
\end{split}\label{fDs}
\end{align}
\end{subequations}
where $\Delta_{H_1H_2}=m_{H_1}-m_{H_2}$, $\hat{m}=\frac{1}{2}(m_u+m_d)$ with $m_{u,d,s}$ the quark masses, while $B'(\Delta,m)= \frac{\partial}{\partial
\Delta}\bar{B}_{00}(m,\Delta)$, and $A_0(m^2)$, $\bar{B}_{00}(m,\Delta)$ can be found
in appendix \ref{app-loops}, Eqs.~\eqref{A_0}, \eqref{B_00}. The formulas \eqref{both-fDs} are valid at the leading order
in $1/m_Q$ \cite{Casalbuoni:1996pg,Boyd:1994pa}. Evaluating expression for $f_{D_s} $ \eqref{fDs} using the tree level values for $g,f$ and the experimental value of $f_{D_s}$ one arrives at $\alpha^{\overline{\text{MS}}}=0.21\pm0.05 \;\text{GeV}^{3/2}$ in $\overline{\text{MS}}$ scheme, and $\alpha^{\text{GL}}=0.24\pm0.05 \;\text{GeV}^{3/2}$ in the Gasser-Leutwyler prescription. The error is equally distributed between experimental errors in $f_{D_s}$, experimental error in $g^{\text{Tree}}$ and variation of unknown counterterms as described below (cf. Eqs.~\eqref{Li}, \eqref{Litwo} and the text below them). The variation of the counterterms introduces relatively large error as they are proportional to $m_K^2/f^2$. Estimated error is only approximate also because $1/m_D$ correction have been neglected. \index{alpha@$\alpha$|)}

\subsubsection{One loop corrections to the $D^*D\pi$ coupling}
\begin{figure}[h]
\begin{center}
\epsfig{file=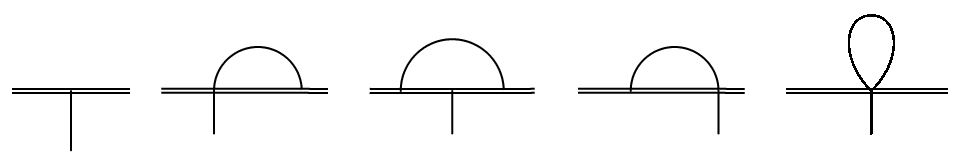, height=1.7cm}
\caption{\footnotesize{Leading and one loop chiral corrections to $g_{D^*D\pi}$ coupling. The double line represents heavy meson, solid line the light pseudoscalar mesons. The second and fourth diagrams vanish.}}\label{gDdiagr}
\end{center}
\end{figure}

\index{g@$g$}
The contributions to the $D^*D\pi$ coupling at one chiral loop are shown on Fig.~\ref{gDdiagr}. They give
\be
g_{D^*D\pi}=g\big[1+\delta g^{\text{1loop}}+\delta g^{\text{c.t.}}\big]\sqrt{Z_DZ_{D^*}Z_\pi}, \label{gDDpi}
\ee
where
\be
\begin{split}
\delta g^{\text{1loop}}=&\frac{g^2}{16 \pi^2 f^2}\Big[\sum_{i}t^i_{aa}t^{i \dagger}_{bb}\left(2 K(m_{P_i},0,\Delta_{D_b^*D_b})-K(m_{P_i},0,0)\right)\Big]\\
&-\frac{1}{6}\frac{1}{16 \pi^2 f^2}\sum_{i}\left(2t^{i}_{aa} t^{i\dagger}_{bb}-(t^it^{i\dagger})_{bb}-(t^it^{i\dagger})_{aa}\right) A_0(m_{P_i}^2),\label{deltag}
\end{split}
\ee
are the 1-loop contributions for $D^*_a\to D_b \pi_{ba}$, $b\ne a$, transition calculated at $v\cdot p=\Delta_{D_a^* D_b}$, where $\Delta_{H_1H_2}=m_{H_1}-m_{H_2}$ as before, while $t^i$ are the $SU(3)$ matrices, corresponding to the pseudoscalars $P_i$ in \eqref{pimatrix}, $\Pi=P_i t^i$. The sum in \eqref{deltag} runs over the light pseudoscalars $K, \pi, \eta$ with masses $m_{P_i}$, while no summation over $a$ and $b$ is assumed. For brevity we have also defined the function
\be
K(m,\Delta_1,\Delta_2)=-\frac{1}{\Delta_1-\Delta_2}\big[\bar{B}_{00}(m,\Delta_1)-\bar{B}_{00}(m,\Delta_1)\big],
\ee
where the limit $K(m,0,0)=-A_0(m^2)$. The $\delta g^{\text{c.t.}}$ term in \eqref{gDDpi} denotes the contribution coming from counterterms \eqref{deltaLg}. For $D^*\to D\pi$ transition the counterterms are either proportional to $m_\pi^2/16\pi^2 f^2$ or are $1/N_c$ suppressed as discussed below Eq.~\eqref{deltaLg}. These counterterms will be thus set to zero in the numerical evaluation. The rough size of $\delta g^{\text{c.t.}}$ can be estimated by comparing the $\overline{\text{MS}}$ and $\text{GL}$ values of g as given below and in Table \ref{tab-koef}.

Comparing with the experimental value \cite{Ahmed:2001xc,Anastassov:2001cw} $g_{D^*D\pi}=g^{\text{Tree}}=0.59\pm 0.08$, we arrive at $g^{\overline{\text{MS}}}=0.41\pm0.10$, $g^{\text{GL}}=0.49\pm0.10$, where the values of $L_{4,5}$ have been used as given in Table \ref{tab-koef}, while other counterterms have been set to zero as discussed above. The error in $g^{\overline{\text{MS}}}$, $g^{\text{GL}}$ is experimental and from the uncertainty on the $L_4$ counterterm (this is proportional to $m_K^2$, see \eqref{Zpi}). The error from the other unknown counterterms can be estimated to be of the same order. 
\index{counterterms!contributions of|)}
\index{corrections, chiral to couplings|)}

\subsubsection{Counterterms} \index{counterterms!values of}
 The values of the counterterms $L_4$ and $L_5$ in \eqref{eqQ:1} are taken from \cite{Gasser:1984gg} and are scaled to $\mu=1\; \text{GeV}$ using
\be
L_i(\mu_2)=L_i(\mu_1) +\frac{\Gamma_i}{16 \pi^2}\ln\frac{\mu_1}{\mu_2},\label{Li}
\ee
where $\Gamma_4=1/8$, $\Gamma_5=3/8$. This yields $L_4^{\text{GL}}=(-0.5\pm0.5)\times 10^{-3}$ and $L_5^{\text{GL}}=(0.6\pm0.5)\times 10^{-3}$. To get values in $\overline{\text{MS}}$ scheme we use the relation
\be
L_i^{\overline{\text{MS}}}=L_i^{\text{GL}}-\Gamma_i\frac{1}{32 \pi^2}.\label{Litwo}
\ee
This gives $L_4^{\overline{\text{MS}}}=(-0.9\pm0.5)\times 10^{-3}$ and $L_5^{\overline{\text{MS}}}=(-0.6\pm0.5)\times 10^{-3}$.

There are no experimental data regarding the sizes of the $k_{1,2}$ and $\varkappa_{1,2}$ counterterms in \eqref{eqQ:3} and \eqref{current}. From large $N_c$ considerations we can conclude that $k_2$ and $\varkappa_2$ are $1/N_c$ suppressed, i.e., the following relations are expected $k_1> k_2$, $\varkappa_1>\varkappa_2$. In the numerical estimates we then set $k_2=\varkappa_2=0$. The approximate size of $k_1$ is determined by observing that $L_5$ term in \eqref{eqQ:1} and $k_1$ term in \eqref{eqQ:3} have similar structure compared to the kinetic term in \eqref{chirallagr} and \eqref{eq-8} respectively. It is thus reasonable to expect that roughly $|k_1|\sim L_5 4 \mu_0 8/f^2$. Similar reasoning applies for $\varkappa_1$, so that in the numerical evaluation we vary $k_1$ and $\varkappa_1$ in the range $- 32 L_5 \mu_0 /f^2<k_1, \varkappa_1< 32 L_5 \mu_0 /f^2$.\index{k12@$k_{1,2}$}\index{kappa12@$\varkappa_{1,2}$}

\begin{table} [h]
\begin{center}
\begin{tabular}{|l|c|c|c|c|} \hline \index{alpha@$\alpha$}\index{beta@$\beta$}\index{g@$g$}
$-$ & Tree & 1-loop $\overline{\text{MS}}$ & 1-loop GL \\ \hline\hline
$\alpha\;[\text{GeV}^{3/2}]$&$0.38\pm0.04 $&$0.21\pm0.05 $& $0.24\pm0.05 $\\ \hline
 $g$&$0.59\pm0.08 $&$0.41\pm0.10 $&$0.49\pm0.10$ \\ \hline
$f \;[\text{MeV}]$&$130.7\pm0.04 $&$120\pm 10 $ & $120\pm10 $\\ \hline
$\beta \;[\text{GeV}^{-1}]$&$2.3\pm0.2 $&$- $ & $- $\\ \hline
$L_4\;[\times 10^{-3}]$& $-$ &$-0.9\pm0.5 $ &$-0.5\pm0.5 $ \\ \hline
$L_5\;[\times 10^{-3}]$& $-$ &$-0.6\pm0.5 $&$0.6\pm0.5 $ \\ \hline
 \end{tabular} 
 \caption[Values of coupling constants in HH$\chi$PT]{\footnotesize{Values of coupling constants used further on in the analysis. The $\overline{\text{MS}}$ and $\text{GL}$ renormalization prescriptions correspond to letting respectively $\bar{\Delta}\to 1$ and $\bar{\Delta}\to0$ in the loop integrals. The values of other ${\cal O}(p^4)$ terms are either set to zero or varied in the ranges discussed in text.}}
\label{tab-koef}\index{coupling constants! values of}
\end{center}
\end{table}

At the end let us summarize the approximations that were made in obtaining the 1-loop values of couplings $g,\alpha, f$ given in Table \ref{tab-koef}\index{alpha@$\alpha$}
\begin{itemize}
\item The $1/m_c$ contributions have been neglected. These are expected to be more important in the determination of $\alpha$, as in it contributions of order $m_K/m_D$ might arise. The $1/m_c$ corrections are expected to be less important in the determination of g, where they are proportional to $m_\pi$.
\item In the determination of $g$ a number of unknown counterterms have been set to zero. Except for the $\tilde{\varkappa}_3$ they are proportional to $m_\pi^2$ which justifies this procedure. The $\tilde{\varkappa}_3$ contribution is proportional to $m_K^2$, while $\tilde{\varkappa}_3$ itself is of order $1/N_c$ and is expected to be suppressed \eqref{deltaLg}. The situation is very similar to the case of $L_4$ contribution, which is proportional to $m_K^2$, while $L_4$ is $1/N_c$ suppressed. Note also that the change of scale and/or renormalization prescription can invalidate the $1/N_c$ argument as can be seen from the relatively large value of $L_4^{\overline{\text{MS}}}$.
\item The uncertainties connected with the couplings in the heavy meson sector do not influence determination of $f$ at one loop. 
\end{itemize}
\index{heavy hadron chiral perturbation theory|)}

\chapter{One loop scalar and tensor functions}\label{scalar-loops} \index{dimensional regularization|(}
In the following we shall present the calculation of the dimensionally regularized one loop scalar functions with one heavy meson propagator, that has been published in \cite{Zupan:2002je}. These will be needed in the evaluation of radiative $D$ meson decays discussed in chapter \ref{rareD}. The expressions for the dimensionally regularized one loop scalar functions within full theory have been know for a long time \cite{'tHooft:1978xw}. The full expressions for scalar functions with one heavy meson propagator, on the other hand, have not been calculated until recently \cite{Zupan:2002je,Boyd:1994pa,Stewart:1998ke,Bouzas:1999ug,Bouzas:2001py,Bouzas:2002xi}. 

The one loop calculations within the heavy quark effective theory are considerably simplified if the light-quark masses are neglected. Very common in the heavy hadron chiral perturbation theory (HH$\chi$PT) is a similar approximation, with the finite contributions omitted, while only the leading logs are retained \cite{Goity:1992tp,Falk:1993fr}. To go beyond the leading log approximation and/or take into account the counterterms appearing at the next order in the chiral expansion, the general solutions for the one loop scalar functions need to be considered. In the context of the HH$\chi$PT a general solution for the one loop scalar two-point function with one heavy quark propagator has been found in \cite{Boyd:1994pa,Stewart:1998ke}. We extend this calculation and find solutions for the scalar three-point and four-point functions with one heavy quark propagator.
\index{tensor functions|(}

The vector and tensor one loop functions can then be expressed in terms of the scalar one loop functions using the algebraic reduction \cite{Passarino:1978jh}. Also, the one loop scalar functions with two or more heavy quark propagators can be expressed in terms of the one loop scalar functions with just one heavy quark propagator. For the case of the equal heavy-quark velocities this can be accomplished using the relation
\begin{equation}
\frac{1}{v\negcdot q-\Delta}\frac{1}{v\negcdot q-\Delta'}=\frac{1}{\Delta-\Delta'}\left(\frac{1}{v \negcdot q -\Delta}-\frac{1}{v \negcdot q-\Delta'}\right).
\end{equation}
For unequal heavy quark velocities techniques developed in \cite{Bouzas:1999ug} can be used.

The scalar one-loop functions with heavy quark propagators can be derived also directly from the scalar functions of the full theory by using the threshold expansion \cite{Beneke:1997zp} (see also appendix B of \cite{Davydychev:2001ui}). This technique has recently been used for the calculation of the scalar and tensor three-point functions with one and two heavy quark propagators \cite{Bouzas:2001py,Bouzas:2002xi}. We will not, however, follow the approach of Bouzas et al. \cite{Bouzas:1999ug,Bouzas:2001py,Bouzas:2002xi} but rather do the calculation from scratch.

This chapter is organized as follows: first we will introduce the notation for scalar and tensor functions that will be used further on. Then we shall proceed to the evaluation of scalar functions. At the beginning we will make some general remarks and list useful relations that will be used further on in the calculation. Then we will review the calculation of one and two point functions. We will continue with the calculation of the three-point and four-point functions in the final sections. 

\section{Notational conventions for loop integrals}\label{Notational-conventions}\index{notation|(}\index{scalar function|(}
In this section we list the definitions of the dimensionally regularized integrals commonly encountered in the evaluation
of the
$\chi$PT and HH$\chi$PT one-loop diagrams. The integrals containing a heavy quark propagator are
\begin{align}
 -\frac{1}{16 \pi^2} \bar{A}_0 (m&)=\frac{i \mu^{\epsilon}}{(2\pi)^{n}} \int d^{n}q \frac{1}{(v \negcdot q -\Delta+i\delta)}=0, \label{HQprop1}\\
 -\frac{1}{16 \pi^2} \bar{B}_{\{0,\mu,\mu\nu\}} (m&,\Delta)=\frac{i \mu^{\epsilon}}{(2\pi)^{n}} \int d^{n}q \frac{\{1,q_\mu,q_{\mu}q_{\nu}\}}{(v \negcdot q -\Delta+i\delta)(q^2-m^2+i \delta)},
\\
\begin{split}
-\frac{1}{16 \pi^2} \bar{C}_{\{0,\mu,\mu\nu\}} (&p, m_1,m_2, \Delta) =\\
&\frac{i \mu^\epsilon}{(2 \pi)^{n}} \int d^{n} q \frac{\qquad \qquad \qquad\quad\{1,q_\mu,q_\mu q_\nu\}\hfill}{(v \negcdot q -\Delta)(q^2-m_1^2)((q+p)^2-m_2^2)}, \label{CBar}
\end{split}
\\
\begin{split}
-\frac{1}{16 \pi^2} \bar{D}_{\{0,\mu,\mu\nu\}}(&p_1,p_2,m_1,m_2,m_3,\Delta)=\\
&\frac{i \mu^\epsilon}{(2\pi)^{n}} \int \frac{\; \; d^{n}q \qquad \qquad \qquad\{1,q_\mu,q_\mu q_\nu\} \hfill }{(v\negcdot q-
\Delta)(q^2-m_1^2)((q+p_1)^2-m_2^2)((q+p_2)^2-m_3^2)},\label{DBar}
\end{split}
\end{align}
where $n=4-\epsilon$. The dependence of scalar and tensor functions on $v^\mu$ is not shown explicitly and also in Eqs.~\eqref{CBar}, \eqref{DBar} the $i\delta$ prescription is not shown. The scalar integrals $\bar{B}_0(m,\Delta)$, $\bar{C}_{0} (p, m_1,m_2, \Delta)$, $\bar{D}_{0}(p_1,p_2,m_1,m_2,m_3,\Delta) $ have been calculated in \cite{Zupan:2002je}. We use the expressions of Ref.~\cite{Zupan:2002je} in the numerical evaluation of the scalar integrals $\bar{B}_0$, $\bar{C}_0$, $\bar{D}_0$. The tensor integrals can be expressed in terms of Lorentz-covariant tensors. The notation we use for the tensor functions resembles closely the notation used in Ref.~\cite{Hahn:1998yk} for the Veltman-Passarino functions \cite{Passarino:1978jh} \index{Veltman-Passarino functions}
\begin{align}
\bar{B}_\mu (m,\Delta)&= v_\mu\bar{B}_1,\label{tensor-beg}\\
 \bar{B}_{\mu\nu}(m,\Delta)&=\eta_{\mu\nu}\bar{B}_{00} +v_\mu v_\nu \bar{B}_{11},\\
\bar{C}_\mu(p,m_1,m_2,\Delta)&= v_\mu\bar{C}_1+p_\mu\bar{C}_2,\\
\bar{C}_{\mu\nu}(p,m_1,m_2,\Delta)&=\eta_{\mu\nu}\bar{C}_{00}+(v_\mu p_\nu+p_\mu v_\nu)\bar{C}_{12}+v_\mu v_\nu\bar{C}_{11}+p_\mu p_\nu \bar{C}_{22},\\
\bar{D}_\mu(p_1,p_2,m_1,m_2,m_3,\Delta)&= v_\mu\bar{D}_1+p_{1\mu}\bar{D}_2+p_{2\mu}\bar{D}_3,\\
\begin{split}
\bar{D}_{\mu\nu}(p_1,p_2,m_1,m_2,m_3\Delta)&=\eta_{\mu\nu}\bar{D}_{00}+v_\mu v_\nu\bar{D}_{11}+(v_\mu p_{1\nu}+p_{1\mu} v_\nu)\bar{D}_{12}+p_{1\mu} p_{2\nu} \bar{D}_{23}\\
&\qquad +p_{1\mu} p_{1\nu} \bar{D}_{22}+(v_\mu p_{2\nu}+p_{2\mu} v_\nu)\bar{D}_{13}+p_{2\mu} p_{2\nu} \bar{D}_{33}.\label{tensor-end}
\end{split}
\end{align}
The tensor functions are calculated using the algebraic reduction \cite{Passarino:1978jh}, i.e., the tensor functions \eqref{tensor-beg}-\eqref{tensor-end} are multiplied by the four-momenta $v^\mu, p^\mu, \dots$ or contracted using $\eta^{\mu\nu}$. Then the identities such as $v\negcdot q=v\negcdot q-\Delta+\Delta$ and/or $q\negcdot p=\frac{1}{2} ((q+p)^2-m^2-(q^2-m^2))$ are used to reduce tensor integrals to a sum of scalar integrals. The result of this procedure has been given explicitly in \cite{Eeg:2001un} for the case of two point functions $\bar{B}_{\{\mu, \mu\nu\}}$ \footnote{Note that different notation is used in Ref.~\cite{Eeg:2001un}, with $\bar{B}_0(m,\Delta)=-I_2(m,\Delta)/\Delta$, $\bar{B}_1(m,\Delta)=-I_2(m,\Delta)-I_1(m)$, $\bar{B}_{00}(m,\Delta)=-\Delta J_1(m,\Delta)$, $\bar{B}_{11}(m,\Delta)=-\Delta J_2(m,\Delta)$.}. For the case of the three and four-point functions $\bar{C}_{\{\mu,\mu\nu\}}$, $\bar{D}_{\{\mu,\mu\nu\}}$ we do not write out explicitly the analytic results of algebraic reductions as the expressions are relatively cumbersome. For instance in the case of $\bar{D}_{\mu\nu}$ the final expression involves the inverse of a $7\times 7$ matrix that corresponds to seven functions $\bar{D}_{00}\dots \bar{D}_{33}$ appearing in the expression of the four-point tensor function \eqref{tensor-end}. Note as well that in this particular case there are ten possible relations between $\bar{D}_{00}\dots \bar{D}_{33}$ and the scalar functions $\bar{B}_0$, $\bar{C}_0$, $\bar{D}_0$ that one gets from algebraic reductions (three equations from each multiplication by $v^\mu$, $p_1^\mu$, $p_2^\mu$ plus one relation from contraction by $\eta^{\mu\nu}$). Obviously not all ten equations can be linearly independent. Using different sets of seven independent equations have to lead to the same results for $\bar{D}_{00}\dots \bar{D}_{33}$ coefficient functions. This fact can then be used as a very useful check in the numerical implementation.

The loophole of the aforementioned procedure is, if the set of equations provided by the algebraic reduction is not invertible. This happens for instance in the calculation of $D^0\to l^+l^-\gamma$ (see section \ref{D0GammaGamma}). Namely, for $p_1=p$ and $p_2=p+k$ appearing in the calculation of $C_0^{4\_4}$ (with $p$ the four-momentum of the lepton pair and $k$ the photon momentum, see section \ref{D0GammaGamma} or appendix \ref{app-D0llBarGamma}, Eqs.~\eqref{eta4_4}, \eqref{kp4_4}) only six out of ten relations following from algebraic reduction are linearly independent. This problem is connected to the special kinematics of $D^0\to l^+l^-\gamma$ decays and has been circumvented by first calculating the tensor four-point functions with the prescription $k\to k+\epsilon a$, where $a$ is some arbitrary four-momentum, and then taking the limit $\epsilon \to 0$ numerically. Similarly, in the calculation of $C_0^{4\_5}$, where $p_1=k$, $p_2=k+p$, see Eqs.~\eqref{eta4_5}, \eqref{kp4_5}, the prescription $p\to p+\epsilon a$ has been used. Because $\bar{D}_{00}\dots \bar{D}_{33}$ are continuous functions of $p_1$ and $p_2$, the outlined limiting procedure leads to an unambiguous result. This has been also checked numerically.

To make the listing of notational conventions self-contained we give in the following also the notation for the Veltman-Passarino functions employed by the LoopTools package \cite{Hahn:1998yk}, \index{LoopTools} that has been used for their numerical evaluation. A general integral is \index{Veltman-Passarino functions}
\begin{equation}
-\frac{1}{16 \pi^2} T^N_{\mu_1\dots\mu_P}=\frac{i \mu^\epsilon}{(2\pi)^n}\int\frac{\; \; d^{n}q \qquad \qquad \qquad q_{\mu_1}\cdots q_{\mu_P} \hfill }{(q^2-m_1^2)((q+p_1)^2-m_2^2)\cdots((q+p_{N-1})^2-m_{N}^2)},
\end{equation}
with two-point functions $T^2$ generally denoted by the letter $B$, the three-point functions $T^3$ by the letter $C$ and the four-point functions $T^4$ by the letter $D$. Thus, e.g., $B_0(p^2,m_1^2,m_2^2)$ and $C_0(p_1^2,(p_1-p_2)^2, p_2^2, m_1^2,m_2^2)$ are two-point and three-point scalar functions respectively. The decomposition of the tensor integrals in terms of the Lorentz-covariant tensors reads explicitly 
\begin{align}
B_\mu&=p_{1\mu}B_1,\\
B_{\mu\nu}&=\eta_{\mu\nu}B_{00}+p_{1\mu}p_{1\nu}B_{11},\\
C_{\mu}&=p_{1\mu}C_1+p_{2\mu}C_2=\sum_{i=1}^2p_{i\mu}C_i,\\
C_{\mu\nu}&=\eta_{\mu\nu}C_{00}+\sum_{i,j=1}^2 p_{i\mu} p_{j\nu}C_{ij},\\
C_{\mu\nu\rho}&=\sum_{i=1}^2(\eta_{\mu\nu}p_{i\rho}+\eta_{\nu\rho} p_{i\mu}+\eta_{\mu\rho} p_{i\nu})C_{00i}+\sum_{i,j,l=1}^2 p_{i\mu}p_{j\nu}p_{l\rho} C_{ijl}.
\end{align}
Note that the tensor-coefficient functions are totally symmetric in their indices.\index{notation|)}\index{tensor functions|)}

\section{General remarks and useful relations}
Let us now turn to the evaluation of the one loop scalar functions, with one heavy quark propagator \eqref{HQprop1}-\eqref{DBar}. Before we start with the actual calculation, let us, however, first list some useful relations and the conventions that are going to be used further on. The greater part of this section is a review of the relations and the conventions used in \cite{'tHooft:1978xw} with certain modifications.
The major difference between the conventional one loop scalar functions and the one loop scalar functions with one heavy quark propagator is the appearance of the propagator linear in the integration variable $q$. Therefore, a modified version of the standard Feynman parameterization is used
\begin{equation}
\begin{split}
\frac{1}{\big(\prod_{i=1}^N A_i\big) B}=& N! \int_0^\infty\negtwo 2 d \lambda \int \prod_{i=1}^N d u_i \; \frac{\delta(1-\sum_i u_i) \;\prod_i \Theta(u_i)}{[\sum_{i=1}^N A_i u_i +2 B \lambda]^{N+1}}\\
=& N!\; \frac{ \int_0^\infty 2 d\lambda \int_0^1 dx_1 \int_0^{x_1} dx_2 \cdots \int _0^{x_{N-2}}dx_{N-1} \hfill}{[A_1(1-x_1)+ \sum_{j=2}^{N-1}A_j (x_{j-1}-x_j) + A_N x_{N-1} +2 B \lambda ]^{N+1}}, \label{eqS:24}
\end{split}
\end{equation}
where $\Theta(u)$ is the Heaviside function $\Theta(u)=1$ for $u>0$ and zero otherwise. In the calculation $A_i$ are going to be ``full'' (inverse) propagators $((q+p_i)^2-m_i^2+i\delta)$ and $B$ the heavy quark propagator $(v\negthincdot q - \Delta + i \delta)$. Note also, that the leading power of $q^2$ in the denominator has increased from the left-hand side's $(q^2)^{N+1/2}$ to the right-hand side's $(q^2)^{N+1}$. The integration over $q$ has been made more convergent, but then another integration over infinite range (integration over $\lambda$) has been introduced through the parameterization.

A very useful identity used in the calculation is
\begin{equation}
\begin{split}
\frac{1}{[(q+p_1)^2-m_1^2+i \delta][(q+p_2)^2-m_2^2+i \delta]}= &\frac{\alpha}{[(q+p_1)^2-m_1^2+i \delta][(q+l)^2-M^2+i \delta]}\\
&+\frac{1-\alpha}{[(q+p_2)^2-m_2^2+i \delta][(q+l)^2-M^2+i \delta]} \label{eqS:9},
\end{split}
\end{equation}
where $\alpha$ is an arbitrary parameter and
\begin{align}
l&=p_1+\alpha(p_2- p_1),\\
M^2&=(1-\alpha)m_1^2+\alpha m_2^2-\alpha(1-\alpha)(p_2-p_1)^2.
\end{align}
The parameter $\alpha$ can then be chosen at will. It is useful to keep it real, though. Then there are no ambiguities connected with the shift of the integration variable $q$, that is performed, as usual, before the Wick rotation. \index{Wick rotation} For instance $\alpha$ can be chosen such that $M^2=0$. If $(p_2-p_1)^2\leqslant (m_1-m_2)^2 $ or $(p_2-p_1)^2\geqslant (m_1+ m_2)^2$, then $\alpha$ is real. If one of the masses is made to be zero, the integration is simplified considerably (as will be seen in the calculation of the four-point function \eqref{eqS:5}). The other option used below is to set $\alpha$ such that $l^2=0$. This can be done for real $\alpha$ if (but not only if) one of $p_1$, $p_2$ or $p_1\pm p_2$ is timelike. This shows, that in general product of propagators at least one internal or one external mass can always be set to zero, even with $\alpha$ restricted to be real.

In doing the integrals the following procedure proves to be very useful. Consider
\begin{equation}
\int_0^\infty \negtwo 2 d\lambda \int_0^1 dx \frac{1}{[a x^2 + b \lambda^2 + c x \lambda+ \cdots]}.
\end{equation}
The integration over $x$ can be simplified by the change of the integration variables $\lambda=\lambda' + \beta x$, where $\beta$ is chosen such, that the coefficient in front of $x^2$ vanishes, i.e., $\beta$ has to solve the equation $b \beta^2+ c\beta + a=0$. Then the integrand is linear in $x$, so the integration over $x$ is trivial. The integration bounds are 
\begin{equation}
\begin{split}
\int_0^\infty \negtwo 2 d\lambda \int_0^1 dx \cdots &= \int_0^1 dx \int_{-\beta x}^\infty 2 d \lambda' \cdots=\\
&= \int_0^\infty \negtwo 2 d\lambda \int_0^1 d x \cdots+ \int_{-\beta}^0 2 d \lambda \int_{-\lambda/\beta}^1 dx \cdots \label{eqS:25}.
\end{split}
\end{equation}

\index{logarithm, branch cuts}
As the results of the integration, the functions such as logarithms, dilogarithms (Spence functions) and hypergeometric functions will appear. Since the arguments of the functions will in general lie in the complex plane it is necessary to discuss the conventions used. The convention used for the logarithms is that they have a cut along the negative real axis. For $x$ exactly negative real we use the prescription $\ln(x)\to \ln(x+i \epsilon)$, where $\epsilon>0$ is a positive infinitesimal parameter. In other words, $\ln(x)=\ln|x|+i\pi$ for $x$ negative real\footnote{Note that this prescription does not change the calculation of the logarithms away from the negative real axis. In particular it does not change the value of a logarithm with an argument that already has an infinitesimal but nonzero imaginary part. For more discussion on this point see text after Eq.~\eqref{eqS:37}. }. In particular $\ln(-1)$ is defined to be $\ln(-1)=i\pi$. Of course this choice is completely arbitrary and at the end of the calculation one has to check that results are independent of this choice. Using this definition for the logs of the negative real arguments the logarithm of an inverse is 
\begin{equation}
\ln\left(\frac{1}{x}\right)=-\ln(x)+2 \pi i \Re^{(-)}(x), \label{eqS:55}
\end{equation}
with
\begin{equation}
\Re^{(-)}(x)=\left\{
\begin{aligned}
1&\; ;\; & x \text{ on negative real axis},\\
0&\; ; \; & \text{otherwise}.
\end{aligned}
\right. \label{eqS:47}
\end{equation}
Note that the change from the usual rule for the logarithm of an inverse is just on the negative real axis. For the arguments away from the negative real axis the function $\Re^{(-)}(x)$ is exactly zero and everything is as usual.
The logarithm of a product is
\begin{equation}
\ln(a b)= \ln(a) +\ln(b) +\eta(a,b),
\label{eqS:17}
\end{equation}
where $\eta$ function is \footnote{Note, that in comparison with \cite{'tHooft:1978xw}, the $\eta$ function has been extended also to the negative real arguments (cf. discussion after Eq.~\eqref{eqS:8} and Eq.~\eqref{eqS:51}). For arguments away from the negative real axis (also if by an infinitesimal amount) it is the same as in \cite{'tHooft:1978xw}.}
\begin{equation}\label{eqS:50}
\eta(a,b)=\left\{
\begin{aligned}
\;&
\begin{aligned}
2 \pi i \big\{ & \Theta(-\Im(a)) \Theta(-\Im (b))\Theta(\Im (ab))\\
&-\Theta(\Im (a)) \Theta(\Im (b))\Theta(-\Im (ab))\big\} 
\end{aligned} \; ; \; &\text{ $a$ and $b$ {\it not} negative real},\\
\;& 
- 2 \pi i \big\{ \Theta(\Im (a))+\Theta(\Im (b)) \big\}
\; ;\; &\text{ either $a$ or $b$ negative real},\\
\;&
-2 \pi i \; ;\; &\text{ $a$ and $b$ negative real}.
\end{aligned}
\right. 
\end{equation}
The normal rule for the logarithm of a product applies for these important cases
\begin{equation}
\begin{split}
\ln(ab)&=\ln(a)+\ln(b) ;\; \text{$\Im (a)$ and $\Im (b)$ have the opposite sign},\\
\ln(a/b)&=\ln(a)-\ln(b) ;\; \text{$\Im (a)$ and $\Im (b)$ have the same sign},
\end{split}\label{eqS:54}
\end{equation}
with $a$, $b$ not negative real.

\index{difficulties! in calculation of scalar functions|(}
A very illuminative example of what kind of problems are encountered when doing the integrals of functions with branch cuts in the complex plane is the following simple calculation taken from \cite{'tHooft:1978xw}. Consider for instance
\begin{equation}
I(a,b)=\int_0^1 \frac{dx}{a x +b}, \label{int-example}
\end{equation}
where $a$ and $b$ are arbitrary complex numbers. The indefinite integral would be $a^{-1}\ln(a x+b)$, but then one has to take care whether for $x\in [0,1]$ the argument of the logarithm crosses the negative real axis or not. It is easier to first divide out $a$
\begin{equation}
I(a,b)=\frac{1}{a} \int_0^1 \frac{dx}{x +b/a}=\frac{1}{a} \ln \left(x +\frac{b}{a}\right)\Big|_0^1.
\end{equation}
Then the argument of the logarithm does not cross the negative real axis as it has the same imaginary part regardless of the value of (real) $x$. So 
\begin{equation}
I(a,b)=\frac{1}{a}\Big[\ln\left(1+\frac{b}{a}\right)-\ln\frac{b}{a}\Big].
\end{equation}
Since the arguments of the logarithms have the imaginary parts of the same sign, the usual rule \eqref{eqS:54} applies and one can write
\begin{equation}
I(a,b)=\frac{1}{a}\ln \frac{a+b}{a},\label{int-example-end}
\end{equation}
which is actually the standard result. The problem with the careless derivation would be, that it would give $a^{-1}(\ln(a+b)-\ln a)$, which is not correct for all choices of $a$ and $b$. For instance, for $a=-1-i \varepsilon$ infinitesimally below the negative real axis and $b=-1+2 i \varepsilon$ infinitesimally above the negative real axis, the integration path lies infinitesimally close to the negative real axis with the starting point below and ending point above the negative real axis. Since the integration path does not cross the pole of the integrand in \eqref{int-example}, the result of integration should be almost real. The naive result $a^{-1}(\ln(a+b)-\ln a)$, however, gives incorrectly $-\ln 2-2\pi i$, while the use of Eq.~\eqref{int-example-end} leads to the correct result $-\ln 2$.
\index{difficulties! in calculation of scalar functions|)}

\section{Dilogarithm and hypergeometric function}\index{hypergeometric function|(}
\label{app:AScal}
In this section we list some properties of the dilogarithm and the hypergeometric function used in the rest of this chapter (for other properties consult, e.g., \cite{'tHooft:1978xw,Gradshteyn}).

The dilogarithm or Spence function is defined as \index{dilogarithm! definition of}
\begin{equation}
\Li(x)=-\int_0^1 dt \; \frac{\ln(1-x t)}{t} \label{eqS:38}.
\end{equation}
The cut for the logarithm along the negative real axis translates into the cut for the dilogarithm along the positive real axis for $x>1$. For $x$ on the positive real axis, $x>1$, dilogarithm is calculated using the following prescription $\Li (x) \to \Li (x- i\epsilon)$. Note as well that $\Li(0)=0$.

Useful identities valid also for the complex arguments (not equal to zero) are \cite{Gradshteyn}\index{dilogarithm!useful relations}
\begin{subequations}
\begin{align}
\Li(x)&=-\Li\Big(\frac{1}{x}\Big)-\frac{1}{6}\pi^2 -\frac{1}{2}\big[\ln(-x)- 2 \pi i \xi(x)\big]^2\label{eqS:19},\\
\Li(x)&=-\Li(1-x)+\frac{1}{6}\pi^2-\ln(x)\ln(1-x), \label{eqS:16}
\end{align}
\end{subequations}
where
\begin{equation}
\xi(x)=
\left\{
\begin{aligned}
\;&1\; ; \; & x\in (0,1),\\
\;&0\; ; \; & \text{ otherwise}.
\end{aligned}
\right.
\end{equation}

\index{hypergeometric function!definition of}
The hypergeometric function for complex argument $|z|<1$ is defined in terms of the series 
\begin{equation}
{}_2F_1(\alpha,\beta;\gamma;z)=1+\sum_{n=0}^{\infty}\frac{\alpha \dots (\alpha+n)\beta\dots (\beta+n)}{\gamma\dots (\gamma+n) (n+1)!}\;z^{n+1}, \label{eqS:15}
\end{equation}
with $\gamma$ not equal to zero or negative integer. Note that the series terminates if $\alpha$ or $\beta$ are equal to negative integer or zero. If either of them is zero then
\begin{equation}
{}_2F_1(0,\beta;\gamma;z)={}_2F_1(\alpha,0;\gamma;z)=1.
\end{equation}
For $z$ outside the unit circle the values of the hypergeometric function can be obtained through \index{analytic continuation}analytic continuation. We make a cut in the $z$ plane along the real axis from $z=1$ to $z=\infty$. Then the series \eqref{eqS:15} will yield, in the cut plain, a single valued analytic continuation that can be obtained using the following identity (other similar transformation formulas can be found in, e.g., \cite{Gradshteyn}) \index{hypergeometric function! useful relations}
\begin{equation}
\begin{split}\label{eqS:14}
{}_2F_1(\alpha,\beta;\gamma;z)=&\frac{\Gamma(\gamma)\Gamma(\beta-\alpha)}{\Gamma(\beta)\Gamma(\gamma-\alpha)} (- z)^{-\alpha} {}_2F_1(\alpha,\alpha+1-\gamma;\alpha+1-\beta;1/z)\\
+&\frac{\Gamma(\gamma)\Gamma(\alpha-\beta)}{\Gamma(\alpha)\Gamma(\gamma-\beta)}(- z)^{-\beta} {}_2F_1(\beta,\beta+1-\gamma;\beta+1-\alpha;1/z).
\end{split}
\end{equation}

The integral representations of the hypergeometric function include
\begin{subequations}
\label{eqS:12}
\begin{align}
\int_z^\infty \frac{x^{\mu-1}dx}{(1+\beta x)^\nu}&=\frac{z^{\mu-\nu}}{\beta^\nu (\nu-\mu)} \; {}_2F_1(\nu,\nu-\mu;\nu-\mu+1;-1/(\beta z)), \label{eqS:22}\\
\int_0^z \frac{x^{\mu-1}dx}{(1+\beta x)^\nu}& = \frac{z^\mu}{\mu} \; {}_2F_1(\nu,\mu;1+\mu;-\beta z), \label{eqS:23}
\end{align}
\end{subequations}
where Eq.~\eqref{eqS:22} is valid for $\Re (\nu)>\Re(\mu)$, while Eq.~\eqref{eqS:23} is valid for the case, when $\arg(1+\beta z)<\pi$ and $\Re(\mu)>0$. \index{hypergeometric function|)}

\section{One- and two-point functions}\label{One-two-point}\index{calculation! scalar functions|(} \index{integration of scalar functions|(}
In this section we will concentrate on the calculation of the dimensionally regularized one-point and two-point functions in the heavy quark effective theory \index{scalar function!one-point}\index{scalar function!two-point}
\begin{align}
-\frac{1}{16 \pi^2} \bar{A}_0 (\Delta)&=\frac{i \mu^{\epsilon}}{(2\pi)^{n}} \int d^{n}q \frac{1}{(v \negcdot q -\Delta+i\delta)},\\
-\frac{1}{16 \pi^2} \bar{B}_0 (m,\Delta)&=\frac{i \mu^{\epsilon}}{(2\pi)^{n}} \int d^{n}q \frac{1}{(v \negcdot q -\Delta+i\delta)(q^2-m^2+i \delta)}, \label{eqS:2}
\end{align}
where $\delta$ is a positive infinitesimal parameter, $n=4-\epsilon$, while $m$ and $\Delta$ are real. The general solution for the two-point function has been found by Stewart in Ref.~\cite{Stewart:1998ke} (see also \cite{Casalbuoni:1996pg} and references therein). In this section we will derive Stewart's result.

We start with the integral
\begin{equation}
I_r=\frac{i \mu^{\epsilon}}{(2\pi)^n }\int d^n q \frac{1}{(v\negcdot q -\Delta+i \delta)(q^2-m^2+i\delta)^r}.
\end{equation}
Using the Feynman parameterization \cite{Manohar:dt}
\begin{equation}
\frac{1}{a^r b^s}=2^s \frac{\Gamma(r+s)}{\Gamma(r)\Gamma(s)} \int_0^\infty \negtwo d\lambda \frac{\lambda^{s-1}}{(a + 2b \lambda)^{r+s}},
\end{equation}
we get
\begin{equation}
I_r=\mu^\epsilon \frac{(-1)^{-r}}{\Gamma(r)}\frac{2 \Gamma(r+1-\frac{n}{2})}{(4 \pi)^{\frac{n}{2}}}\int_0^\infty \negtwo d\lambda (\lambda^2+2 \lambda \Delta+m^2 -i \delta) ^{\frac{n}{2} -r -1}.
\end{equation}
 The integral can be expressed in terms of the hypergeometric function. Introducing the new variable $\lambda'=\lambda +\Delta$ and then splitting the integration interval for negative $\Delta$ we get
\begin{equation}
\int_\Delta^\infty\frac{d \lambda'}{({\lambda'}^2-\Delta^2+m^2 -i \delta)^N}=\int_{-|\Delta|}^{|\Delta|} \frac{d \lambda}{(\lambda^2 -\Delta^2+m^2-i\delta)^N} \Theta(-\Delta) +\int_{|\Delta|}^{\infty}\frac{d\lambda}{(\lambda^2-\Delta^2+m^2-i\delta)^N},
\end{equation}
where we write $N=r+1-n/2$ for short. Another change of variables $u=\lambda^2$ leads to
\begin{equation}
\begin{split}
(m^2-\Delta^2-i\delta)^{-N}\Big\{&\int_{\Delta^2}^\infty \frac{du}{2 \sqrt{u}}\Big[\frac{u}{m^2-\Delta^2-i\delta}+1 \Big]^{-N}\;+ \\
\;&+ 2 \int_0^{\Delta^2} \frac{d u}{2 \sqrt{u}} \Big[\frac{u}{m^2-\Delta^2-i\delta}+1 \Big]^{-N} \Theta(-\Delta)\Big\}\label{eqS:3}.
\end{split}
\end{equation}
These integrals can be expressed in terms of the hypergeometric functions ${}_2F_1(\alpha, \beta;\gamma; z)$ through the identities \eqref{eqS:22}, \eqref{eqS:23} listed in section \ref{app:AScal}. Using the transformation formula \eqref{eqS:14} together with ${}_2F_1(0,\beta;\gamma;z)={}_2F_1(\alpha,0;\gamma;z)=1$ we arrive at
\begin{equation}
\begin{split}
I_r=\mu^\epsilon\frac{(-1)^{-r}}{\Gamma(r)} \frac{ 2 \Gamma(N)}{(4\pi)^{\frac{n}{2}}}\bigg[ &-\Delta (m^2 -\Delta^2-i\delta)^{-N} {}_2F_1\negthickspace \left(N,\frac{1}{2};\frac{3}{2};\frac{-\Delta^2}{m^2-\Delta^2-i \delta}\right)\\
&+ \frac{\Gamma\left(N+\frac{1}{2}\right)\Gamma\left(\frac{1}{2}\right)}{(2N-1) \Gamma(N)} (m^2-\Delta^2-i \delta)^{\frac{1}{2}-N}\bigg] \label{eqS:1},
\end{split}
\end{equation}
where $N=r+1-n/2$. 

Let us first discuss the case when $r$ is equal to zero or negative integer, i.e., when the integrand, apart from the heavy quark propagator, is a polynomial. The integrals for the physical case $n\to 4$ are divergent, but we can make sense of it through the \index{analytic continuation}analytic continuation. At fixed $r$ the integral $I_r$ is taken to be an analytic function of the complex dimension $n$. For $r$ equal to zero or negative integer and $n/2 \ne \mathbb{Z}$ all functions appearing in \eqref{eqS:1} are finite, apart from $\Gamma(r)$ that is infinitely large. Thus $I_r$ vanishes for $r$ zero or negative integer everywhere in the $n$ complex plane apart from the points on the real axis with integer $n/2 \geqslant r+1 $. Analytic continuation of $I_r$ \eqref{eqS:1} is then equal to zero in the whole $n$ plane. Integrals over polynomials (and one heavy quark propagator) are in the dimensional regularization thus equal to zero. In particular, the one-point scalar function $\bar A_0=0$. \index{scalar function!one-point}

For the two-point function we have $r=1$ and therefore $N=\epsilon/2$. So the two point function is
\begin{equation}
\begin{split}
-\frac{2}{16 \pi^2} \left(4 \pi \mu^2 \right)^\frac{\epsilon}{2}\; &\Big\{\Gamma \negthickspace\left(\frac{\epsilon}{2}\right) (-\Delta) (m^2-\Delta^2-i\delta)^{-\frac{\epsilon}{2}}\; {}_2F_1\negthickspace \left(\frac{\epsilon}{2},\frac{1}{2};\frac{3}{2};\frac{- \Delta^2}{m^2-\Delta^2-i\delta}\right)\\
&- \pi (m^2-\Delta^2-i\delta)^{\frac{1}{2}}\Big\}.
\end{split}
\end{equation}
Since $\Gamma(\epsilon/2)\to 2/\epsilon -\gamma+{\cal O}(\epsilon)$ we have to expand hypergeometric function around $\epsilon/2=0$ in order to get the finite terms correctly
\begin{equation}
{}_2F_1\negthickspace\left(\frac{\epsilon}{2},\frac{1}{2};\frac{3}{2};z\right)=1+ \frac{\epsilon}{2} \frac{\partial}{\partial N}\left. {}_2F_1\negthickspace\left(N,\frac{1}{2};\frac{3}{2};z\right)\right|_{N=0}+\dots.
\end{equation}
The partial derivative can be found using the series expansion \eqref{eqS:15} and is
\begin{equation}
\frac{\partial}{\partial N} \; \left. {}_2F_1\negthickspace\left(N,\frac{1}{2};\frac{3}{2};z\right)\right|_{N=0}=-\ln(1-z)-z^{-\frac{1}{2}} \ln\Big(\frac{1+\sqrt{z}}{1-\sqrt{z}}\Big)+2.
\end{equation}
This leads us to the final result for the two-point function \index{scalar function!two-point}
\begin{equation}
\begin{split}
\frac{i \mu^{\epsilon}}{(2\pi)^{4-\epsilon}}& \int d^{4-\epsilon}q \frac{1}{(v \negcdot q -\Delta+i \delta)(q^2-m^2+i \delta)}=\\
& \qquad \qquad \frac{2\Delta}{(4 \pi)^2} \Big\{ \frac{2}{\epsilon}-\gamma+\ln 4\pi -\ln\Big(\frac{m^2}{\mu^2}\Big)+2 -2 F\left(\frac{m}{\Delta}\right)\Big\}\label{eqS:33},
\end{split}
\end{equation}
where F(x) is a function as defined in \cite{Stewart:1998ke} valid for both positive and negative $\Delta$ (while $m$ is always taken to be positive real)
\begin{equation} \label{F_def}
F\left(\frac{1}{x}\right)=\left\{
\begin{aligned}
\;&\frac{1}{x}\sqrt{x^2-1} \; \ln(x+\sqrt{x^2-1}+i \delta) \; ; &\;|x|>1,\\
-&\frac{1}{x}\sqrt{1-x^2}\left[\frac{\pi}{2}-\tan^{-1} \left(\frac{x}{\sqrt{1-x^2}}\right)\right]\; ;& \;|x|\le 1,
\end{aligned}\right.
\end{equation}
with $\delta$ an infinitesimal positive parameter. Note that for $x<-1$, $ F(1/x)$ has an imaginary part that corresponds to the particle creation. Also, the two point function has to be a continuous function of $\Delta$, as can be seen from \eqref{eqS:2} or \eqref{eqS:3}. It is easy to check that for $|x|=1$ the two-point function is continuous as then $F(\pm 1)=0$. The two-point function is also continuous for $\Delta\to 0$. Even though $F(1/x)$ diverges as $x\to0$, the two point function \eqref{eqS:33} is finite and equal to $m/8 \pi$.

Finally, for $r \geqslant 2$ both $\Gamma(r)$ and $\Gamma (N)$ in \eqref{eqS:1} are finite in the limit $n\to 4$, so that Eq.~\eqref{eqS:1} can be used directly, with $n$ set to $n=4$.

\section{Three-point scalar function}
The one loop scalar three-point function with one heavy quark propagator is given by \index{scalar function!three-point}
\begin{equation}
-\frac{1}{16 \pi^2} \bar{C}_0 (v,k, \Delta, m_1,m_2)=\frac{i \mu^\epsilon}{(2 \pi)^{n}} \int d^{n} q \frac{1}{(v \negcdot q -\Delta+i\delta)(q^2-m_1^2 +i \delta)((q+k)^2-m_2^2+i\delta)}.\label{eqS:Three}
\end{equation}
This integral is finite in 4 dimensions, so that $\epsilon$ can be set to zero. Using the Feynman parameterization \eqref{eqS:24} we get
\begin{equation}
\bar{C}_0=- \int_0^\infty \negtwo 2 d\lambda \int_0^1 dx \frac{1}{[\lambda^2 +k^2 x^2 +2 v\negcdot k x \lambda -(k^2+m_1^2-m_2^2)x + 2\Delta \lambda+m_1^2-i\delta]}\label{eqS:42}.
\end{equation}
The integration over $x$ can be made trivial through the change of variables $\lambda=\lambda'+\alpha x$. We choose $\alpha$ to be the solution of 
\begin{equation}
(k+\alpha v)^2=0, \label{eqS:49}
\end{equation}
 as then the term quadratic in $x$ is zero. The solution is $\alpha_{1,2}=(-v \negthincdot k\pm \sqrt{(v\negthincdot k)^2-k^2})$ and is real for any real four-vector $k^\mu$ (this can be easily seen by going in the frame, where $v^\mu=(1,0,0,0)$ with the square root then equal to $\sqrt{\vec{k}^2}$). Changing the integration order as in \eqref{eqS:25}, and then integrating over $x$, we get
\begin{equation}
\begin{split}
\bar{C}_0=- \bigg[ &\int_0^\infty \frac{2 d\lambda}{(A\lambda +B)}\ln\bigg(\frac{\lambda^2+C\lambda +D+(A\lambda +B)}{\lambda^2+C \lambda +D}\bigg)\\
&+\int_0^1\frac{2 \alpha d\lambda}{(-A\alpha \lambda +B)}\Big\{ \ln[\alpha^2 \lambda^2-C\alpha\lambda+D+(-A\alpha \lambda+B)]\\
&\qquad\qquad\qquad \qquad\quad-\ln [\alpha^2 \lambda^2-C\alpha \lambda +D+(-A \alpha \lambda +B)\lambda]\Big\}\bigg]\label{eqS:4},
\end{split}
\end{equation}
where
\begin{equation*}
\begin{matrix}
&A=2(v\negcdot k +\alpha), &B=2 \Delta \alpha +m_2^2-m_1^2-k^2, \\
&C=2\Delta, &D=m_1^2-i\delta.
\end{matrix}
\end{equation*}
and $\alpha$ one of the solutions $\alpha_{1,2}$ of the quadratic equation \eqref{eqS:49}. In \eqref{eqS:4} we have used the fact that D is the only complex parameter and split the logarithm in the second integrand. The integrands of both the first and the second integral have vanishing residues. In the second integral we then add and subtract the values of the logarithms for $\lambda=B/A\alpha$ and write
\begin{equation}
\begin{split}
\bar{C}_0=-\frac{1}{A}\Bigg[&\int_0^\infty\negtwo\frac{2 d\lambda}{\lambda +B/A} \ln \bigg(\frac{\lambda^2+(A+C)\lambda +(B+D)}{\lambda^2+C\lambda+D}\bigg)\\
-&\int_0^1 \negtwo\frac{2 d\lambda}{\lambda-B/(A\alpha)} \Big(\ln \big[\alpha^2\lambda^2-(A+C)\alpha \lambda+(B+D)\big]\\
&\qquad\qquad \qquad\qquad\qquad\qquad-\ln\big[\alpha^2\lambda_0^2-(A+C)\alpha \lambda_0+(B+D)\big]\Big)\\
+&\int_0^1\negtwo \frac{2 d\lambda}{\lambda-B/(A\alpha)}\Big(\ln\big[\alpha(\alpha-A)\lambda^2+(B-C\alpha)\lambda +D\big]\\
&\qquad\qquad\qquad\qquad\qquad\qquad -\ln\big[\alpha(\alpha-A)\lambda_0^2+(B-C\alpha)\lambda_0+D\big]\Big)\bigg]\label{eqS:21},
\end{split}
\end{equation}
with $\lambda_0=B/(A\alpha)$. Note, that all three integrals in \eqref{eqS:21} have integrands with vanishing residues. These integrals can be reduced into the sums of the dilogarithms, where care has to be taken regarding the imaginary parts of the arguments of the logarithms. The solutions of the integrals can be found in section \ref{app:BScal}. The solution of the first integral can be found in \eqref{eqS:27}, \eqref{eqS:41}, with the definitions \eqref{eqS:36}, \eqref{eqS:48} (where $a_1=a_2=0$, note also the minus sign), while the solutions to the last two integrals can be found in \eqref{eqS:28}, \eqref{eqS:29}. Using the functions $S_3$ and $I_2$ defined in section \ref{app:BScal} the three-point function \eqref{eqS:Three} finally reads \index{scalar function!three-point}
\begin{equation}
\begin{split}
\bar{C}_0= \frac{2}{A}\Big[\;& I_2\left(0,0,A+C,B+D,C,D,-B/A\right)\\
+&S_3\left(\alpha^2,-(A+C)\alpha,B+D,B/A\alpha\right)\\
-& S_3\left(\alpha(\alpha-A),B-C\alpha,D,B/A\alpha\right)\Big].\label{ThreePoint}
\end{split}
\end{equation}
Note that the value of the three-point function in \eqref{ThreePoint} does not depend on which of the solutions $\alpha_{1,2}$ of the equation \eqref{eqS:49} is used. This can be used as a useful check in the numerical implementation.

The solution is simplified considerably if $k^2=0$. Then the $x$ integration in \eqref{eqS:42} is trivial. Proceeding similarly as above we arrive at
\begin{equation}
\bar{C}_0(v,k,\Delta,m_1,m_2)\Big|_{k^2=0}=-\frac{1}{v \negcdot k} \sum_i \rho(\varkappa_i) \left[ \Li\left(\frac{\lambda_0}{\lambda_0-\varkappa_i}\right)+\frac{1}{2} \ln^2(\lambda_0-\varkappa_i)\right],
\end{equation}
where $\lambda_0=(m_1^2-m_2^2)/(2 v \cdot k)$, while $\varkappa_i$ are the solutions of
\begin{equation}
\begin{split}
&\lambda^2+ 2 (v \negcdot k+\Delta)\lambda +m_2^2-i \delta=(\lambda-\varkappa_1)(\lambda-\varkappa_3),\\
& \lambda^2+2 \Delta \lambda +m_1^2-i \delta=(\lambda-\varkappa_2)(\lambda-\varkappa_4),\label{eqS:43}
\end{split}
\end{equation}
and $\rho(\varkappa_i)=(-1)^{i+1}$.

The solution is even further simplified if besides $k^2=0$ also $m_1=m_2=m$. Then
\begin{equation}
\bar{C}_0(v,k,\Delta,m,m)\Big|_{k^2=0}=-\frac{1}{v \negcdot k} \sum_i \rho(\varkappa_i) \left[ \frac{1}{2} \ln^2(-\varkappa_i)\right],
\end{equation}
with $\varkappa_i$ and $\rho(\varkappa_i)$ given in \eqref{eqS:43}. The three point function in this limit has been calculated before and is given explicitly in \cite{Fajfer:2001ad} (see Eq. (A10) of \cite{Fajfer:2001ad}). The two expressions agree completely. A number of numerical checks between numerically integrated expression \eqref{eqS:4} and final expression \eqref{ThreePoint} have been performed as well. 

\section{Four-point function}
The scalar four-point function with one heavy quark propagator is defined as \index{scalar function!four-point}
\begin{equation}
\begin{split}
-\frac{1}{16 \pi^2} \bar{D}_0(v,p_1,&p_2,\Delta,m_1,m_2,m_3)=\\
&\frac{i \mu^\epsilon}{(2\pi)^{n}} \int \frac{\; \; d^{n}q \hfill }{(v\negcdot q-
\Delta)(q^2-m_1^2)((q+p_1)^2-m_2^2)((q+p_2)^2-m_3^2)}, \label{eqS:10}
\end{split}
\end{equation}
where the $i\delta$ prescription has been omitted in the notation. Again, the integral is convergent and $\epsilon$ can be set to zero. Since the Feynman parameterization \eqref{eqS:24} is not symmetric in $A_i$ and $B$, the elegant transformation used in the calculation of the conventional four-point function \cite{'tHooft:1978xw} and further improved in \cite{Denner:qq} unfortunately cannot be applied. Instead, one repeatedly uses the propagator identity \eqref{eqS:9} to solve the integral \eqref{eqS:10}. Since the parameter $\alpha$ in the propagator identity \eqref{eqS:9} has to be real, the calculation differs depending on the values of the external momenta $p_1$ and $p_2$. 

First we take up the case, when one of the following inequalities is true $p_{1,2}^2\geqslant (m_1+m_{2,3})^2$ or $p_{1,2}^2\leqslant (m_1-m_{2,3})^2$. If necessary, we renumber the momenta and reshuffle the propagators in \eqref{eqS:10} in such a way that either $p_1^2\geqslant (m_1+m_{2})^2$ or $p_{1}^2\leqslant (m_1-m_{2})^2$ in order to simplify the discussion. Then we use the propagator identity \eqref{eqS:9} on the second and the third propagators of \eqref{eqS:10} 
\begin{equation}
\begin{split}
\frac{1}{(q^2-m_1^2+i \delta)((q+p_1)^2-m_2^2+i \delta)}= &\frac{1-\alpha}{[(q+p_1)^2-m_2^2+i \delta][(q+l)^2-M^2+i \delta]}\\
&+\frac{\alpha}{[q^2-m_1^2+i \delta][(q+l)^2-M^2+i \delta]},
\end{split}
\end{equation}
where $\alpha$ is an arbitrary parameter and
\begin{align}
l&=\alpha p_1,\\
M^2&=(1-\alpha)m_1^2+\alpha m_2^2-\alpha(1-\alpha)p_1^2.
\end{align}
We choose $\alpha$ such that $M^2=0$. This is satisfied by real $\alpha$ if either $p_1^2\geqslant (m_1+m_{2})^2$ or $p_{1}^2\leqslant (m_1-m_{2})^2$ as has been assumed above. The scalar four point function is then
\begin{equation}
\begin{split}
\frac{i}{(2\pi)^{4}} &\int d^{4} q\frac{1-\alpha}{(v\negcdot q -\Delta)[(q+p_1)^2-m_2^2][(q+p_2)^2-m_3^2][(q+l)^2+i\delta]}\;+\\
\frac{i }{(2\pi)^{4}} &\int d^{4} q\frac{\alpha}{(v\negcdot q -\Delta)[q^2-m_1^2][(q+p_2)^2-m_3^2][(q+l)^2+i\delta]}, \label{eqS:5}
\end{split}
\end{equation}
with $\alpha$ being the solution of
\begin{equation}
p_1^2 \alpha^2 + (m_2^2-m_1^2-p_1^2)\alpha +m_1^2=0. \label{eqS:32}
\end{equation}
To calculate the two integrals in \eqref{eqS:5} it suffices to consider 
\begin{equation}
\begin{split}
-\frac{1}{16\pi^2}\tilde{D}_0&(v,k_1,k_2,k_3,\Delta,M_1,M_2)=\\
&\frac{i}{(2\pi)^{4}}\int d^{4}q\frac{1}{(v\negcdot q-\Delta)[(q+k_1)^2-M_1^2][(q+k_2)^2-M_2^2][(q+k_3)^2+i\delta]},\label{eqS:6}
\end{split}
\end{equation}
where the $i\delta$ prescription has not been written out explicitly in the first three propagators.

Using the Feynman parameterization \eqref{eqS:24} and integrating over $q$ we arrive at
\begin{equation}
\begin{split}
\tilde{D}_0= \int_0^\infty \negtwo 2d\lambda \int_0^1 \negtwo dx \int_0^x\negtwo dy [&p_{23}x^2+p_{12}y^2+(p_{13}-p_{23}-p_{12})xy+\lambda^2+(P_2-P_3)x\lambda+P_3\lambda \\
&+(P_1-P_2)y\lambda +(-p_{23}+M_2^2)x+(p_{23}-p_{13}+M_1^2-M_2^2)y- i \delta]^{-2},
\end{split}\label{eqS:52}
\end{equation}
with
\begin{equation}
\begin{split}
p_{ij}&=(k_i-k_j)^2,\\
P_i&=2(v\negcdot k_i+\Delta).\\
\end{split}\label{eqS:44}
\end{equation}
To simplify the integration we introduce new variables $y=x y'$ and $\lambda=x\lambda'$. The integration limits are then $\int_0^1 dx \int_0^x dy \int_0^\infty 2 d\lambda \to \int_0^1 dx\int_0^1 x dy' \int_0^\infty x 2 d\lambda'$. Since $x$ is positive and $\delta$ an infinitesimal parameter of which only the sign matters, the extra factor of $x^2$ in the numerator can be canceled against the similar factor in the denominator. After the cancellation the denominator is linear in $x$. The integration over $x$ is now trivial and yields
\begin{equation}
\begin{split}
\tilde{D}_0=\int_0^\infty \negtwo 2 d\lambda \int_0^1 dy &[(p_{23}-p_{13}+M_1^2-M_2^2)y +P_3 \lambda -p_{23}+M_2^2-i\delta]^{-1} \times\\
& [p_{12}y^2+\lambda^2+(P_1-P_2) y\lambda +P_2\lambda +(-p_{12}+M_1^2-M_2^2)y+M_2^2-i\delta]^{-1}.\label{eqS:53}
\end{split}
\end{equation}
To cancel the $y^2$ term in the integral above a new variable $\lambda=\lambda'+\beta y$ is introduced, with $\beta$ chosen to solve
\begin{equation}
\beta^2+(P_1-P_2)\beta+ p_{12}=0. \label{eqS:45}
\end{equation}
The solutions are real, if $(k_2-k_1)^\mu$ is real (cf. Eq.~\eqref{eqS:44}). Since both $k_1$ and $k_2$ are taken to be real four-vectors, this is always the case. We then get
\begin{equation}
\tilde{D}_0=\big(\int_0^\infty \negtwo 2 d\lambda \int_0^1 \negtwo dy+\int_{-\beta}^0 \negtwo 2 d\lambda \int_{-\lambda/\beta}^1 \negtwo \negtwo dy\;\big) \frac{1}{[a_1 y+b_1 \lambda +c_1] [a_2y +b_2\lambda +c_2+d_2y\lambda +\lambda^2]}, \label{eqS:7}
\end{equation}
with
\begin{equation}
\begin{matrix}
&
\begin{aligned}
a_1&=\beta P_3+p_{23}-p_{13}+M_1^2-M_2^2,\\
b_1&=P_3,\\
c_1&=M_2^2-p_{23}-i\delta,\\
{}&{}
\end{aligned}
&\begin{aligned}
 a_2&=\beta P_2+M_1^2-M_2^2-p_{12},\\
b_2&=P_2,\\
c_2&=M_2^2-i\delta,\\
d_2&=2\beta+P_1-P_2.
\end{aligned}
\end{matrix}\label{eqS:46}
\end{equation}
After $y$ integration we arrive at
\begin{equation}
\begin{split}
\tilde{D}_0=& \frac{1}{(\lambda_1-\lambda_2)}\frac{1}{(b_1d_2-a_1)}\Bigg[\\
&\int_0^\infty \negtwo\frac{2 d \lambda}{\lambda-\lambda_1} \bigg\{ \ln\left(\frac{b_1\lambda+c_1}{b_1\lambda+a_1+c_1}\right)-\ln\left(\frac{\lambda^2+b_2\lambda+c_2}{\lambda^2+(b_2+d_2)\lambda+a_2+c_2}\right)\bigg\}\\
-&\int_0^\infty \negtwo\frac{2 d \lambda}{\lambda-\lambda_2} \bigg\{ \ln\left(\frac{b_1\lambda+c_1}{b_1\lambda+a_1+c_1}\right)-\ln\left(\frac{\lambda^2+b_2\lambda+c_2}{\lambda^2+(b_2+d_2)\lambda+a_2+c_2}\right)\bigg\}\\
-&\int_0^1\negtwo\frac{2 d\lambda}{\lambda+\lambda_1/\beta} \bigg\{ \ln\left(\frac{(a_1-\beta b_1)\lambda+c_1}{-\beta b_1\lambda+a_1+c_1}\right) -\ln\left(\frac{\beta(\beta -d_2)\lambda^2+(a_2-b_2\beta)\lambda+c_2}{\beta^2\lambda^2-\beta (b_2+d_2)\lambda+a_2+c_2}\right)\bigg\}\\
+&\int_0^1 \negtwo\frac{2 d\lambda}{\lambda+\lambda_2/\beta} \bigg\{ \ln\left(\frac{(a_1-\beta b_1)\lambda+c_1}{-\beta b_1\lambda+a_1+c_1}\right) -\ln\left(\frac{\beta(\beta -d_2)\lambda^2+(a_2-b_2\beta)\lambda+c_2}{\beta^2\lambda^2-\beta (b_2+d_2)\lambda+a_2+c_2}\right)\bigg\}\Bigg], \label{eqS:8}
\end{split}
\end{equation}
with $\lambda_{1,2}$ the solutions of 
\begin{equation}
(b_1d_2-a_1)\lambda^2+(a_2b_1-a_1b_2+c_1d_2)\lambda+a_2c_1-a_1c_2=(b_1d_2-a_1)(\lambda-\lambda_1)(\lambda-\lambda_2).\label{lam12}
\end{equation}
Note that the integrands above have vanishing residues, i.e., arguments of the two logarithms in the integrands are the same for $\lambda=\lambda_{1,2}$. Note as well, that the infinitesimal imaginary parts of $c_1$ and $c_2$, the $-i \delta$ in \eqref{eqS:46}, have to be equal. They originate from the same infinitesimal parameter in \eqref{eqS:52} that after the integration over $x$ appears twice in \eqref{eqS:53}. The size of the infinitesimal parts of $\lambda_{1,2}$ compared to $-i\delta$ are thus unambiguously defined and have to be kept track of until the end of the calculation.

 The integrals \eqref{eqS:8} can be expressed in terms of the dilogarithms. This has been done in section \ref{app:BScal}. The solution for the first two integrals can be found in \eqref{eqS:27}, with the definitions in \eqref{eqS:36}, \eqref{eqS:48}, while the solution for the last two integrals can be found in \eqref{eqS:31}, \eqref{eqS:26}. Together with Eqs.~\eqref{eqS:5}, \eqref{eqS:6} and the cascade of abbreviations \eqref{eqS:46}, \eqref{eqS:45}, \eqref{eqS:44} this gives the complete solution of the four point function with at least one external momentum $p_1$, $p_2$ satisfying $p_{1,2}^2\geqslant (m_1+m_{2,3})^2$ or $p_{1,2}^2\leqslant (m_1-m_{2,3})^2$. Collecting the terms and rearranging the last two propagators in \eqref{eqS:10} if necessary, the four point function for $p_{1}^2\geqslant (m_1+m_{2})^2$ or $p_{1}^2\leqslant (m_1-m_{2})^2$ finally reads \index{scalar function!four-point}
\begin{equation}
\begin{split}
\bar{D}_0(v,p_1,p_2,\Delta,m_1,m_2,m_3)=&(1-\alpha) \tilde{D}_0(v,p_1,p_2,l,\Delta,m_2,m_3)\\
&+\alpha \;\tilde{D}_0(v,0,p_2,l,\Delta,m_1,m_3),\label{eqS:Tildefun}
\end{split}
\end{equation}
with $l=\alpha p_1$ and $\alpha$ the solution of $p_1^2\alpha^2+(m_2^2-m_1^2-p_1^2)\alpha +m_1^2=0$, while
\begin{equation}
\begin{split}
\tilde{D}_0&(v, k_1,k_2,k_3,\Delta,M_1,M_2)=\frac{1}{(\lambda_1-\lambda_2)}\frac{2}{(b_1d_2-a_1)}\Big[\\
 &I_2\left(-c_1/b_1,-(a_1+c_1)/b_1,b_2,c_2,b_2+d_2,a_2+c_2,\lambda_1\right)\\
-&I_2(-c_1/b_1,-(a_1+c_1)/b_1,b_2,c_2,b_2+d_2,a_2+c_2,\lambda_2)\\
-&I_1(\beta^2-\beta d_2,\beta^2,a_2-\beta b_2,-\beta(b_2+d_2),a_1-\beta b_1,-\beta b_1,c_2,a_2+c_2,c_1,a_1+c_1,-\lambda_1/\beta)\\
+&I_1(\beta^2-\beta d_2,\beta^2,a_2-\beta b_2,-\beta(b_2+d_2),a_1-\beta b_1,-\beta b_1,c_2,a_2+c_2,c_1,a_1+c_1,-\lambda_2/\beta)\Big], \label{eqS:Ifun}
\end{split}
\end{equation}
with $a_1 \dots d_2$ defined in \eqref{eqS:46}, $\beta$ defined in \eqref{eqS:45}, $p_{ij}$, $P_i$ defined in \eqref{eqS:44} and $\lambda_{1,2}$ solutions of \eqref{lam12}. Note that parameters $\alpha$ and $\beta$ are solutions of quadratic equations \eqref{eqS:32} and \eqref{eqS:45} that in general have two real solutions each. The value of the four point function $\bar{D}_0(v, k_1,k_2,k_3,\Delta,M_1,M_2)$ does not depend on which of the solutions are chosen in the evaluation of \eqref{eqS:Tildefun}, \eqref{eqS:Ifun}. This fact can be used as a useful test in the numerical implementation of the expressions given above. 

Now we take up the special case of $p_1$ on the light-cone, i.e., $p_1^2=0$. From \eqref{eqS:32} it follows that $\alpha=m_1^2/(m_1^2-m_2^2)$. If $m_1^2\ne m_2^2$ then $\alpha$ is finite and the calculation proceeds as before, \eqref{eqS:10}-\eqref{eqS:8}. For equal masses $m_1$ and $m_2$, we evaluate the integrals by first taking $m_1^2\ne m_2^2$ and then performing the limit $m_1^2\to m_2^2$ and thus $|\alpha|\to \infty$. The external momenta in the last propagators of \eqref{eqS:5} are both equal to $l=\alpha p_1$. Thus the last external momentum in $\tilde{D}_0$ of \eqref{eqS:6} is going to be $k_3=l=\alpha p_1$ for both of the integrals in \eqref{eqS:5}. In the limit $\alpha\to \infty$ the following leading order values \eqref{eqS:46} are obtained: 
$a_1\to 2 \alpha p_1\negthincdot(\beta v- p_2)$,
$b_1\to 2 \alpha v\negthincdot p_1$,
$c_1\to 2 \alpha p_1\negthincdot p_2$
, where we have used also the fact that $p_1^2=0$. The other coefficients $a_2$, $b_2$, $c_2$, $d_2$ do not depend on $\alpha$. The first term in the denominator of the integrand in \eqref{eqS:7} is then proportional to $\alpha$, while the second term in the denominator does not depend on $\alpha$ at all. The $\alpha $ in the denominator cancels against the $\alpha$ in the numerator of \eqref{eqS:5}. For the case of equal masses $m_1^2=m_2^2$ and $p_1^2=0$, the solution is then the same as for the case $m_1^2 \ne m_2^2$ except that (i) one has to replace $(1-\alpha)$ and $\alpha$ in \eqref{eqS:5} with $-1$ and $1$ respectively, and that (ii) $a_1$, $b_1$, $c_1$ in \eqref{eqS:46}, \eqref{eqS:8} have to be replaced by their limiting values (divided by $\alpha$)
\begin{equation}
\begin{split}
a_1&\to a_1^l= 2 p_1\negthincdot (\beta v-p_2),\\
b_1&\to b_1^l= 2 v\negthincdot p_1,\\
c_1&\to c_1^l=2 p_1\negthincdot p_2- i \delta, \label{eqS:34}
\end{split}
\end{equation}
where $\beta$ is the solution to \eqref{eqS:45} (with $P_i$ defined in \eqref{eqS:44}), and is different for the two integrals in \eqref{eqS:5}. In the limiting value of $c_1$ coefficient given in \eqref{eqS:34} an additional $- i\delta$ prescription has been added. As will be shown in the next paragraph, this does not have any effect on the value of the four-point function. It does make possible, however, to express the integrals in \eqref{eqS:8} in terms of the functions $I_1$ and $I_2$ as in \eqref{eqS:Ifun}.

It is easy to see, that the limiting procedure as explained above does lead to an unambiguous result. One might in principle worry that limits $m_1^2\to m_2^2$ taken from above and below, corresponding to the limits $\alpha\to \infty$ and $\alpha\to -\infty$ respectively, would lead to different results. The question is most conveniently settled if the $\tilde{D}_0$ functions in Eq.~\eqref{eqS:Tildefun} are replaced by the expressions given in Eq.~\eqref{eqS:53}. Once the limit $m_1\to m_2$ is taken, the first factors of the integrands have the same limiting value. For $\alpha$ large, thus the leading term is
\begin{equation}
\begin{split}
\bar{D}_0\to -\alpha \int_0^\infty \negtwo 2 d\lambda \int_0^1 dy &[-2 \alpha p_1\negcdot p_2 y +2 \alpha v\negcdot p_1\lambda +2 \alpha p_1\negcdot p_2 - i\delta]^{-1} \times\\
& [(p_1-p_2)^2y^2+\lambda^2+2 v\cdot (p_1-p_2) y\lambda +\\
& \qquad +P_2\lambda +(-(p_1-p_{2})^2+m_2^2-m_3^2)y+m_3^2-i\delta]^{-1}\\
+\alpha \int_0^\infty \negtwo 2 d\lambda \int_0^1 dy &[-2 \alpha p_1\negcdot p_2 y +2 \alpha v\negcdot p_1\lambda +2 \alpha p_1\negcdot p_2 - i\delta]^{-1} \times\\
& [p_2^2y^2+\lambda^2-2 v\cdot p_2 y\lambda +\\
& \qquad +P_2\lambda +(-p_{2}^2+m_1^2-m_3^2)y+m_3^2-i\delta]^{-1}, \label{limit}
\end{split}
\end{equation}
with $P_2=2 (v\negthincdot p_2+\Delta)$. After collecting the two integrands in \eqref{limit} the first factor in the integrands cancels and one finds 
\begin{equation}
\begin{split}
\bar{D}_0\to - \int_0^\infty \negtwo 2 d\lambda \int_0^1 y dy &[(p_1-p_2)^2y^2+\lambda^2+2 v\cdot (p_1-p_2) y\lambda +\\
& \qquad +P_2\lambda +(-(p_1-p_{2})^2+m_2^2-m_3^2)y+m_3^2-i\delta]^{-1}\times \\
& [p_2^2y^2+\lambda^2-2 v\cdot p_2 y\lambda +\\
& \qquad +P_2\lambda +(-p_{2}^2+m_1^2-m_3^2)y+m_3^2-i\delta]^{-1},
\end{split}
\end{equation}
This result exhibits clearly the fact that (i) the limit $m_1^2\to m_2^2$ is independent of whether it is taken from above or below and (ii) the limit is independent of the size (or even the sign) of the infinitesimal parameter in the first terms of the integrands in \eqref{limit}.

When the momenta $p_1$, $p_2$ satisfy $(m_1-m_{2,3})^2<p_{1,2}^2< (m_{1}+ m_{2,3})^2$, rendering a complex $\alpha$, the procedure outlined above in \eqref{eqS:10}-\eqref{eqS:Tildefun} cannot be applied directly. Starting from \eqref{eqS:10}, we then use the propagator identity \eqref{eqS:9} on the last two propagators in \eqref{eqS:10}, where we set $\alpha$ such that $l^2=(p_1+\alpha(p_2-p_1))^2=0$. This has a real solution for $\alpha$ since $p_1$ and $p_2$ are timelike as has been assumed at the beginning of this paragraph. Changing the notation slightly we then have for the scalar four point function (omitting the $- i\delta$ prescription in the notation)
\begin{equation}
\begin{split}
\frac{i}{(2\pi)^{4}} &\int d^{4} q\frac{\alpha'}{(v\negcdot q -\Delta)[q^2-m_1^2][(q+p_1)^2-m_2^2][(q+l)^2-M^2]}\; +\\
\frac{i }{(2\pi)^{4}} &\int d^{4} q\frac{1-\alpha'}{(v\negcdot q -\Delta)[q^2-m_1^2][(q+p_2)^2-m_3^2][(q+l)^2-M^2]}, \label{eqS:11}
\end{split}
\end{equation}
with $\alpha' $ the (real) solution of 
\begin{equation}
(p_1+\alpha'(p_2-p_1))^2=0, \label{eqS:quadalphaPr}
\end{equation}
 and 
\begin{subequations}
\begin{align}
l&=p_1+\alpha'(p_2-p_1),\\
M^2&=(1-\alpha')m_2^2+\alpha' m_3^2-\alpha'(1-\alpha') (p_2-p_1)^2.
\end{align}
\end{subequations}
The integrals in \eqref{eqS:11} can now be solved using the procedure outlined above \eqref{eqS:10}-\eqref{eqS:34}, once we permute the last two propagators with $l$ taking the role of $p_1$ in \eqref{eqS:10}. Note also, that $\alpha'$ solves quadratic equation \eqref{eqS:quadalphaPr} that in general has two solutions. The final result for the four point function $\bar{D}_0$ does not depend on which of the two solutions is taken in \eqref{eqS:11}. This fact can be exploited in the numerical implementation as a useful check.

	The four point function has already been calculated before for a special case of $m_1=m_2=m_3$, $p_1^2=p_2^2=0$ and $p_1^\mu-p_2^\mu=Mv^\mu$ (see Eq. (A11) of \cite{Fajfer:2001ad}). It has been checked numerically that the two solutions, the one given here and the solution of \cite{Fajfer:2001ad}, agree for this special case. A number of other numerical tests have been performed. The direct numerical integration of \eqref{eqS:52} and the evaluation of analytical result given above have been found to agree numerically. It has been also checked that the results do not depend on which of the two solutions for $\alpha$, $\beta$ or $\alpha'$ is taken. The solution for the four-point function calculated above has also been checked numerically to have the branch cuts as required by analyticity \index{analyticity} and \index{unitarity} unitarity.

The five-point as well as the higher-point functions can be expressed in terms of the scalar functions given above using the standard procedure \cite{'tHooft:1978xw,vanNeerven:1983vr}. Consider for instance the case of five-point scalar function. This is a function of four vectors, $v$ and $p_1,p_2,p_3$. The five-point function is first multiplied by $v_\mu \epsilon_{\alpha\beta\gamma\delta}$ and then antisymmetrized in all five indices. The resulting tensor is zero, because there is no antisymmetrical tensor with five indices in four dimensions. Then the tensor is multiplied first with $p_{1\alpha}p_{2\beta}p_{3\gamma}q_\delta$ and finally with $v^\mu \epsilon_{\alpha'\beta'\gamma'\delta'}p_1^{\alpha'}p_2^{\beta'}p_3^{\gamma'}q^{\delta'}$. Using the decomposition of the product of two Levi-Civita tensors in terms of the Kronecker delta functions and expressing the scalar products $q \negthincdot p_i$ in terms of the propagators $((q+p_i)^2- m_{i+1}^2)$ and $(q^2-m_1^2)$, the five-point function can be expressed in terms of the four point functions. The tensor functions can also be expressed in terms of the scalar functions using the algebraic reduction \cite{Passarino:1978jh}.

In the numerical implementation of the expressions as given here, further care has to be taken regarding the numerical instabilities. Such numerical instabilities can for instance arise, if one of the solutions of the quadratic equation is much smaller than its coefficients. There is also a possibility of a cancellation between the dilogarithmic functions, when the values of the dilogarithms separately are much larger then their sum. These difficulties can be dealt with along the lines of Ref.~\cite{vanOldenborgh:1989wn}.
\index{integration of scalar functions|)}

\section{Reduction to dilogarithms}\index{calculation! scalar functions|)}
\label{app:BScal} \index{dilogarithm!useful relations|(}
In this section we will express the integrals appearing in \eqref{eqS:21}, \eqref{eqS:8} in terms of the dilogarithms.
First we review the derivations given in \cite{'tHooft:1978xw}. Consider
\begin{equation}
\begin{split}
R(\lambda_1,\lambda_0)& =\int_0^1 d\lambda \frac{1}{\lambda -\lambda_0}[\ln(\lambda-\lambda_1)-\ln(\lambda_0-\lambda_1)]\\
&=\int_{-\lambda_1}^{1-\lambda_1} d\lambda \frac{1}{\lambda -\lambda_0 +\lambda_1}[\ln\lambda-\ln(\lambda_0-\lambda_1)], \label{eqS:18}
\end{split}
\end{equation}
where $\lambda_{0,1}$ may be complex. The residue of the pole of the integrand is zero. The cut of the logarithm is along the negative real axis, so for $\lambda_1$ not real, the cut is outside the triangle $0$, $-\lambda_1$, $1-\lambda_1$. The integration path can thus be deformed to (for $\lambda_1$ real, this statement is trivial)
\begin{equation*}
\int_{-\lambda_1}^{1-\lambda_1} d\lambda =\int_0^{1-\lambda_1}d\lambda-\int_0^{-\lambda_1} d\lambda.
\end{equation*}
Making the substitutions $\lambda=(1-\lambda_1)\lambda'$ and $\lambda=\lambda_1\lambda'$ we obtain
\begin{equation}
\begin{split}
R(\lambda_1, \lambda_0)&=\int_0^1 d\lambda\Big[\frac{d}{d\lambda}\ln\Big(1+\lambda \frac{1-\lambda_1}{\lambda_1-\lambda_0}\Big)\Big][\ln \lambda(1-\lambda_1)-\ln(\lambda_0-\lambda_1)]\\
&-\int_0^1 d\lambda\Big[\frac{d}{d\lambda}\ln\Big(1-\lambda \frac{\lambda_1}{\lambda_1-\lambda_0}\Big)\Big][\ln(-\lambda\lambda_1)-\ln(\lambda_0-\lambda_1)]. \label{eqS:35}
\end{split}
\end{equation}
Since $\lambda$ is positive real, none of the arguments of the logarithms crosses the negative real axis. After integration per parts 
\begin{equation}
\begin{split}
R(\lambda_1,\lambda_0)=\Li\left(\frac{\lambda_1-1}{\lambda_1-\lambda_0}\right) +&\ln\left(\frac{1-\lambda_0}{\lambda_1-\lambda_0}\right)\left[ \ln(1-\lambda_1)-\ln(\lambda_0-\lambda_1)\right]\\
-\Li \left(\frac{\lambda_1}{\lambda_1-\lambda_0}\right)-&\ln\left(\frac{-\lambda_0}{\lambda_1-\lambda_0}\right)\left[ \ln(-\lambda_1)-\ln(\lambda_0-\lambda_1)\right].
\end{split}
\end{equation}
This can be further simplified using \eqref{eqS:16} 
\begin{equation}
\begin{split}
R(\lambda_1,\lambda_0)=&\Li\Big(\frac{\lambda_0}{\lambda_0-\lambda_1}\Big) +\left[\eta\Big(-\lambda_1, \frac{1}{\lambda_0-\lambda_1}\Big)+ 2 \pi i \Re^{(-)}(\lambda_0-\lambda_1)\right] \ln\frac{\lambda_0}{\lambda_0-\lambda_1}\\
-&\Li\Big(\frac{\lambda_0-1}{\lambda_0-\lambda_1}\Big)-\left[\eta\Big(1-\lambda_1, \frac{1}{\lambda_0-\lambda_1}\Big)+2\pi i\Re^{(-)}(\lambda_0-\lambda_1)\right] \ln\frac{\lambda_0-1}{\lambda_0-\lambda_1}, \label{eqS:51}
\end{split}
\end{equation}
with $\eta$ defined in \eqref{eqS:50} and $\Re^{(-)}(x)$ defined in \eqref{eqS:47}. Note that this result differs slightly from the one in \cite{'tHooft:1978xw} as it is defined also for the arguments lying on the negative real axis. The extension to negative real arguments was not necessary in \cite{'tHooft:1978xw} as then the $\lambda_0$ was always real. This is not the case in the calculation of the four point function with one heavy quark propagator, as the $\lambda_1$ and $\lambda_2$ in \eqref{eqS:8} can have nonzero imaginary parts. The momenta and the masses in the calculation can then be chosen such, that one of the arguments appearing in \eqref{eqS:51} can lie on the negative real axis.

Next we turn to the integral 
\begin{equation}
S_3(a,b,c,\lambda_0)=\int_0^1 d\lambda \frac{1}{\lambda-\lambda_0}[\ln(a \lambda^2+b\lambda+c)-\ln(a \lambda_0^2+b\lambda_0+c)], \label{eqS:28}
\end{equation}
with $a$ real, while $b$, $c$, $\lambda_0$ may be complex but such, that the imaginary part of the argument of the logarithm does not change sign for $x\in [0,1]$ (also $\Im (c) \ne 0$). 

Let $\epsilon$ and $\delta$ be infinitesimal quantities that have the {\it opposite} sign from the imaginary part of first and second argument of the logarithm respectively. That is, the signs of the arguments are as given by $-i\epsilon$ and $-i \delta$. Then
\begin{equation}
\begin{split}
S_3=&\int_0^1 d\lambda \frac{1}{\lambda-\lambda_0}[\ln(\lambda-\lambda_1)(\lambda-\lambda_2)-\ln(\lambda_0-\lambda_1)(\lambda_0-\lambda_2)]\\
&-\eta\Big(a-i\epsilon,\frac{1}{a-i\delta}\Big)\ln\Big(\frac{\lambda_0-1}{\lambda_0}\Big),\label{eqS:S3}
\end{split}
\end{equation}
with $\lambda_{1,2}$ the solutions of $a\lambda^2+b\lambda+c=a(\lambda-\lambda_1)(\lambda-\lambda_2)$. Next we split up the logarithms, use the fact that the imaginary part of $(\lambda-\lambda_1)(\lambda-\lambda_2)$ has the same sign as the imaginary part of $c/a$ and use the definitions of $R(\lambda_1,\lambda_0)$ \eqref{eqS:18} to get
\begin{equation}
\begin{split}
S_3(a,b,c,\lambda_0)=&R(\lambda_1,\lambda_0)+R(\lambda_2,\lambda_0)\\
&+\Big[\eta(-\lambda_1,-\lambda_2)-\eta(\lambda_0-\lambda_1,\lambda_0-\lambda_2)-\eta\Big(a-i \epsilon, \frac{1}{a-i\delta}\Big)\Big]\ln\frac{\lambda_0-1}{\lambda_0}, \label{eqS:29}
\end{split}
\end{equation}
with $\epsilon$ and $\delta$ defined before Eq.~\eqref{eqS:S3}.

For future reference we also define 
\begin{equation}
S_2(b,c,\lambda_0)=\int_0^1 d\lambda \frac{1}{\lambda-\lambda_0} [\ln(b \lambda+c)-\ln(b \lambda_0+c)],
\end{equation}
with $b$ real and $c$, $\lambda_0$ possibly complex ($\Im (c) \ne 0$). Defining as above infinitesimal parameters $\epsilon'$ and $\delta'$ to have signs {\it opposite} to the imaginary parts of the first and the second argument of the logarithms respectively, we obtain
\begin{equation}
S_2(b,c,\lambda_0)=R(-c/b, \lambda_0)-\eta\Big(b-i\epsilon',\frac{1}{b-i\delta'}\Big)\ln\frac{\lambda_0-1}{\lambda_0}.
\end{equation}

Next we turn to the integrals appearing in the calculations of the three-point and four-point functions with one heavy quark propagator. Consider first
\begin{equation}
\begin{split}
I_1(a_1,a_2,b_1,b_2,b_3,b_4,&c_1,c_2,c_3,c_4,\lambda_0)=\\
&\int_0^1d\lambda \frac{1}{\lambda-\lambda_0} \Big[\ln\frac{b_3 \lambda+c_3}{b_4\lambda+c_4}-\ln\frac{a_1 \lambda^2+b_1\lambda+c_1}{a_2\lambda^2+b_2\lambda +c_2}\Big], \label{eqS:31}
\end{split}
\end{equation}
with $a_{1,2}$ and $b_{1,\cdots,4}$ real, while $\lambda_0$, $c_{1\cdots 4}$ may be complex but such that $\Im (c_1) \Im (c_2)>0$ and $\Im (c_3)\Im (c_4)>0$. Also, the coefficients are such, that for $\lambda=\lambda_0$ the two logarithms are equal, so that the residue of the integrand is equal to zero. Such an integral appears in the calculation of the four-point scalar function \eqref{eqS:8}. 
To reduce the integral $I_1$ to the integrals $S_2$, $S_3$ we add and subtract the values of the logarithms at the pole. Since the numerators and the denominators of the logarithms in \eqref{eqS:31} have imaginary parts of the same sign, we can split the logarithms. Additional $\eta$ terms appear, however, when we split the logarithms with $\lambda$ set to $\lambda_0$. As the result we get
\begin{equation}
\begin{split}
I_1&=S_2(b_3,c_3,\lambda_0)-S_2(b_4,c_4,\lambda_0)-S_3(a_1,b_1,c_1,\lambda_0)+S_3(a_2,b_2,c_2,\lambda_0)\\
&+\Big[\eta\Big(a_1\lambda_0^2+b_1\lambda_0+c_1,\frac{1}{a_2\lambda_0^2+b_2\lambda_0+c_2}\Big)+2 \pi i \Re^{(-)}(a_2 \lambda_0^2+b_2 \lambda_0+c_2)\\
&\qquad -\eta\Big(b_3\lambda_0+c_3,\frac{1}{b_4\lambda_0+c_4}\Big) -2 \pi i \Re^{(-)}(b_4\lambda_0+c_4)\Big]\ln\frac{\lambda_0-1}{\lambda_0}. \label{eqS:26}
\end{split}
\end{equation}
with $\eta$ defined in \eqref{eqS:50} and $\Re^{(-)}$ in \eqref{eqS:47}.

Next consider the integral 
\begin{equation}
\begin{split}
I_2(a_1,a_2,g_1,f_1,g_2,&f_2,\lambda_0)=\\
&\int_0^\infty \negtwo d\lambda\frac{1}{\lambda-\lambda_0} \left\{ \ln\frac{\lambda-a_1}{\lambda-a_2}-\ln\frac{\lambda^2+g_1\lambda+f_1}{\lambda^2+g_2\lambda+f_2}\right\}, \label{eqS:36}
\end{split}
\end{equation}
with $g_{1,2}$ real, while $\lambda_0$, $a_{1,2}$, $f_{1,2}$ may be complex with the restriction $\Im (a_1) \Im (a_2)>0$, $\Im (f_1)\Im (f_2)>0$. Then the logarithms can be split without introducing $\eta$ terms, independent of the value of $\lambda$ as long as this is real. Also the arguments of the logarithms in \eqref{eqS:36} are taken to be the same for $\lambda=\lambda_0$, so that the residue of the integrand is zero. Such integrals appear in the calculation of the three-point function \eqref{eqS:21} and in the calculation of the four-point function \eqref{eqS:8}. We rewrite the integral \eqref{eqS:36} as
\begin{equation}
I_2=\int_0^\infty \negtwo d\lambda\frac{1}{\lambda-\lambda_0} \Big\{ \ln\frac{\lambda-a_1}{\lambda-a_2}-\ln\frac{(\lambda-b_1)(\lambda-b_2)}{(\lambda-c_1)(\lambda-c_2)}\Big\}, \label{eqS:30}
\end{equation}
with
\begin{equation}
\begin{split}
\lambda^2+g_1\lambda+f_1&=(\lambda-b_1)(\lambda-b_2),\\
\lambda^2+g_2\lambda+f_2&=(\lambda-c_1)(\lambda-c_2), \label{eqS:48}
\end{split}
\end{equation}
where $\Im (b_1)\Im (b_2)<0$, $\Im (c_1)\Im (c_2)<0$, $\Im(b_1b_2)\Im(c_1c_2)>0$ as can be seen from the constraints on $g_{1,2}$, $f_{1,2}$. Then the logarithms can be split up in the sum of the logarithms with arguments linear in $\lambda$. 

To the integral \eqref{eqS:30} we add logarithms with $\lambda$ set to $\lambda_0$ and then split the logarithms
\begin{equation}
\begin{split}
0&=\ln\frac{\lambda_0-a_1}{\lambda_0-a_2}-\ln\frac{(\lambda_0-b_1)(\lambda_0-b_2)}{(\lambda_0-c_1)(\lambda_0-c_2)}\\
&=\sum_i \rho(\varkappa_i) \ln(\lambda_0-\varkappa_i) -\eta', \label{eqS:39}
\end{split}
\end{equation}
where $\varkappa_i$ are the coefficients $a_{1,2}$, $b_{1,2}$, $c_{1,2}$ with $\rho(\varkappa_i)=1$ for $a_1,c_{1,2}$ and $\rho(\varkappa_i)=-1$ for $a_2,b_{1,2}$. There is also a sum of $\eta$ terms that we do not write out explicitly, but just denote by $\eta'$, as it will be reabsorbed in the final result. Note also, that in the case of $\lambda_0 -\varkappa_i$ real and negative the logarithm is calculated using the prescription $\lambda_0-\varkappa_i\to \lambda_0-\varkappa_i + i \delta$, with $\delta$ a positive infinitesimal parameter (see also \eqref{eqS:55}-\eqref{eqS:54}). The integral is then
\begin{equation}
I_2=\sum_i \rho(\varkappa_i) \int_0^\infty \negtwo d\lambda \frac{1}{\lambda-\lambda_0}[\ln(\lambda-\varkappa_i)-\ln(\lambda_0-\varkappa_i)] + \eta' \int_0^\infty \frac{d\lambda}{\lambda-\lambda_0}. \label{eqS:20}
\end{equation}
The separate integrals are divergent so they have to be regulated. We use the cutoff $M$ that is sent to infinity at the end of the calculation. Note also, that there is no problem with the pole in the last term even if $\lambda_0$ is real, as then $\eta'$ is zero. 

The regulated integrals are then
\begin{equation}
\int_0^M d \lambda \frac{1}{\lambda-\lambda_0} [\ln(\lambda-\varkappa_i)-\ln(\lambda_0-\varkappa_i)].
\end{equation}
Let us from here on first assume, that $\lambda_0-\varkappa_i$ is not negative real. Changing the variable $\lambda=M\lambda'$ and using the calculation of $R(\lambda_1,\lambda_0)$ \eqref{eqS:18}, \eqref{eqS:51} we get
\begin{equation}
\begin{split}
\int_0^1 &d\lambda \frac{1}{\lambda-\frac{\lambda_0}{M}} \left[\ln\left(\lambda-\frac{\varkappa_i}{M}\right)-\ln\left(\frac{\lambda_0}{M}-\frac{\varkappa_i}{M}\right)\right]= \\
&\Li\frac{\lambda_0}{\lambda_0-\varkappa_i}-\Li \frac{\lambda_0-M}{\lambda_0-\varkappa_i}+\eta\Big(-\varkappa_i, \frac{1}{\lambda_0-\varkappa_i}\Big) \ln \frac{\lambda_0}{\lambda_0-\varkappa_i} -\eta\Big(1-\frac{\varkappa_i}{M}, \frac{M}{\lambda_0-\varkappa_i}\Big) \ln\frac{\lambda_0-M}{\lambda_0-\varkappa_i}.
\end{split}
\end{equation}
For $M$ big enough the last term is zero. The $M$ dependent dilogarithm can be transformed using relation \eqref{eqS:19}
\begin{equation}
\Li\frac{-M}{\lambda_0-\varkappa_i}=-\Li\frac{\lambda_0-\varkappa_i}{-M} -\frac{1}{6}\pi^2 -\frac{1}{2} \ln^2\Big(\frac{M}{\lambda_0-\varkappa_i}\Big).
\end{equation}
The argument of the dilogarithm on the right-hand side goes toward zero as $M\to \infty$, so that in that limit the dilogarithm vanishes. Next we split the logarithm in the last term and write
\begin{equation}
\ln^2\Big(\frac{M}{\lambda_0-\varkappa_i}\Big)=\ln^2 M-2 \ln M\ln (\lambda_0-\varkappa_i)+\ln^2(\lambda_0-\varkappa_i).
\end{equation}
The first term gives zero once summed over in \eqref{eqS:20}, while the second term cancels against the $\eta'$ term in \eqref{eqS:20}. Leaving the case of $\lambda_0-\varkappa_i$ negative real to the reader, the final result is
\begin{equation}
\begin{split}
I_2=\sum_i \rho(\varkappa_i) \bigg\{ &\Li \frac{\lambda_0}{\lambda_0-\varkappa_i} +\left[\eta\Big(-\varkappa_i, \frac{1}{\lambda_0-\varkappa_i}\Big)+ 2 \pi i \Re^{(-)}(\lambda_0-\varkappa_i)\right] \ln\frac{\lambda_0}{\lambda_0-\varkappa_i}\\
+&\frac{1}{2}\ln^2(\lambda_0-\varkappa_i)-\ln(\lambda_0-\varkappa_i)\ln(-\lambda_0)\bigg\},\label{eqS:27}
\end{split}
\end{equation}
with
\begin{equation}
\rho(\varkappa_i)=\left\{
\begin{aligned}
\;&+1; & \varkappa_i=a_1, c_{1,2},\\
\;& -1; & \varkappa_i=a_2,b_{1,2}. \label{eqS:37}
\end{aligned}
\right.
\end{equation}
Note that this solution applies also for the case encountered in the calculation of the three point function \eqref{eqS:21}, when $a_1=a_2=0$. Then the terms containing $a_{1,2}$ cancel each other, so they can be dropped altogether for the case of Eq.~\eqref{eqS:21}.
 
There is one more point worth mentioning regarding the expression \eqref{eqS:27}. One might think that problems could arise for $\lambda_0-\varkappa_i$ negative real or $\lambda_0/(\lambda_0-\varkappa_i)$ real as then one has to deal with the cuts in the logarithm and the dilogarithmic function\footnote{Note that there exists such a combination of parameters $v$, $p_{1,2}$, $\Delta$ and $m_{1,2,3}$ in \eqref{eqS:10} that $\lambda_0-\varkappa_i$ in \eqref{eqS:27} is negative real for some $i$, as can be seen from definition of $a_1,\dots,d_2$ \eqref{eqS:46}, definition of $\lambda_{1,2}$ \eqref{lam12} and the expression for the four-point function \eqref{eqS:Ifun}.}. We use the prescription for the arguments lying exactly on the cuts of the functions as described before Eq.~\eqref{eqS:55} and after Eq.~\eqref{eqS:38}. One could as well use a different prescription, with infinitesimal parameter $\epsilon$ in \eqref{eqS:55}, \eqref{eqS:38} taken to be negative, and with appropriately adjusted $\eta$ and $\Re^{(-)}$ functions. It has been checked numerically, that the result \eqref{eqS:27} does not change, if the alternative prescription is used. Thus the result \eqref{eqS:27} is valid for any complex $\lambda_0$, $\varkappa_i$ independent of the prescription used for the arguments lying on the cut.

For the special case of $\lambda_0$ real the result \eqref{eqS:27} simplifies considerably. The $\eta$ term is then zero. Also the last term in \eqref{eqS:27}, that arises from the $\eta'$ term in \eqref{eqS:20}, then sums up to zero. For $\lambda_0$ real we have
\begin{equation}
I_2=\sum_i \rho(\varkappa_i) \Big[ \Li \frac{\lambda_0}{\lambda_0-\varkappa_i}+ \frac{1}{2}\ln^2(\lambda_0-\varkappa_i)\Big], \label{eqS:41}
\end{equation}
with $\varkappa_i$ and $\rho(\varkappa_i)$ as in \eqref{eqS:37}. 

Of special interest is the case of $\lambda_0$ almost real, i.e., $\lambda_0=\lambda_0^{re}+i \delta' $, where $\lambda_0^{re}$ is the real part of $\lambda_0$ and $\delta'$ an infinitesimal (not necessarily positive) parameter. One can of course still use the solution \eqref{eqS:27}. The problem is, however, that for both $\varkappa_i$ and $\lambda_0$ almost real one has to keep track of the relative sizes of the infinitesimal imaginary parts. This complication can be avoided by the following procedure. First we set $\lambda_0$ in the second line of \eqref{eqS:39} equal to its real part. By doing this, the arguments of the logarithms can cross the negative real axis, which is compensated by a new sum of $\eta$ functions, $\eta'$. Then instead of \eqref{eqS:20} we have
\begin{equation}
\sum_i \rho(\varkappa_i) \int_0^\infty \negtwo d\lambda \frac{1}{\lambda-\lambda_0^{re}- i\delta'}[\ln(\lambda-\varkappa_i)-\ln(\lambda_0^{re}-\varkappa_i)] +\eta' \int_0^\infty \frac{d\lambda}{\lambda-\lambda_0}.
\end{equation}
In the first integral $\delta'$ can be safely put equal to zero as the resulting integrand has vanishing residue, with the logarithm in the numerator being an analytic function in some neighbourhood of the pole (since $\Im(\varkappa_i)\ne0)$. The integral thus does not depend on how we avoid the pole (i.e. $\delta'$ can be positive, negative or zero). In the second integral one has to keep the imaginary part of $\lambda_0$.

The final result for almost real $\lambda_0=\lambda_0^{re}+i\delta'$ is then
\begin{equation}
I_2=\sum_i \rho(\varkappa_i) \Big[ \Li \frac{\lambda_0^{re}}{\lambda_0^{re}-\varkappa_i} 
+\frac{1}{2}\ln^2(\lambda_0^{re}-\varkappa_i)-\ln(\lambda_0^{re}-\varkappa_i)\ln(-\lambda_0)\Big], \label{eqS:40}
\end{equation}
where in the last logarithm $\lambda_0$ is kept together with its infinitesimal imaginary part. \index{dilogarithm!useful relations|)}\index{dimensional regularization|)}\index{scalar function|)}

\chapter{Weak interactions in the effective theory approach}\label{weak-int}
In this chapter we will briefly review the standard methods used in the phenomenology of weak interactions, the operator product expansion (OPE) and the renormalization group (RG) equations. These are used to arrive at a set of local operators describing weak interactions at low energies. At the end the factorization approximation, that is used to evaluate hadronic matrix elements of the current-current local operators, is described. 

\section{Operator Product Expansion}\label{OPE}\index{effective weak interaction|(}\index{OPE|(}
In this section we will briefly review the ideas behind the Operator Product Expansion (OPE) and its application to weak interactions. The original idea dates back to Wilson \cite{Wilson:zs}, who conjectured that the divergent part of a product of two operators could be described by a sum of local operators $Q_n(x)$
\be
A(x) B(y) \xrightarrow[x\to y]{} \sum_{n=1}^{\infty} C_n^{AB}\!(x-y)\; Q_n(x), \label{OPE-beg}
\ee
where the coefficients $C_n^{AB}\!(x-y)$ contain the divergences. There are two major advantages to the expansion as written above. From simple dimensional analysis it immediately follows that if operators $A$ and $B$ have dimensions $d_{A}$ and $d_{B}$ respectively (in terms of mass), then $\dim(C_n)= d_A+d_A- d_{n}$ with $d_{n}$ the dimension of operator $Q_n$. We then expect the highest divergence in $C_n^{AB}(x-y)$ to be of the form $x-y$ to the power $d_{n} -d_A-d_B$. So only those operators $Q_n$ with a small enough dimensions $d_n$ are expected to have divergent coefficients and be relevant in the $x\to y$ limit. Since dimensions of operators grow with the number of fields and with the number of derivatives on the fields that they are constructed from, only a finite number of terms in \eqref{OPE-beg} have to be retained. This discussion will be changed only slightly by quantum effects, where anomalous dimensions come into play. The other surprising thing about the OPE \eqref{OPE-beg} is, that it is an operator identity. It holds regardless of what the states it acts on are. 

Let us note on passing that the perturbative proof of OPE has been given by Zimmerman \cite{Zimmerman:1970}, while a nonperturbative proof can be found in \cite{Weinberg:kr}. The operator product expansion in general reads
\be
T\{A_1(x_1) A_2(x_2)\dots A_k(x_k)\} \xrightarrow[x_i\to x]{} \sum_n C_n^{A_1,\dots,A_k}(x_1-x,\dots, x_k-x) Q_n(x),
\ee
with $T$ being the time ordering operator.

The application of the OPE to weak interactions comes from the observation that the distances at which weak interactions occur are set by the mass of the intermediate $W$ and $Z$ bosons, i.e., $x-y\sim 1/m_W$. If one is interested in the processes at energy scales $\mu$ much smaller than the weak scale, $\mu\ll m_W$, or, in other words, in the processes, that effectively occur at typical distances $1/\mu$ that are much larger than $x-y\sim 1/m_W$, we can take the limit $x\to y$ (or equivalently $m_W\to \infty$) and use the operator product expansion. 

Let us formulate this in some more detail. We start from the charged current part of the weak Lagrangian
\be
{\cal L}_{\text{CC}}=\frac{g_2}{2 \sqrt{2}} \big(J_\mu^+ W^{+\mu}+J_\mu^-W^{-\mu}\big),
\ee
where 
\be
\begin{split}
J_\mu^+=&(\bar{u}d')_{V-A}+(\bar{c}s')_{V-A}+(\bar{t}b')_{V-A}\\
&+(\bar{\nu}_e e)_{V-A}+(\bar{\nu}_\mu \mu)_{V-A}+(\bar{\nu}_\tau \tau)_{V-A},
\end{split}
\ee
with $J_\mu^-=(J_\mu^+)^\dagger$, $q_i'=V_{ij}^{CKM} q_j$ are the rotated weak states, while $(\bar{q} q')_{V-A}=(\bar{q}\gamma^\mu(1-\gamma_5)q')$, $W^{\pm\mu}$ are the $W$ vector boson fields, and $g_2$ is the weak isospin coupling constant.\index{V-A}\index{weak isospin}

\begin{figure}
\begin{center}
\epsfig{file=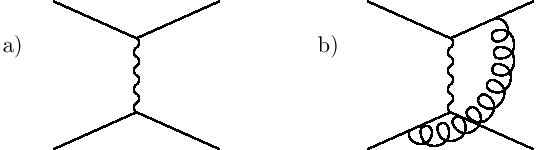, height=1.8cm}
\caption{\footnotesize{The tree level diagram in the full theory a), that is translated to the $Q_2$ operator in the effective theory. The operator $Q_1$ arises from the QCD interactions b). In addition to the diagram b) also the diagrams with gluons attached to different legs appear.}}\label{treeQCD}\index{Q12@$Q_{1,2}$}
\end{center}
\end{figure}

The scattering matrix $S_{fi}$ to the first nonzero order in perturbation theory for a process involving four quarks is
\be
S_{fi}=-\frac{1}{2!}\int d^4 x\int d^4 y \langle f| T\{ {\cal L}_{\text{CC}}(x) {\cal L}_{\text{CC}} (y) \}|i \rangle. \label{weakpert}
\ee
Taking as an example $cd\to us$ scattering process shown on Fig.~\ref{treeQCD}, the tree level amplitude is \index{amplitude!$cd\to us$ scattering}
\be
M_{fi}=-\Big(\frac{g_2}{2 \sqrt{2}}\Big)^2 V_{cs}^* V_{ud} \bar{u}_u \gamma^\mu(1-\gamma_5) u_d \frac{\eta_{\mu\nu}-\frac{p_\mu p_\nu}{m_W^2}}{p^2-m_W^2+i\epsilon} \bar{u}_s\gamma^\nu (1-\gamma_5) u_c. \label{weaktree}
\ee
In this process, there are two relevant scales. One is determined by the momentum exchange $p$, the other by the $W$ boson mass $m_W$. The effective size of the weak interaction region $x-y$ in \eqref{weakpert} is of the order $1/m_W$ and shrinks to a point when the limit $m_W \to \infty$ is taken. This limit corresponds to the limit $p\ll m_W$, so that the $p$ dependence in \eqref{weaktree} can be neglected. It then follows immediately, that the amplitude \eqref{weaktree} can be obtained with an effective Lagrangian
\be
{\cal L}_{\text{eff}}=-\frac{G_F}{\sqrt{2}}V_{cs}^* V_{ud} (\bar{u}d)_{V-A}(\bar{s}c)_{V-A}=-\frac{G_F}{\sqrt{2}} V_{cs}^* V_{ud} C_2 Q_2, \label{eff-tree}
\ee
where the definition of the Fermi constant \index{Fermi constant} \index{GF@$G_F$} $G_F/\sqrt{2}=g_2^2/(8m_W^2)$ has been used and tentatively a local four-quark operator $Q_2=(\bar{u}d)_{V-A}(\bar{s}c)_{V-A}$ with a Wilson coefficient $C_2=1$ has been defined. The important feature of the effective weak Lagrangian \eqref{eff-tree} is that it is valid also for other final states, mesons, baryons,... and not only for the free quarks for which it has been calculated. As advertised before, this is one of the salient features of the operator product expansion. The expansion in a series of local operators is calculated using quark initial and final states, but is then valid also for hadron initial and final states. \index{V-A}\index{Wilson coefficients}

Actually, the effective Lagrangian \eqref{eff-tree} is valid only in the absence of QCD interactions. Once these are taken into account, as shown on Fig.~\ref{treeQCD}, another four-quark operator appears in the OPE, $Q_1=(\bar{u}^\alpha d^\beta)_{V-A} (\bar{s}^\beta c^\alpha)_{V-A}$, where summation over color indices $\alpha,\beta$ is understood. To arrive at this operator, the following identity for the $SU(N_c)$ generators is used 
\be
T^a_{\alpha\gamma}T^a_{\beta\delta}=-\frac{1}{2N_c}\delta_{\alpha\gamma}\delta_{\beta\delta}+\frac{1}{2} \delta_{\alpha\delta} \delta_{\beta\gamma},\label{SU3Iden}
\ee
where $N_c=3$ is the number of colors, while $T^a=\frac{1}{2}\lambda^a$, with $\lambda^a$ the Gell-Mann $SU(3)$ matrices, \index{Ta@$T^a$}for which $\tr(\lambda^a\lambda^b)=2 \delta^{ab}$. The effective weak Lagrangian for $\Delta C=1$, $\Delta S=1$ is then
\be
{\cal L}_{\text{eff}}=-\frac{G_F}{\sqrt{2}} V_{cs}^* V_{ud}\big(C_1 Q_1+C_2 Q_2\big), \label{weak-cs}
\ee
with
\be
Q_1=(\bar{u}^\alpha d^\beta)_{V-A} (\bar{s}^\beta c^\alpha)_{V-A}, \qquad Q_2=(\bar{u}d)_{V-A}(\bar{s}c)_{V-A}.\label{weak-cs-oper} \index{Q12@$Q_{1,2}$}
\ee
The coefficient $C_1$ is proportional to $\alpha_s$, while $C_2\sim 1$ is nonzero already at tree level in the perturbative QCD expansion, as discussed above.\index{Wilson coefficients}

\begin{figure}
\begin{center}
\epsfig{file=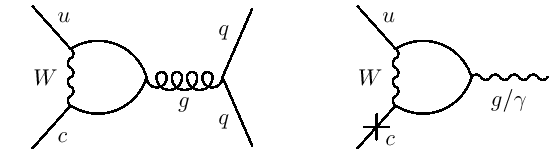, height=1.8cm}
\caption{\footnotesize{The diagrams that give rise to the penguin operators. The diagram on the left gives rise to the operators $Q_{3.\dots, 6}$, while the operator to the right gives rise to the magnetic penguin operators $Q_{7,8}$. Cross denotes mass insertion.}}\label{Penguins}
\end{center}
\end{figure}

\index{operators! four-fermion}\index{operators! penguin}\index{penguin operators|(}\index{penguin operators! definition of}
If one is interested in the processes involving a slightly different change of flavors, other operators can appear as well. For instance, if the $s$-quark in \eqref{weak-cs}, \eqref{weak-cs-oper} is replaced by a $d$ quark (or vice versa), an additional set of diagrams, shown on Fig.~\ref{Penguins}, is possible at ${\cal O}(\alpha_s)$ order and gives rise to the ``penguin'' operators $Q_{3,\dots 6}$. The complete set of operators for $\Delta C=1$, $\Delta S=0$ (neglecting electroweak penguins) is then \index{Q12@$Q_{1,2}$}
\begin{subequations}\label{list-oper}
\begin{align}
Q_1^d&=(\bar{u}^\alpha d^\beta)_{V-A} (\bar{d}^\beta c^\alpha)_{V-A},& Q_2^d=&(\bar{u}d)_{V-A}(\bar{d}c)_{V-A},\\
Q_1^s&=(\bar{u}^\alpha s^\beta)_{V-A} (\bar{s}^\beta c^\alpha)_{V-A}, & Q_2^s=&(\bar{u}s)_{V-A}(\bar{s}c)_{V-A},\\
Q_3&=(\bar{u}c)_{V-A}\sum_q (\bar{q}q)_{V-A}, & Q_4=&(\bar{u}^\alpha c^\beta)_{V-A}\sum_q (\bar{q}^\beta q^\alpha)_{V-A},\\
Q_5&=(\bar{u}c)_{V-A}\sum_q (\bar{q}q)_{V+A}, & Q_6=&(\bar{u}^\alpha c^\beta)_{V-A}\sum_q (\bar{q}^\beta q^\alpha)_{V+A},
\end{align}
\end{subequations}
where we have suppressed the color indices in the currents of the form $(\bar{q}q')=(\bar{q}^{\alpha}q'^\alpha)$, while the sum over $q$ runs over the active quark-flavors. At the scale $\mu\simeq m_c$ these are $q=u,d,s,c$. In the following chapters, we will be interested also in the final states involving photons and lepton pairs. For these decays the set of the relevant operators is enlarged by the magnetic penguins (corresponding to the diagrams on Figure \ref{Penguins})\index{Q7@$Q_7$}
\begin{align}
Q_{7}&=\frac{e}{4\pi^2} m_c F_{\mu\nu} \bar{u} \sigma^{\mu\nu} P_R c, &Q_{8}=\frac{g_s}{4\pi^2} m_c G^a_{\mu\nu} \bar{u} \sigma^{\mu\nu} T^a P_R c, \label{magn-peng}
\end{align}
with $g_s$ the strong coupling constant, and the semileptonic operators (corresponding to the diagrams on Fig.~\ref{InamiLim})\index{operators! semileptonic}\index{penguin operators! semileptonic}\index{Q9@$Q_9$}\index{Q99@$Q_{10}$}
\begin{align}
Q_9&=\frac{e^2}{16 \pi^2} (\bar{u}_L \gamma^\mu c_L) (\bar{l} \gamma_\mu l),
&Q_{10}&=\frac{e^2}{16 \pi^2} (\bar{u}_L \gamma^\mu c_L) (\bar{l} \gamma_\mu \gamma_5 l),\label{semilep-peng}
\end{align}
where $q_L=P_L q$ and $P_{L,R}=\frac{1}{2}(1\pm \gamma_5)$ are the chirality projection operators.
\index{effective weak interaction|)}

\section{Renormalization Group and OPE}\label{RG-OPE}\index{anomalous dimension|(}\index{renormalization group|(}\index{renormalization group! general framework|(}\index{Wilson coefficients|(}
In the calculation of the Wilson coefficients typically expressions of the form $\alpha_s \ln(\mu/m_W)$ appear. Here $\mu$ is a typical scale at which the processes occur. For processes involving the decay of $c$-quark a typical scale is of order $1$ GeV. The ratio of scales in the argument of the logarithm is thus very large, of order $100$, and consequentially the factor $\alpha_s \ln\big(\mu/m_W\big)$ is of order ${\cal O}(1)$. Even though the QCD coupling $\alpha_s$ is not terribly large at the scales of around $1$ GeV and could be used as a perturbative expansion parameter, the appearance of large logarithms prevents the straightforward application of perturbation theory. All large logarithms of the form $\big(\alpha_s \ln\big(\mu/m_W\big)\big)^n$ have to be summed up using renormalization group equations, if one wants to get the correct leading order expression for the Wilson coefficients at lower energy scales. 

The RG evolution is done in several steps \cite{Buchalla:1995vs}. First the Wilson coefficients $C_i$ are calculated at the weak scale $\mu\sim m_W$ to some given order in the perturbative QCD expansion. For instance, at the leading order $C_2(m_W)=1$, while $C_1=C_{3,\dots,6}=0$ (the coefficients $C_{7,9,10}$ will be discussed later on). Then to the same order anomalous dimensions of the four-quark operators in the effective theory are calculated (at the leading order this is to the order $\alpha_s$). These are then used to evolve the Wilson coefficients to lower energy scales.

Let us first introduce the notion of the \index{anomalous dimension! definition of|(}anomalous dimensions\footnote{We will follow closely the introduction given in \cite{Buchalla:1995vs}.}. To do so, consider first an invariant amplitude $A$ for a given process. Assume that the calculation of the invariant amplitude $A$ in the full theory is known. This has to be the same to the value of $A$ obtained in the effective theory, i.e., by using the operator product expanded effective Lagrangian. Using LSZ theorem the amplitude for the four-quark scattering is proportional to $Z_q^2 \langle Q_i^{(0)}\rangle_0$, where $Z_q$ is the renormalization constant for the quark field $q^{(0)}=Z_q^{1/2} q$, while $\langle Q_i^{(0)}\rangle_0$ is the amputated Green function of the unrenormalized operator\index{amputated Green function}.\index{bare operators}\index{renormalized operators} However, $Z_q^2 \langle Q_i^{(0)}\rangle_0$ is still divergent, so that additional multiplicative operator renormalization has to be introduced \index{LSZ}
\be
Q_i^{(0)}=Z_{ij} Q_j,
\ee
with operator $Q_j$ a function of renormalized quark fields $q$.
 The renormalization constants $Z_{ij}(\mu)$ are scale dependent, and so are the renormalized operators $Q_j$. This dependence cancels in the product, so that the bare operators $Q_i^{(0)}$ \index{bare operators} are scale independent, as they should be. The unrenormalized amputated Green function $ \langle Q_j^{(0)} \rangle_0$ and renormalized one $\langle Q_j\rangle$ are connected by
\be
Z_q^2 \langle Q_i^{(0)} \rangle_0= Z_{ij} \langle Q_j \rangle. \label{oper-renorm}
\ee
The renormalized amputated Green functions $\langle Q_j \rangle$ are now finite and can be used to define the Wilson coefficients $C_i$ by matching the calculation in the effective theory to the full theory calculation
\be
A=-\frac{G_F}{\sqrt{2}} C_i \langle Q_i\rangle, \label{Cidef}
\ee
where the dependence on the CKM matrix elements has not been written out explicitly for the sake of simplicity. With $A$ known from the full theory calculation, Eq.~\eqref{Cidef} then defines the values of the Wilson coefficients $C_i$.

It is illuminating to consider also a different point of view, closer to the conventional renormalization in terms of the coupling constants. Instead of absorbing the divergences in the renormalizations of operators, these can be absorbed in the ``coupling constants'', the Wilson coefficients $C_i$. The renormalized Wilson coefficients are thus\index{renormalized operators}
\be
C_i^{(0)}=Z_{ij}^c C_j, \label{c-renorm}
\ee
while the effective Lagrangian is
\be
\begin{split}
{\cal L}_{\text{eff}}&=-\frac{G_F}{\sqrt{2}} C_i^{(0)} Q_i (q^{(0)})=-\frac{G_F}{\sqrt{2}} Z_{ij}^c Z_q^2 C_j Q_i\\
&=-\frac{G_F}{\sqrt{2}} \big[ C_i Q_i+\left(Z_{ij}^c Z_q^2-\delta_{ij}\right) C_j Q_i\big],
\end{split}\label{lagr-eff}
\ee
where $Q_i(q^0)$ denotes that the four-quark operator is constructed from the bare quark fields. The amplitude is then using Lagrangian \eqref{lagr-eff} \index{bare operators}
\be
A=-\frac{G_F}{\sqrt{2}} C_i^{(0)} Z_q^2 \langle Q_i^{(0)} \rangle_0=-\frac{G_F}{\sqrt{2}} C_i \langle Q_i\rangle, \label{ampl-eff}
\ee
where the last equality in \eqref{ampl-eff} has been obtained using the effective Lagrangian in the last line of \eqref{lagr-eff} \cite{Buchalla:1995vs}. Note that the divergences are absorbed by the $(Z_{ij}^c Z_q^2-\delta_{ij})$ part of the effective Lagrangian \eqref{lagr-eff}. Using \eqref{oper-renorm} and \eqref{c-renorm} in \eqref{ampl-eff} it immediately follows that
\be
Z_{ij}^c=Z_{ji}^{-1}. \label{Z-connect}
\ee
The renormalization constants of the operators are directly connected to the renormalization constants of the Wilson coefficients.

The evolution of the Wilson coefficient is now easily determined. Following the usual notation \cite{Buchalla:1995vs}, first the anomalous dimensions matrix $\gamma$ is introduced
\be
\gamma=Z^{-1} \frac{d Z}{d\ln\mu}.
\ee
The RG equations for Wilson coefficients then immediately follow from Eqs.~\eqref{c-renorm}, \eqref{Z-connect} and the fact that the bare Wilson coefficients $C_i^{(0)}$ are scale independent
\be
\frac{d \vec{C}}{d\ln\mu}=\gamma^\top \vec{C}. \label{evol-eq}
\ee
Using the evolution of the QCD coupling constant $g_s$ (in the $\overline{\text{MS}}$ renormalization scheme)
\be
\frac{d g_s(\mu)}{d\ln \mu}=\beta(\epsilon, g_s(\mu))=-\epsilon g_s +\beta(g_s),
\ee
together with the expansion of the beta functions $\beta(g_s)$ and the anomalous dimension matrix $\gamma$ in terms of the strong coupling constant \index{anomalous dimension!expansion of}
\begin{align}
\beta(g_s)&=-\beta_0 \frac{g_s^3}{16 \pi^2} -\beta_1 \frac{g_s^5}{(16 \pi^2)^2}+\dots\\
\gamma(\alpha_s)&=\gamma^{(0)} \frac{\alpha_s}{4\pi}+\gamma^{(1)}\big(\frac{\alpha_s}{4\pi}\Big)^2+\dots \label{gamma-expand}
\end{align}
where $\alpha_s=g_s^2/(4\pi)$, while for the $f$ active flavors and $N_c$ colors
\be
\beta_0=\frac{11N_c-2f}{3}, \qquad \qquad \beta_1=\frac{34}{3}N_c^2 -\frac{10}{3} N_c f - \frac{N_c^2-1}{N_c} f,
\ee
the evolution equation \eqref{evol-eq} can be solved to any given order. Below we will consider the solutions $U_f(m_1,m_2)$ of \eqref{evol-eq} for the evolution $m_2\to m_1$ at the NLO. The form of the evolution matrices will, however, not be written out explicitly. They can be found in \cite{Buchalla:1995vs,Buras:1998ra} (see, e.g., Eqs. (3.92)-(3.98) of \cite{Buchalla:1995vs}), where also the expressions for the anomalous dimension matrices to next-to-leading order for a number of processes can be found. \index{anomalous dimension! definition of|)} 

To get some flavor for the effects of the RGE, we show the leading order RG evolution of a single operator
\be
C(\mu) = \Big(\frac{\alpha_s(m_W)}{\alpha_s(\mu)}\Big)^{\gamma^{(0)}/2\beta_0} C(m_W). \label{c-mu-dep}
\ee
The strong coupling constant $\alpha_s(\mu)$ appearing in \eqref{c-mu-dep} is at the two-loop order \index{running, strong coupling constant}
\be
\alpha_s(\mu)=\frac{4\pi}{\beta_0 \ln(\mu^2/\Lambda^2)}\Bigg[1-\frac{\beta_1}{\beta_0^2}\frac{\ln\ln(\mu^2/\Lambda^2)}{\ln(\mu^2/\Lambda^2)}\Bigg], \label{alpha-two-loop}
\ee
where $\Lambda $ is the QCD scale (the value of which depends on the number of active flavors). Using the precisely measured value of the strong coupling constant at the $Z$ boson mass, $\alpha_s(m_Z)=0.1172\pm0.002$ \cite{Hagiwara:pw} one arrives at $\Lambda^{(5)}=216\pm25$ MeV, while for $\mu<m_b$ with four active flavors the matching at $m_b=4.25$ GeV gives $\Lambda^{(4)}=311\pm33$ MeV. The important observation about Eq.~\eqref{c-mu-dep} is that it contains {\it all} the terms of the form $\big(\alpha_s \ln(\mu/m_W)\big)^n$ as has been announced at the beginning of this section. \index{QCD scale}

\index{renormalization group! calculation of Wilson coefficients|(}
In general, the Wilson coefficients at lower scale are calculated through the following steps \cite{Buchalla:1995vs}. First the Wilson coefficients $C_i(m_W)$ at weak scale are calculated by matching the effective theory with five active flavors $q=u,d,s,c,b$ onto the full theory. Then the anomalous dimensions $\gamma^{(5)}$ are calculated in the effective theory with five flavors. Using $\gamma^{(5)}$, Wilson coefficients are evolved down to the scale of b-quark, obtaining $C_i(m_b)$. If one is interested in the processes at lower scales, e.g at the charm quark scale, $b$-quark is integrated out as an effective degree of freedom. This is accomplished by matching the effective theory with five flavors onto the effective theory with four flavors. The remaining Wilson coefficients are then evolved down to the charm scale using the anomalous dimension matrices of the four-flavor effective theory. Thus
\be
\vec{C}(m_c)=U_4(m_c,m_b)M_5(m_b) U_5(m_b,m_W)\vec{C}(m_W),
\ee
where $U_{4,5}(\mu_1,\mu_2)$ are the evolution matrices from the scale $\mu_2$ to the scale $\mu_1$ in four and five-flavor effective theories respectively, while $M_5$ is the threshold matrix that matches the two effective theories at the scale $\mu\sim m_b$. \index{renormalization group! general framework|)}

\index{renormalization group! in charm decays|(}
Let us be more specific and discuss the case of $\Delta C=1$, $\Delta S=0$ charm decays in some more detail. The effective Lagrangian at the weak scale $\mu\sim m_W$ is
\be
\begin{split}
{\cal L}_{\text{eff}}=-\frac{G_F}{\sqrt{2}} \bigg[ &V_{cd}^* V_{ud} \big(\sum_{i=1,2}C_i Q_i^d+\sum_{i=3,\dots,10}C_i Q_i\big)+\\
+&V_{cs}^* V_{us} \big(\sum_{i=1,2}C_i Q_i^s+\sum_{i=3,\dots,10} C_iQ_i\big)+\\
+&V_{cb}^* V_{ub} \big(\sum_{i=1,2}C_i Q_i^b+\sum_{i=3,\dots,10}C_i Q_i\big)\bigg]\\
=-\frac{G_F}{\sqrt{2}} \bigg[ &V_{cd}^* V_{ud} \sum_{i=1,2}C_i \big(Q_i^d-Q_i^b\big)
+V_{cs}^* V_{us} \sum_{i=1,2}C_i \big(Q_i^s-Q_i^b\big)\bigg], \label{lagr-at-weak}
\end{split}
\ee
where $Q_{1,2}^{d,s}$ are defined in \eqref{list-oper}, while $Q_{1,2}^b$ are obtained through the replacement $d\to b$ from $Q_{1,2}^d$. The penguin and electromagnetic operators $Q_i$, $i=3,.., 10$ are defined in Eqs.~\eqref{magn-peng}, \eqref{semilep-peng}, while the contributions of the electromagnetic penguins have been neglected as they are suppressed by additional powers of $\alpha$ in the processes considered. In the last line of \eqref{lagr-at-weak} the unitarity of the CKM matrix has been used $V_{cd}^* V_{ud}+V_{cs}^* V_{us}+V_{cb}^* V_{ub}=0$. Above, also the masses of $d,s,b$ quarks have been neglected compared to the weak scale, so that the Wilson coefficients $C_i$ are the same regardless of the flavor of down quark flowing in the loop in the full theory (i.e. regardless of the CKM structure in front of the parenthesis in \eqref{lagr-at-weak}). Thus the penguin operators do not appear in the effective Lagrangian at the weak scale as long as the mass of the $b$-quark can be neglected compared to $m_W$. This is in contrast to the case of $\Delta B=1$ decays, where the up-type quarks flow in the loops in the full theory. Since the top quark is very heavy, its mass cannot be neglected in the loops. This induces penguin operators already at the weak scale.

\index{Q99@$Q_{10}$}
Regarding the RG evolution of the operators $Q_{1,..,10}$ there are several important things to note. First of all $Q_{10}$ does not mix with other operators due to chirality. Furthermore, it has vanishing anomalous dimension, so that $C_{10}(\mu_c)=C_{10}(m_W)$. Next, the dimension five operators $Q_{7,8}$ do not mix into the dimension six operators $Q_{1,\dots,6}$ and $Q_{9}$. If one is interested in these operators solely, the dimension five operators can be dropped from the RG analysis. We will follow this procedure and evaluate $C_7$ separately. Note also, that (i) $Q_9$ operator does not mix into the operators $Q_{1,\dots,6}$ and (ii) the penguin operators $Q_{3,\dots,6}$ do not mix into the operators $Q_{1,2}$. One can thus consider the RG evolution of the reduced operator basis $Q_{1,2}$, $Q_{1, \dots,6}$ or $Q_{1,\dots,9}$, if one is interested in smaller sets of the Wilson coefficients $C_{1,2}$, $C_{1,\dots,6}$, or $C_{1,\dots,9}$, without introducing any error in the calculation. Finally, it is convenient to introduce a rescaled operator $\tilde{Q}_9=\alpha/\alpha_s (\bar{u}c)_{V-A} (\bar{l}l)_V$, as then the anomalous dimension depends only on the strong coupling, and can be expanded as in \eqref{gamma-expand}. The calculation of the Wilson coefficients then proceed as outlined above. 

It is instructive to do the $\alpha_s$ counting. At the leading order the RG evolution sums terms of the form $\alpha_s \ln(m_c^2/m_W^2)$, which are numerically of the order ${\cal O}(1)$. At the leading order one thus has to start with the initial values $C_i(m_W)$ calculated at $\alpha_s^0$, and then evolve them using the 1 loop anomalous dimensions (i.e. of order $\alpha_s$) to get the order ${\cal O}(1)$ values $C_i(\mu)$ at lower scales. Going to higher orders, an additional power of $\alpha_s$ is added at each step. We thus have
\be
C_i(\mu)={\cal O}(1)+{\cal O}(\alpha_s)+\dots
\ee
This expansion is valid also for $\tilde{C}_9$ multiplying the rescaled operator $\tilde{Q}_9$. Since $Q_9=\alpha_s/(8\pi) \tilde{Q}_9$, then $C_9=8 \pi/\alpha_s \tilde{C}_9$, so that the expansion is
\be
C_9(\mu)={\cal O}(1/\alpha_s)+{\cal O}(1)+{\cal O}(\alpha_s)+\dots \label{C9exp}
\ee
Thus it is only the NLO term, that is of the order ${\cal O}(1)$ in the calculation of the $C_9$ Wilson coefficient. It is then consistent in the $\alpha_s$ counting to work with $C_9$ determined at the NNLO and with the other Wilson coefficients at the NLO (if one wishes to work to ${\cal O}(\alpha_s)$). Partial calculations at the NNLO became available in the literature recently \cite{Bobeth:1999mk,Asatrian:2001de,Asatryan:2001zw,Lunghi:2002qw,Ghinculov:2002pe}, however, the three-loop calculation of the NNLO dimensional matrix has still not been performed. For this reason we will work in the following with both $C_9$ and $C_{1,\dots,6}$ determined at the NLO.

We start a more quantitative discussion with the values of the Wilson coefficients to the order ${\cal O}(\alpha_s)$ at the weak scale. These are known for quite some time and are in the naive dimensional regularization scheme (NDR)\footnote{In the naive dimensional regularization the Dirac matrices are assumed to obey $\{\gamma_\mu,\gamma_\nu\}=2 g_{\mu\nu}$, where $g_{\mu\nu}$ is a $4-\epsilon$ dimensional metric tensor. The $\gamma_5$ matrix is assumed to commute with the Dirac matrices $\{\gamma_\mu,\gamma_5\}=0$ \cite{Buras:1998ra}.} \cite{Buras:1989xd}
\begin{align}
C_1(m_W)&=\frac{11}{2} \frac{\alpha_s(m_W)}{4 \pi}, &C_2(m_W)&=1-\frac{11}{6}\frac{\alpha_s(m_W)}{4 \pi},\label{C1C2mW}
\end{align}
while $C_{3,\dots,9}(m_W)=0$. Since above $\mu_b$ the penguin operators do not enter the effective Lagrangian due to the unitarity of the CKM matrix, the Wilson coefficients $C_{1,2}$ in \eqref{lagr-at-weak} can be evolved down to $\mu\sim \mu_b$ using the $2\times 2$ anomalous dimension matrix (which can be found in \cite{Buras:1989xd} or in Eq. (5.12) of \cite{Buchalla:1995vs}). At the scale $\mu_b$ the $b$-quark is integrated out, i.e., the five-flavor effective theory \eqref{lagr-at-weak} is matched onto the four-flavor theory given by \index{effective weak interaction}
\be
\begin{split}
{\cal L}_{\text{eff}}=-\frac{G_F}{\sqrt{2}} \bigg[ &V_{cd}^* V_{ud} \Big(\sum_{i=1,2}C_i Q_i^d+\sum_{i=3,\dots,6,9}C_i Q_i\Big)+ V_{cs}^* V_{us} \Big(\sum_{i=1,2}C_i Q_i^s+\sum_{i=3,\dots,6,9} C_iQ_i\Big)\bigg]\\
=-\frac{G_F}{\sqrt{2}} \bigg[ &V_{cd}^* V_{ud} \sum_{i=1,2}C_i Q_i^d+ V_{cs}^* V_{us} \sum_{i=1,2}C_i Q_i^s-V_{cb}^* V_{ub} \sum_{i=3,\dots,6,9} C_iQ_i\bigg]. \label{lagr-at-bandc}
\end{split}
\ee
The penguin operator Wilson coefficients $C_{3,\dots,6,9}$ arise from the matching procedure. This is the only nontrivial step in the application of the formulas from the literature, as these were calculated for the down-type quark transitions. We use the expressions for the $K_L\to \pi^0 e^+ e^-$ decay \cite{Buras:qa}, where a similar procedure has to be done at the charm mass, with the $c$-quark being integrated out. For the gluonic penguins there are no changes, when going to the case of $b$-quark being integrated out, while the semileptonic Wilson coefficient has to be multiplied by $e_b/e_c=-1/3 \cdot 3/2=-1/2$. We then have (see Eqs. (6.20), (8.9) of \cite{Buchalla:1995vs})
\begin{align}
Z_1(m_b)&=C_1(m_b), &Z_2(m_b)&=C_2(m_b),\label{Z-begin}\\
Z_3(m_b)&=-\frac{\alpha_s}{24 \pi} F_s(m_b), &Z_4(m_b)&=\frac{\alpha_s}{8 \pi} F_s(m_b),\\
Z_5(m_b)&=-\frac{\alpha_s}{24 \pi} F_s(m_b), &Z_6(m_b)&=\frac{\alpha_s}{8 \pi} F_s(m_b),\\
Z_9(m_b)&=-\frac{1}{2} Z_{7V}'(m_b)=\frac{\alpha_s}{4 \pi} F_e(m_b), &{}&\label{Z-end}
\end{align}
with $Z_{7V}'$ defined as in Eq. (8.9) of \cite{Buchalla:1995vs}, while the functions
\begin{align}
F_s(\mu)&=-\frac{2}{3} \Big[\ln\Big(\frac{m_b^2}{\mu^2}\Big)+1\Big]Z_2(\mu),\\
F_e(\mu)&=-\frac{4}{9} \Big[\ln\Big(\frac{m_b^2}{\mu^2}\Big)+1\Big]\big(3Z_1(\mu)+ Z_2(\mu)\big),
\end{align}
are again calculated in the NDR.

 The sets of operators $\{ Q_{1,2}^d, Q_{3,\dots,6}, Q_9\}$ and $\{ Q_{1,2}^s, Q_{3,\dots,6}, Q_9\}$ from the first line of \eqref{lagr-at-bandc} are then evolved to the charm scale $\mu\sim m_c$ using the $7\times 7$ anomalous dimension matrices $\gamma^{(4)}$ for the four quark effective theory. The $6\times 6$ LO and NLO submatrices involving the gluonic penguins are listed in Eqs. (6.25), (6.26) of Ref.~\cite{Buchalla:1995vs} and have been calculated in \cite{Buras:1991jm,Ciuchini:1993vr}. The remaining entries are listed in Eqs (8.11), (8.12) of Ref.~\cite{Buchalla:1995vs} and have been calculated in \cite{Buras:qa}. 

In summary, the RG evolution from $\mu_W\sim m_W$ to $\mu_c$ for the $\Delta C=1$ transitions is described by the following procedure \index{summary}
\begin{align}
m_b< \mu< m_W:& \qquad \vec{C}(\mu)=U_5(\mu,m_W) \vec{C}(m_W),\label{RG-flow-beg}\\
\mu=m_b:&\qquad \vec{C}(m_b)\to \vec{Z}(m_b),\\
m_c< \mu< m_b:& \qquad \vec{C}(\mu)=U_4(\mu,m_b) \vec{Z}(m_b),\label{RG-flow-end}
\end{align}
with $U_5$ and $U_4$ the $2\times 2$ and $7\times 7$ evolution matrices for five and four active flavors respectively. They can be found in Eqs. (3.93)-(3.98) of \cite{Buchalla:1995vs}. The $Z(m_b)$ are given in \eqref{Z-begin}-\eqref{Z-end}. The values of the Wilson coefficients are listed in Table \ref{tab-Wilson}. For a comparison the values of the Wilson coefficients at the leading order are given as well, but calculated with the two-loop evolution of the strong coupling constant \eqref{alpha-two-loop}. The values are given for the central value of $\Lambda^{(5)}=216\pm25$ MeV and $m_b=4.25$ GeV. The one sigma change in $\Lambda^{(5)}$ corresponds to a change of about 10\% in $C_{1,\dots,6}$. We find a pronounced scale dependence for the $C_9$ coefficient below 1.5 GeV, as a consequence of the large cancelations in the RG evolution equations. The situation is very similar to the case of the coefficient $Z_{7V}$ in $K_L \to \pi^0 e^+e^-$ \cite{Buchalla:1995vs}. The LO value of $C_9$ even changes sign near $\mu \sim 1$ GeV, being positive for $\mu >1$ GeV. \index{anomalous dimension! values of}\index{renormalization group! effect on $C_9$}

\begin{table} [h] \index{tables of results}
\begin{center}
\begin{tabular}{|l|c|c|c|c|c|c|c|c|c|} \hline\index{C1@$C_1$}\index{C2@$C_2$}\index{C9@$C_9$}\index{calculation! results of}
-&$\mu$(GeV) & $C_1$& $C_2$ & $C_3$ & $C_4$ & $C_5$ & $C_6$ & $\tilde{C}_9$ & $C_9$\\ \hline\hline
LO&$1.0$ & $-0.64$& $1.34$ & $0.016$ & $-0.036$ & $0.010$ & $-0.046$ & $-0.0013$ & $-0.07$\\ \hline
NLO&$1.0$ & $-0.49$& $1.26$ & $0.024$ & $-0.060$ & $0.015$ & $-0.060$ & $-0.011$ & $-0.60$\\ \hline
NLO&$1.5$ & $-0.37$& $1.18$ & $0.013$ & $-0.036$ & $0.012$ & $-0.033$ & $-0.0018$ & $-0.13$\\ \hline
NLO&$2.0$ & $-0.30$& $1.14$ & $0.009$ & $-0.025$ & $0.009$ & $-0.021$ & $-0.0016$ & $-0.13$\\ \hline
 \end{tabular} 
 \caption[Values of Wilson coefficients at scales $\mu=1,1.5, 2$ GeV]{\footnotesize{Values of Wilson coefficients at scales $\mu=1,1.5, 2$ GeV, calculated at the next-to-leading order (NLO) as explained in the text. For a comparison in the first line the LO values are given at the scale $\mu=1$ GeV, but calculated with the two loop evolution of the strong coupling constant \eqref{alpha-two-loop}. In the last column the properly scaled $C_9=8\pi/\alpha_s(\mu) \tilde{C}_9$ Wilson coefficient is given. }}
\label{tab-Wilson}
\end{center}
\end{table}

\index{renormalization group! calculation of Wilson coefficients|)}
\index{operators!penguin}\index{penguin operators!contribution in charm decays}
 Note, however, that the penguin operators are proportional to the $V_{cb}^* V_{ub}$ matrix elements \eqref{lagr-at-bandc}. In the Wolfenstein parametrization \eqref{VCKM-Wolf} this is $\sim \lambda^5$, which has to be compared to the CKM suppression of the $Q_{1,2}$ operators, $V_{cs}^* V_{us}\sim \lambda$, where $\lambda=\sin{\theta_c}=0.22$. Penguin operators are thus suppressed by $\lambda^4\sim 10^{-3}$ in the $\Delta C=1$ transitions, even more so because the penguin Wilson coefficients are of the order $C_{3,\dots,6}(m_c)\leq 10^{-1} C_{1,2}(m_c)$ as shown in Table \ref{tab-Wilson} (see also \cite{Buccella:1994nf,Pham:1986cj}). The penguin operators in the $\Delta C=1$ transitions are not relevant numerically, except in special observables such as $CP$ \index{CP@$CP$}asymmetries \cite{Buccella:1994nf}. They are thus neglected in the following. 

\index{Q7@$Q_7$}\index{Q9@$Q_9$}
Note, that also the $Q_7$, $Q_9$ operators are suppressed by a factor $\lambda^4\sim 10^{-3}$ compared to the $Q_{1,2}$ operators. They will, however, be kept in the analysis, because of possibly large non-SM contributions that will be discussed in more detail in chapter \ref{rareD}. In the SM they are, however, negligible. Incidentally this also means, that the uncertainties in the value of the $C_9$ coefficient, observed above, will not propagate into the decay rates. 

It is interesting to compare the $\Delta C=1$ transition discussed above with the $\Delta B=1$, $b\to s $ transition. The relevant effective Lagrangian at the $\mu_b$ scale is \cite{Buchalla:1995vs}
\be
{\cal L}_{\text{eff}}=-\frac{G_F}{\sqrt{2}} \bigg[ V_{ub}^* V_{us} \sum_{i=1,2}C_i Q_i^{u(b\to s)}+V_{cb}^* V_{cs} \sum_{i=1,2}C_i Q_i^{c(b\to s)} - V_{tb}^* V_{ts} \sum_{i=3,\dots,6,9} C_iQ_i^{(b\to s)}\bigg].
\ee
The CKM factors are here $V_{ub}^* V_{us}\sim \lambda^4$, $V_{cb}^* V_{cs}\sim \lambda^2$, $V_{tb}^* V_{ts}\sim \lambda^2$. Because of this CKM structure, the penguin operators are the dominant operators in the charmless $b\to s$ decays, quite contrary to the case of the $\Delta C=1$ transition discussed above. 
\index{renormalization group! in charm decays|)}

\begin{figure}
\begin{center}
\epsfig{file=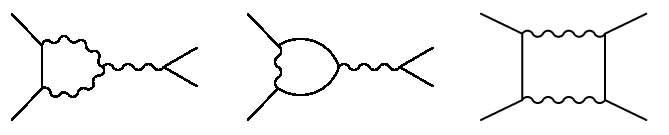, height=2.cm}
\caption{\footnotesize{ The penguin and box diagrams contributing to $ c \to u l^+ l^-$ decay at the quark level.}}\label{InamiLim}
\end{center}
\end{figure}

Let us now conclude the discussion of the $\Delta C=1$ transitions by turning to the magnetic penguin operator $Q_7$ and the semileptonic operator $Q_{10}$. The value of the $C_7$ Wilson coefficient is obtained following the same procedure as outlined in Eqs.~\eqref{RG-flow-beg}-\eqref{RG-flow-end}, using the operator basis $Q_{1,\dots,7}$. The major difference compared to the case of RG evolution with the $Q_9$ operator is, that the leading order mixing of the operators $Q_{7,8}$ with the operators $Q_{1,\dots,6}$ vanishes. It is only at the two-loop level, that the anomalous dimension matrix has nonzero values mixing $C_{1,\dots,6}$ into $C_7$. The expansion of $C_7$ in powers of $\alpha_s$ then begins at the order ${\cal O}(1)$, contrary to the case of $C_9$ \eqref{C9exp}. The factor $e$ in $Q_7$ also insures that the expansion of $\gamma(\alpha_s)$ in \eqref{gamma-expand} is unchanged, with the difference, that $\gamma^{(0)}_{i7}$ receive contributions from the two-loop calculation. Since the two-loop results are scheme dependent, so is $\gamma^{(0)}$. It is then customary to introduce the effective anomalous dimension matrix $\gamma^{(0)\text{eff}}$ \cite{Buras:xp}, which is scheme independent, as is the case for the leading order results. Using the LO anomalous dimension matrix $\gamma^{(0)\text{eff}}$, the NLO evolution for $\alpha_s$, $m_b=4.25$ GeV, the result is (see also \cite{Greub:1996wn})\index{C7@$C_7^{\text{eff}}$}
\be
C_7^{\text{eff}}(1.0 \;\text{GeV})=0.13, \qquad C_7^{\text{eff}}(1.5\;\text{GeV})=0.087, \qquad C_7^{\text{eff}}(2.0 \;\text{GeV})=0.066, \label{C7numbers}
\ee

\index{Inami-Lim calculation of $C_{7,9,10}$|(}\index{neglected QCD in $C_{7,9,10}$|(}
It is instructive to compare the values of $C_{7,9}$ Wilson coefficients obtained from the RG analysis with the invariant amplitudes that one would get from the full electroweak theory, but by neglecting the QCD interactions (i.e. by evaluating the diagrams of Fig.~\ref{InamiLim}). The invariant amplitudes of the QCD neglected calculation have the same structure as is obtained from the effective Lagrangian \eqref{effective_lagr} when used {\it at tree level}. The parameters of the invariant amplitudes obtained by neglecting the QCD contributions will be denoted by $C_{7,9}^{\text{IL}}$ (with IL standing for Inami, Lim \cite{Inami:1980fz}). It is important to stress that these are {\it not the Wilson coefficients}, as they only parametrize the invariant amplitudes. However, based on the (unproved) expectations, that $C_9$ is not much changed by the QCD corrections, $C_9^{\text{IL}}$ has been often used in the literature as an estimate for $C_9(\mu)$ \cite{Burdman:2001tf,Fajfer:2001sa}.

 The values of the parameters $C_{7,9,10}^{\text{IL}}$ are easily obtained from the calculation of Ref.~\cite{Inami:1980fz} for the $b\to s l^+l^-$ transitions. Following \cite{Ho-Kim:1999bs} we find, that the coefficients are of the form
\be
C_i^{\text{IL}}= I_q F_I^{(i)}(x_j)+Q_q F_Q^{(i)}(x_j),\label{Cnmass-dep}
\ee
where $i=7,9,10$, while $I_q$ is connected to the weak isospin \index{weak isospin} of the quarks in the loops of Fig.~\ref{InamiLim} and $Q_q$ is their charge. For the up-type quarks in the loops, as is the case in the $b\to s l^+l^-$ transition, we have $I_q=+1$, $Q_q=2/3$. For the case of the $c\to u l^+l^-$ transition, that we are interested in, $I_q=-1$, $Q_q=-1/3$, as then the down-type quarks appear as the intermediate states in the loops. $F_I(x_i)$ and $F_Q(x_i)$ are functions of the CKM matrix elements and the masses of the quarks running in the loops, $x_i=m_{q_i}^2/m_W^2$. The functions $F_I^{(i)}(x_j)$ and $F_Q^{(i)}(x_j)$ have been determined by Inami and Lim \cite{Inami:1980fz}. Using their definitions one arrives at
\begin{align}
C_9^{\text{IL}}&=-\frac{\tilde{C}}{2}\frac{1}{\sin^2\theta_W} - \frac{\tilde{H}_1}{4},\\
C_{10}^{\text{IL}}&=\frac{\tilde{C}}{2} \frac{1}{\sin^2\theta_W},\label{C10IL}
\end{align}
with
\be
\tilde{C}=-4 \sum_{j=s,b} \lambda_j \bar{C}(x_j,x_d)I_q, \qquad
\tilde{H}_1= 16\sum_{j=s,b}\lambda_j \big[ \bar{F}_1(x_j,x_d)+2 \bar{\Gamma}_z(x_j,x_d) I_q\big],
\ee
where $\lambda_j=V_{cj}^* V_{uj}/(V_{cb}^* V_{ub})$, the function $\bar{C}(x_j, x_d)$ is defined in Eq. (2.14) of Ref.~\cite{Inami:1980fz}, the function $\bar{\Gamma}_z(x_j, x_d)$ in Eq. (2.7) of Ref.~\cite{Inami:1980fz}, while $\bar{F}_1(x_j, x_d)$ is defined in Eq. (B.2) of Ref.~\cite{Inami:1980fz} (note also the errata), the latter function being changed slightly, as now
\be
\bar{F}_1=Q_q \big\{\dots\big\}+I_q \dots
\ee
(i.e. the last two lines of Eq. (B.2) in Ref.~\cite{Inami:1980fz} are to be multiplied with $I_q$). 

The important thing to note is, that the variable $x_j=m_{q_j}^2/m_W^2$ is very small for $q_j=d,s,b$. The functions $\bar{C}$ and $\bar{\Gamma}_z$ are proportional to $\bar{C},\bar{\Gamma}_z\propto x_j $ and are thus very small. The function $\bar{F_1}$, on the other hand, is to the leading order $\bar{F}_1(x_j, x_d)\sim \frac{2}{3} Q_q \ln(x_j/x_d)$ which is of the order ${\cal O}(1)$. We thus arrive at
\be
V_{cb}^*V_{ub} C_9^{\text{IL}}\simeq -V_{cs}^*V_{us} 16/9 \ln \big(m_s/m_d)=-1.13 \pm 0.06,\label{C9IL}
\ee\index{renormalization group! effect on $C_9$}
where the value $m_s/m_d=17-22$ \cite{Hagiwara:pw} has been used. The value $V_{cb}^*V_{ub} C_9^{\text{IL}}$ should be compared with $V_{cb}^*V_{ub} C_9(\mu)\sim 10^{-4}$. The value of the Wilson coefficient is four magnitudes smaller than the corresponding parameter obtained without RG resummation and by neglecting QCD interactions! The reason for this discrepancy lies in the appearance of large logarithms $\ln(m_{d,s}/m_W)$ in the perturbative calculation. These have to be resummed using RG. Also, in the calculation very small scales $m_{d,s}$ appear. Neglecting QCD effects, as in the calculation outlined above, is thus not justified. Let us note, however, that the logarithm appearing in \eqref{C9IL} will be reproduced in the calculation of the inclusive modes $c\to u l^+l^-$ (cf. section \ref{ctoullbar}), as is expected, if one uses mass-independent renormalization (see appendix C of \cite{Buras:1991jm}). In the calculation of the inclusive mode at the quark level, the logarithm appears from the application of the $Q_{1,2}$ operators at the one-loop level. This contribution then dominates the rate due to the $V_{cb}^*V_{ub}$ suppression of the $Q_9$ operator. A similar situation occurs in the calculation of the exclusive modes, where inclusions of the $Q_{1,2}$ operators also dominate the rate as will be discussed in more detail in the following chapters.

Similarly one could determine the value of $C_7$ at the weak scale, arriving at the leading order expression
\be
C_7^{\text{IL}}\sim -\frac{5}{24} \sum_j \lambda_j x_j.
\ee
Using $|V_{cb}^* V_{ub}|=(1.3 \pm 0.4)\times 10^{-4}$ this leads to the value $|C_7^{\text{IL}}|\sim 10^{-3}$, which is an order of magnitude smaller than the RG improved Wilson coefficient $C_7$. Namely, the RG evolution lifts the hard GIM mechanism \index{GIM suppression} of $C_7\sim \sum_j \lambda_j x_j $ and replaces it with the logarithmic dependence on the scales $\mu\sim m_c, m_b, m_W$ involved in the RG evolution. Since in the RGE the masses $m_{s,d}$ in the QCD loops are neglected, the suppression due to the small value of $V_{cb}^* V_{ub}=-V_{cs}^* V_{us}-V_{cd}^*V_{ud}$ is still present. Again, the inclusive rate $c\to u\gamma$ \index{ctougamma@$c\to u\gamma$} is dominated by the inclusion of operators $Q_{1,2}$ at one-loop level. 

Finally, the leading order expression in terms of $x_j=m_{q_j}^2/m_W^2$ for the $C_{10}^{\text{IL}}$ coefficient \eqref{C10IL}, is 
\be
C_{10}^{\text{IL}}\simeq 2\sum_j \lambda_j x_j \frac{1}{\sin^2 \theta_W}.
\ee
Using the Wolfenstein CKM parameters $\rho=0.4, \eta=0.45, A=0.83$ and the quark masses $m_d=6$ MeV, $m_s=130$ MeV, $m_b=4.25$ GeV, we arrive at $C_{10}^{\text{IL}}=(3.9+1.7 i)\times 10^{-2}$. Note that (i) if the masses of $d,s,b$ quarks can be neglected compared to $m_W$, then $C_{10}=0$ and, that then (ii) the low energy QCD and QED interactions cannot induce a nonzero value of the $C_{10}$ Wilson coefficient. It is thus consistent with the assumptions of the OPE to set $C_{10}=0$, which will be done in the following.\index{C10@$C_{10}$}

\index{Inami-Lim calculation of $C_{7,9,10}$|)}\index{neglected QCD in $C_{7,9,10}$|)}
\index{approach, used in thesis|(} \index{effective weak interaction}
For the sake of completeness we write down at the end the effective Lagrangian of the weak interactions induced at the scales $\mu\sim m_c$, containing the operators relevant for the processes, that will be considered in chapters \ref{D0K0K0bar}, \ref{rareD} \index{C1@$C_1$}\index{C2@$C_2$}\index{C9@$C_9$}\index{C7@$C_7^{\text{eff}}$}\index{C10@$C_{10}$}\index{operators! four-fermion}\index{Q12@$Q_{1,2}$}\index{Q7@$Q_7$}\index{Q9@$Q_9$}
\be
\begin{split}
{\cal L}_{\text{eff}}= &-\frac{G_F}{\sqrt{2}} \sum_{q=d,s} V_{uq}
 V_{cq}^* \big[ C_1 Q_1^{(q)} +C_2 Q_2^{(q)}\big]\\
&+ \frac{G_F}{\sqrt{2}} V_{cb}^* V_{ub}\big[ \sum_{i=7,9,10}C_i Q_i+\sum_{i=7,9,10} C_i'Q_i'\big], \label{effective_lagr}
\end{split}
\ee
where 
\begin{subequations}\label{Qall}
\begin{align}
Q_1^{(q)}&=(\bar{u}^\alpha\Gamma_\mu q^\beta) (\bar{q}^\beta\Gamma^\mu c^\alpha), & Q_2^{(q)}&=(\bar{u}^\alpha\Gamma_\mu q^\alpha) (\bar{q}^\beta\Gamma^\mu c^\beta),\label{Q1Q2}\\
Q_7&=\frac{e}{4 \pi^2} F_{\mu \nu} m_c \bar{u}\sigma ^{\mu \nu} P_R c, & Q_7'&=\frac{e}{4 \pi^2} F_{\mu \nu} m_c \bar{u}\sigma ^{\mu \nu} P_L c,\\
Q_9&=\frac{e^2}{16 \pi^2} (\bar{u}_L \gamma^\mu c_L) (\bar{l} \gamma_\mu l), &Q_9'&=\frac{e^2}{16 \pi^2} (\bar{u}_R \gamma^\mu c_R) (\bar{l} \gamma_\mu l),\\
Q_{10}&=\frac{e^2}{16 \pi^2} (\bar{u}_L \gamma^\mu c_L) (\bar{l} \gamma_\mu \gamma_5 l), &Q_{10}'&=\frac{e^2}{16 \pi^2} (\bar{u}_R \gamma^\mu c_R) (\bar{l} \gamma_\mu\gamma_5 l), \index{Q99@$Q_{10}$}
\end{align}
\end{subequations}
 with $\Gamma^\mu=\gamma^\mu(1-\gamma_5)$, $\alpha, \beta$ the color indices written out explicitly in $O_{1,2}^{(q)}$, while the summation over the repeated indices is understood. In \eqref{effective_lagr} also the operators $Q_i'$ have been introduced. The $Q_7'$ operator is $m_u/m_c$ suppressed compared to the $Q_7$ operator in the SM and is usually neglected. The $Q_9',Q_{10}'$ operators do not arise from the SM interactions. They will be relevant, however, for the discussion of the SM extensions. In the SM calculation we set $C_7'=C_9'=C_{10}'=C_{10}=0$, while the other Wilson coefficients are listed in Table \ref{tab-Wilson} and Eq.~\eqref{C7numbers}. \index{anomalous dimension|)}\index{OPE|)}\index{penguin operators|)}\index{renormalization group|)}\index{Wilson coefficients|)}

\section{Factorization approximation}\label{Factorization}\index{factorization approximation|(}
As discussed in the previous section, the weak interactions can be described at low energies by means of an effective Lagrangian obtained through the operator product expansion and the renormalization group evolution. The effective Lagrangian for the Cabibbo allowed transitions \index{Cabibbo allowed} is\index{Q12@$Q_{1,2}$}
\be
\begin{split}
{\cal L}_{\text{eff}}=&-\frac{G_F}{\sqrt{2}} V_{cs}^* V_{ud} \big[ C_1 (\bar{u}^\alpha d^\beta)_{V-A} (\bar{s}^\beta c^\alpha)_{V-A} +C_2 (\bar{u} d)_{V-A} (\bar{s} c)_{V-A} \big]=\\
&=-\frac{G_F}{\sqrt{2}} V_{cs}^* V_{ud} \big[ C_1 (\bar{u}c)_{V-A} (\bar{s}d)_{V-A} +C_2 (\bar{u} d)_{V-A} (\bar{s} c)_{V-A} \big] \\
=&-\frac{G_F}{\sqrt{2}} V_{cs}^* V_{ud} \big[ C_1 Q_1 +C_2 Q_2 \big],\label{fact-start}
\end{split}
\ee
where the second line has been obtained using the Fierz transformation, $\alpha,\beta$ denote color indices as before, and are written out explicitly only in the first term of \eqref{fact-start}. The Wilson coefficients contain contributions from hard gluon exchanges and can be calculated perturbatively as described in the previous section. They are scale and at the NLO also renormalization scheme dependent. This dependence is canceled by the scale and renormalization scheme dependence of the local four-quark operators $Q_{1,2}$. The matrix elements in the hadronic weak transitions
\be
M_{fi}=\frac{G_F}{\sqrt{2}} V_{cs}^* V_{ud} \big[ C_1 \langle f|Q_1|i\rangle +C_2 \langle f| Q_2 |i \rangle \big],
\ee
are thus scale and scheme independent, as they should be. \index{nonperturbative effects} The nonperturbative nature of these transitions is hidden in the matrix elements $\langle f|Q_{1,2}|i\rangle$ between hadronic final and initial states. Evaluation of these elements is a very hard problem and lies at the core of all the difficulties connected with the weak transitions between hadronic states. In principle the only exact way to estimate them is to calculate them on lattice. However the problem is so involved, especially for the heavy-to-light hadron transitions, that even the ``exact'' calculations on lattice have to resort to a number of approximations. One of them, the quenching approximation, will be discussed later on, in chapter \ref{quenching-errors}. The other option is to try to use some phenomenologically or theoretically motivated approximation to calculate the weak elements in question.

The factorization approximation is a very simple but extremely useful and quite successful approximation \cite{Bauer:1986bm}. In this approach the currents appearing in the operators $Q_{1,2}$ are assumed to factor. Each of the currents is proportional to interpolating stable or quasistable hadronic fields. The approximation comes in, when these interpolating full hadronic fields are approximated in one or both of the currents by an asymptotically free hadronic field, i.e., by the ``in'' and ``out'' fields. The effective interaction \eqref{fact-start} is then \index{effective weak interaction}
\be
{\cal L}_{\text{eff}}=
-\frac{G_F}{\sqrt{2}} V_{cs}^* V_{ud} \big[ a_1 (\bar{u} d)_{V-A}^H (\bar{s} c)_{V-A}^H+a_2 (\bar{u}c)_{V-A}^H (\bar{s}d)_{V-A}^H \big]. \label{fact-lagr-allowed}
\ee 
where $(\bar{q}q)_{V-A}^H$ are the hadronized $V-A$ currents. For instance
\be
 (\bar{u} d)_{V-A}^H =-f \partial_\mu \pi^-+\cdots
\ee
with the dots representing other hadronic fields with the same quantum numbers. In the approach adopted here, the hadronized current containing charm quark is obtained using the heavy quark symmetry and is given in \eqref{current} plus terms coming from fields with the same quantum numbers. The {\it effective} Wilson coefficients $a_{1,2}$ in \eqref{fact-lagr-allowed} are in principle unknown coefficients that have to be estimated from the experimental data. For instance, for the case of the $D$ meson two-body nonleptonic decays $D\to P_1 P_2$, the decay amplitudes in the (naive) factorization approximation read \index{amplitude!in factorization approximation}
\begin{equation}
\begin{split}
M_{P_1P_2,D}^W =\hskip15mm&\\
\frac {G_F}{\sqrt 2}V_{cs}^*V_{ud} \bigg[&a_1\Big(\langle P_2| 
(\bar{u}d)_\mu | 0\rangle \langle 
P_1| (\bar{s} c)^\mu |D\rangle +\langle P_1| (\bar{u}d)_\mu | 
0\rangle\langle 
P_2| (\bar{s} c)^\mu |D\rangle+ \\
&\langle P_1P_2| (\bar{u}d)_\mu | 0\rangle\langle 0| (\bar{s} c)^\mu 
|D\rangle\Big)+ a_2\Big(\langle P_2| (\bar{s}d)_\mu | 0\rangle\langle P_1| 
(\bar{u} c)^\mu |D\rangle +\\
&\langle P_1| (\bar{s}d)_\mu | 0\rangle\langle P_2| (\bar{u} c)^\mu 
|D\rangle 
+\langle P_1P_2| (\bar{s}d)_\mu | 0\rangle\langle 0| (\bar{u} c)^\mu 
|D\rangle\Big)\bigg]. \label{fakt-2}
\end{split}
\end{equation} 
This is then compared with the experimental data on the Cabibbo allowed \index{Cabibbo allowed} decays $D\to K\pi$, arriving at the values $a_1\approx 1.3 \pm 0.1$, $a_2\approx -0.55\pm0.1$ \cite{Bauer:1986bm,Kamal:1994ep}. In this analysis the final state interactions \index{final state interactions} have to be taken into account. The naive factorization approximation as explained above, is taken to be valid only in the weak vertex, for the so called {\it bare} amplitudes. The outgoing hadronic states then interact strongly, which can lead to elastic and inelastic rescattering effects. Because $D$ mesons lie close to the resonance region, a number of $s$ and $t$ channel resonances can in principle contribute. These effects are especially important for the $K\pi$, $K\eta^{(')}$ states because of the presence of $S=1$ scalar meson resonance $K^0(1950)$ (for more details see \cite{Buccella:1994nf,Buccella:1996uy,Fajfer:1999hh,Cheng:2002wu,Cheng:2000fd,Elaaoud:1999pj,Zenczykowski:1996bk,Rosner:1999xd,Rosner:1999zm,Gronau:1999zt}, for the final state interactions in the $D\to PV$ decays see also \cite{Li:2002pj,Kamal:gu,Sharma:1995xt} and for the doubly Cabibbo suppressed \index{Cabibbo doubly suppressed}  $D\to K\pi$ decays \cite{Chiang:2001av}). \index{resonant contributions}

On the other hand, starting from the weak Lagrangian \eqref{fact-start} one can expect that
\be
a_1\simeq C_2+\frac{1}{N_c} C_1, \qquad a_2\simeq C_1+\frac{1}{N_c} C_2. \label{coeff-Nc}
\ee
To see this, consider for instance the $Q_1$ operator in \eqref{fact-start}\index{Q12@$Q_{1,2}$}
\be
\begin{split}
Q_1&=(\bar{u}^\alpha c^\alpha)_{V-A} (\bar{s}^\beta d^\beta)_{V-A}\\
&=(\bar{s}^\beta c^\alpha)_{V-A} (\bar{u}^\alpha d^\beta)_{V-A}\\
&=\frac{1}{N_c} (\bar{s}^\alpha c^\alpha)_{V-A} (\bar{u}^\beta d^\beta)_{V-A}+2 (\bar{s}T^a c)_{V-A} (\bar{u}T^a d)_{V-A},
\end{split}
\ee
where first Fierz transformation and then the $SU(N_c)$ completeness relation \index{completeness relation for $SU(N_c)$}
\be
\delta_{\alpha \beta} \delta_{\gamma \delta}=\frac{1}{N_c} \delta_{\alpha \delta}\delta_{\beta \gamma} +2 T^a_{\alpha \delta} T^a_{\gamma \beta},
\ee
have been used (cf. \eqref{SU3Iden}). The weak Lagrangian \ref{fact-start} can thus be written in two equivalent ways\index{Q12@$Q_{1,2}$}
\be
\begin{split}
{\cal L}_{\text{eff}}=&-\frac{G_F}{\sqrt{2}} V_{cs}^* V_{ud} \left[ \left(C_1+\frac{1}{N_c}C_2\right) Q_1 +2 C_2 Q_1^{(8)} \right]\\
=&-\frac{G_F}{\sqrt{2}} V_{cs}^* V_{ud} \left[ 2 C_1 Q_2^{(8)} +\left(C_2+\frac{1}{N_c}C_1\right) Q_2 \right],
\label{fact-two-ways}
\end{split}
\ee
with
\be
\begin{split}
Q_1^{(8)}& = (\bar{u}T^a c)_{V-A} (\bar{s}T^a d)_{V-A},\\
 Q_2^{(8)}&= (\bar{s}T^a c)_{V-A} (\bar{u}T^a d)_{V-A},\label{Q8}
\end{split}
\ee
the products of two colored currents. In the factorization approximation these are then expected to give vanishing contributions between the colorless hadronic states, leading to the naive estimates for the $a_{1,2}$ effective Wilson coefficients as given in \eqref{coeff-Nc}. Using the NLO values for the Wilson coefficients $C_1(m_c)=-0.49\pm 0.15$, $C_2=1.26\pm0.10$ (see Table \ref{tab-Wilson} and also Ref.~\cite{Buras:1994ij}), where the naive dimensional regularization has been used, one arrives at $a_1^{\text{NDR}}\approx 1.10\pm 0.05$ and $a_2^{\text{NDR}}\approx -0.07\pm0.12$. The value of $a_2^{\text{NDR}}$ is phenomenologically unacceptable. One gets a phenomenologically much more satisfactory description in the $N_c\to\infty$ limit, where the $1/N_c$ terms are dropped. Then $a_1=C_2$ and $a_2=C_1$, so that
\be
a_1=1.26\pm0.10, \qquad a_2=-0.49\pm0.15. \label{coeff-a1a2}
\ee
Some arguments for this modified approach to the factorization can be given on the theoretical side, either in the $1/N_c$ expansion \cite{Buras:1985xv} or by using the QCD sum rules \cite{Blok:1992na,Blok:1992hw,Blok:1992he}, however the situation is not yet completely clear. Nevertheless, given the simplicity of the approach and its relative phenomenological success, we will use in the following the factorized weak Lagrangian with the effective Wilson coefficients given in \eqref{coeff-a1a2}. The effective Lagrangian for the Cabibbo once suppressed decays is thus \index{Cabibbo once suppressed}
\be
{\cal L}_{\text{eff}}=
-\frac{G_F}{\sqrt{2}}\sum_{q=d,s} V_{cq}^* V_{uq} \big[a_1 (\bar{u} q)_{V-A}^H (\bar{q} c)_{V-A}^H+ a_2 (\bar{u}c)_{V-A}^H (\bar{q}q)_{V-A}^H\big], \label{fact-lagr}
\ee
with the effective Wilson coefficients $a_{1,2}$ given in \eqref{coeff-a1a2}. Possible penguin contributions have been neglected as explained in the previous section. Note also, that the $a_{1,2}$ effective Wilson coefficients obtained from $C_{1,2}(\mu)$ in \eqref{coeff-Nc} (with $N_c$ terms dropped) are scale and scheme dependent. The hadronized operators appearing in \eqref{fact-lagr} on the other hand are not. At the end one thus ends up with the matrix elements, and consequentially observables such as decay widths, that are scale and scheme dependent, which is a serious theoretical downsize of the approach. This problem has been discussed in detail in Ref.~\cite{Buras:1994ij}. It can be, however, circumvented entirely, if one views the effective coefficients \eqref{coeff-a1a2} as phenomenologically determined parameters, which are then of course scale and scheme independent.

The idea of factorization has recently received a lot of attention due to the theoretical work of two groups, the approach of the QCD factorization \index{QCD factorization} \cite{Beneke:1999br,Beneke:2000ry,Beneke:2001ev} and the pQCD approach \cite{Li:1994ck,Li:1994iu,Yeh:1997rq,Cheng:1999gs,Chang:1996dw}. The underlying physical picture of these approaches is the idea of color transparency \cite{Bjorken:kk,Dugan:1990de,Politzer:au}, which is effective in the heavy meson decays with energetic final decay products. As an example consider the case of $B\to \pi \pi$. The fast moving final mesons produced by the point-like source (the local operators in the OPE expansion)\index{OPE} decouple from the soft QCD interactions. Contributions of the soft gluons are suppressed by $\Lambda_{\text{QCD}}/m_b$. The QCD factorization approach gives rigorous results valid in the heavy quark limit to the leading power in $\Lambda_{\text{QCD}}/m_b$, but to all orders in the perturbation theory. These ideas have been further developed for the heavy-to-heavy transitions in \cite{Bauer:2001cu}. The application of the above formalism is, however, not possible for the $D$ meson nonleptonic decays as here the energy release is much smaller than in the case of $B$ mesons. Nevertheless, the factorization procedure has been applied to the nonleptonic $D$ decays in a phenomenologically successful way as discussed above. In this sense nonleptonic $D$ decays are halfway between $B$ and $K$ nonleptonic decays. Namely, it
is well known that the factorization does not work in the nonleptonic $K$
decays \cite{Buras:1993dy,Buras:1996dq,BEF,Cheng:1999dj}. 
\index{approach, used in thesis|)}

\chapter{Nonfactorizable contributions to the decay mode $D^0\to K^0 \bar{K}^0$}\label{D0K0K0bar}\index{D0K0K0Bar@$D^0\to K^0 \bar{K}^0$|(}\index{decay!$D^0\to K^0 \bar{K}^0$|(}
The decay mechanism of the weak nonleptonic $D^0$ decays has motivated
numerous studies, e.g., \cite{Bauer:1986bm,Pham:1987rj,Dai:1999cs,Lipkin:es,Gerard:1998un,Kamal:1995fr,Zenczykowski:1999cu,Terasaki:1998tf}. For the nonleptonic decays of $D$ mesons, as well as for $K$'s and $B$'s,
the {\em factorization} hypothesis explained in section \ref{Factorization} has commonly been used. In this section we discuss nonfactorizable contributions to $D$ decays,
in particular in the decay mode $D^0\to K^0\bar{K^0}$. This decay mode has been advertised as an interesting probe of the nonperturbative physics in weak decays long time ago \cite{Pham:1987rj}. Additional
motivation to consider this decay mode
 comes from the recent
experimental searches for the CP violating asymmetry in
 $D^0 \to K_S K_S$ \cite{Bonvicini:2000qm}. \index{CP@$CP$}

In $D$ decays the factorization hypothesis works
reasonably well, if one is interested in an order of magnitude estimate,
 but it does not reproduce experimental data completely. For example, a
naive
application of the factorization in the charm decays
leads to the rates for the $D^0 \to \pi^0
\bar K^0$,
$D^0 \to \pi^0 \pi^0$, $D^0 \to K^+ K^-$, $D^0 \to \pi^+ \pi^-$
decays which are
too strongly suppressed (see, e.g., \cite{Kamal:gu,Alexander:1990nf,Ablikim:2002ep,Buccella:1994nf}). Consideration of either the final state interactions through \index{final state interactions}resonant or nonresonant rescattering and/or of other nonfactorizable mechanisms is thus mandatory. 
Moreover, and this is the important point of the present chapter,
 in $D^0\to K^0\bar{K^0}$ a naive application of factorization misses
completely, predicting a vanishing branching ratio, in contrast with the
experimental situation. \index{resonant contributions}

To see this, note that at tree level the $D^0\to K^0\bar{K^0}$ decay
might
occur through two \index{annihilation diagrams} annihilation diagrams \cite{Bauer:1986bm} with either $c\to s$ or $c\to d$ transition. However, they cancel each other by the \index{GIM suppression}
GIM mechanism. Moreover, in the factorization limit, the amplitude \index{amplitude! in $D^0\to K^0\overline{K}^0$} is
proportional to
\begin{equation}
\langle K^0 \bar K^0| V_\mu |0\rangle \langle 0 | A^\mu | D^0\rangle
\simeq (p_{K^0} - p_{\bar K^0})_\mu \, f_D p_{D}^\mu= 0.
\label{eq-9}
\end{equation}
In many of the studies (e.g.
\cite{Pham:1987rj,Dai:1999cs,Lipkin:es,Gerard:1998un,Zenczykowski:1999cu}) this decay has been understood as a
result of the final state interactions (FSI). \index{final state interactions}
In the analysis of Ref.~\cite{Pham:1987rj}
the rescattering mechanism included $K^+ K^-$ and $\pi^+ \pi^-$
states leading to a branching ratio
$\text{Br}(D^0 \to K^0 \bar K^0)$ $= \frac{1}{2} \text{Br}(D^0 \to K^+ K^-)$.
Experimental data on the other hand are \cite{Hagiwara:pw} $\text{Br}(D^0 \to K^0
\bar K^0)
= (7.1\pm 1.9)\times 10^{-4}$ and $\text{Br}(D^0 \to K^+ K^-)=
 (4.12\pm 0.14)\times 10^{-3}$.
 A recent investigation of the $D^0\to K^0\bar {K^0}$ decay mode
performed in \cite{Dai:1999cs}
 has focused on the $s$ channel and $t$ channel one
particle exchange contributions. The $s$ channel contribution has been
taken into account through a poorly known scalar meson $f_0(1710)$
and was found to be very small, while the one particle $t$-exchanges
yielded higher contributions, with pion exchange being the highest. In
the approach of \cite{Terasaki:1998tf} the $D^0 \to K^0 \bar K^0$ decay was
realized through the scalar glueball
or glue-rich scalar meson.

\index{approach, used in thesis|(}
We will adopt the approach of the effective Lagrangians as explained in chapter \ref{HQET}. The subsequent analysis has been published in \cite{Eeg:2001ip,Eeg:2001un}. Because the ${\cal O}(p)$ (factorizable) contribution is zero
\eqref{eq-9},
we will try to approach to the $D^0 \to K^0 \bar K^0$
decay
systematically to ${\cal O}(p^3)$. We do this by including first
 the nonfactorizable
 contributions coming from the chiral loops. In the weak vertex the factorization hypothesis will be used, leading to the weak transitions of the type $D^0 \rightarrow \pi^+ \pi^-$ and $D^0 \rightarrow K^+ K^-$ (see Figs. \ref{fig-2}-\ref{fig-5}). In this sense the approach is similar to the factorization hypothesis as put forward in Ref.~\cite{Bauer:1986bm} for the Cabibbo allowed \index{Cabibbo allowed} $D$ decays. In Ref.~\cite{Bauer:1986bm} factorization was assumed for the weak vertex, leading to the {\it bare} amplitudes, that are then modified by the FSI. The final state interactions correspond to the diagram $F_4$ on Figure \ref{fig-5}. \index{final state interactions} A number of additional chiral loop diagrams will be considered in this approach. In addition, we consider the gluon condensate contributions, also of
${\cal O}(p^3)$, which we calculate within the Heavy-Light Chiral Quark Model (HL$\chi$QM) framework. The HL$\chi$QM is an extension of the effective Lagrangian approach of chapter \ref{HQET}, as it models also the interactions of light pseudoscalars and heavy mesons with quarks. By integrating out the quark degrees of freedom one is able to reproduce the effective Lagrangians of chapter \ref{HQET}. A brief introduction to the HL$\chi$QM will be given in section \ref{HeavyLight}, while a more detailed description can be found in \cite{Hiorth:2002pp}.

We emphasize \index{difficulties!of approach}
that one cannot a priori expect for the chiral expansion to work to a good precision in the process $D \rightarrow K
\bar{K}$,
 because the energy release is $p =788$
MeV and hence $p/\Lambda_\chi$ (for $\Lambda_\chi\geq 1$ GeV)
is close to unity. However, the
leading contributions, that we will consider, do turn out to
describe the data reasonably well.
The next to leading ${\cal O}(p^5)$ terms might
be almost
of the same order of magnitude compared to the ${\cal O}(p^3)$ terms,
with a weak suppression of the order
$p^2/\Lambda_\chi^2$.
On the other hand, the inclusion of
${\cal O}(p^5)$ order in this framework is not straightforward.
Before doing loop calculations at that order,
one has to find a reliable framework to include
 light resonances $\rho$, $K^*$, $a_0(980)$, $f_0(975)$, etc.\index{resonant contributions}
Usually the light resonances
are treated using hidden gauge symmetry (see, e.g., \cite{Casalbuoni:1996pg}). This is not easily reconciled with the chiral perturbation theory.
Even if the light resonances were included in the effective 
Lagrangian, one would face the problem of determining their
couplings
to the rest of the heavy and light states. The poorly
known scalar resonances would introduce a rather
large uncertainty \cite{Dai:1999cs}. Right now, the
consistent calculation of this or higher orders does not seem to be
possible.
Still, the amplitude of the $D^0 \to K^0 \bar K^0$ decay, calculated
within our framework to the order ${\cal O}(p^3)$ turns out to be in agreement with the experimental result. 
Note also, that $1/m_Q$ terms have been omitted in the calculation.

\index{approach, used in thesis|)}\index{factorization approximation|)}

\section{Chiral loop contributions}\label{chiral-loops}\index{contributions!chiral loop|(}\index{nonfactorizable contributions|(}
As discussed above, in the factorization limit there are no contributions \index{factorization approximation}
to the $D^0\to K^0 \bar{K^0}$ decay at tree level \eqref{eq-9}. The observation of a partial
decay width $\text{Br}(D\to K^0\bar{K^0})=(7.1\pm 1.9)\times 10^{-4}$ on the
other hand implies, that we can expect sizable contributions at the one
loop level. Calculations to one loop in the framework of combined
chiral
perturbation theory and heavy quark symmetry, the Heavy Hadron Chiral Perturbation Theory (HH$\chi$PT), involve a
construction of the most general effective Lagrangian, that has the
correct
symmetry properties, in order to make the renormalization work. This construction together with the chiral counting has been explained in chapter \ref{HQET}. 

\begin{figure}
\begin{center}
\epsfig{file=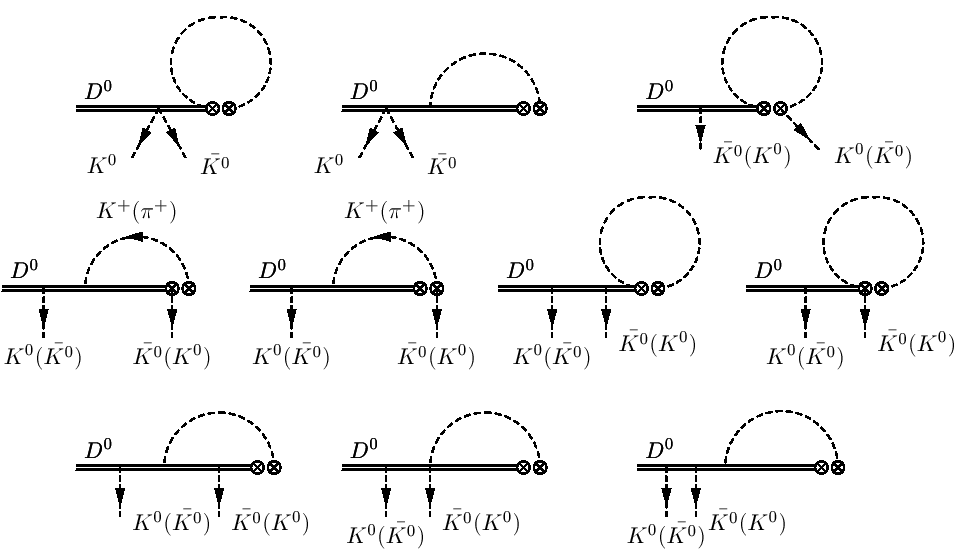, height=6cm}
\caption{\footnotesize{Diagrams, that give zero contribution, since the relevant
vertices
appearing in the heavy meson chiral Lagrangian \eqref{eq-8} are zero.
 The double line represents
heavy meson $D$ or $D^*$, while dashed lines denote pseudo-Goldstone
bosons. }}\label{fig-2}
\end{center}
\end{figure}

\begin{figure}
\begin{center}
\epsfig{file=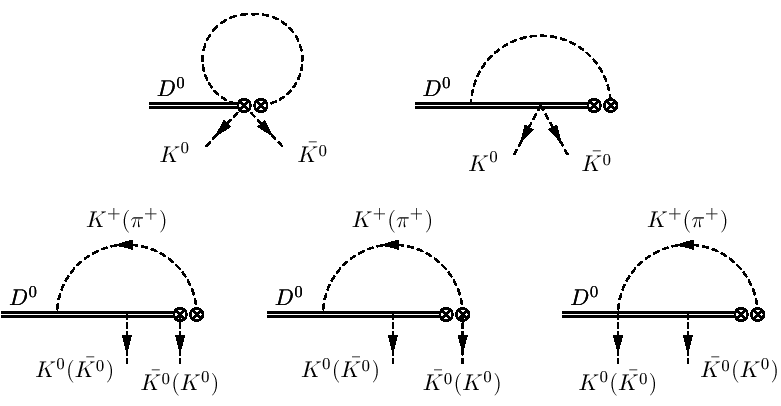, height=4cm}
\caption{\footnotesize{Diagrams, that give zero contributions, since the loop
integrals are zero. The double line represents the
heavy meson $D$ or $D^*$, while the dashed lines denote the pseudo-Goldstone
bosons.}}\label{fig-3}
\end{center}
\end{figure}

\begin{figure}
\begin{center}
\epsfig{file=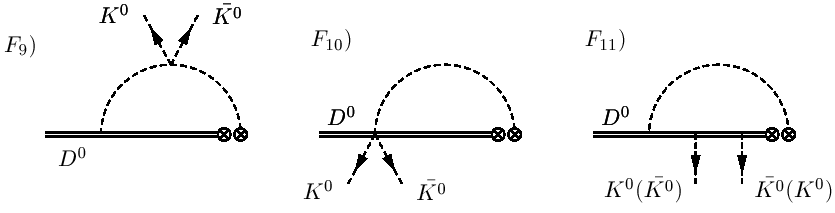, height=2.3cm}
\caption{\footnotesize{Power suppressed diagrams (neglected in the
calculation).}}\label{fig-4}
\end{center}
\end{figure}

\begin{figure}
\begin{center}
\epsfig{file=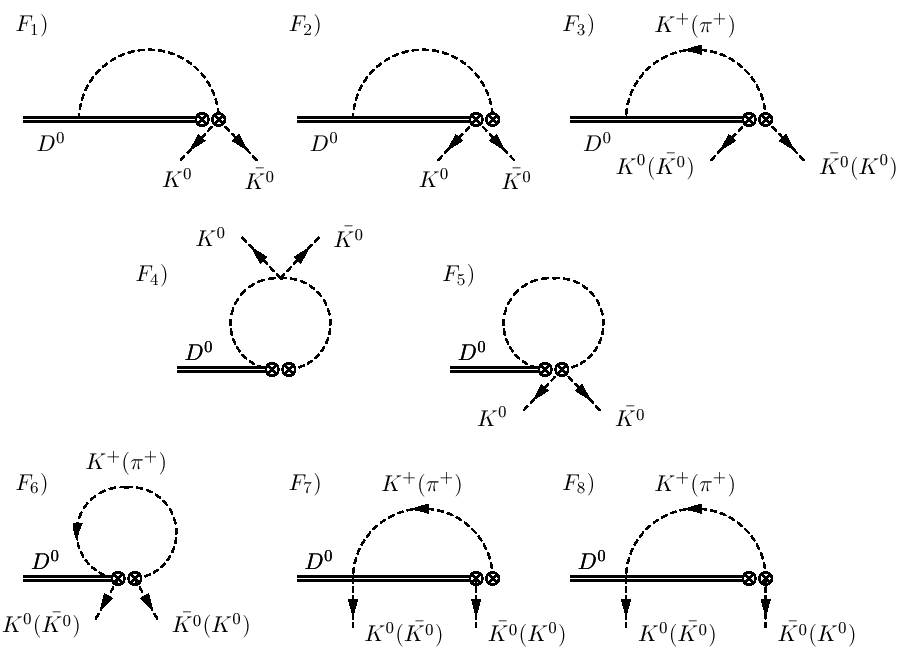, height=6.5cm}
\caption{\footnotesize{The nonzero diagrams in the $D^0 \to K^0 \bar{K^0}$
decay.}}\label{fig-5}
\end{center}
\end{figure}

The weak Lagrangian relevant for the $D^0\to K^0 \bar{K}^0$ decay is \index{Q12@$Q_{1,2}$}
\be
\begin{split}
{\cal L}_{\text{eff}}=&-\frac{G_F}{\sqrt{2}}\sum_{q=d,s} V_{cq}^* V_{uq} \left[ \left(C_1+\frac{1}{N_c}C_2\right) Q_1^{(q)} +2 C_2 Q_1^{q(8)} \right]\\
=&-\frac{G_F}{\sqrt{2}}\sum_{q=d,s} V_{cq}^* V_{uq} \left[ 2 C_1 Q_2^{q(8)} +\left(C_2+\frac{1}{N_c}C_1\right) Q_2^{(q)} \right],
\label{fact-two-ways-Cabibbo-suppressed}
\end{split}
\ee
with
\be
\begin{split}
Q_1^{q(8)}& = (\bar{u}T^a c)_{V-A} (\bar{q}T^a q)_{V-A},\\
 Q_2^{q(8)}&= (\bar{q}T^a c)_{V-A} (\bar{u}T^a q)_{V-A},\label{Q8-supp}
\end{split}
\ee
while $Q_1^{(q)}$ and $Q_2^{(q)}$ are defined in \eqref{Q1Q2}, and consist of a product of two color-singlet currents. The operators $Q_{1,2}^{(q)}$ in the weak vertex will be evaluated using factorization.\index{factorization approximation} The nonfactorizable contributions will then arise from the chiral loops shown on Figs. \ref{fig-2}-\ref{fig-5}. The other nonfactorizable contributions come from the operators with the colored currents $Q_{1,2}^{q(8)}$, which will be evaluated in the next section using chiral quark model. For convenience we define $a_1^{\text{NDR}}=(C_2+C_1/N_c)\sim 1.10$ and $a_2^{\text{NDR}}=(C_1+C_2/N_c)\sim -0.07 $, with $N_c=3$ the number of colors, while the numerical values are given using $C_{1,2}$ calculated at $m_c$ in the NDR (cf. footnote to the text above \eqref{C1C2mW}). Note, that these are not the same as the phenomenologically motivated values of $a_{1,2}$ \eqref{coeff-a1a2} of the factorization approximation, where the contributions of the $Q_{1,2}^{q(8)}$ operators in the weak vertex are then neglected. On the contrary, we do take the contributions of colored currents into account, and therefore use the $a_{1,2}^{\text{NDR}}$ values of Eq.~\eqref{coeff-Nc}.

The loop diagrams are divergent and have to be regulated. We work both in the strict $\overline{\text{MS}}$ renormalization scheme, where we \index{loop! renormalization}\index{MS@$\overline{\text{MS}}$}\index{renormalization scheme!$\overline{\text{MS}}$}
put $\bar{\Delta}=\frac{2}{\epsilon}-\gamma+\ln(4\pi)+1\to 1$ in the loop calculations as well as in the Gasser-Leutwyler (GL) renormalization scheme $\bar{\Delta}\to0$.\index{renormalization scheme!Gasser-Leutwyler} The first choice is the same as the one made by Stewart in \cite{Stewart:1998ke}, while the other was made by the authors of Ref.~\cite{Casalbuoni:1996pg}. The renormalization prescription determines
the appropriate renormalization of couplings in the ${\cal O}(p^3)$
effective
Lagrangian as discussed in section \ref{coupling}. Using two prescriptions makes possible to estimate the size of the counterterms, that are otherwise neglected. Further, we consider
 only contributions coming from the $a_1^{\text{NDR}}$ part of the weak Lagrangian,
 as $a_2^{\text{NDR}}$ is suppressed compared to $a_1^{\text{NDR}}$ \eqref{coeff-a1a2}(see also \cite{Buras:1994ij}).\index{counterterms!contributions of}\index{counterterms!values of}

Writing down the most general one loop graphs with two outgoing
Goldstone
bosons, $K^0$ and $\bar{K^0}$, one arrives at 26 Feynman
diagrams. A number of these give zero contributions or are suppressed and are shown on
Figures \ref{fig-2}, \ref{fig-3}, \ref{fig-4}. The graphs that do
contribute to the $D^0\to K^0\bar{K^0}$ decay are shown on Fig.~\ref{fig-5}.
Note that the factorizable loops, which renormalize vertices are omitted, as they contribute only at higher order in the chiral expansion
(they do appear, however, in the loop determination of the $\alpha$
coupling
related to $f_D$. For more details see section \ref{coupling}.)

To shorten the notation, the common factors in the $S$ matrix have been factored out, so that the amplitude is written as \index{amplitude! in $D^0\to K^0\overline{K}^0$} 
\begin{equation}
M(D^0 \rightarrow K^0 \bar{K^0}) \; = \; - \,
\frac{G_F}{\sqrt{2}} \, a_1^{\text{NDR}} \, V_{us}\,V^*_{cs} \,
\frac{F}{8\pi^2}\sqrt{m_D},\label{eq-12}
\end{equation}
where
$F=\sum_n F_n$ is the sum of the amplitudes corresponding to the graphs
on Fig.
 \ref{fig-5}. In \eqref{eq-12} we have also neglected the contributions of order $V_{ub} V_{cb}^*$, so that we use $V_{us}V_{cs}^*=-V_{ud}V_{cd}^*$. The partial decay
width
for the decay $D^0 \to K^0\bar{K^0}$ is then
\begin{equation}
\Gamma_{D^0\to
K^0\bar{K^0}}=\frac{1}{2\pi}\frac{G_F^2}{8m_D}
\left(a_1^{\text{NDR}}\right)^2 |V_{us} \, V^*_{cs}|^2
\frac{|F|^2}{(8\pi^2)^2}
|\vec{p}|,\label{eq-2}
\end{equation}
where $\vec{p}$ is the $K^0$ three-momentum in the $D^0$ rest frame
\begin{equation}
|\vec{p}|=\frac{1}{2}\sqrt{m_D^2-4m_K^2} \; \,.
\end{equation}
The nonzero amplitudes corresponding to the graphs on Fig.~\ref{fig-5}
are \index{amplitude! in $D^0\to K^0\overline{K}^0$} 
\begin{align}
F_1&+F_2+F_3=\frac{g \alpha}{f^2}\frac{13}{4} \big[ \bar{B}_{00}(m_\pi
,\Delta_d^*)-\bar{B}_{00} (m_K,\Delta_s^*)\big],\label{eq-3}\\
\begin{split}
F_4&=-\frac{\alpha}{3 f^2}\frac{m_D}{2} \Bigl\{\big(m_D^2-2
m_K^2\big)\big[B_0(m_D^2,m_K^2, m_K^2)-B_0(m_D^2,m_\pi^2,m_\pi^2)\big]\Bigr.+\\
&\qquad\qquad+m_D^2\big[B_{11}(m_D^2,m_\pi^2,m_\pi^2)-B_{11}(m_D^2, m_K^2, m_K^2)\big]+\\
&\qquad\qquad\Bigl.+\big[B_{00}(m_D^2,m_\pi^2, m_\pi^2)-B_{00}(m_D^2, m_K^2, m_K^2)\big]+(m_\pi^2
-m_K^2)B_0(m_D^2,m_\pi^2,m_\pi^2)\Bigr\},\label{eq-5}
\end{split}\\
F_5&+F_6=-\frac{\alpha m_D}{f^2}\frac{7}{24}\big[A_0(m_\pi^2)-A_0(m_K^2)\big],\\
\begin{split}
F_7&+F_8=-\frac{\alpha}{4
f^2}\Bigl\{\Bigr.\bar{B}_{00}(m_K,\tilde{\Delta}_d)+
\bar{B}_{11}(m_K,\tilde{\Delta}_d)-
\bar{B}_{00}(m_\pi,\tilde{\Delta}_s)-\bar{B}_{11}(m_\pi,\tilde{\Delta}_s)\\
&\qquad\qquad+m_D\Delta_d\bar{B}_0(m_K,\tilde{\Delta}_d)-
m_D\Delta_s\bar{B}_0(m_\pi,\tilde{\Delta}_s)+
\frac{m_D}{2\tilde{\Delta}_d}A_0(m_K^2)-\frac{m_D}{2
\tilde{\Delta}_s}A_0(m_\pi^2)\Bigl.\Bigr\},
\end{split}\label{eq-4}
\end{align}
where $\Delta_q^{(*)}=m_{D_q^{(*)}}-m_{D^0}$ and
$\tilde{\Delta}_q=m_D/2+\Delta_q$ for $q=d,s$.
Note that $\tilde{\Delta}_q$ are
of the order $m_D/2$, a consequence of relatively high momenta
flowing in the loops of graphs $F_7, F_8$. The one and two point functions $A_0(m^2)$,$B_{0,00,11}(k^2,m^2,m^2)$,
$\bar{B}_{0,00,11}(m,\Delta)$ appearing in the amplitudes
\eqref{eq-3}-\eqref{eq-4} were defined in section \ref{Notational-conventions}. Explicit expressions can be found in section \ref{One-two-point} and in appendix \ref{app-loops}.

It should be noted that in Eqs.~\eqref{eq-3}-\eqref{eq-4} all the
expressions vanish in the exact $SU(3)$ limit, where $m_K \to
m_{\pi}$ and $\Delta_s \to \Delta_d$, $\tilde{\Delta}_s \to
\tilde{\Delta}_d$. This shows explicitly, that the $D^0 \to K^0
\bar{K^0}$
decay mode is a manifestation of the $SU(3)$ breaking effects (as already
noted by H. Lipkin \cite{Lipkin:es}, if $U$ symmetry is exact, then
$\Gamma(D^0\to K^0\bar K^0)=0$).

The amplitudes shown on Figs.~\ref{fig-2}, \ref{fig-3}, \ref{fig-4} are
either exactly zero or are suppressed by powers of $1/m_D$ and
$g$. The amplitudes corresponding to the diagrams on
 Figs.~\ref{fig-2}, \ref{fig-3} are zero due to symmetry reasons
(because there are no such
couplings in the heavy sector chiral Lagrangian \eqref{eq-8}, or because
of Lorentz covariance), while the amplitudes $F_9$, $F_{10}$ and
$F_{11}$ shown on Fig.~\ref{fig-4} are power suppressed.
An analysis of the loop integrals leads to the conclusion
 that $F_9\sim g \bigl(\tilde{q}/m_D\bigr)^2 F_4$, $F_{10}\sim g
\bigl(\tilde{q}/m_D\bigr) F_4$ and $F_{11}\sim g^3
\bigl(\tilde{q}/m_D\bigr) F_4$,
where $\tilde{q}$ is a typical loop momentum less than $m_D/2$, so that the
suppression need not be substantial. However,
a direct evaluation of the amplitude
$F_{10}$
shows, that it is about 10 times smaller than $F_4$. 
Therefore, in our numerical calculation we neglect contributions
of $F_9$, $F_{10}$ and $F_{11}$. Numerical results are listed in Table \ref{tab-1}, section \ref{Results-nonfact}.

\begin{figure}
\begin{center}
\epsfig{file=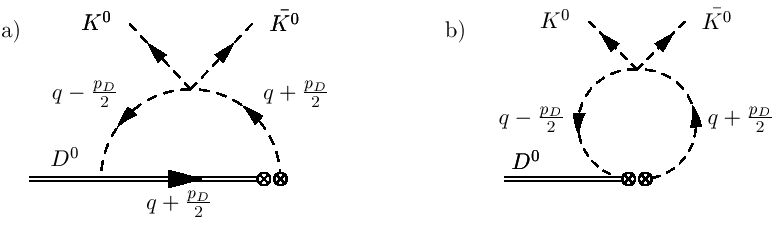, height=3cm}
\caption{\footnotesize{The momenta flowing in the graphs corresponding to a) the power
suppressed $F_9$ amplitude
 and b) the leading contribution $F_4$
amplitude. }}\label{fig-1}
\end{center}
\end{figure}
\index{contributions!chiral loop|)}

\section{The nonfactorizable color-current contributions}\label{HeavyLight} \index{color-currents, contributions of|(}\index{gluon condensate|(}
In this section we will estimate the contributions of $Q_{1,2}^{q(8)}$ operators in the weak Lagrangian \eqref{fact-two-ways-Cabibbo-suppressed}. In the factorization limit the product of colored currents does not contribute at the
meson level, as mesons are color singlet objects. At quark level, however, the colored currents can contribute through the gluon
 condensate. In order to
 estimate this contribution, we have to establish the connection between
the underlying quark-gluon dynamics and the meson level picture. This is
 done through the use of the Heavy-Light Chiral Quark Model (HL$\chi$QM).
 
\subsection{The Heavy-Light Chiral Quark Model
}\label{Quark-model}\index{Heavy-Light Chiral Quark Model}\index{HL$\chi$QM}
In the $\chi$QM \cite{Weinberg:1978kz,Cronin:1967jq,Manohar:1983md,Bijnens:1985ry,Diakonov:1987ty,Espriu:1989ff},
the light quarks ($u, d, s$) couple to the would-be Goldstone octet
mesons ($K, \pi, \eta$) in a chiral invariant way. All effects
are in principle calculable in terms of physical quantities and a few
model dependent parameters, the quark condensate, the gluon
condensate
and the constituent quark mass \cite{BEF,Pich:1990mw,epb}. 
 Among many approaches the Chiral Quark Model
 ($\chi$QM)\cite{Pich:1990mw} was shown to be able to accommodate the
intriguing $\Delta I = 1/2$ rule in the
$K \to \pi \pi$ decays, as well
as the CP violating parameters, by systematic involvement of the soft
gluon emission forming gluon condensates and chiral
loops at the order ${\cal O} (p^4)$ \cite{BEF}. Also, in the ``generalized factorization'' it was shown \cite{Cheng:1999dj}, that
 the inclusion of gluon condensates is important
 in order to understand the $\Delta I = 1/2$ rule in the $K \rightarrow 2 \pi$ decays. 

As the $\chi$QM approach successfully indicated the main
 mechanisms in the $K \to \pi \pi$
decays, it seems worthwhile to investigate the decays of charm mesons
within a similar framework. In the case of $D$ meson decays one
has to extend the ideas of the $\chi$QM to the sector involving a
heavy quark ($c$) using the chiral symmetry of the light degrees of freedom
as well as the heavy quark symmetry. This leads to the formulation
of the Heavy-Light Chiral Quark Models (HL$\chi$QM) \cite{Hiorth:2002pp,Bardeen:1993ae,Ebert:1994tv,Deandrea:1998uz,Polosa:2000ym}.

The Lagrangian of the HL$\chi$QM is 
\begin{equation}
{\cal L}={\cal L}_{\text{HQ}}+{\cal L}_{\chi \text{QM}} + {\cal L}_{\text{Int}},
\label{totlag}
\end{equation}
where 
\begin{equation}
{\cal L}_{\text{HQ}}=\overline{Q}_v \, i v \cdot D \, Q_v
+ {\cal O}(m_Q^{-1}),
\label{LHQET}
\end{equation}
is the leading order Lagrangian of the heavy quark effective theory \cite{Neubert:1993mb} (cf. Eq.~\eqref{heavyquark}), with $v^\mu$ the heavy quark velocity and $D_\mu$ the covariant derivative containing gluon field. The chiral quark model Lagrangian ${\cal L}_{\chi \text{QM}}$ is 
\be
{\cal L}_{\chi \text{QM}}=\sum_n \bar{\Psi}^{(n)}\big(i\gamma^\mu D_\mu-m^{(n)}\big) \Psi^{(n)}-m_\chi \big(\bar{\Psi}_R \Sigma \Psi_L +\bar{\Psi}_L \Sigma^\dagger \Psi_R\big).
\ee
In comparison to the usual QCD Lagrangian \eqref{QCD}, the chiral quark model has an additional term proportional to the constituent quark mass $m_\chi$. This is taken to be of the order $m_\chi\simeq 200$ MeV. As in the discussion of the chiral perturbation theory in section \ref{CHPTsection}, the Goldstone boson degrees of freedom are factored out from the quark fields
\be
\Psi(x)=e^{i \gamma_5 \phi(x)} \tilde{\Psi}(x),
\ee
with $\Psi=\{ \Psi^{(n)}\}$ a vector of quark fields. 
The $\chi$QM Lagrangian is then
\be
{\cal L}_{\chi \text{QM}}=\bar{\tilde{\Psi}}i\gamma^\mu\big(D_\mu +{\cal V}_\mu -i \gamma_5 {\cal A}_\mu\big) \tilde{\Psi} -m_\chi \bar{\tilde{\Psi}}\tilde{\Psi} -\bar{\tilde{\Psi}}\widetilde{\cal M}_+ \tilde{\Psi},\label{chqmR}
\ee
where 
\be
\widetilde{\cal M}_+=\xi m\xi P_R+\xi^\dagger m\xi^\dagger P_L,
\ee
with $m=\diag(m^{(n)})$ the diagonal matrix of quark masses and $P_{R,L}=\frac{1}{2}(1\pm \gamma_5)$ the chiral projection operators. The vector and the axial vector currents ${\cal V}_{\mu}$ and
${\cal A}_\mu$ are defined in \eqref{defVA}. 

In the heavy-light case, the generalization of the
 meson-quark interactions in the pure light sector $\chi$QM
is given by the following $SU(3)_V$
invariant Lagrangian \cite{Hiorth:2002pp,Bardeen:1993ae,Ebert:1994tv,Deandrea:1998uz}
\begin{equation}
{\cal L}_{\text{Int}} =
- G_H \, \left[ \bar{\tilde{\Psi}}_a \, \overline{H}_{va} \, Q_v \,
 + \, \overline{Q}_v \, H_{va} \, \tilde{\Psi}_a \right],
\label{Int}
\end{equation}
with $H_{va}$ the heavy meson field \eqref{heavyfield}. The dependence on heavy-quark velocity $v$ is denoted explicitly, while $a$ is the flavor index. The unknown constant $G_H$ can be related to constants $\alpha$, $g$ of HH$\chi$PT as described below (cf. Eqs.~\eqref{GHlog}-\eqref{beta}).

The weak currents have the usual form, except that the Goldstone bosons are factored out. The weak current with two light quarks is
\begin{equation}
\bar q_L \gamma^\mu \lambda^a q_L \, = \, \,
\bar{\tilde{\Psi}}_L \gamma ^\mu \Lambda^a \, \tilde{\Psi}_L,
\qquad
\Lambda^a \, \equiv \, \xi^\dagger \lambda^a \, \xi.
\label{leftcur}
\end{equation}
The weak current with one heavy quark is as given by HQET \cite{Neubert:1993mb}, except that as before, the Goldstone bosons are factored out
\begin{equation}
J_c^\mu
=C_{\gamma}(\mu) \bar{\tilde{\Psi}}_b \xi^{\dagger}_{bc}\gamma^\mu
P_L Q_v \, +
\, C_v(\mu) \bar{\tilde{\Psi}}_b \xi^{\dagger}_{bc} v^\mu P_L Q_v,
\label{modcur}
\end{equation}
The coefficients $C_{\gamma,v}$ are determined
from the QCD renormalization for $\mu < m_c$.
However, for $\mu \simeq \Lambda_\chi$, $C_\gamma \simeq 1$
 and $C_v \simeq 0$. When quark fields will be integrated out, this will lead to the leading order term of the current \eqref{current}. 

\subsection{Estimate of the color-current contributions}\label{color-current}
We are now able to outline the strategy used in \cite{Eeg:2001un} to estimate the contribution of the nonfactorizable colored currents to $D^0\to K^0\bar{K}^0$. We will not discuss all the details, for which we refer the reader to \cite{Eeg:2001un,Hiorth:2002pp}. The key observation is, that once the quark degrees of freedom are integrated out, one has to end up with the most general effective Lagrangian containing the meson fields, i.e., the HH$\chi$PT Lagrangian \eqref{totlag}. By integrating out the quark fields one can thus (i) connect the unknown couplings $m_\chi$, $G_H$ in \eqref{totlag} to the constants of the HH$\chi$PT, that are fixed from the experiment, (ii) calculate (in a model dependent way) the higher order constants of the HH$\chi$PT, (iii) relate the constants of the HH$\chi$PT to each other. 
For instance, the lowest order chiral
 Lagrangian in the light pseudoscalar 
sector \eqref{chirallagr} 
can be obtained
by coupling two axial fields to a quark loop using the Lagrangian
in Eq.~\eqref{chqmR}:
\begin{equation}
 i {\cal{L}}^{(2)}_{\text{str}} \,
= - N_c \, \int \frac{d^dp}{(2\pi)^d} \,
\Tr \left[ \left(\gamma_\sigma \gamma_5 {\cal{A}}^\sigma \right)
 \, S(p) \,
\left(\gamma_\rho \gamma_5 {\cal{A}}^\rho \right) \, S(p) \right]
\sim
\Tr \left[{\cal{A}}_\mu {\cal{A}}^\mu \right],
\label{L2str}
\end{equation}
where $S(p) = (\; \sls p - m_\chi)^{-1}$, and the trace is
both in flavor and Dirac spaces.
The result on the right-hand side of Eq.~\eqref{L2str} is the standard form of the lowest order chiral Lagrangian
\eqref{chirallagr}, as can easily be seen by using the relations
\begin{equation}
{\cal{A}}_\mu \; = \; - \frac{1}{2 i} \xi \,
 (\partial_\mu \Sigma^\dagger) \, \xi
\; = \; \frac{1}{2 i} \xi^\dagger \, (\partial_\mu \Sigma) \,
\xi^\dagger.
\label{ASigma}
\end{equation}
Similarly one obtains
 the lowest order ${\cal O}(p)$
strong chiral Lagrangian \eqref{eq-8} in the heavy sector.

\begin{figure}[bt]
\begin{center}
 \epsfig{file=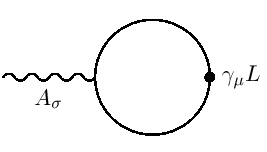,width=4cm}
\caption{\footnotesize{Feynman diagram for the bosonization of the left-handed current
to the order ${\cal O}(p)$}.}\label{fig:va}
\end{center}
\end{figure}

In a similar way, i.e., by integrating out the quark fields, we can dress-up the quark weak currents with mesonic fields. This has been called the process of bosonization in \cite{Eeg:2001un}. Let us consider the bosonization of the light weak current. \index{bosonization}
The lowest order term ${\cal O}(p)$ is obtained, when the vertex
$\Lambda^a$
from
(\ref{leftcur}) and the axial vertex ($ \sim {\cal A}_\mu$) from
(\ref{chqmR})
are combined with the quark loops (see Fig.~\ref{fig:va}):
\begin{equation}
 j^a_\mu({\cal{A}}) \, = - \, i N_c \int \frac{d^dp}{(2\pi)^d} \,
\Tr \left[ \left(\gamma_\mu L \, \Lambda^a \right) S(p) \,
\left(\gamma_\sigma \gamma_5 {\cal{A}}^\sigma \right) \, S(p) \,
\right]
 \; \sim \; \Tr \left[\Lambda^a \, {\cal{A}}_\mu \right].
\label{jXA}
\end{equation}
This coincides with (\ref{jX}) when (\ref{ASigma})
is used.

Note that the proportionality factors in \eqref{L2str}, \eqref{jXA} contain divergent integrals. These can be regulated in different ways, but in this context they are treated as free parameters. They are used to relate different model parameters and chiral Lagrangian constants to each other. For instance to get a leading order estimate of coupling $G_H$, one uses the self-energy diagrams of heavy mesons and light-pseudoscalars. 
 A logarithmically divergent integral is contained in both calculations and is used to relate $f_\pi$ to $G_H$
\begin{equation}
G_H \; \simeq \; \frac{2 \sqrt{m_\chi}}{f_\pi}.
\label{GHlog}
\end{equation}
Furthermore, the quadratic
divergence contained in the loop integral of 
the diagram in Fig.~\ref{fig:heavylight} (left) is related to the quark condensate of the light
quark, which is also quadratically divergent. 
To the leading order 
\begin{equation}
G_H \; \simeq \; -2 \frac{m_\chi \, \alpha}{\langle \bar{q} q\rangle
},
\label{GHquad}
\end{equation}
Combining (\ref{GHlog}) and (\ref{GHquad}) we obtain
\begin{equation}
\alpha \; \simeq \; - \, \frac{\langle \bar{q}q\rangle }{f_\pi \,
\sqrt{m_\chi}},
\label{alphaH}
\end{equation}
which for the values $m_\chi$= 200 MeV, $f_\pi$ = 131 MeV and
 $\langle \bar{q}q\rangle $ = (-240MeV$)^3$ gives $\alpha=0.24 \; \text{GeV}^{3/2}$ in agreement with the values for
$\alpha$ cited in section \ref{coupling}. The fact that (\ref{alphaH}) works
well numerically, gives some support to the leading order estimates
in (\ref{GHlog}) and (\ref{GHquad}). These relations are 
slightly modified, when other contributions such as soft gluon emission are taken into account \cite{Hiorth:2002pp}. In the following we will, however, use the simple relations (\ref{GHlog}) - (\ref{alphaH}). For the future reference we define the ratio
\be \label{beta-def}
\tilde\beta = \frac{f^2 G_H}{2 \alpha}.
\ee
Using (\ref{GHlog}) and (\ref{alphaH})
we obtain
\begin{equation}
\tilde \beta \; \simeq \; - \, \frac{m_\chi \, f_\pi^2}{\langle \bar{q}
q\rangle } \; \simeq
\; \frac{1}{4}.
\label{beta}
\end{equation}
This then gives the approximate value of the unknown coupling $G_H$ \eqref{Int}, \eqref{beta-def}, that will be needed in Eqs. (\ref{amp}) and (\ref{DivRel}) below.

\begin{figure}[bt]
\begin{center}
 \epsfig{file=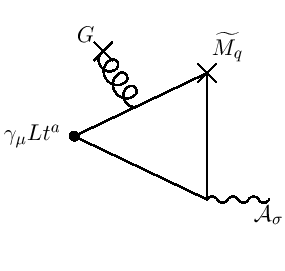,width=4cm}
\caption{\footnotesize{Diagram for the bosonization of the colored light current to ${\cal O}(p^3)$}.}
\label{fig:colcur}
\end{center}
\end{figure}

\begin{figure}[bt]
\begin{center}
\epsfig{file=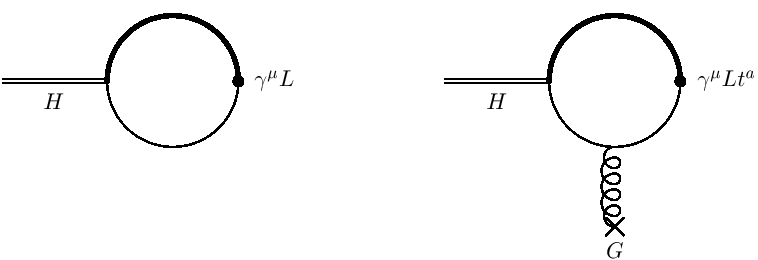,width=11cm}
\caption{\footnotesize{Diagrams representing bosonization of the heavy-light weak current.
The boldface line represents the heavy quark, the solid line the light
quark.}}
\label{fig:heavylight}
\end{center}
\end{figure}

\begin{figure}[bt]
\begin{center}
 \epsfig{file=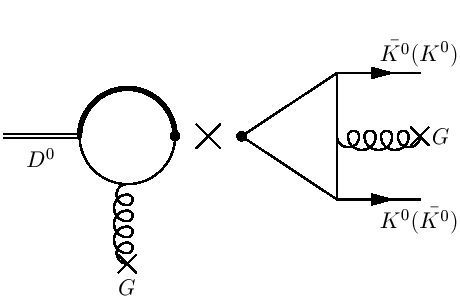,width=7cm}
\caption{\footnotesize{Diagram representing the calculation of nonfactorizable contributions of the colored currents to $D^0\to K^0\bar{K}^0$. The gluons contribute to the gluon condensate.}}
\label{fig:condensate}
\end{center}
\end{figure}

Finally, we switch to the estimation of the nonfactorizable contribution of the colored currents in \eqref{alphaH}. First the light and heavy-to-light colored currents are bosonized in a very similar way to the light current \eqref{jXA}. To get a nonzero contribution one has to work in a gluonic background. The bosonization of the colored light current is depicted on Fig.~\ref{fig:colcur}, while the bosonization of the heavy-light colored current is shown on Fig.~\ref{fig:heavylight} (right). In the calculation gluons are first treated as external fields. When the product of the two colored currents is taken at the end (see Fig. \ref{fig:condensate}), the external gluons are assumed to contribute to the gluon condensate \index{bosonization}

Details of the calculation can be found in \cite{Eeg:2001un}, while here we write down only the final results. We find for the nonfactorizable gluon condensate contribution:
\begin{equation}
\begin{split}
 {\cal{L}}_{\text{eff}}(D^0 \mbox{decay})_{\langle G^2\rangle }
\; &= \; 8 \frac{G_F}{\sqrt{2}} \, C_2
\left(\frac{g_s \, G_H}{16 \pi^2} \right)
\left(\frac{g_s}{12 m_\chi} \frac{1}{16 \pi^2} \right) \langle
G^2\rangle \\
&\hskip12mm\times \, v_\rho
\, \left[ B_\varepsilon \, T^{a,\rho}_\varepsilon
\; + \, B_g \, T^{a,\rho}_g \right] \,
 \big(D \xi^\dagger\big)_c,
\label{effLag}
\end{split}
\end{equation}
where $B_{\varepsilon,g}$ are numerical factors that come from integration, $T^{a,\rho}_\varepsilon$, $T^{a,\rho}_g$ are functions of the light-pseudoscalar field and correspond to the light current Eq.~\eqref{leftcur}, $D$ is the $D$ meson field, with the last term corresponding to the heavy current \eqref{current}, while $\langle G^2\rangle $ is the gluon condensate, obtained by the
prescription
\begin{equation}
G_{\mu \nu}^a G_{\rho\sigma}^a \; \rightarrow \; \frac{1}{12} \,
 (\eta_{\mu \rho} \, \eta_{\nu \sigma} \, -
\, \eta_{\mu \sigma} \, \eta_{\nu \rho}) \, \langle G^2\rangle.
\end{equation}
Expanding functions $T^{a,\mu}_\varepsilon$, $T^{a,\mu}_g$ to the lowest order to get the $D^0\to K^0\bar{K}^0$ amplitude, one finds that they contain terms proportional to $(p_{K^0} + p_{\bar K^0})^\mu$. Unlike in the factorization approximation \eqref{eq-9}, the effective Lagrangian \eqref{effLag} contains terms that do not vanish once they are contracted with $v^\mu = p_D^\mu /M_D$, where $p_D^\mu=(p_{K^0} + p_{\bar K^0})^\mu$. They will
give a nonfactorizable contribution to the $D^0 \rightarrow K^0 \overline{K}^0$ decay, 
 proportional to $\langle G^2\rangle $
\begin{equation}
M(D^0 \rightarrow K^0 \bar{K^0})_{\langle G^2\rangle }
\; = \; - \, C_2 \frac{G_F}{\sqrt{2}}V_{us}V_{cs}^* m_D^2 \, \frac{(m_s - m_d)}{m_\chi} \;
\frac{\tilde\beta \, \delta_G}{12 N_c} \pi \, f_D
\label{amp}
\end{equation}
where:
\begin{equation}
\delta_G = N_c \frac{\langle \alpha_s \, G^2/\pi\rangle }{8 \pi^2
f^4},
\label{DivRel}
\end{equation}
while $\tilde \beta$ is defined in \eqref{beta-def} and is estimated to be $\tilde \beta \simeq 1/4$.

It should be noted, that in principle other terms than the
 one in (\ref{effLag}) could contribute. There is one possible term where the field
$\widetilde{M}_q$ inserted in Fig.~\ref{fig:colcur} may instead be attached to
the light quark line in the diagram of Fig.~\ref{fig:heavylight} (right). However, this term
will
not give contributions to $D^0 \rightarrow K^0 \bar{K^0}$. Another possible term could arise by attaching the 
 field ${\cal A_\sigma}$ to the light quark line in Fig.~\ref{fig:heavylight} (right) instead of attaching it to the quark lines on 
 Fig.~\ref{fig:colcur}. This term
is identically zero.

In the language of the chiral perturbation theory, the term (\ref{effLag})
can be interpreted as a counterterm. To be more specific, the
(divergent part of the) counterterm
has the Lorentz and flavor structure of the second line of
(\ref{effLag})
and is multiplied with a (divergent) coefficient adjusted to cancel the
loop divergences obtained in section \ref{chiral-loops}.
 \index{color-currents, contributions of|)}\index{gluon condensate|)}

\section{Results}\label{Results-nonfact}\index{calculation! results of|(}\index{nonresonant contributions}\index{results}

In the numerical evaluation of the chiral loop contributions we use the values of $\alpha$, $g$ and
$f$ that were obtained in section \ref{coupling} and are given in Table \ref{tab-koef}. We present results both for the $\overline{\text{MS}}$ ($\bar{\Delta}\to 1$) and for the $\text{GL}$ ($\bar{\Delta}\to 0$) renormalization prescriptions. The difference between the two predictions reflects the relative importance of the counterterms, that were neglected in our approach. We put everywhere $\mu = 1$ $\rm{GeV}$. For the effective Wilson coefficients $a_{1}^{\text{NDR}}$, we use the value $a_1^{\text{NDR}}=C_2+C_1/N_c=1.10\pm0.05$ calculated at the scale $\mu=1$, while $a_2^{\text{NDR}}$ is neglected as discussed at the beginning of section \ref{chiral-loops}.\footnote{Even if the ``new factorization'' values \eqref{coeff-a1a2} had been used,
the $a_2$ part of weak interaction would be suppressed by a factor of $1/3$
compared to the $a_1$ one.}
We present the numerical results \footnote{Note that the numerical results differ somewhat from \cite{Eeg:2001ip,Eeg:2001un}, as slightly different values for $\alpha$, $g$ couplings have been used.} for the nonzero one chiral loop
amplitudes in
 Table \ref{tab-1}.

\begin{table} [h] \index{tables of results}
\begin{center}
\begin{tabular}{|l|c|c|} \hline \index{amplitude! in $D^0\to K^0\overline{K}^0$} \index{contributions!chiral loop}
$-$&$M_i^{\overline{\text{MS}}}[\times 10 ^{-7}{\rm \;GeV}]$ & ${\cal
M}_i^{\text{GL}}[\times 10 ^{-7}{\rm \;GeV}]$ \\ \hline\hline
$M_1$ & $-0.58$& $-0.66$ \\ \hline
$M_2$ & $-0.43$ & $-0.49$\\ \hline
$M_3$ &$ -0.87$ & $-0.98$ \\ \hline
$M_4$ & $0.69 -2.33 i$ & $1.37 -2.66 i$ \\ \hline
$M_5$ &$-0.75$ & $-0.46$ \\ \hline
$M_6$ &$-0.56$ & $-0.35$ \\ \hline
$M_{7} $&$-0.91$ & $-2.94$ \\ \hline
$M_{8}$ &$ 0.84$ & $0.97$\\ \hline\hline
$\sum_i M_i$& $-2.58 -2.32 i$ & $ -4.42 -2.65 i$\\ \hline
\end{tabular}
\caption[Table of one chiral loop amplitudes in $D^0\to K^0\overline{K}^0$]{\footnotesize{Table of the one chiral loop amplitudes (see
Fig.
\ref{fig-5}), where $M=\sum_n M_n$ is defined in
\eqref{eq-12}. The second column shows the amplitudes calculated using
$\bar{\Delta}\to 1$, while the third column
amplitudes have been calculated using $\bar{\Delta}\to 0$. The values of the coupling constants are taken from Table \ref{tab-koef}.
In the last line the sum of all amplitudes is presented. It can be
compared
with the experimental result
$|M_{\rm Exp}|=3.80\times 10^{-7}{\rm \;GeV}$.}}
\label{tab-1}
\end{center}
\end{table}

The imaginary
part of the amplitude comes from the $F_4$ graph, when the $\pi$'s
or the $K$'s in the loops are on-shell. All other
 graphs contribute only to the
real part of the amplitude.
The imaginary part of the amplitude is scale and scheme independent to the first order in the 
chiral expansion\footnote{The imaginary parts given in Table \ref{tab-1} are not the same because of slightly different values of $g,\alpha$ couplings in the two schemes. The difference is obviously of higher order in the chiral expansion}.
This amplitude is also obtainable from unitarity considerations, and is valid beyond
the chiral loop expansion.
We also mention, that the rescattering contribution, considered in
\cite{Buccella:1994nf,Buccella:1996uy,Pham:1987rj},
is the same contribution, as the one we calculate from graphs on
Fig.~\ref{fig-1}.

\index{counterterms! contributions of}
In order to cancel the divergences one has to construct counterterms.
In the framework of the HL$\chi$QM they can be calculated model dependently, as discussed at the end of section \ref{HeavyLight}. In the effective Lagrangian approach they are taken to be free parameters, while the form of the terms is determined by the symmetry arguments. We estimated the size of the counterterms by using two renormalization prescriptions. The difference between the two indicates that the contributions coming from the counterterms can be substantial. However, the values one obtains by considering the chiral loop contributions solely, are in fair agreement with the experimental data.

The final expression for the nonfactorizable contribution of the color-currents is given in Eq.~\eqref{amp}. Using the values \cite{BEF}
 $\langle \frac{\alpha_s}{\pi} G^2\rangle \simeq (334 {\mbox \;{\rm
MeV}})^4$, $m_\chi=200$
\;MeV,
and $m_s \simeq$ 150 \;MeV, we obtain the numerical value:\index{color-currents, contributions of}
\begin{equation}
M(D^0 \rightarrow K^0 \bar{K^0})_{\langle G^2\rangle }
\; \simeq \; 0.43 \times 10^{-7} \text{\;GeV},
\label{ampNum}
\end{equation}
which is comparable in size to the chiral loop contributions
in Table \eqref{tab-1}.

Adding both the chiral loops and the gluon \index{errors, estimation}
condensate \eqref{ampNum} contributions, we obtain the total amplitude \index{amplitude! in $D^0\to K^0\overline{K}^0$} 
 to ${\cal
O}(p^3)$
\begin{equation}
\begin{split}
GL (\bar{\Delta}\to0) ; \qquad &
M_{\text{Th}}=(-3.99 - 2.65 \, i)\times 10^{-7}{\text{\;GeV}}, \\
\overline{\text{MS}}(\bar{\Delta}\to1) ; \qquad &
M_{\text{Th}}=(-2.15 -2.32 \, i)\times 10^{-7}{\text{\;GeV}}.
\end{split}
\end{equation}
or in terms of the branching ratio
\begin{equation}
\begin{split}
GL(\bar{\Delta}\to0) ; \qquad &
\text{Br}(D^0\to K^0\bar{K^0})_{\text{Th}}=(10 \pm 3.5)\times 10^{-4}, \\
\overline{\text{MS}}(\bar{\Delta}\to1) ; \qquad &
\text{Br}(D^0\to K^0\bar{K^0})_{\text{Th}}=(4.3\pm 2.5)\times 10^{-4}.
\end{split} \label{result-nonfact}
\end{equation}
where the estimated uncertainties reflect the uncertainties
 in the couplings $\alpha, g,f$. These results should be compared with the
experimental value \cite{Hagiwara:pw} $\text{Br}(D^0 \to K^0
\bar K^0)
= (7.1\pm 1.9)\times 10^{-4}$.

\index{difficulties!of approach}
Note that, as described above, in the calculation of $D^0\to K^0\bar{K}^0$ we have neglected the order ${\cal O}(p^3)$ counterterms, as the entire set cannot be fixed from experiment (see discussion in section \ref{coupling}). The counterterms would absorb the leading scheme dependence of the final result \eqref{result-nonfact}. In other words, the scheme dependence of the final result \eqref{result-nonfact} indicates the sizes of counterterm contributions to the decay. These are as large as the chiral-loop contributions. This is not surprising and is similar to the case of the chiral corrections to the $f_K$ decay constant. Nonetheless, the important notion is, that the chiral loop contributions do lie in the right ballpark.

There are also other possible contributions to the decay mode $D^0\to K^0\bar{K}^0$, apart from the ones discussed above. For instance, around the charm mesons mass region there are many resonances.
One might think that their contribution will appear in this decay mode,
 either as scalar resonance exchange \cite{Dai:1999cs}
or as $K^*$ exchanges \cite{Dai:1999cs,Zenczykowski:1999cu,Buccella:1994nf,Buccella:1996uy}.
Within our framework they would appear as the next
order contribution in the chiral expansion.
This is, however, beyond the present scope of our investigations.
It is interesting to point out that the effects we calculate,
both from chiral loops and from the gluon condensate,
 are results of the $SU(3)$ flavor
symmetry breaking. In the limit of exact symmetry both contributions
will disappear. \index{resonant contributions}

We can summarize that we indicate the leading
 nonfactorizable
contributions to $D^0 \to K^0 \bar K^0$. Even though the use of the chiral
perturbation theory in this decay mode could be questioned, some of 
the calculated chiral loops
can be considered as part of the final state
interactions. \index{final state interactions}
Although the next to leading ${\cal O}(p^5)$ order terms
might give sizable contributions
to this decay, we have demonstrated that contributions due to
the chiral loops and gluon
condensates are of the same order of magnitude
as the amplitude extracted from the experimental result.\index{calculation! results of|)} \index{D0K0K0Bar@$D^0\to K^0 \bar{K}^0$|)}\index{decay!$D^0\to K^0 \bar{K}^0$|)}\index{nonfactorizable contributions|)}

\chapter{Rare $D$ decays} \label{rareD}\index{rare decays|(}
\section{Why rare $D$ decays?}\index{rare decays! motivation for}
Why should one be interested in the rare decays in the first place? It takes an incredible amount of hard work to do the rare-event experiments, so one better be sure it is worth the effort. First of all, rare decays are interesting if they are associated with a conservation law. Prominent examples of such decays are the proton decay and $\mu\to e \gamma$. These processes are completely forbidden within the Standard Model. Their observation would thus signal new physics, for instance the Grand Unified Theories. \index{signatures, new physics}

But also processes, that are not completely forbidden in the Standard Model, can be extremely useful probes of New Physics. For instance, precision measurements of the rare $K, D, B$ meson decays offer studies of the flavor changing neutral currents (FCNCs), i.e., transitions of the following type \index{FCNC}
\be
q_i\to q_j +\left\{ 
\begin{aligned}
&\nu \bar{\nu}\\
&e^+e^-\\
&\gamma
\end{aligned}\right.,
\ee
where $q_i$, $q_j$ are quarks of different flavor (but of the same electromagnetic charge). In the SM these transition are forbidden at tree level because of the unitarity of the CKM matrix (this is the so called Glashow-Iliopoulos-Maiani (GIM) mechanism).\index{GIM suppression} They can, however, occur at the 1-loop level. They are additionally suppressed because of the hierarchical structure of the CKM matrix. FCNCs are naturally suited as probes of the quark-flavor mixing dynamics in the SM and beyond. Actually, some of these processes (e.g., $K_L\to \mu^+\mu^-$) were very important in the historical construction of the SM \cite{Isidori:2001nd}.

Note also, that the information used at present to constrain the parameters of the CKM matrix, comes either from the charged currents (that occur already at tree level) or from the $\Delta F=2$ amplitudes. These data are in a very good agreement with the SM. However, new physics could affect $\Delta F=1$ and $\Delta F=2$ processes in a substantially different manner. The consideration of the FCNCs thus constitutes an invaluable test of the SM in the quark sector. \index{FCNC} \index{signatures, new physics}

\index{nonperturbative effects}
Of special interest are the so-called {\it golden modes}. For the decay mode to be golden-plated it has to fulfill the following requirements: (i) the SM amplitude has to be small or forbidden entirely, (ii) it has to be theoretically clean, i.e., contributions of long distance physics (nonperturbative QCD effects) have to be well under control, (iii) it has to allow for potentially large contributions from new physics. An example of such a golden mode is $K\to \pi \nu \bar{\nu}$. In the SM it proceeds at the one loop level through the same diagrams as shown on Fig.~\ref{InamiLim}. The quarks flowing in the loop are from the up sector, $u,c,t$. As discussed in section \ref{OPE} (see Eq.~\eqref{Cnmass-dep}), the amplitudes are proportional to $x_q=m_q^2/m_W^2$, with $m_q$ the mass of the up-type quark in the loop
\be
M(s\to d\nu \bar{\nu})=\sum_{q=u,c,t} V_{qs}^* V_{qd} {\cal M}_q\sim \left\{
\begin{aligned}
{\cal O}&(\lambda^5 m_t^2)+i {\cal O}(\lambda^5 m_t^2)&: t-\text{quark},\\
{\cal O}&(\lambda m_c^2)+i {\cal O}(\lambda^5 m_c^2)&: c-\text{quark},\\
{\cal O}&(\lambda \Lambda_{\text{QCD}}^2)&: u-\text{quark},
\end{aligned}\right.
\ee
where the Wolfenstein parametrization of the CKM matrix has been used with $V_{us}=\lambda=0.22$ and with the standard CKM phase convention $\Im(V_{us})=\Im (V_{ud})=0$. In the last line we have tacitly written the scale of QCD instead of the $u$-quark mass, indicating the size of the nonperturbative QCD effects. Even though the top contribution is suppressed by the CKM hierarchy, it still dominates the $s\to d\nu\bar{\nu}$ transition. This has two important consequences (i) the $s\to d \nu \bar{\nu}$ transition is dominated by the short distance physics, so that the QCD corrections are small and calculable in the perturbation theory and (ii) the $s\to d\nu \bar{\nu}$ transition is very suppressed within the SM and thus sensitive to new physics. \index{signatures, new physics}Especially clean theoretically is $K_L\to \pi^0 \nu \bar{\nu}$. Because of the CP \index{CP@$CP$} structure only the imaginary part of the CKM matrix contributes, so that the charm contribution is completely negligible and thus the theoretical error is below 3\% \cite{Isidori:2001nd}. The theoretical error in the $K^+\to \pi^+ \nu \bar{\nu}$ decays, on the other hand, is $\sim 10\%$. One has to pay the price, however, on the experimental side, where one has to search for the events with the probability of $10^{-10}$. Recently the second event in the $K^+\to \pi^+ \nu \bar{\nu}$ decay channel has been observed, giving \cite{Adler:2001xv}\index{rare decays! experimental searches}
\be
\text{Br}(K^+\to \pi^+ \nu \bar{\nu})_{\text{Exp}}=\left(1.57 \genfrac{}{}{0pt}{}{+1.75}{-0.82}\right)\times 10^{-10}.
\ee
This is to be compared with
\begin{align}
\text{Br}(K^+\to \pi^+ \nu \bar{\nu})_{\text{Th}}&=(0.8\pm0.3)\times 10^{-10},\\
\text{Br}(K_L\to \pi^0 \nu \bar{\nu})_{\text{Th}}&=(2.8\pm1.1)\times 10^{-10}.
\end{align}

A very similar situation occurs in the $B$ physics, which is in many respects even more favorable theoretically. For instance, the $b\to s\nu\bar{\nu}$ transition is dominated completely by the top quark. Namely, in contrast to $s\to d \nu\bar{\nu}$ transition discussed above, charm and $u$-quark are not CKM enhanced. At the moment the most significant information about the $\Delta B=1$ FCNCs is, however, coming from the $B\to X_s\gamma$ decay. Here QCD corrections play an important role, but are well under control \cite{Isidori:2001nd}. \index{FCNC}

In the case of $c\to u$ transitions the situation is not so favorable. Instead of the up-type quarks with a very distinct mass hierarchy, now the down-type quarks run in the loops. One can thus in general expect the following contributions
\be
M(c\to u)=\sum_{q=d,s,b}V_{uq}^* V_{cq} {\cal M}_q \sim \left\{
\begin{aligned}
{\cal O}&(\lambda^5 m_b^2)&: b-\text{quark},\\
{\cal O}&(\lambda m_s^2)&: s-\text{quark},\\
{\cal O}&(\lambda \Lambda_{\text{QCD}}^2)&: d-\text{quark},
\end{aligned}\right.
\ee
Since the $b$ quark is much lighter than the top quark, it cannot surpass the $\lambda^4$ suppression. In a sense the situation is just the opposite to the $s\to d$ FCNC, where now the contributions from the heaviest, $b$-quark, are expected to be the least important. One can thus expect that in the rare $D$ decays, the nonperturbative long distance (LD) effects coming from the lighter two down quarks, $d,s$, will dominate. Thus, there are no golden modes in the rare $D$ decays. \index{FCNC}\index{long-distance contributions}\index{nonperturbative effects}

Because the LD effects dominate $D$ decays, no extraction or tests of the CKM matrix are possible in these decays. Also, in order to be able to probe new physics, new effects, if present, have to be large. However, there is an important sidepoint to the whole story. Namely, $D$ physics probes the flavor structure of the up-quark sector, in contrast to the $K$ and $B$ decays discussed in the beginning of this section. The potential new physics effects in the two sectors can be very different. In this sense rare $D$ meson decays can prove a valuable probe of the new physics effects. However, to be able to use it, firm limits on the LD contributions are needed.

\section{Rare radiative $D$ decays}\label{rare-review}\index{rare decays! short review of charm}
 As discussed in the previous section, FCNC rare \index{FCNC} $D$ decays are dominated by the LD effects. Nevertheless they constitute an interesting probe of the quark-flavor dynamics in the up-quark sector and have as such received an ongoing theoretical and experimental attention. In this section we will review recent theoretical work on the FCNCs in $D$ decays and move on in the next sections to estimate $D^0\to \gamma\gamma$ and $D^0\to l^+l^- \gamma$ decays in the framework of the effective Lagrangians discussed in chapters \ref{HQET} and \ref{weak-int}. 

Since LD effect are difficult to control theoretically one would like to either find the decay modes, where the LD effects are as small as possible, and/or find observables where the LD effects cancel. Such an observable was constructed in \cite{Fajfer:2000zx}, where $D^0\to \rho, \omega \gamma$ decays were considered. The important observation is that the $(\bar{d}d)\gamma$ final state arises through the nonleptonic $W$-exchange $c\bar{u}\to d\bar{d}$, and is thus LD dominated, while the $(u\bar{u})\gamma$ final state is mainly due to the electromagnetic penguin $c\to u\gamma$ \index{ctougamma@$c\to u\gamma$} transition. The LD contributions and thus also decay amplitudes of $D^0\to \rho \gamma$, $D^0\to \rho \gamma$ are then almost equal. Subtracting the two one obtains a quantity which is small in the SM 
\be
D^{\omega-\rho}=\frac{\hat{\Gamma}(D^0\to \omega \gamma)-\hat{\Gamma}(D^0\to \rho^0\gamma)}{\hat{\Gamma}(D^0\to \omega\gamma)}\label{omega-rho},
\ee
where $\hat{\Gamma}$ is the decay width divided by the phase space. Since most of the LD effects cancel in $D^{\omega-\rho}$, it is sensitive to the SD contributions that can be altered by the new physics effects. The SM prediction is $D^{\omega-\rho}\sim 6\pm 15\%$. A larger difference would be a smoking gun \index{smoking gun} for new physics. Note, that generic supersymmetric scenarios can lead to $D^{\rho-\omega}\sim {\cal O}(1)$. \index{Drhoomega@$D^{\rho-\omega}$} \index{signatures, new physics}\index{supersymmetry}

Another interesting analysis is connected with the decay modes $D\to (P, V) l^+l^-$ estimated both in the SM and in the Minimal Supersymmetric Standard Model (MSSM) \cite{Fajfer:2001sa,Fajfer:1998rz,Burdman:2001tf,Schwartz:he,Singer:1996it}, where $P=\pi, K, \eta$ are the light pseudoscalar mesons, $V=\rho, \omega, \dots$ are the light vector mesons and $l^+l^-$ is the electron or muon lepton pair. The decay widths are estimated either by employing the vector meson dominance \cite{Burdman:2001tf} or an effective Lagrangian approach similar to the framework we follow, with the vector mesons treated using hidden symmetry \cite{Fajfer:2001sa,Fajfer:1998rz}. In Ref.~\cite{Burdman:2001tf} analysis of $D\to (P,V) l^+l^-$ decays in the $R$ parity violating MSSM was performed. \index{R parity@$R$ parity violating terms} It was found, that using constraints on trilinear $R$ parity violating couplings obtained prior to this analysis from different decay modes, the predicted branching ratios for $D^+\to \pi^+ \mu^+\mu^-$, $D^0\to \rho^0 \mu^+\mu^-$ already saturate the experimental bounds \footnote{In \cite{Burdman:2001tf} previous bound $\text{Br}(D^+\to \pi^+ \mu^+\mu^-)<1.5 \times 10^{-5}$ \cite{Aitala:1999db,Aitala:2000kk} has been used.} $\text{Br}(D^+\to \pi^+ \mu^+\mu^-)<8.6 \times 10^{-6}$ \cite{Johns:2002hd}, $\text{Br}(D^0\to \rho^0 \mu^+\mu^-)<2.2 \times 10^{-5}$ \cite{Aitala:1999db,Aitala:2000kk}. Measurements of the rare $D$ meson decays can thus already now constrain the new physics scenarios in the up-type quark sector. \index{signatures, new physics}

The radiative decay modes $D\to V\gamma$ are dominated by the LD effects \cite{Fajfer:1998dv,Fajfer:1997bh,Burdman:1995te,Bajc:1994ui,Lebed:1999kq}. They can be used as probes of new physics only through observables such as $D^{\rho-\omega}$ \eqref{omega-rho}, where the LD effects cancel. A more direct possibility for probing $c\to u\gamma$ \index{ctougamma@$c\to u\gamma$} is through at present rather exotic $B_c\to B_u^*\gamma$ decay \cite{Fajfer:1999dq,Poljsak:2001hh,Fajfer:2001hj,Prelovsek:2001ny}. In this decay the SM long distance and short distance contributions are found to be of comparable size \cite{Fajfer:1999dq}, giving a branching ratio $\sim 10^{-8}$.

The leptonic decays $D^0\to \mu^+ \mu^-$, $D^0\to e^+e^-$ are very rare in the SM as they are helicity suppressed. This suppression is lifted slightly by the LD effects, mainly through the two-photon unitarity contribution, giving branching ratios of the order $\sim 10^{-13}$ \cite{Burdman:2001tf}. These decay modes can, however, receive large contributions from the physics beyond the SM. For instance, a modest improvement on the experimental upper limit $\text{Br}(D^0\to \mu^+\mu^-)<3.3 \times 10^{-6}$ would already yield a new bound on the product of trilinear $R$ parity violating couplings $\tilde{\lambda}'_{11k}\tilde{\lambda}'_{12k}$ \cite{Burdman:2001tf,Golowich:2002ac}. Note, that the $c \to u$ equivalents of the golden modes $K\to \pi \nu \bar{\nu}$, the $D\to P \nu \bar{\nu}$ do receive comparable contributions from the SD as from the nonperturbative LD physics, but are of the order $\sim 10^{-15}$ and are out of reach experimentally. \index{nonperturbative effects} \index{R parity@$R$ parity violating terms}

The radiative decay $D^0\to \gamma \gamma $ has been considered in Ref.~\cite{Fajfer:2001ad} and will be discussed at length in section \ref{D0GammaGamma}. It is dominated by the LD effects with a branching ratio of $\sim 10^{-8}$. A separate analysis of the same decay mode has been performed by Ref.~\cite{Burdman:2001tf}, arriving at the compatible result. Possible contributions to the $D^0\to \gamma\gamma$ decay coming from the MSSM give results comparable to the SM branching ratio. More sensitive to the possible new physics contributions is the decay mode $D^0\to l^+ l^- \gamma $ \cite{Fajfer:2002gp}, which will be discussed in section \ref{llbarGamma}. \index{signatures, new physics}

\index{rare decays! experimental searches}
Parallel to the theoretical work, there is also a considerable experimental effort devoted to the charm physics. The high statistics and an excellent quality of data at FOCUS \index{FOCUS}experiment now allow, among others, for high precision studies of charm semileptonic decays \cite{Link:2002ev}, determination of $D^{0,\pm}$ decay times below $1\%$ error level \cite{Link:2002bx}, as well as for the searches of CP violation and rare $D$ decays \cite{Link:2000aw,Link:2000kr}. There is a very rich potential for the charm physics at the B-factories, \index{B-factories} with both Belle \index{Belle} and Babar having an active program in the charm studies \cite{Yabsley:2002,Williams:2002}. For instance, more than 120 milion charm pairs have already been produced at BaBar. \index{BaBar} This corresponds to more than 220~000 $D^*$-tagged $D^0$ decays, which will allow for the precision lifetime and $D^0$ mixing analyses as well as for the searches of rare charm decays \cite{Williams:2002}. An exciting charm physics program is under way also at CLEO, \index{CLEO} that was recently able to measure $\Gamma(D^*)$ for the first time \cite{Ahmed:2001xc,Anastassov:2001cw}. Among the rare D
decays, the decays $D\to V \gamma$ and $D \to V(P) l^+ l^-$ are subjects of CLEO\index{CLEO} and FERMILAB \index{FERMILAB} searches \cite{E791,Johns:2002hd}. In the following years a great phenomenological impact is expected from proposed CLEO-c \index{CLEO-c} physics programme. Next year more than 6 million tagged $D$ decays are expected to be measured. This will allow for precision charm branching ratio measurements and consequently improved measurements of the CKM matrix elements also in the $b$-sector, as well as for extensive studies of $D$-mixing, CP violation and rare decays in the charm sector \cite{Pedlar}.

\index{cll@$c\to u l^+l^-$|(}\index{rare decays! calculation of|(}\index{Wilson coefficients}
\section{Inclusive $c\to u l^+l^-$ decay}\label{ctoullbar}\markright{INCLUSIVE $c\to u l^+l^-$ DECAY}
Before we turn to the estimates of the experimentally more tractable exclusive rare decays $D^0\to \gamma\gamma$, $D^0\to l^+l^-\gamma$, we will discuss the calculation of the inclusive $c\to u l^+l^-$ decay. By doing this exercise, we will further clarify several points regarding the relative importance of Wilson operators, that have been brought forward in section \ref{RG-OPE}. \index{OPE}

The effective Lagrangian for the $c\to u l^+l^-$ decay mode can be found in Eqs.~\eqref{lagr-at-bandc}, \eqref{effective_lagr}, with the Wilson coefficients listed in Table \ref{tab-Wilson}. Note that in the literature \cite{Burdman:2001tf,Fajfer:2001sa,Geng:2000if} as an estimate for the $C_9(\mu_c)$ Wilson coefficient, the result from electroweak theory without QCD, $C_9^{\text{IL}}$, has been used. The leading order expression in terms of $m_{d,s}^2/m_W^2$ is \footnote{For further details about the calculation see section \ref{RG-OPE}, where also the discussion regarding $C_{7,10}$ is presented.}
\be
C_9^{\text{IL}}\simeq -\lambda_s 16/9 \ln \big(m_s/m_d), \label{C9ILsecond}
\ee
where $\lambda_j=V_{cj}^* V_{uj}/(V_{cb}^* V_{ub})$. Using $m_s/m_d=17-22$ \cite{Hagiwara:pw} we arrive at the value $V_{cb}^*V_{ub} C_9^{\text{IL}}\simeq -V_{cs}^* V_{us} 16/9 \ln \big(m_s/m_d)=-1.13 \pm 0.06 $, which should be compared to the value obtained using the QCD corrected Wilson coefficient $C_9$ in Table \ref{tab-Wilson}. This gives $V_{cb}^*V_{ub} C_9(\mu)\sim 10^{-4}$, which is four magnitudes smaller than the corresponding parameter obtained by neglecting QCD interactions! The reason for this discrepancy lies in the appearance of large logarithms $\ln(m_{d,s}/m_W)$, that avoid the GIM suppression \index{GIM suppression} otherwise present in $C_9$. It is exactly these large logarithms that RG evolution sums correctly \cite{Buchalla:1995vs}. Since small scales of the order $m_{d,s}$ lie in the nonperturbative region of QCD, the calculation leading to \eqref{C9ILsecond} is not valid at all. 

The logarithm appearing in \eqref{C9IL} is exactly reproduced in the calculation of the inclusive channel $c\to u l^+ l^-$, if mass-independent renormalization is used (see appendix C of \cite{Buras:1991jm}). To show this explicitly, we consider the calculation of $c\to u l^+l^-$ in the naive dimensional regularization (NDR). The amplitude can be parametrized as \index{amplitude!in $c\to u l^+l^-$}
\be
M=- \frac{G_F}{\sqrt{2}}V_{cb}^* V_{ub}\left [ \hat{C}_7^{\text{eff}}\langle Q_7\rangle^{0}+\hat{C}_9^{\text{eff}}\langle Q_9\rangle^{0}+\hat{C}_{10}^{\text{eff}}\langle Q_{10}\rangle^{0}\right],
\ee
with $\langle Q_{7,9,10}\rangle^0$ the tree level matrix elements of the operators. Note that $\hat{C}_{7,9,10}^{\text{eff}}$ are not Wilson coefficients, but merely parametrize the invariant amplitude. The $\hat{C}_9^{\text{eff}}$ coefficient is dominated by the 1-loop contributions coming from insertion of the $Q_{1,2}^q$ operators, $q=d,s$. The virtual photon is emitted from the intermediate $d,s$ quarks. Note that this contribution is of order $\alpha_s^0$ and proportional to $V_{cq}^* V_{uq}$, $q=d,s$, and is thus only once Cabibbo suppressed\index{Cabibbo once suppressed}. Using existing results for $b\to s\l^+l^-$ at the NLO \cite{Grinstein:1988me,Misiak:bc,Buras:1994dj} we arrive at
\be
V_{cb}^* V_{ub} \hat{C}_9^{\text{eff}}=2 V_{cs}^* V_{us} \left(h(z_s,\hat s)-h(z_d, \hat s)\right) \left(3 C_1(m_c) +C_2(m_c)\right), \label{hatc9}
\ee
with $z_q=m_q/m_c$, $\hat s=(m_{l^+l^-}/m_c)^2$ is the reduced mass of the lepton pair, while
\be
\begin{split}
h(z,s)=&-\frac{8}{9} \ln z+\frac{8}{27}+\frac{4}{9}x\\
&-\frac{2}{9}(2+x)\sqrt{|1-x|}
\left\{ 
\begin{aligned}
 \ln\left|\frac{\sqrt{1-x}+1}{\sqrt{1-x}-1}\right|-i \pi,\qquad &\text{for\;}x<1,\\
2 \Arctan\left(\frac{1}{\sqrt{x-1}}\right),\qquad &\text{for\;}x \ge 1,
\end{aligned}
\right.
\end{split}
\ee
where $x=4z^2/s$. In \eqref{hatc9} the contributions suppressed by $V_{cb}^* V_{ub}$ are neglected. These include the tree level contribution from $Q_9$ as well as the 1-loop contributions coming from insertions of the QCD penguin operators $Q_{3,\dots 6}$. From expression \eqref{hatc9} one should reproduce Inami-Lim result \eqref{C9IL}, when momenta and masses of the external particles are set to zero. Taking the limit $m_{l^+l^-}\ll m_{d,s}$, one gets 
\be
\lim_{\hat{s}\to 0}\left(h(z_s,\hat s)-h(z_d, \hat s)\right)\to -\frac{8}{9} \ln\left(\frac{m_s}{m_d}\right).
\ee
Taking the values of the $C_{1,2}$ Wilson coefficients at the weak scale $C_1\simeq 0$, $C_2 \simeq 1$, one arrives at the Inami-Lim result \eqref{C9IL}, as expected. Note, that the logarithm $\ln(m_d/m_s)$ in \eqref{C9IL} arises from insertion of the $Q_{1,2}$ operators. Phenomenologically more interesting is the limit $m_{l^+l^-}\sim m_c\gg m_{d,s}$. In the limit $m_{l^+l^-}\to \infty$ the difference $\left(h(z_s,\hat s)-h(z_d, \hat s)\right)$ vanishes, while for $m_{l^+l^-}\sim m_c$ it is at a level of few percent! Using $C_9^{\text{IL}}$ \eqref{C9IL} instead of $\hat{C}_9^{\text{eff}}$ \eqref{hatc9} as in \cite{Burdman:2001tf,Fajfer:2001sa} then overestimates the $d\text{Br}(c\to u l^+l^-)/d\hat s$.

Explicitly, the branching ratio is \cite{Fajfer:2001sa}
\be
\begin{split}
\frac{\text{Br}(c\to ul^+l^-)}{ds}=&\frac{G_F^2 \alpha_{\text{QED}}^2 m_c^5}{768 \pi^5\Gamma(D^0)}|V_{cb}^* V_{ub}|^2 (1-\hat s)^2\bigg[\bigg\{4\left(1+\frac{2}{\hat s}\right) |\hat{C}_7^{\text{eff}}|^2\\
&+\frac{1+2\hat s}{16}\left(|\hat{C}_9^{\text{eff}}|^2+|\hat{C}_{10}^{\text{eff}}|^2\right)+3 \Re\left(\hat{C}_7^{\text{eff}*}\hat{C}_9^{\text{eff}}\right)\bigg\}\\
&+\bigg\{(\hat{C}_{7,9,10}^{\text{eff}}\to \hat{C}_{7,9,10}^{'\text{eff}}\bigg\} \bigg],
\end{split}
\ee
where we write $\hat s=(m_{l^+l^-}/m_c)^2$ as before, while $\hat{C}_{7,9,10}^{'\text{eff}}$ correspond to the tree level matrix elements of the $Q_{7,9,10}'$ operators \eqref{Qall}. For the value of $\hat{C}_7^{\text{eff}}$ we use the two-loop result of Ref.~\cite{Greub:1996wn}, $\hat{C}_7^{\text{eff}}=\lambda_s (0.007+0.020 i)(1\pm 0.2)$, with $\lambda_s$ defined after Eq.~\eqref{C9ILsecond}. The dominant contribution to $\hat{C}_7^{\text{eff}}$ comes from the insertion of $Q_2^q$ operator, while the contributions from the insertion of $Q_1^q$ operators vanish because of the color structure. The coefficient $\hat{C}_{10}^{\text{eff}}\simeq 0$ in the Standard Model.

Using $m_c=1.4$ GeV one arrives at \index{results}
\begin{equation}
\begin{split}
\text{Br}(c\to u e^+e^-)&=2.4 \times 10^{-10},\\
\text{Br}(c\to u \mu^+\mu^-)&=0.5 \times 10^{-10},
\end{split}
\end{equation}
where the dominant contribution comes from the $\hat{C}_7^{\text{eff}}$ part of the amplitude. This is in contrast to Refs. \cite{Burdman:2001tf,Fajfer:2001sa}, where $\hat{C}_9^{\text{eff}}$ was estimated using $C_9^{\text{IL}}$. This lead to the branching ratios of one (for $e^+e^-$) to two (for $\mu^+ \mu^-$) orders of magnitude higher, with $\hat{C}_9^{\text{eff}}$ contribution dominating the branching ratio.

The suppression of QCD corrected $\hat{C}_9^{\text{eff}}$ \eqref{hatc9} compared to $C_9^{\text{IL}}$ \eqref{C9IL} comes from two sources. The cancellation of $s$ and $d$ quark contributions in \eqref{hatc9} is very strong even at moderate values of $\hat{s}$, with $\left(h(z_s,\hat s)-h(z_d, \hat s)\right)\le 10\% $ for $\hat{s}\ge 0.3$. There is also a sizable cancelation between $C_1(m_c)$ and $C_2(m_c)$ in \eqref{hatc9}. This cancelations could in principle be modified by the two-loop QCD corrections to the $Q_{1,2}$ matrix elements\footnote{The existing two-loop calculations of $Q_{1,2}$ matrix elements in $b\to s l^+l^-$ \cite{Asatrian:2001de,Asatryan:2001zw} have been done for small $\hat s$, where no substantial increase in $c\to u l^+l^-$ is expected.}. If the cancelations are completely lifted, one can estimate the possible effect by $\hat{C}_9^{\text{eff}}\sim \alpha_s(m_c) C_9^{\text{IL}}$. This leads to roughly the same prediction for $\text{Br}(c\to u e^+e^-)$, while it can increase $\text{Br}(c\to u \mu^+\mu^-)$, as $\hat{C}_9^{\text{eff}}$ affects mostly the higher $\hat s$ part of the decay width distribution.

Note, that the calculation of $c\to u l^+ l^-$ is in many respects different than the calculation of $b\to s l^+ l^-$. The operators $Q_{1,2}^{u,c(b\to s)}$ in $b\to s l^+ l^-$ are equivalent to the $Q_{1,2}^{d,s}$ operators in the $c\to u l^+l^-$ transition, but with different CKM factors multiplying the operators in the effective Lagrangian. In $b\to s l^+ l^-$ then only the $Q_{1,2}^{c(b\to s)}$ operators contribute, as $Q_{1,2}^{u(b\to s)}$ operators are $V_{ub}$ suppressed. Hence, there is no approximate cancellation of the type $\left(h(z_s,\hat s)-h(z_d, \hat s)\right)$ as found above. Note also, that in $b\to s l^+ l^-$ the penguin operators $Q_{3,\dots,10}$ are not CKM suppressed relative to $Q_{1,2}$ and have to be taken into account, contrary to the $c\to u l^+ l^-$ case, where the penguin operators are $V_{ub}$ suppressed.

The $V_{ub}$ suppression of the penguin operators $Q_{3,\dots,10}$ is present also in the calculation of the exclusive charm decays, where the insertions of $Q_{1,2}$ operators again dominate the rate. This will be discussed in more detail for the case of $D^0 \to l^+l^- \gamma$ decay in section \ref{llbarGamma}. Before we proceed with the calculation, let us mention the commonly used terminology of long distance (LD) and short distance (SD) contributions. These are usually separated in the discussion of weak radiative decays \index{long-distance contributions} \index{short-distance contributions} 
$q' \to q \gamma \gamma$ or $q' \to q \gamma $ decays.
The SD contribution in these transitions is a result of the
penguin-like transition induced by the operators $Q_{7,9,10}$, while the long distance contribution
arises from the insertions of $Q_{1,2}$ operators, when the off- or on-shell photon is emitted
from the
quark legs. We will follow this classification in the following.

\begin{figure}
\begin{center}
\epsfig{file=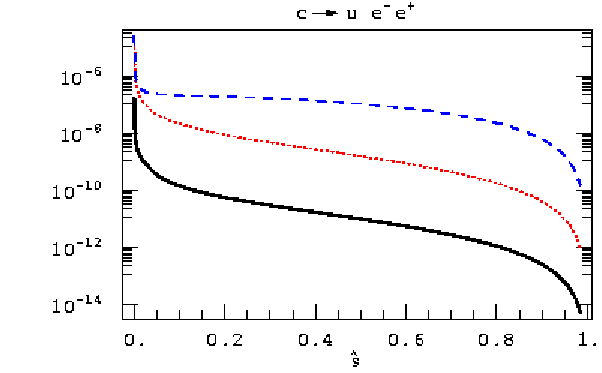}
\caption{\footnotesize{The branching ratio distribution $d \text{Br}(c\to u e^+e^-)/d \hat s$ with $\hat s=(m_{e^+e^-}/m_c)^2$. The Standard Model prediction is represented by the solid line, largest MSSM prediction is denoted by red,dotted line, while largest possible $R$ parity violating contribution is denoted by blue,dashed line.}} \label{dGammaElIncl}
\end{center}
\end{figure}

Finally, we study possible enhancements of the $c\to u l^+l^-$ transitions in the extension beyond the Standard Model. At lower energies the contributions of new physics show up as enhancements of the Wilson coefficients $C_{7,9,10}$, compared to the Standard Model values. The upper bounds on the possible enhancements will be discussed in more detail in section \ref{Beyond}. Here we will only list the effects on inclusive modes. Taking the largest possible effects due to the Minimal Supersymmetric Standard Model, as given in Eq.~\eqref{MSSMbounds} below, one arrives at \index{MSSM} \index{results}\index{signatures, new physics}\index{supersymmetry}
\be
\text{Br}(c\to u e^+e^-)_{\text{MSSM}}= 3.5\times {10}^{-8}, \qquad \text{Br}(c\to u \mu^+\mu^-)_{\text{MSSM}} =0.65\times{10}^{-8},
\ee
where the largest contribution is due to the $Q_7$ and $Q_7'$ operator insertions, similarly to the SM case. If the assumption of $R$ parity violation is relaxed, then the possible enhancement of inclusive modes can be even more drastic(cf. Eq.~\eqref{bounds} below), giving \index{R parity@$R$ parity violating terms}
\be
\text{Br}(c\to u e^+e^-)_{\;\;\sls R\;}= 1.4\times{10}^{-7}, \qquad \text{Br}(c\to u \mu^+\mu^-)_{\;\;\sls R\;}= 1.7\times {10}^{-6},
\ee
On Fig.~\ref{dGammaElIncl} we also show the decay width distribution for $c\to u e^+e^-$ in the SM and beyond. The $c\to u \mu^+\mu^-$ decay width is similar to the $c\to u e^+e^-$ one. In the SM the largest contribution to the inclusive decay come from the low $\hat s$ region, as this is dominated by the $Q_7$ operator. This feature can be changed drastically in the SM extensions. For instance in the MSSM with $R$ parity breaking the largest part of the transition probability comes from the intermediate range of $\hat s$, i.e., with the lepton pair mass of around $m_{l^+l^-}\sim 1$ GeV. Even though possible enhancements due to the non-SM physics can completely dominate in $c\to u l^+l^-$, one has to keep in mind, that inclusive modes are difficult to measure experimentally. From the experimental point of view far more interesting are exclusive rare decays, to which we turn in the next sections.

\index{cll@$c\to u l^+l^-$|)}

\section{The rare $D^0\to \gamma\gamma$ decay}\label{D0GammaGamma}\index{D0gammagamma@$D^0\to \gamma\gamma$|(}\index{decay!$D^0\to \gamma\gamma$|(}
In this section we will present an update to the first detailed calculation of the rare decay mode
$ D^0 \to \gamma \gamma $ in the context of the Standard Model, that has been published in \cite{Fajfer:2001ad}. At the end of section, we will provide also an estimate of possible new physics effects in this decay mode.

The decays which are of some relevance to the $D^0\to \gamma \gamma$
mode, like $D^0\to \rho^0 \gamma$, $D^0\to \omega \gamma$, are
expected to have the 
branching ratios in the $10^{-6}$ range \cite{Fajfer:2000zx}. It is then
hard to believe, that the
branching ratio of the $D^0\to \gamma \gamma$ decay mode could be as high as
$10^{-5}$
in the Standard Model (SM), as found by \cite{Routh:1995np}. Apart from this
estimation, there
was no other detailed work on $D^0\to \gamma \gamma$ in the literature prior to our analysis \cite{Fajfer:2001ad}, to the
best of
our knowledge. Recently a separate analysis of the same decay mode appeared in Ref.~\cite{Burdman:2001tf}. 

 On the other hand, in the B and K meson systems there are numerous
studies
of the two photon decays. For example, the $B_s\to \gamma \gamma$
decay
has been studied with various approaches within the SM and beyond. In the SM,
the
short distance (SD) contribution \cite{LSH} leads to a branching ratio
$\text{Br}(B_s \to \gamma \gamma) \simeq
3.8 \times 10^{-7}$. The QCD corrections enhance this rate to $5 \times
10^{-7}$ \cite{RRS}.
On the other hand, in some of the SM extensions the branching ratio can
be
considerably larger. The two Higgs doublet scenario, for example, could
enhance this branching ratio by an order of magnitude \cite{Boz:1999pr}. Such
"new physics"
effects could at least in principle be dwarfed by the long distance (LD)
effects.
However, existing calculations show, that these are not larger than the
SD
contribution \cite{Choudhury:1998rb,Liu:1999qz}, which is typical of the situation in the 
radiative B decays \cite{Eilam}. In the $K^0$ system the situation is
rather different. Here, the SD
contribution is too small to account for the observed rates of $K_S\to \gamma\gamma$,
$K_L\to \gamma\gamma$ by factors of $\sim 3-5$ \cite{Gaillard:1974hs}, although it
could be of relevance in the
mechanism of CP-violation. Many detailed calculations of these processes
have
been performed over the years (see recent Refs.
 \cite{Gaillard:1974hs,Goity:1986sr,D'Ambrosio:ze,Kambor:1993tv} and Refs. therein),
especially using the chiral approach to account for the pole diagrams
and
the loops. These LD contributions lead to rates which are compatible
with
existing measurements.

\subsection{The theoretical framework}\label{theoretical-frame}
\index{approach, used in thesis|(}
 Motivated by the experimental efforts to observe rare D meson decays (cf. section \ref{rare-review}),
as well as by the lack of detailed theoretical treatments, we undertook
an investigation of the $D^0\to \gamma \gamma$ decay \cite{Fajfer:2001ad}. The short
distance
contribution
is expected to be rather small, as already discussed in the introductory section to this chapter, hence the main contribution would come from the long
distance
interactions. In order to treat the long distance contributions, we use
the
heavy quark effective theory combined with the chiral perturbation theory,
HH$\chi$PT (see chapter \ref{HQET}). This approach was used before for $D^*$ strong and
electromagnetic
decays \cite{Casalbuoni:1996pg,Stewart:1998ke,Guetta:1999vb}. The leptonic and semileptonic decays of
D meson were also treated
within the same framework (see \cite{Casalbuoni:1996pg} and references therein).

In the calculation we will use the following classification of the short distance (SD) and long distance (LD) contributions. The SD contributions arise from the insertion of penguin like operators that describe either $c\to u\gamma$ \index{ctougamma@$c\to u\gamma$} or $c\to u \gamma\gamma$ \index{cgammagamma@$c\to u\gamma\gamma$} FCNC transitions, that occur through weak-scale interactions (exchanges of $W$ bosons). The seven dimensional Wilson operators describing $c\to u \gamma\gamma$ transition are expected to be suppressed by additional powers of weak scale compared to the five dimensional $c\to u\gamma$ Wilson operator $Q_7$. For instance, for a very similar $b \to s \gamma \gamma$ decay, one obtains that
without QCD corrections the ratio
$\Gamma (b \to s \gamma \gamma)/ \Gamma (b \to s \gamma)$
is about $10^{-3}$ \cite{Herrlich:1991bq}. The largest SD contribution to $D^0\to \gamma\gamma$ is thus expected from the SD $c\to u\gamma$ transition, i.e., from the insertion of $Q_7$ operator \eqref{effective_lagr}, while the other photon is emitted from initial meson leg. The value of $C_7$ Wilson coefficient is given in \eqref{C7numbers}, where we use the scale $\mu=1$ GeV. 

The LD contributions on the other hand arise from the processes, where ${\it both}$ photons are emitted through purely electromagnetic transitions, that occur at low scales. The main LD contribution will arise from the insertion of $Q_{1,2}$ operators. To estimate the matrix elements of these operators, we use the factorization approximation (explained in section \ref{Factorization}). \index{factorization approximation} \index{nonfactorizable contributions} The resulting effective four quark
nonleptonic $\Delta C = 1$
 weak Lagrangian is given in Eq.~\eqref{fact-lagr}, with the effective Wilson coefficients $a_{1,2}$ given in \eqref{coeff-a1a2}. The hadronic degrees of freedom are described using the HH$\chi$PT, explained in chapter \ref{HQET}. The heavy quark and chiral symmetries also provide us with the form of weak currents, given in \eqref{jX}, \eqref{current}. In a similar way, also the hadronic matrix element of $Q_7$ is determined \eqref{Q7hadr}.

The photon couplings are obtained by gauging the Lagrangians
\eqref{chirallagr}, \eqref{eq-8} and the light current \eqref{jX}
with the $U(1)$ photon field $B_\mu$.
 Following \cite{Stewart:1998ke} we add in addition the electromagnetic interaction \eqref{eq-100} with
an unknown coupling $\beta$
of dimension -1,
which is needed to account, for example, for $D^*\to D \gamma$. Even though the
Lagrangian \eqref{eq-100} is formally $1/m_Q \sim m_q$ suppressed,
we
do not neglect it, as it has been found that it gives a sizable
contribution to $D^*(B^*)\to D(B) \gamma\gamma$ decays \cite{Guetta:1999vb}. In the
case of $D^0\to \gamma\gamma$ it gives largest contribution to the
parity conserving part
of the
amplitude, however, it does not contribute to the decay rate
 by more than $10\%$, as
 will be shown later.

 The approach of the HH$\chi$PT introduces several coupling constants that
have
to be determined from experiment. The recent measurement of the $D^*$
decay
width \cite{Ahmed:2001xc,Anastassov:2001cw} has determined the $D^*D\pi$ coupling, which is
related to $g$, the basic
strong coupling of the Lagrangian. Further discussion on this point can be found in section \ref{coupling}. There is more ambiguity, however,
concerning
the value of the anomalous electromagnetic coupling $\beta$, which is responsible for
the $D^*D\gamma$ decays \cite{Stewart:1998ke,Guetta:1999vb} (see section \ref{coupling}).

 Let us address now some issues concerning the theoretical framework \index{difficulties!of approach}
 used in our treatment. The
typical energy of the intermediate pseudoscalar mesons is of order $m_D/2$,
so that
the chiral expansion $p/\Lambda_\chi$ (for $\Lambda_\chi \gtrsim 1$
GeV) is rather close to
unity. Thus, for the decay under study here, we extend the possible range
of
applicability of the chiral expansion of HH$\chi$PT, compared to the 
previous treatments
like $D^*\to D \pi$, $D^*\to D \gamma$ \cite{Stewart:1998ke} or $D^*\to D
\gamma \gamma$ \cite{Guetta:1999vb}, in which a heavy meson
appears in the final state, making the use of the chiral perturbation theory
rather natural. The suitability of our undertaking here must be
confronted
with the experiment, and possibly other theoretical approaches.

 At this point we also remark that the contribution of the order ${\cal
O}(p)$ does
not exist in the $D^0\to \gamma \gamma$ decay, and the amplitude starts
with
the contribution of the order ${\cal O}(p^3)$. At this order the amplitude
receives an \index{annihilation diagrams}
annihilation type contribution proportional to the $a_2$ Wilson
coefficient, with the Wess-Zumino anomalous term coupling light
pseudoscalars
to two photons. As we will show, the total amplitude is dominated by
terms
proportional to $a_1$ that contribute only through loops with Goldstone
bosons. Loop contributions proportional to $a_2$ vanish at this order.
We point out that any other model which does not involve intermediate
charged states cannot give this kind of contribution. Therefore, the
chiral
loops naturally include effects of the intermediate meson exchange.

 The chiral loops of order ${\cal O}(p^3)$ are finite, as they are in
the similar
case of $K\to \gamma \gamma$ decays \cite{Gaillard:1974hs,Goity:1986sr,D'Ambrosio:ze,Kambor:1993tv}. The
next to leading terms might be
almost of the same order of magnitude compared to the leading ${\cal
O}(p^3)$ term,
the expected suppression being approximately $p^2/\Lambda^2_\chi$. The
inclusion of next order terms in the chiral expansion is not
straightforward
in the present approach. As already mentioned earlier, we do include, however, terms which contain the
anomalous
electromagnetic coupling $\beta$ \eqref{eq-100}, and appear as next to leading order terms in
the
chiral expansion, in view of their potentially large contribution (as in
$B^*(D^*)\to B(D) \gamma \gamma$ decays considered in \cite{Guetta:1999vb}). As it
turns out, these
terms are suppressed compared to the leading loop effects, which at
least
partially justifies the use of HH$\chi$PT for the decay under
consideration.
Contributions of the same order could arise from the light resonances like
$\rho$,
$K^*$, $a_0(980)$, $f_0(975)$. Such resonances are sometimes treated
with hidden
gauge symmetry (see, e.g., \cite{Casalbuoni:1996pg}), which is not compatible with
the chiral perturbation
symmetry. Therefore, a consistent calculation of these terms is beyond
our
scheme and we disregard their possible effect. 

\index{approach, used in thesis|)}\index{resonant contributions}

\subsection{Results}\index{contributions!chiral loop|(}
\label{results}
The invariant amplitude for the $D^0 \to \gamma \gamma $ decay can
 be written using gauge and Lorentz invariance in the following
 form: \index{amplitude!in $D^0\to \gamma\gamma$}
\begin{equation}
M = \left[ i M^{(-)} \left(g^{\mu \nu} -\frac{k_2^\mu k_1^\nu}{k_1
\negcdot k_2} \right)+
 M^{(+)} \epsilon^{\mu \nu\alpha\beta}k_{1\alpha}k_{2\beta}\right]
 \epsilon_{1\mu}\epsilon_{2\nu},\label{eq-104}
\end{equation}
where $M^{(-)}$ is the parity violating and $M^{(+)}$
the parity conserving part of the amplitude,
while $k_{1(2)}$, $\epsilon_{1(2)}$ are respectively the
four momenta and the polarization vectors of the outgoing
photons.

\begin{figure}
\begin{center}
\epsfig{file=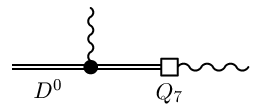, height=1.5cm}
\caption{\footnotesize{The short distance diagram with the insertion of the $Q_7$ operator. The blob $\bullet$ denotes the $\beta$-like vertex \eqref{eq-100}.}}\label{SDgammagamma}
\end{center}
\end{figure}

\begin{figure}
\begin{center}
\epsfig{file=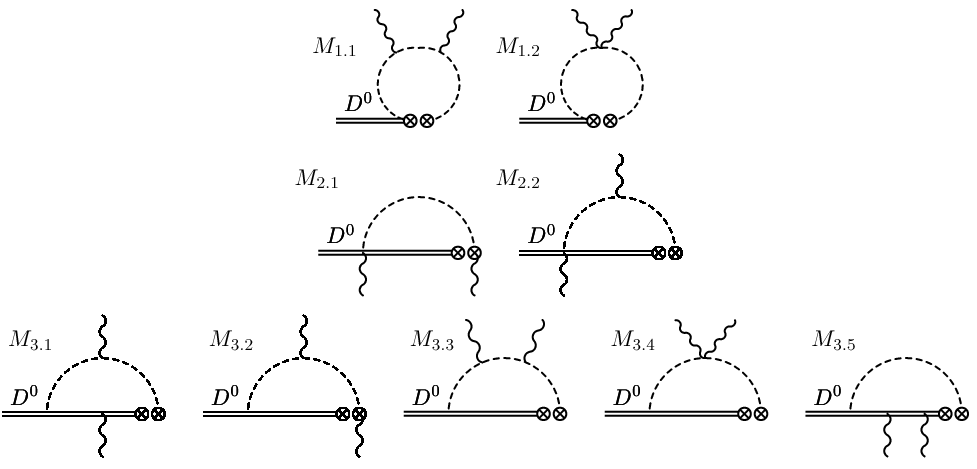, height=5.cm}
\caption{\footnotesize{One loop diagrams, not containing beta-like terms
\eqref{eq-100}, that give nonvanishing contributions to the
$D^0\to
\gamma \gamma$ decay amplitude. Each sum of the amplitudes corresponding to the
diagrams in one row $M_i=\sum_j
M_{i.j}$ is gauge invariant and finite. The numerical values are
listed in
Table \ref{tab-1D0GammaGamma}. }}\label{fig2D0GammaGamma}
\end{center}
\end{figure}

\begin{figure}
\begin{center}
\epsfig{file=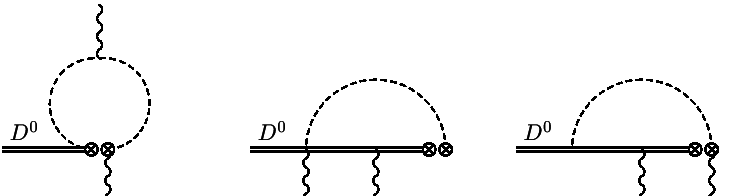, height=2.1cm}
\caption{\footnotesize{One loop diagrams (not containing $\beta$-like terms
\eqref{eq-100}) that give vanishing contributions. The dashed line
represents
charged
Goldstone bosons flowing in the loop ($K^+,\pi^+$), while the
double line represents the heavy mesons, $D$ and $D^*$.}}\label{fig1D0GammaGamma}
\end{center}
\end{figure}

Using the amplitude decomposition \eqref{eq-104}, the decay width for the $D^0\to \gamma \gamma$ decay is
\begin{equation}
\Gamma_{D^0\to \gamma \gamma}= \frac{1}{16 \pi m_D} \left(\left|M^{(-)}\right|^2+
\frac{1}{4} \left|M^{(+)}\right|^2 m_D^4\right). \label{eq-106}
\end{equation}
The short distance contribution to the $D^0\to \gamma\gamma$ decay
width is
estimated using the $c\to u \gamma$ transition induced by the $Q_7$ operator in \eqref{effective_lagr}, with one photon emitted from the $D^0$ leg via ${\cal
L}_\beta$ term \eqref{eq-100}, as shown on Figure \ref{SDgammagamma}. The parity violating part of the
short distance amplitude is
\begin{equation}\label{MSD-}
 M_{\text{SD}}^{(-)}= -\frac{m_D^{3/2}}{12 \pi^2} \frac{G_F}{\sqrt{2}}
V_{ub} V_{cb}^* C_{7}^{\text{eff}} e^2
(\beta m_c+1)\alpha \frac{1}{1+2 \Delta^*/m_D},
\end{equation}
while the parity conserving part of the amplitude is
\begin{equation}\label{MSDplus}
 M_{\text{SD}}^{(+)}=-\frac{\sqrt{m_D}}{12 \pi^2} \frac{G_F}{\sqrt{2}}
V_{ub} V_{cb}^* C_{7}^{\text{eff}}
e^2
(\beta m_c+1) \alpha \frac{2}{m_D+2 \Delta^*},
\end{equation}
where $\Delta^{*}=m_{D^{0*}}-m_{D^0}$.
Turning now to the long distance contributions, we depict in Figs.
\ref{fig1D0GammaGamma} and \ref{fig2D0GammaGamma} the loop diagrams
arising to the leading order ${\cal O }(p^3)$ by using
Eqs.~\eqref{chirallagr}, \eqref{eq-8}, \eqref{jX}, \eqref{current}. The circled
crosses indicate the currents in the weak
interaction, with the left circled cross representing the heavy current \eqref{current}, and the right circled cross indicating the light current \eqref{jX}. In Figure \ref{fig1D0GammaGamma} we grouped all the diagrams which vanish
by
 the symmetry considerations. All
nonvanishing contributions are assembled in Fig.~\ref{fig2D0GammaGamma}\footnote{Note that in \cite{Fajfer:2001ad} the diagram $M_{3.5}$ had been erroneously left out of the figure. The results presented in \cite{Fajfer:2001ad} are, however, correct.}. We denote
the gauge invariant sums corresponding
to the nonvanishing diagrams of Fig.~\ref{fig2D0GammaGamma} by $M_i^{(\pm)}=\sum_j
M_{i.j}^{(\pm)}$ (the gauge invariant sums are sums of the diagrams in each
row of Fig.~\ref{fig2D0GammaGamma}), where $+ (-)$ denotes parity conserving
(violating) part of the amplitude, as in \eqref{eq-106}. Note that the gauge invariant sets of diagrams satisfy the theorem given in section \ref{gauge}\footnote{In each row actually a sum of two smaller gauge invariant sets is given. This is done for the sake of convenience, as otherwise the $k_1\leftrightarrow k_2$ symmetry would not be present explicitly in the results. Compare also with Figure \ref{fig-1D0llBarGamma} in the next section.}. The parity
violating sums, which
arise from the $a_1$ term in \eqref{fact-lagr} are \\
\begin{align}
M_1^{(-)}&= -\frac{(m_D)^{3/2}}{4 \pi^2} \frac{G_F}{\sqrt{2}}
a_1
\alpha e^2 \left[V_{us} V_{cs}^*\;
M_4\!\!\left(m_K,-\tfrac{m_D^2}{2}\right)+V_{ud}V_{cd}^*\;
M_4\!\!\left(m_\pi, -\tfrac{m_D^2}{2}\right)\right]\label{eq-1},\\
\begin{split}
M_2^{(-)}&= - \sqrt{m_D} \frac{G_F}{\sqrt{2}} a_1 e^2 g \alpha
\frac{1}{8 \pi^2} \Big[\\
&\qquad V_{us} V_{cs}^*\left(
\bar{B}_0\left(m_K,\tfrac{m_D}{2}+\Delta_s^*\right) +
2G_3\left(m_K,m_D+\Delta_s^*,-\tfrac{m_D}{2}\right)\right)+\\
&\qquad V_{ud} V_{cd}^*\left(
\bar{B}_0\left(m_\pi,\tfrac{m_D}{2}+\Delta_d^*\right) +
2G_3\left(m_\pi,m_D+\Delta_d^*,-\tfrac{m_D}{2}\right)\right)\Big],
\end{split}\\
M_3^{(-)}&= \sqrt{m_D} \frac{G_F}{\sqrt{2}} a_1 g e^2 \alpha
\frac{1}{2 \pi^2} \Big[
V_{us}V_{cs}^* f(m_K,\Delta_s^*,m_D) +V_{ud} V_{cd}^*
f(m_\pi,\Delta_d^*,m_D)\Big], \label{M3-}
\end{align}
with
\begin{equation}
\begin{split}
f(m,&\Delta,m_D)= \frac{m^2}{m_D} \Big[
G_0\big(m,\Delta+\tfrac{m_D}{2},\tfrac{m_D}{2}\big)-\frac{1}{2}
G_0(m,\Delta,\tfrac{m_D}{2})\Big]+\frac{5 m_D}{8} + \frac{\Delta}{2}\\
& +\frac{(m^2-\Delta^2)}{2} \Big[
\Big(\frac{1}{2}+\frac{\Delta}{m_D}\Big)\; \overline{G}_0\big(m,
\Delta+\tfrac{m_D}{2},m_D\big)+m^2\;
\overline{M}_0\big(m,\Delta+\tfrac{m_D}{2},m_D\big)\\
&\qquad\qquad\qquad-\frac{1}{m_D}B_0(m_D^2,m^2,m^2)\Big]\\
&+ \Big(\Delta-\frac{m_D}{2}\Big)
M_2(m,-\tfrac{m_D^2}{2})+\frac{1}{4}\Big(\frac{m_D}{2}-\Delta\Big)
B_0(m_D^2,m^2,m^2)\\
&- \frac{(2\Delta+m_D)}{4 m_D }
\bar{B}_0(m,\Delta+m_D)+\frac{(3m_D^2/2+3
\Delta m_D +2 \Delta^2-2 m^2)}{2 m_D^2 }
\bar{B}_0(m,\Delta+\tfrac{m_D}{2})\\
&-\frac{(m_D \Delta -2 m^2 +
2 \Delta^2)}{4 m_D^2} \bar{B}_0(m,\Delta).
\end{split}
\end{equation}
The parity conserving parts of the amplitude $M_i^{(+)}$ vanish
for the diagrams on Fig.~\ref{fig2D0GammaGamma}.
We denoted $\Delta_q^{(*)}=m_{D_q^{(*)}}-m_{D^0}$,
while the definitions of the scalar integrals 
$\bar{B}_0(m,\Delta)$, $B_0(k^2,m^2,m^2)$ can be found in section \ref{Notational-conventions}, with the explicit expressions given in \eqref{eqS:33}, \eqref{B_0Veltman} The abbreviations for the tensor integrals $G_0(m,\Delta,v\negcdot k)$,
$G_3(m,\Delta,v\negcdot k)$,
$\overline{G}_0(m,\Delta,v\negcdot k)$,
$\overline{M}_0(m,\Delta,v \negcdot k)$, $M_2(m,k_1\negcdot k_2)$,
 $M_4(m,k_1 \negcdot k_2)$ are presented in appendix \ref{Notational-glossary}.

 Note that the sums of the amplitudes \eqref{eq-1}-\eqref{M3-}
 are gauge invariant and finite.
This is expected, since one cannot generate counterterms at \index{counterterms! contributions of}
 this order. There is no $\mu$ dependence apart
 from the one hidden in $a_1$ (cf. discussion after Eq.~\eqref{fact-lagr}), even though $\mu$ appears in
 the above functions, but it
 cancels out completely. Note also, that the one
 loop chiral corrections vanish in the exact $SU(3)$ limit,
 i.e., when $m_K \to m_\pi$, as it is expected.
One should note that taking the chiral limit (i.e. $m_s, m_d \to 0$)
is not unambiguous. Namely, in the combined heavy quark effective theory
and
the chiral perturbation theory, beside chiral logarithms
there are also functions of
the form $F(m_q/\Delta)$ \eqref{F_def} whose value depend on the way one takes the
limit (see, e.g., Ref.~\cite{Boyd:1994pa}).

We remark that there exist additional diagrams of the same order in
the chiral
expansion as the ones given on Fig.~\ref{fig2D0GammaGamma}, but proportional to
the $a_2$ part of the effective
weak Lagrangian \eqref{fact-lagr}. In these additional diagrams, the chiral
loop is attached to the light current in the factorized vertex, while
the photons are
 emitted from the
pseudoscalars in the loop, or they come from the weak vertex.
However, the amplitudes of these diagrams
 vanish due to Lorentz symmetry.

\begin{figure}
\begin{center}
\epsfig{file=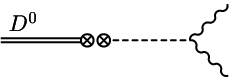, height=1cm}
\caption{\footnotesize{Anomalous contributions to $D^0\to \gamma\gamma$ decay.
The intermediate pseudoscalar mesons propagating from the weak
vertex are $\pi^0,
\eta,\eta'$. }}\label{fig3D0GammaGamma}
\end{center}
\end{figure}

The contribution coming from the anomalous coupling \index{anomalous interaction}
$\pi^0 \gamma \gamma$, $\eta\gamma \gamma$, $\eta'\gamma \gamma$ \index{Wess-Zumino term}
(Fig.~\ref{fig3D0GammaGamma}) is
\begin{equation}
\begin{split}
M_{\text{Anom.}}^{(+)}=-&\sqrt{m_D} \frac{G_F}{\sqrt{2}}a_2 \alpha
\frac{e^2}{4 \pi^2}
\sum_{P=\pi^0,\eta,\eta'} \frac{m_D}{m_D^2-m_P^2} K_P\\
&K_{\pi^0}=V_{ud}V_{cd}^*\\
& K_\eta=\big[V_{ud}V_{cd}^*\big(\tfrac{\sin
\Theta}{\sqrt{3}}-\tfrac{\cos
\Theta}{\sqrt{6}}\big)+V_{us}V_{cs}^*\big(\tfrac{\sin
\Theta}{\sqrt{3}}+\tfrac{\sqrt{2}\cos
\Theta}{\sqrt{3}}\big)\big]\big[\tfrac{\sqrt{2}\cos
\Theta}{\sqrt{3}}-\tfrac{4\sin
\Theta}{\sqrt{3}}\big]\\
& K_\eta'=\big[-V_{ud}V_{cd}^*\big(\tfrac{\sin
\Theta}{\sqrt{6}}+\tfrac{\cos
\Theta}{\sqrt{3}}\big)+V_{us}V_{cs}^*\big(\tfrac{\sqrt{2}\sin
\Theta}{\sqrt{3}}
-\tfrac{\cos
\Theta}{\sqrt{3}}\big)\big]\big[\tfrac{\sqrt{2}\sin\Theta}{\sqrt{3}}
+\tfrac{4\cos \Theta}{\sqrt{3}}\big],
\end{split}
\end{equation}
where $\theta=-20^o \pm 5^o $ is the $\eta -\eta'$ mixing angle
and we have set $f_\pi=f_{\eta_8}=f_{\eta_0}$. This choice of the 
parameters reproduces the experimental results for the
$\pi^0\to \gamma \gamma$, $\eta \to \gamma \gamma$, and
$\eta'\to \gamma \gamma$ decay width \cite{Hagiwara:pw}.
In the numerical evaluation we use the values of $\alpha=0.38\pm 0.04 \; \text{GeV}^{3/2}$ and $g=0.59\pm0.08$
obtained in the beginning of section \ref{coupling} and summarized in Table \ref{tab-koef}.
 Note, that the $SU(3)$ breaking effects in the form of chiral loops
and the counterterms can change the extracted value of $\alpha$.
One chiral loop corrections can amount to about
$40\%$ when $g$ is taken to be $0.59$.
This value might be changed
by the finite part of the counterterms. \index{counterterms! contributions of}
However, the contributions coming from the 
counterterms are not known and due to the lack of experimental data
they cannot be fixed yet. In our
calculation we take $\alpha=0.38\; {\rm GeV}^{3/2}$, keeping in
mind that
the chiral corrections might be important. 
For the effective Wilson coefficients $a_1$ we take $1.26$ \eqref{coeff-a1a2}.
We present the numerical results\footnote{Note, that in \cite{Fajfer:2001ad} slightly different value of $\alpha$ has been used. Also the two-loop QCD improved $\hat{C}_7^{\text{eff}}$ from section \ref{ctoullbar}, instead of the Wilson coefficient $C_7$, has been used. }
 for the one loop amplitudes in Table \ref{tab-1D0GammaGamma}. Short distance contribution is negligible as expected.

\begin{table} [h] \index{tables of results}
\begin{center}
\begin{tabular}{|l|r l|r l|} \hline\index{amplitude!in $D^0\to \gamma\gamma$}
 &$M_{ i}^{(-)}$&$ [\times 10 ^{-10}{\rm \;GeV}]$& $M_{i}^{(+)}$&$
[\times10
^{-10}{\rm \;GeV^{-1}}]$\\ \hline\hline
Anom. & $0$ & & $-0.65$& \\ \hline
SD & $-0.004$&$-0.005 i$ & $-0.002$&$ -0.003 i$ \\ \hline
$1$ & $4.35$&$+11.5i$ & $0$&\\ \hline
$2$ &$2.05$ & & $0$& \\ \hline
$3$ & $-0.66$&$+3.43i$&$0$& \\ \hline
\hline
$\sum_i M_i^{(\pm)}$& $5.73$&$+14.95 i$ &$-0.65$&$ -0.003 i$\\ \hline

\end{tabular}
\caption[Table of the nonvanishing finite
amplitudes in $D^0\to \gamma\gamma$]{\footnotesize{Table of the nonvanishing finite
amplitudes. The amplitudes coming from the
anomalous and short distance
($C_{7}^{\text{eff}}$) Lagrangians are presented. The finite
and gauge invariant sums of the
one-loop amplitudes are listed in the next three lines
($M_i^{(\pm)}=\sum_j M_{i.j}^{(\pm)}$).
The numbers $1,2,3$ denote the row of diagrams on the
Fig.~\ref{fig2D0GammaGamma}.
In the last line the sum of all the amplitudes is given.}}
\label{tab-1D0GammaGamma}
\end{center}
\end{table}

\begin{figure}
\begin{center}
\epsfig{file=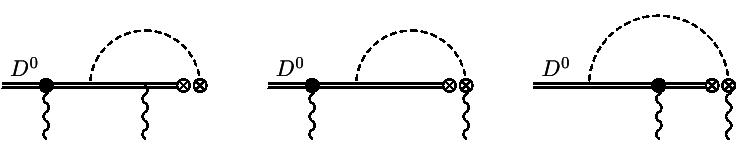, height=2cm}
\caption{\footnotesize{The diagrams with one $\beta$-like \eqref{eq-100} coupling
(described by $\bullet$), which give
vanishing amplitudes. }}\label{fig4D0GammaGamma}
\end{center}
\end{figure}

\begin{figure}
\begin{center}
\epsfig{file=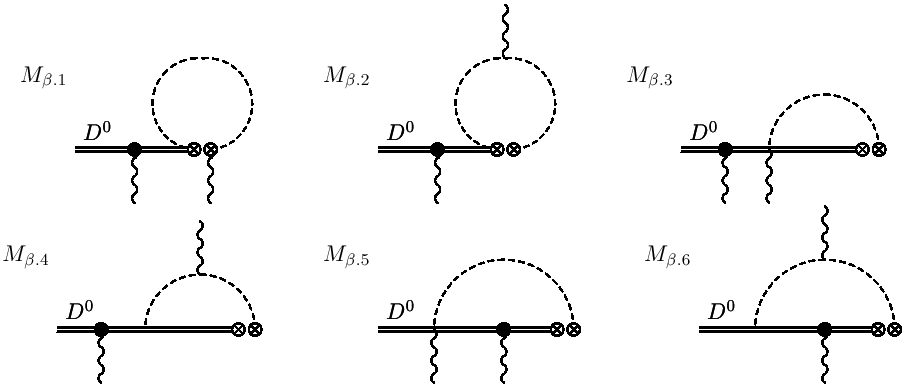, height=4.5cm}
\caption{\footnotesize{The diagrams which give nonzero amplitudes
with one $\beta$-like coupling. }}\label{fig5D0GammaGamma}
\end{center}
\end{figure}

In the determination of $D^*\to D \gamma\gamma $ and
$B^* \to B \gamma \gamma$ a sizable contribution from
$\beta$-like electromagnetic terms \eqref{eq-100} has been
found \cite{Guetta:1999vb}. Therefore we have to investigate their
effect
in the $D^0 \to \gamma \gamma $ decay amplitude. The terms in
Eq.~\eqref{eq-100} lead to an additional set of diagrams, which is given in Figs.
\ref{fig4D0GammaGamma} and \ref{fig5D0GammaGamma}, where the $\beta$ vertex is indicated by
$\bullet$.
The nonzero parity violating parts of the one loop
diagrams containing
$\beta$
coupling are (Fig.~\ref{fig5D0GammaGamma})
\begin{align}
\begin{split}
M_{\beta.4}^{(-)}&= \sqrt{m_D}\frac{G_F}{\sqrt{2}}a_1 e^2 g
\alpha
\left(\beta
+\tfrac{1}{m_c}\right) \frac{1}{(m_D +2 \Delta^*)} \frac{1}{16 \pi^2}
\frac{m_D^2}{3}\times\\
&\times\Big[ V_{us}V_{cs}^*
G_3(m_K,m_D+\Delta_s^*,-\tfrac{m_D}{2})+
V_{ud}V_{cd}^*G_3(m_\pi,m_D+\Delta_d^*,-\tfrac{m_D}{2})\Big],
\end{split}\\
\begin{split}
M_{\beta.6}^{(-)}&= \frac{G_F}{\sqrt{2}} a_1 e^2 g \alpha
\left(\beta-\tfrac{2}{m_c}\right)
\frac{(m_D)^{\frac{3}{2}}}{48 \pi^2}\Big\{ V_{us}V_{cs}^*
\Big[G_3(m_K,\tfrac{m_D}{2}+\Delta_s^*,\tfrac{m_D}{2})-G_3(m_K,\Delta_s^*,\tfrac{m_D}{2})\Big]\\
&\qquad \qquad \qquad \qquad + V_{ud}V_{cd}^*
\Big[G_3(m_\pi,\tfrac{m_D}{2}+\Delta_d^*,
\tfrac{m_D}{2})-G_3(m_\pi,\Delta_d^*,\tfrac{m_D}{2})\Big]\Big\},
\end{split}
\end{align}
while the parity conserving parts of the amplitudes arising from the
one loop diagrams with $\beta$
coupling are
\begin{align}
M_{\beta.1}^{(+)}&=-\frac{G_F}{\sqrt{2 m_D}}a_1 e^2 \alpha
\left(\beta+\tfrac{1}{m_c}\right)
\frac{1}{m_D+2 \Delta^*} \frac{1}{12 \pi^2}\Big[ V_{us}V_{cs}^*
A_0(m_K^2)+V_{ud}V_{cd}^* A_0(m_\pi^2)\Big],\\
\begin{split}
M_{\beta.2}^{(+)}&=\frac{G_F}{\sqrt{2 m_D}}a_1 e^2 \alpha
\left(\beta+\tfrac{1}{m_c}\right)\frac{1}{(m_D+2\Delta^*)}\frac{1}{12\pi^2}
\Big[V_{us}V_{cs}^* A_0(m_K^2)+V_{ud}V_{cd}^*
A_0(m_\pi^2)\Big]\\
&=-M_{\beta.1}^{(+)},
\end{split}
\\
\begin{split}
M_{\beta.3}^{(+)}&=\frac{G_F}{\sqrt{2 m_D}} a_1 e^2 g
\alpha\left(\beta+\tfrac{1}{m_c}\right)
\frac{1}{m_D+ 2 \Delta^*} \frac{1}{12\pi^2}\Big\{ V_{us}V_{cs}^*\big[
(m_D+\Delta_s) \bar{B}_0(m_K,m_D+\Delta_s) +\\
&+A_0(m_K^2)\big]+
+V_{ud}V_{cd}^*\big[(m_D+\Delta_d)\bar{B}_0(m_\pi,m_D+\Delta_d)+A_0(m_\pi^2)\big]\Big\},
\end{split}\\
\begin{split}
M_{\beta.4}^{(+)}&=\frac{G_F}{\sqrt{2 m_D}}a_1 e^2 g \alpha
\left(\beta
+\tfrac{1}{m_c}\right)
\frac{1}{m_D+2 \Delta^*} \frac{1}{6\pi^2}\Big\{ V_{us}V_{cs}^* \big[(m_D+\Delta_s) G_3(m_K,m_D+\Delta_s,-\tfrac{m_D}{2})-\\
&-\frac{1}{2}
A_0(m_K^2)\big]+ V_{ud}V_{cd}^* \big[(m_D+\Delta_d)G_3(m_\pi,m_D+
\Delta_d,-\tfrac{m_D}{2})-\frac{1}{2}
A_0(m_\pi^2)\big]\Big\},
\end{split}
\end{align}
\begin{align}
\begin{split}
M_{\beta.5}^{(+)}&=\frac{G_F}{\sqrt{2m_D}} a_1 e^2
g \alpha\left(\beta
-\tfrac{2}{m_c}\right)
\frac{1}{48 \pi^2}\Big\{ V_{us}V_{cs}^* \frac{1}{\tfrac{m_D}{2}+
(\Delta_s-\Delta_s^*)}\big[(m_D+\Delta_s)\bar{B}_0(m_K,m_D+\Delta_s)\\
&-\left(\tfrac{m_D}{2}+\Delta_s^*\right)\bar{B}_0\!\left(m_K,\tfrac{m_D}{2}+\Delta_s^*\right)\big]+ V_{ud}V_{cd}^*\frac{1}{\tfrac{m_D}{2}+
(\Delta_d-\Delta_d^*)}\big[(m_D+\Delta_d)\times\\
& \times\bar{B}_0(m_\pi,m_D+\Delta_d)-
\left(\tfrac{m_D}{2}+\Delta_d^*\right)\bar{B}_0\left(m_\pi,\tfrac{m_D}{2}+\Delta_d^*\right)\big]\Big\},
\end{split}\\
\begin{split}
M_{\beta.6}^{(+)}&=\frac{G_F}{\sqrt{2 m_D}}a_1 e^2 g \alpha
\left(\beta-\tfrac{2}{m_c}\right)
\frac{1}{24 \pi^2}\Big\{
V_{us}V_{cs}^* \Big[-\frac{(\frac{m_D}{2}+
\Delta_s^*)}{(\frac{m_D}{2}+\Delta_s-\Delta_s^*)}G_3\big(m_K,
\tfrac{m_D}{2}+\Delta_s^*,-\tfrac{m_D}{2}\big)+\\
&+\frac{(m_D+\Delta_s)}{(\frac{m_D}{2}+\Delta_s-
\Delta_s^*)}G_3\big(m_K,m_D+\Delta_s,-\tfrac{m_D}{2}\big)\Big]
+V_{ud}V_{cd}^* \Big[-\frac{(\frac{m_D}{2}+
\Delta_d^*)}{(\frac{m_D}{2}+\Delta_d-
\Delta_d^*)}\times \\
&\times G_3\big(m_\pi,\tfrac{m_D}{2}+\Delta_d^*,-
\tfrac{m_D}{2}\big)+\frac{(m_D+\Delta_d)}{(\frac{m_D}{2}+\Delta_d-\Delta_d^*)}G_3\big(m_\pi,m_D+\Delta_d,-
\tfrac{m_D}{2}\big)\Big]\Big\}.
\end{split}
\end{align}
The amplitudes with $\beta$ coupling are not finite and have to be
regularized. We use the strict $\overline{\text{MS}}$ prescription
 $\bar{\Delta}=1$ and take $\mu = 1$ $\rm{GeV}
\simeq
\Lambda_\chi$ as in \cite{Stewart:1998ke}.

 Results are gathered in Table \ref{tab-2}
using
$\beta=2.3 \; {\rm GeV}^{-1}$ and $m_c=1.4 \; {\rm GeV}$.
Inspection of Tables \ref{tab-1D0GammaGamma} and \ref{tab-2} reveals that for the
real
parts of the
amplitudes, the
contributions of Figs. \ref{fig2D0GammaGamma}, \ref{fig3D0GammaGamma} and of Fig.~\ref{fig5D0GammaGamma}
are
comparable in size. However, the
decay rate is dominated
by the contribution of the imaginary part of the parity-violating
amplitude, which arises from the
one loop diagrams of Fig.~\ref{fig2D0GammaGamma}. For the parity-conserving
amplitude, the
contributions of anomaly
and $\beta$-like terms are comparable in magnitude. Due to the suppression of $a_2$ in comparison to $a_1$, we do not
include diagrams proportional to $a_2$ in the calculation of terms with
$\beta$.

\begin{table} [h] \index{tables of results}
\begin{center}
\begin{tabular}{|l|c|c|c|} \hline \index{amplitude! in $D^0\to \gamma\gamma$}
Diag. &$M_{i}^{(-)} [\times 10 ^{-10}{\rm \;GeV}]$& $M_{i}^{(+)}
[\times 10
^{-10}{\rm \;GeV^{-1}}]$\\ \hline\hline
$\beta.1$ & $0$ & $-3.30$ \\ \hline
$\beta.2$ & $0$ & $3.30$\\ \hline
$\beta.3$ &$0$ & $2.60$ \\ \hline
$\beta.4$ & $1.08$ &$-0.01$ \\ \hline
$\beta.5$ &$0$ & $0.63$ \\ \hline
$\beta.6$ &$-3.53 $&$-0.64$ \\ \hline
\hline
$\sum_i M_i^{(\pm)}$& $-2.45$ &$2.57$\\ \hline

\end{tabular}
\caption[Table of nonzero 
amplitudes corresponding to the diagrams
 with $\beta$ coupling in $D^0\to \gamma\gamma$]{\footnotesize{Table of nonzero 
amplitudes corresponding to the diagrams
 with $\beta$ coupling (Fig.~\ref{fig5D0GammaGamma}).
In the last line the sums of the contributions are presented.
We use $\beta=2.3 $ GeV${}^{-1}$, $m_c=1.4$ GeV. }}
\label{tab-2}
\end{center}
\end{table}

Using short distance contributions, the finite one loop diagrams and the
anomaly parts of the
amplitudes (shown in Figs. \ref{fig2D0GammaGamma}, \ref{fig3D0GammaGamma}
and with the numerical values of the amplitudes listed in Table
\ref{tab-1D0GammaGamma}),
one obtains
\begin{equation}
\text{Br}(D^0\to \gamma\gamma)=1.7 \times 10^{-8}.
\end{equation}
This result is slightly changed when one takes into account the terms
dependent on $\beta$ \eqref{eq-100}. The branching ratio obtained, when
we
sum all the contributions is
\begin{equation}
 \text{Br}(D^0\to \gamma\gamma)=1.6 \times 10^{-8}.
\end{equation}

By varying $\beta$ within $1 \;{\rm GeV}^{-1}
\le \beta \le 5 \; {\rm GeV}^{-1}$ and keeping
$g=0.59\pm 0.08$, the branching ratio
is changed by at most 10\%. On the other hand, one has to keep in mind
that
the loop contributions
involving beta are not finite and have to be regulated. We have used
$\overline{\rm MS}$ scheme, with the
 divergent parts being absorbed by the counterterms. \index{counterterms! contributions of}
 The size of these is not
known, so they might influence the error in our
 prediction of the branching ratio. Note, however, that the use of a different renormalization scheme, e.g., the GL renormalization scheme, would not change the result considerably. Namely, the leading order results coming from the chiral loops on Fig. \ref{fig2D0GammaGamma} are finite and scheme independent, while the contributions from scheme dependent diagrams \ref{fig5D0GammaGamma} are suppressed by an order of magnitude.
Note also, that changing $\alpha$ would
affect the predicted
branching ratio. For instance, if the chiral corrections do
decrease the value of $\alpha$ by $30\%$ this would decrease
 the predicted
branching ratio down to $0.8\times 10^{-8}$.
\index{contributions!chiral loop|)}

\subsection{Summary of $D^0\to \gamma\gamma$}\index{calculation! results of|(}
\label{Summary} \index{errors, estimation|(}\index{nonresonant contributions}\index{results}\index{summary}
We have presented a detailed calculation of the decay amplitude
$D^0\to \gamma \gamma$, which
includes short distance and long distance contributions, by making use
of
the theoretical
tool of the HH$\chi$PT. Within this
framework, the leading
contributions are found
to arise from the charged $\pi$ and $K$ mesons running in the chiral
loops, and
are of the order ${\cal O}(p^3)$.
These terms are finite and contribute only to the parity violating part
of
the amplitude. The
inclusion of higher order terms in the chiral expansion is
unfortunately
 plagued
 with the uncertainty caused
by the lack of knowledge of the counterterms. As to the parity \index{counterterms! contributions of}
conserving
part of the decay, it is given
by the terms coming from the anomaly and
from the loop terms containing
the beta coupling, the latter giving most of the amplitude. The size of
this
part of the amplitude
is approximately one order of magnitude smaller than the parity
violating
amplitude, thus
contributing less than 20\% to the decay rate. Therefore, our
calculation
predicts that the $D\to 2 \gamma$
decay is mostly a parity violating transition.

 In addition to the uncertainties we have mentioned, there is the 
question
of the suitability of
the chiral expansion for the energy involved in this process; the size
of
the uncertainty related
to this is difficult to estimate. Altogether, our estimate is that the
total
uncertainty is not larger
than 50\%. Accordingly, we conclude that the predicted branching ratio
is\footnote{This result is higher then the one published in \cite{Fajfer:2001ad} due to new experimental data on $f_{D_s}$ \cite{Heister:2002fp} and consequently higher $\alpha$. The two results agree within error-bars.}
\begin{equation}
\text{Br}(D^0\to \gamma \gamma)^{\text{SM}}= (1.6 \pm 0.8)\times 10^{-8}. \label{fin-res}
\end{equation}
 The reasonability of this result can be deduced also from a comparison
with the calculated decay
rates for the $D^0\to \rho(\omega)\gamma$, which are found to be
expected
 with
a branching ratio of
approximately $10^{-6}$ \cite{Fajfer:1998dv,Burdman:1995te,Fajfer:2000zx}.
\index{errors, estimation|)}

Recently an independent analysis of the $D^0\to \gamma\gamma$ decay width has been performed \cite{Burdman:2001tf}. To estimate the LD effect, authors of Ref.~\cite{Burdman:2001tf} use vector meson dominance (VMD) \index{VMD}for the lightest vector meson contributions as well as the single higher pseudoscalar resonance and two light-pseudoscalar the intermediate states. The largest contribution is found to come from the VMD estimate which gives $\text{Br}(D^0\to \gamma \gamma)=(3.5 \genfrac{}{}{0pt}{}{+4.0}{-2.6})\times 10^{-8} $. The higher pseudoscalar resonances give negligible contribution. The contributions from intermediate two light-pseudoscalar states are similar to the contributions considered in our approach and give $\text{Br}(D^0\to \gamma \gamma)=0.7\times 10^{-8} $ in good agreement with our estimate. \index{resonant contributions}

Note, however, that both the approach presented here and the approach of Ref.~\cite{Burdman:2001tf} include only the contributions of lowest lying bound states and the resonances. Recent analysis \cite{Bosch:2002bv,Bosch:2002bw} based on the QCD factorization \index{QCD factorization}in the heavy-quark limit $m_c\gg \Lambda_{\text{QCD}}$ and on the quark-hadron duality, \index{quark-hadron duality} suggests that large cancellations between LD contributions in the duality sum could occur. Whether the quark-hadron duality as well as the QCD factorization do take place in the charm decays, is however not clear, so that the question of large cancellations between the LD effects remains open.

It is interesting to estimate what sizes of new physics \index{signatures, new physics} effects one can expect in the $D^0\to \gamma\gamma$ decay mode \cite{Zupan:Dubrovnik}. In \cite{Fajfer:2001sa,Bigi:1989hw,Prelovsek:2000xy} it has been found, that the $c\to u\gamma$ \index{ctougamma@$c\to u\gamma$} short distance contribution (corresponding to the insertion of $Q_7, Q_7'$ operators) can get considerably enhanced, if one takes into consideration the MSSM spectrum. The leading contribution comes from the gluino exchange and can be at most as large as $|V_{cb} V_{ub}^*C_7^{\text{MSSM}}|\lesssim 0.04 $, $|V_{cb} V_{ub}^*C_7^{'\text{MSSM}}|\lesssim 0.04 $ \eqref{MSSMbounds}. Making the replacements $ C_{7}^{\text{eff}}\to C_{7}^{\text{eff}}- C_{7}^{'\text{eff}}$ in \eqref{MSD-} and $ C_{7}^{\text{eff}}\to C_{7}^{\text{eff}}+C_{7}^{'\text{eff}}$ in \eqref{MSDplus} and using the values of $C_{7}^{\text{eff}},C_{7}^{'\text{eff}}$ on the upper bounds, one finds that the MSSM contributions might increase the Standard Model prediction for the branching ratio up to \index{MSSM}
\begin{equation}
 \text{Br}(D^0\to \gamma\gamma)^{\text{MSSM}}=4.6 \times 10^{-8},\label{gammagammaMSSM}
\end{equation}
where largest contribution is found for $C_{7}^{\text{eff}}\simeq -C_{7}^{'\text{eff}}$. Note, that no additional contribution to the $c\to u\gamma$ transition arises, if $R$ parity conservation is lifted. Thus the $D^0\to \gamma\gamma$ branching ratio will not be enhanced above \eqref{gammagammaMSSM}, even if $R$ parity violating terms are introduced in the construction of MSSM (cf. Eq.~\eqref{bounds}). \index{calculation! results of|)}\index{D0gammagamma@$D^0\to \gamma\gamma$|)}\index{decay!$D^0\to \gamma\gamma$|)}\index{R parity@$R$ parity violating terms}

From the experimental side, CLEO has quite recently set the first experimental upper limit  on this decay mode with $Br(D^0\to \gamma\gamma)<2.9 \times 10^{-5}$ at the 90\% confidence level \cite{Coan:2002te}. The upcoming CLEO-c experiment is expected to be sensitive to  this decay mode at the level of approximately $\sim 10^{-6}$ \cite{Selen}.

\section{The $D^0\to l^+l^- \gamma$ decays}\label{llbarGamma}\index{D0llgamma@$D^0\to l^+l^- \gamma$|(}\index{decay!$D^0\to l^+l^- \gamma$|(}
\markright{THE $D^0\to l^+l^-\gamma$ DECAYS}
In this section we study the rare decays $D^0\to e^+e^-\gamma$, $D^0\to \mu^+\mu^-\gamma$ both in the Standard Model and in the MSSM \cite{Fajfer:2002gp}. A Standard Model analysis of $D^0\to l^+l^- \gamma$ branching ratios neglecting QCD and LD effects, has been made in Ref.~\cite{Geng:2000if}, giving $\text{Br}(D^0\to l^+l^- \gamma)=6.3 \times 10^{-11}$. However, LD effects are expected to dominate the SM prediction similarly to the $D\to P, V l^+l^-$ decays. To evaluate the nonresonant LD effects, we use the heavy quark effective theory combined with the chiral perturbation theory as explained in chapter \ref{HQET}. We also include the contributions of vector resonances in our analysis.

Another expectation based on the experience from the $D\to (P, V) l^+l^-$ decays is, that there are possibly large contributions in the $D^0\to l^+l^- \gamma$ decays coming from the SM extensions such as the MSSM with R-parity violation. These expectations make $D^0\to l^+l^- \gamma$ channels interesting from both experimental as well as from theoretical side.

\subsection{Standard Model prediction}
We will devote the first part of this section to the estimation of the $ D^0 \to l^+ l^-\gamma$ decay width in the context of the Standard Model. At the quark level, this decay mode cannot proceed through tree diagrams and is thus induced only at the one loop level in the Standard Model. Possible diagrams at the quark level are shown on Fig.~\ref{InamiLim}. These than translate into an effective weak Lagrangian at the scale of $m_c$ \eqref{effective_lagr}.

\subsubsection{Nonresonant contributions}
\label{nonresonant}\index{nonresonant contributions|(}

First we turn to estimating the nonresonant contributions in $D^0\to l^+l^-\gamma$. As we will see later on, it is in the nonresonant contributions that the extensions of the Standard Model can show up.

The most general invariant amplitude for the $D^0 \to l^+l^- \gamma $ decay, that one obtains from the effective Lagrangian \eqref{effective_lagr}, can
 be written in the following form \index{amplitude! in $D^0\to l^+l^-\gamma$}
\be
\begin{split}
M &=M_0^{\mu\nu}\epsilon_\mu^*(k)\frac{1}{p^2} \bar{u}(p_1)\gamma_\nu v(p_2)+M_5^{\mu\nu}\epsilon_\mu^*(k)\frac{1}{p^2} \bar{u}(p_1)\gamma_\nu\gamma_5 v(p_2)+\\
&\qquad+ M_{\text{BS}}(p^2)\Big[ \bar{u}(p_1)\Big(\frac{\sls{\epsilon}^* \sls{p}_D}{p_1 \cdot k}-\frac{\sls{p}_D \sls{\epsilon}^*}{p_2 \cdot k}\Big) \gamma_5 v(p_2)\big],
\end{split} \label{inv_ampl}
\ee
where
\be
M_{0,5}^{\mu\nu}=C_{0,5}(p^2) \big(\eta^{\mu\nu}-\frac{p^\mu k^\nu}{p\negcdot k}\big)+ D_{0,5}(p^2) \epsilon^{\mu \nu \alpha \beta} k_\alpha p_\beta, \label{inv_amp}
\ee
with $p_{1,2}$ the four-momenta of the lepton and antilepton respectively, $p=p_1+p_2$ the momentum of lepton pair, $k$ the photon momentum and $\epsilon_\mu$ its polarization vector. The form factors $C_{0,5}(p^2)$, $D_{0,5}(p^2)$, $M_{\text{BS}}(p^2)$ are functions of $p^2$ only and in particular do not depend on $k\cdot p_1$ or $k\cdot p_2$. The $C_0,D_5$ terms are parity violating, while
$C_5,D_0$, and the bremsstrahlung part of the amplitude, $M_{\text{BS}}$, are parity conserving. \index{bremsstrahlung} \index{form factors! in $D^0\to l^+l^-\gamma$}

The partial decay width is then
\begin{equation}
\begin{split}
\frac{d\Gamma}{dp^2}=&\frac{1}{16 \pi^3 m_D^3} \Bigg\{ \frac{k \negcdot p}{3 p^2}\sqrt{1-4\hat{\mu}_p^2}\bigg[\big(|C_0|^2+|D_0|^2(k\negcdot p)^2\big) (1+2\hat{\mu}_p^2)+\\
&+\big(|C_{5}|^2+|D_{5}|^2(k\negcdot p)^2\big) (1-4\hat{\mu}_p^2)\bigg]\\
&+\frac{|M_{\text{BS}}|^2}{k\negcdot p} \Big[\left((p^2)^2+m_D^2(m_D^2-4 m^2)\right) \; \ln\Big(\frac{1+\sqrt{\quad}}{1-\sqrt{\quad}}\Big)-2 p^2 m_D^2\sqrt{\quad}\Big]\\
&+4 \Im(D_0 M_{\text{BS}}^*) \frac{m}{p^2} (k\negcdot p)^2 \; \ln\Big(\frac{1+\sqrt{\quad}}{1-\sqrt{\quad}}\Big)\Bigg\},
\end{split}
\end{equation}
where $\hat{\mu}_p^2=m^2/p^2$, with $m$ the lepton mass, $\sqrt{\quad}=\sqrt{1-4 \hat{\mu}_p^2}$, while $k\negcdot p=(m_D^2-p^2)/2$. This expression agrees with the expression for the partial decay width $K_L\to l^+l^- \gamma$ as given in \cite{D'Ambrosio:1994ae,D'Ambrosio:1996sw}, as well as with the $B\to l^+l^-\gamma$ decay width as given in \cite{Aliev:2001bw}. 


\begin{figure}
\begin{center}
\epsfig{file=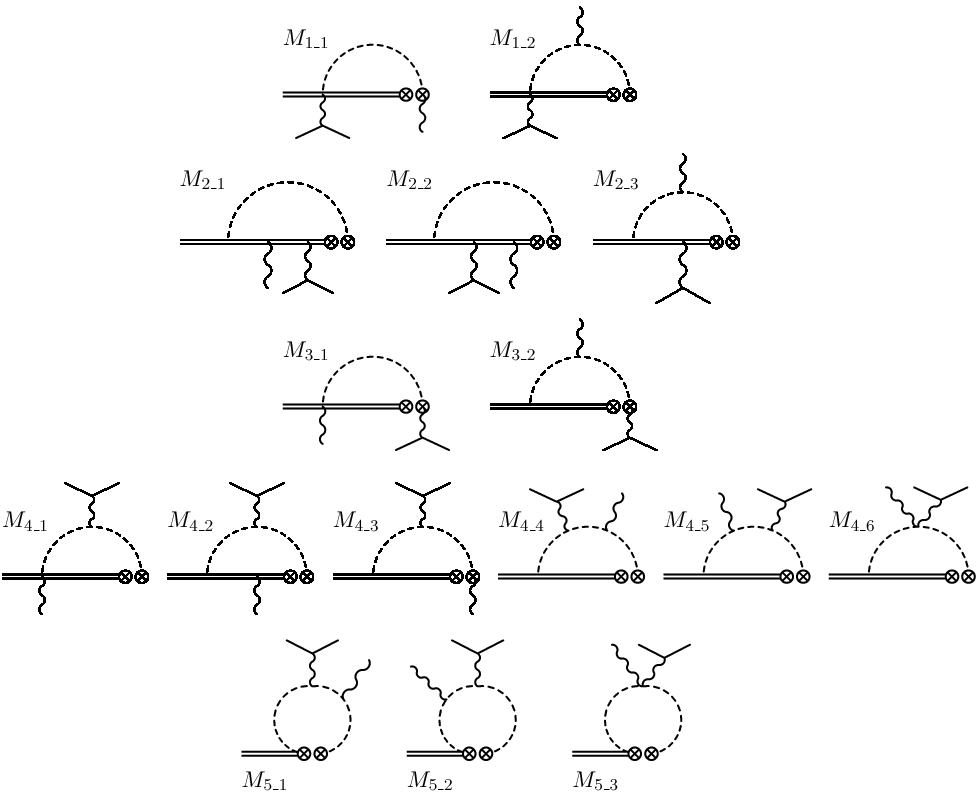, height=7cm}
\caption{\footnotesize{Nonresonant contributions to $D^0\to l^+l^-\gamma$ decay at one chiral loop order. Sum of diagrams in each row is gauge invariant and finite.}}\label{fig-1D0llBarGamma}
\end{center}
\end{figure}
\index{contributions!chiral loop|(}
The nonresonant LD contributions will arise from the chiral loop contributions shown on Figure \ref{fig-1D0llBarGamma}. The weak vertices receive contributions from the $Q_{1,2}$ operators in the effective Lagrangian \eqref{effective_lagr}. The sizes of these contributions are estimated using the factorization approximation \eqref{fact-lagr}. \index{factorization approximation} We use the phenomenologically motivated values $a_1=1.26$, $a_2=-0.49$ of the ``new factorization'' \cite{Buras:1994ij}. As in the $D^0\to \gamma\gamma$ decay the long distance interactions will
contribute only if the $SU(3)$
flavor symmetry is broken, i.e., if $m_s\neq m_d$. Note also that in the diagrams of Fig.~\ref{fig-1D0llBarGamma} only the term proportional to $a_1$ contributes. The $a_2$ part of effective Lagrangian \eqref{fact-lagr} gives rise to the resonant LD contributions and will be discussed later on.

We will calculate the nonresonant LD contributions in the framework of HH$\chi$PT that has been explained in chapter \ref{HQET} (see also section \ref{theoretical-frame}). The values of coupling constants are listed in Table \ref{tab-koef}. Note, that the $\beta$ coupling that describes $D^* D\gamma$ transition will not be taken into account in the chiral loop contributions of Fig.~\ref{fig-1D0llBarGamma}, as it has been found to give negligible contribution in a very similar case of the $D^0\to \gamma\gamma$ analysis. The $D\to D^{*}\gamma$ transition will be, however, needed to estimate the short distance contributions shown on Fig. \ref{SDdiagr}. These will give numerically irrelevant contributions for the SM predictions, but will be important later on, when we extend the analysis to the MSSM case. 

Using HH$\chi$PT one arrives at the set of nonzero ${\cal O }(p^3)$ diagrams listed in Fig.~\ref{fig-1D0llBarGamma}. Each row of diagrams on Fig.~\ref{fig-1D0llBarGamma} is a gauge invariant set. Note that each row actually represents the smallest gauge invariant set of diagrams, that can be obtained using the theorem stated in section \ref{gauge}. The sum of the diagrams in each row is also finite. Separate diagrams are in general divergent and are regulated using the dimensional regularization. Further details on this subject can be found in chapter \ref{scalar-loops}. The explicit expressions of the corresponding amplitudes can be found in appendix \ref{app-D0llBarGamma}. Note that the chiral loop contributions of Fig.~\ref{fig-1D0llBarGamma} contribute only to the $M_0^{\mu\nu}$ part of the invariant amplitude \eqref{inv_ampl}. Namely, the $l^+l^-$ pair couples to the charged mesons in the loop only via electromagnetic current. This leads to the $1/p^2$ photon pole in the amplitude ($p$ being the momentum of the lepton pair). The LD nonresonant contributions coming from Fig.~\ref{fig-1D0llBarGamma} thus exhibit a pole behaviour at small lepton momenta. This pole is either cut off by the phase space because of the nonzero lepton masses ($p^2=4 m^2$), or by the experimental limitations due to Dalitz conversion \cite{Burdman:2001tf}.

\begin{figure}
\begin{center}
\epsfig{file=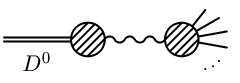}
\caption{\footnotesize{One particle reducible diagrams with photon connecting initial (pseudo)scalar and final state particles are zero.}}\label{TwoBlobs}
\end{center}
\end{figure}

\begin{figure}
\begin{center}
\epsfig{file=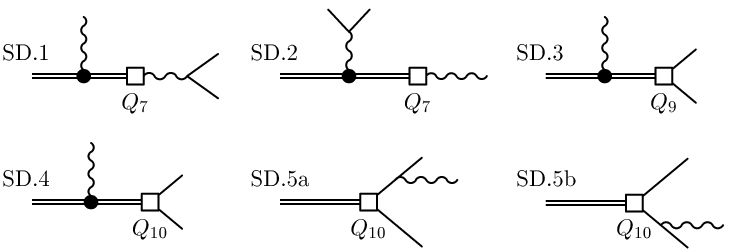, height=3.5cm}
\caption{\footnotesize{The nonresonant short distance diagrams. The effective weak Lagrangian vertex is denoted by square. The relevant operator is denoted as well. The blob $\bullet$ denotes the $\beta$-like vertex \eqref{eq-100}.}}\label{SDdiagr}
\end{center}
\end{figure}

Note that there is no photon bremsstrahlung off the final lepton pair in the chiral loop contributions. Namely, diagrams of the type shown on Fig.~\ref{TwoBlobs}, with the initial meson being a (pseudo)scalar, and with a photon connecting the two blobs, vanish due to gauge invariance. \index{bremsstrahlung}

The diagrams of Fig.~\ref{fig-1D0llBarGamma} are evaluated in the minimal subtraction ($\overline{\text{MS}}$) renormalization scheme. However, the sum of the diagrams is finite and scheme independent. We use the values of coupling constants listed in Table \ref{tab-koef}. Integrating over the whole available phase space one arrives at the estimates
\be
\text{Br}(D^0\to e^+e^-\gamma)_{\text{nonres}}=1.29 \times 10^{-10}, \qquad \text{Br}(D^0\to \mu^+\mu^-\gamma)_{\text{nonres}}=0.21 \times 10^{-10}.\label{nonres-res}
\ee
Due to a photon pole, the larger part of the electron channel branching ratio comes from the region of the phase space with $p^2\sim 0$. The phase space is cut off by the muon masses at much higher $p^2$, giving a smaller contribution of nonresonant LD effects to this decay channel.

Note that the short distance (SD) contributions coming from the penguin operators $Q_{7,9,10}$ are indeed very small in the Standard Model due to the CKM suppression \eqref{effective_lagr}. Evaluating the expectation values of the operators $Q_{7,9,10}$ using heavy quark symmetry as described in Eq.~\eqref{eq-105} (see also explicit expressions in appendix \ref{app-C}) and using the values of Wilson coefficients listed in Table \ref{tab-Wilson}, one arrives at the corresponding branching ratios of order $10^{-17}-10^{-18}$. This is negligible compared to the LD nonresonant contributions.

\index{nonresonant contributions|)}

\index{contributions!chiral loop|)}
\index{contributions!vector resonance|(}
\subsubsection{Resonant contributions}
\begin{figure}
\begin{center}
\epsfig{file=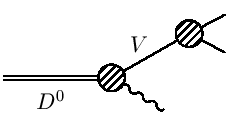}
\caption{\footnotesize{The mechanism of $D^0\to l^+l^-\gamma$ decay through intermediate vector resonance state $V$.}}\label{Reson}
\end{center}
\end{figure}

\index{resonant contributions|(}
The mechanism of the decay $D^0\to l^+l^-\gamma$ through a resonant intermediate state is depicted on Fig.~\ref{Reson}. The $D^0$ meson first decays into a vector meson and a photon, $D^0\to V \gamma$. The vector meson than decays into a lepton pair, completing the cascade $D^0\to V\gamma\to l^+l^- \gamma$. The decay width coming from this mechanism can be written as \cite{Lichard:1998ht}
\be
\frac{d \Gamma_{D^0\to V \gamma\to l^+ l^- \gamma}}{d p^2}=\Gamma_{D^0\to V\gamma} \frac{1}{\pi}\frac{\sqrt{p^2}\; }{(M_V^2-p^2)^2+M_V^2 \Gamma^2}\Gamma_{V\to l \bar{l}}, \label{factor}
\ee
where $p$ is the momentum of the lepton pair, while $M_V$ and $\Gamma$ are the mass and the decay width of the vector meson resonance. Several assumptions go into the derivation of the simple, but physically well motivated formula \eqref{factor}. First of all the interference with other channels is neglected. Under this approximation the formula is generally valid for the case of scalar resonances. Following the reasoning of Ref.~\cite{Lichard:1998ht} it is easy to show, that Eq.~\eqref{factor} is valid also for the case of the electromagnetic decay of vector resonance into a lepton pair.

Since vector resonances $\rho, \omega, \phi$ are relatively narrow Eq.~\eqref{factor} can be further simplified using the narrow width approximation $\Gamma\ll M_V$
\be
\text{Br}(D^0\to V\gamma\to l^+ l^- \gamma)= \text{Br}(D^0\to V \gamma) \text{Br}(V\to l^+l^-). \label{narrow_width}
\ee
The narrow width approximation is valid at $5\%$ level for $\rho$, and below $1\%$ for $\omega, \phi$ mesons. To obtain numerical estimates, the experimental data on the branching ratios $\text{Br}(V\to l^+l^-)$ \cite{Hagiwara:pw} can be used. On the other hand none of the decays $D^0\to V\gamma$ have been measured yet. We thus use the theoretical predictions of branching ratios $\text{Br}(D^0\to V \gamma)$. As the central values we use the recent predictions of Ref.~\cite{Prelovsek:2000rj}, where a reanalysis of Ref.~\cite{Fajfer:1998dv} using the quark model to determine relative phase uncertainties has been performed. As a comparison we also list in Table \ref{table-input2} the predictions of Ref.~\cite{Burdman:1995te}. Note that for the upper limit predictions in \cite{Burdman:1995te} VMD model was used, with the main numerical input the experimental value of $\text{Br}(D^0\to \rho^0 \phi)$. However, the central value of this branching fraction as cited in \cite{Hagiwara:pw} has decreased by a factor of three between 1994-2002. Thus the upper limits on predictions of \cite{Burdman:1995te} should be divided by three, bringing the values in fair agreement with \cite{Prelovsek:2000rj}.\index{VMD}

\begin{table}[h] \index{tables of results}
\begin{center}
\begin{tabular}{|c|r||c|r|}\hline
Decay& Exp. \cite{Hagiwara:pw} & Decay & Exp. \cite{Hagiwara:pw}\\ \hline\hline
$\text{Br}(\rho^0\to e^+e^-)$ & $(4.54\pm0.10)\times 10^{-5}$ & $\text{Br}(\rho^0\to \mu^+\mu^-)$ & $(4.60\pm0.28)\times 10^{-5}$\\ \hline
$\text{Br}(\omega \to e^+e^-)$ & $(6.95\pm0.15)\times 10^{-5}$ & $\text{Br}(\omega \to \mu^+\mu^-)$ & $(9.0\pm 3.1)\times 10^{-5}$\\ \hline
$\text{Br}(\phi\to e^+e^-)$ & $(2.96\pm 0.04)\times 10^{-4}$ & $\text{Br}(\phi\to \mu^+\mu^-)$ & $(2.87\genfrac{}{}{0pt}{}{+0.18}{-0.22})\times 10^{-4}$\\\hline
\end{tabular}
\caption[Experimental values for branching ratios of $V\to l^+l^-$]{\footnotesize{Branching ratios of vector mesons decaying to a lepton pair as compiled in Ref.~\cite{Hagiwara:pw}. }}
\label{table-input1}
\end{center}
\end{table}

\begin{table}[h]
\begin{center}
\begin{tabular}{|c|c|c|c|}\hline
 Decay & Theor. \cite{Prelovsek:2000rj}& Theor. \cite{Burdman:1995te} & Exp. \cite{Hagiwara:pw}\\ \hline\hline
 $\text{Br}(D^0\to \rho^0 \gamma)$ & $1.2\times 10^{-6} $ & $(1 -5)\times 10^{-6} $ & $ <2.4 \times 10^{-4}$\\ \hline
 $\text{Br}(D^0\to \omega \gamma)$ & $1.2\times 10^{-6} $ & $ \simeq2 \times 10^{-6} $& $ <2.4 \times 10^{-4}$\\ \hline
 $\text{Br}(D^0\to \phi \gamma)$ & $3.3 \times 10^{-6} $& $(1-34) \times 10^{-6} $ & $ <1.9 \times 10^{-4}$\\ \hline
\end{tabular}
\caption[Theoretical predictions for decays $D^0\to V\gamma$]{\footnotesize{Theoretical predictions for decays $D^0\to V\gamma$ \cite{Prelovsek:2000rj,Burdman:1995te}. Predictions of Ref.~\cite{Prelovsek:2000rj} are used as central values (see also comments in text). 
In the last column the experimental upper limits are listed.}}
\label{table-input2}
\end{center}
\end{table}

Using the values compiled in Tables \ref{table-input1}, \ref{table-input2} together with Eq.~\eqref{narrow_width} one immediately arrives at 
\begin{align}
\text{Br}(&D^0\to\rho\gamma\to l^+l^- \gamma) \sim 5 \times 10^{-11},\label{reson1}\\
\text{Br}(&D^0\to\omega\gamma\to l^+l^- \gamma) \sim 8 \times 10^{-11},\\
\text{Br}(&D^0\to\phi\gamma\to l^+l^- \gamma) \sim 10^{-9},\label{reson3}
\end{align}
 with $l^+l^-=e^+e^-, \mu^+\mu^-$. Above we have used the fact, that the differences between $e^+e^-$ and $\mu^+\mu^-$ decay modes in the Standard Model come from the phase space differences only. These are relatively small compared to other theoretical and experimental uncertainties entering predictions \eqref{reson1}-\eqref{reson3}, and are as such neglected.

\begin{figure}
\begin{center}
\epsfig{file=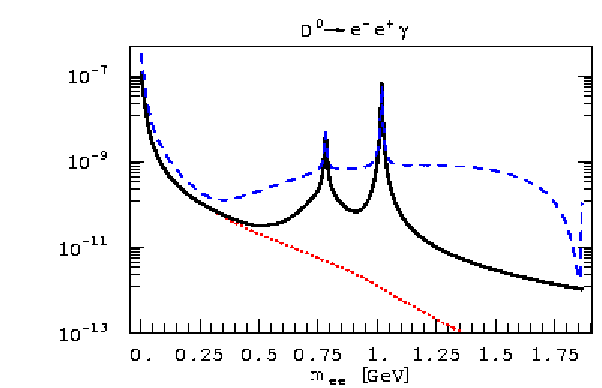, height=5.2cm}
\epsfig{file=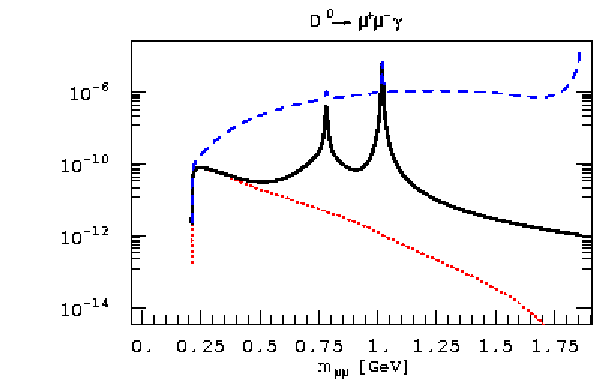, height=5.2cm}
\caption{\footnotesize{The normalized decay width distribution $(d\Gamma/d p^2)/\Gamma$ as function of effective lepton pair mass $m_{l^+ l^-}$ (where $m_{l^+ l^-}^2=p^2$) for $e^+e^-$ (left plot) and $\mu^+\mu^-$ (right plot) final lepton pair. The dotted line denotes SM nonresonant contribution, solid black line denotes full SM prediction, while dashed line denotes largest possible MSSM contribution with $R$ parity violation.}}\label{fig-res}
\end{center}
\end{figure}

As seen from the estimates \eqref{reson1}-\eqref{reson3} the largest contribution to $D^0\to l^+l^-\gamma$ comes from the intermediate $\phi$ resonance, being approximately one order of magnitude larger than the other two contributions. Note also, that in the region of $p^2$, where vector resonances are important, the nonresonant contribution calculated in the previous section is several orders of magnitude smaller. We can thus safely neglect possible interference between the nonresonant and the resonant contributions and simply add the resonant contributions \eqref{reson1}-\eqref{reson3} to the nonresonant ones \eqref{nonres-res}. The decay width distribution is plotted on Fig.~\ref{fig-res}, while the predicted branching ratios are
\be
\text{Br}(D^0\to e^+e^-\gamma)_{\text{SM}}=1.2 \times 10^{-9}, \qquad \text{Br}(D^0\to \mu^+\mu^-\gamma)_{\text{SM}}=1.1 \times 10^{-9}. \label{SM-pred}
\ee
Note that, if the values of Ref.~\cite{Burdman:2001tf} had been used, the predicted branching ratios could be utmost a factor of three higher.

Incidentally Fig.~\ref{fig-res} also explains, why the $D^0\to V\gamma\to \gamma^*\gamma$ cascade could be neglected in the $D^0\to \gamma\gamma$ decay rate calculation. Namely, for $\gamma^*$ almost on-shell the decay width is dominated by the nonresonant contributions. These are described using the HH$\chi$PT as in the calculation of $D\to \gamma\gamma$ presented in section \ref{D0GammaGamma}.

\index{contributions!vector resonance|)}\index{resonant contributions|)}
\subsection{Beyond the Standard Model}\label{Beyond} \index{beyond the Standard Model|(}
We turn now to the possible effects of physics beyond the Standard Model, that could enhance the predicted branching ratios \eqref{SM-pred}. The effects of new physics show up in the values of the Wilson coefficients \index{signatures, new physics}
\be
C_i^{\text{new}}=C_i+\delta C_i,
\ee
where $C_i$ are the SM values of Wilson coefficients listed in Table \ref{tab-Wilson} and in Eq.~\eqref{C7numbers}, while $\delta C_i$ denote the changes due to new physics effects. Note that the general feature of all the SM extension is to overcome the $V_{cb}^* V_{ub}$ suppression of the penguin operators $Q_{7,9,10}$ \eqref{effective_lagr}. Another general feature is that the new physics effects will extend the basis of penguin operator \eqref{list-oper} by operators $Q_{7,9,10}'$ with quark chiralities switched \eqref{Qall}.

\subsection{Minimal Supersymmetric Standard Model}\index{MSSM|(} \index{supersymmetry|(}
We start with the simplest supersymmetric extension of the SM, the Minimal Supersymmetric Standard Model (MSSM). It is constructed by putting the SM fermions in the chiral multiplets and the SM gauge bosons in the vector multiplets, thus in effect doubling the spectrum of the Standard Model fields. If no particular SUSY breaking mechanism is assumed, the MSSM Lagrangian contains well over 100 unknown parameters. It is thus very useful to adopt the so-called mass insertion approximation. In this approximation the basis of fermion and sfermion states is chosen such, that all the couplings of these particles to the neutral gauginos are flavor diagonal, but then the squark mass matrices are not diagonal. The squark propagators are then expanded in terms of nondiagonal elements, where mass insertions induce changes of squark flavor \cite{Hall:1985dx}. The mass insertions are parametrized as
\be
(\delta^u_{ij})_{AB}=\frac{(M_{ij}^u)^2_{AB}}{M_{\tilde{q}}^2},
\ee
where $i \ne j$ are flavor indices, $A,B$ denote chirality, $(M_{ij}^u)^2$ are the off-diagonal elements of the up-type squark mass matrices and $M_{\tilde{q}}$ is the average squark mass.

The largest contribution to the $c\to ul^+l^-$ transition is expected from gluino-squark exchanges \cite{Burdman:2001tf,Fajfer:2001sa,Lunghi:1999uk}. Allowing for only one insertion, the contributions from gluino-squark exchange diagrams are\index{Q7@$Q_7$}
\begin{subequations}\label{gluino}
\begin{align}
V_{cb}^* V_{ub} \delta C_7 &= \frac{8}{9} \frac{\sqrt{2}}{G_F M_{\tilde q}^2} \pi \alpha_s \left[(\delta_{12}^u)_{LL} \frac{P_{132}(z)}{4}+ (\delta^u_{12})_{LR} P_{122}(z) \frac{M_{\tilde{g}}}{m_c} \right],\\
V_{cb}^* V_{ub} \delta C_9&=\frac{32}{27} \frac{\sqrt{2}}{G_F M_{\tilde{q}}^2}\pi \alpha_s (\delta_{12}^u)_{LL} P_{042}(z),\\
 V_{cb}^* V_{ub} \delta C_{10}&\simeq 0,
\end{align}
\end{subequations}
where $z=M_{\tilde g}^2/M_{\tilde q}^2$, while the functions $P_{ijk}(z)$ are
\be
P_{ijk}(z)=\int_0^1 dx \frac{x^i (1-x)^j}{(1-x+z x)^k}.
\ee
The Wilson coefficient $C_{10}$ receives the first nonzero contributions from double mass insertions, therefore we neglect it in the following. The Wilson coefficients $C_{7, 9, 10}'$ corresponding to the operators with ``wrong chirality'' receive contributions from gluino-squark exchanges that are of the same form as expressions \eqref{gluino}, but with the interchange $L\leftrightarrow R$.

In the numerical evaluation of possible MSSM effects we use the gluino mass $M_{\tilde g}=250$ GeV and the average squark mass $M_{\tilde q}=250$ GeV, that are given by the lower experimental bounds \cite{Hagiwara:pw}. For the bounds on the mass insertions we use the analysis of \cite{Fajfer:2001sa,Prelovsek:2000xy}. The strongest bounds on mass insertion parameters $(\delta^u_{12})_{LR}$ are obtained by requiring that the minima of the scalar potential do not break charge or color, and that they are bounded from below \cite{Prelovsek:2000xy,Casas:1996de}, giving
\be
|(\delta^u_{12})_{LR}|, |(\delta^u_{12})_{RL}|\le 4.6 \times 10^{-3}, \qquad \text{for} \quad M_{\tilde{q}}=250\; \text{GeV}.
\ee
The bounds on mass insertions $(\delta_{12}^u)_{LL}$ and $(\delta_{12}^u)_{RR}$
can be obtained from the experimental upper bound on the mass difference in the neutral D system. Saturating the experimental bound $\Delta m_D<4.5 \times 10^{-14}$ GeV \cite{Godang:1999yd,Link:2000cu} by the gluino exchange, gives \cite{Fajfer:2001sa,Prelovsek:2000xy,Gabbiani:1996hi}
\be
|(\delta^u_{12})_{LL}|\le 0.03, \qquad \text{for} \quad M_{\tilde{g}}=M_{\tilde{q}}=250\; \text{GeV}, 
\ee
where $(\delta^u_{12})_{RR}$ has been set to zero. These translate into
 \begin{subequations}\label{MSSMbounds}
\begin{align}
|V_{cb}^* V_{ub} \delta C_7| & \le 0.04, &|V_{cb}^* V_{ub} \delta C_7'| & \le 0.04, \label{c7bound}\\
|V_{cb}^* V_{ub} \delta C_9|&\le 0.0016, 
 &|V_{cb}^* V_{ub} \delta C_9'|&\simeq 0.
\end{align}
\end{subequations}
Note that both $C_7$ and $C_7'$ receive largest contributions from $(\delta^u_{12})_{LR}$ insertions. Note also that the upper limit on $C_{7}$ coefficient is three orders of magnitude larger than the Standard Model value, while for $C_9$ is an order of magnitude larger than the SM value. However, as discussed in the previous section, SM prediction is dominated by the $Q_{1,2}$ insertions and therefore by the long distance effects. 

 The contributing diagrams are shown on Fig.~\ref{SDdiagr}, to which the diagrams with $Q_i\to Q_i'$ should be added. In the mass insertion approximation the coefficient $C_{10}$ is small and will be neglected in the following. When the $Q_{7,9}$ operators are inserted, the photon bremsstrahlung off the final lepton pair is not possible. In the case of the $Q_7$ operator this is because the diagrams are of the type shown in Fig.~\ref{TwoBlobs}, while in the case of the $Q_9$ operator the bremsstrahlung is prohibited because of the vector current conservation. \index{bremsstrahlung}

Taking the values of induced Wilson coefficients at the upper bounds we obtain 
\be
\text{Br}(D^0\to e^+e^-\gamma)_{\text{MSSM}}=1.4 \times 10^{-9}, \qquad \text{Br}(D^0\to \mu^+\mu^-\gamma)_{\text{MSSM}}=1.2 \times 10^{-9}. \label{MSSM-pred}
\ee
The MSSM contribution to the decay rate is entirely due to the gluino exchange enhancement of the $C_7,C_7'$ coefficients. The decay rate is thus enhanced in low $p^2$ region, which also explains larger increase of the $D^0\to e^+e^- \gamma$ decay rate. The increase is, however, not significant enough to dominate over the resonant contributions \eqref{reson1}-\eqref{reson3}. The MSSM effects, if any, are thus too small to be unambiguously detected
experimentally in the decays $D^0\to l^+l^- \gamma$. \index{MSSM|)}

\subsection{R parity violation}\index{Q99@$Q_{10}$}\index{Q9@$Q_9$} \index{R parity@$R$ parity violating terms|(}
The situation is quite different once the assumption of $R$ parity conservation is relaxed and the soft symmetry breaking terms are introduced. We follow the analysis of Ref.~\cite{Burdman:2001tf}. The tree level exchange of down squarks results in the effective interaction
\be
{\cal L}_{\text{eff}}=\frac{\tilde{\lambda}_{i2k}' \tilde{\lambda}_{i1k}'}{2 M_{\tilde{d}_R^k}^2} (\bar{u}_L \gamma^\mu c_L) (\bar{l}_L \gamma^\mu l_L), \label{R-inter}
\ee
where $\tilde{\lambda}_{ijk}'$ are the coefficients of the lepton--up-quark--down-squark $R$ parity breaking terms of the superpotential in the quark mass basis. The effective interaction \eqref{R-inter} translates into the additional contributions $\delta C_i$ to the $C_{9,10}$ Wilson coefficients
\be
V_{cb}^* V_{ub} \delta C_9=-V_{cb}^* V_{ub} \delta C_{10}=\frac{2 \sin^2 \theta_W}{\alpha_{\text{QED}}^2}\left(\frac{m_W}{M_{\tilde{d}_R^k}}\right)^2 \tilde{\lambda}_{i2k}' \tilde{\lambda}_{i1k}',
\ee
while no contributions are generated to the $C_{9,10}'$ Wilson coefficients \cite{Burdman:2001tf}. For electrons in the final states we use bounds on $\tilde{\lambda}_{i2k}', \tilde{\lambda}_{i1k}'$ from charged current universality \cite{Allanach:1999ic}
\be
\tilde{\lambda}_{11k}'\le 0.02 \left(\frac{M_{\tilde{d}_R^k}}{100 \text{ GeV}}\right),\qquad \tilde{\lambda}_{12k}'\le 0.04 \left(\frac{M_{\tilde{d}_R^k}}{100 \text{ GeV}}\right).\label{el-tri}
\ee
 For muons in the final state, the limits come from $D^+\to \pi^+\mu^+\mu^-$ \cite{Burdman:2001tf}. Using the new experimental bound 
$\text{Br}(D^+\to \pi^+\mu^+\mu^-)<8.8 \times 10^{-6}$ \cite{Johns:2002hd}, this gives
\be
\tilde{\lambda}_{22k}', \tilde{\lambda}_{21k}'\le 0.003 \left(\frac{M_{\tilde{d}_R^k}}{100 \text{ GeV}}\right)^2.\label{mu-tri}
\ee
The bounds on trilinear couplings \eqref{el-tri}, \eqref{mu-tri} then give the following bounds on possible enhancements of the $C_{9,10}$ Wilson coefficients for the electron or muon channel
\begin{subequations}\label{bounds}
\begin{align}
|V_{cb}^* V_{ub} &\delta C_{9,10}^e|\le 4.4,\\
|V_{cb}^* V_{ub} &\delta C_{9,10}^\mu|\le 17,
\end{align}
\end{subequations}
with $\delta C_9^{e,\mu}=-\delta C_{10}^{e,\mu}$. Note that in \eqref{bounds} the squark mass cancels. The enhancements $\delta C_{9,10}^{e,\mu}$ are then added to the Standard Model values. The diagrams are listed on Figure \ref{SDdiagr}. The possible enhancement over the SM branching ratio predictions is quite striking and is in the case of muons in the final state by almost two orders of magnitude, if the values of $C_{9,10}$ Wilson coefficients are taken to be the upper bounds in \eqref{bounds}. The diagrams on Fig.~\ref{SDdiagr} with photon bremsstrahlung off the final lepton pair and the insertion of the $Q_{10}$ operator are IR divergent.\index{infrared divergences} We take cutoff energy to be $E_{\gamma}\ge 50$ MeV or $E_{\gamma}\ge 100$ MeV. The contributions from various sources, the nonresonant \eqref{nonres-res} and resonant \eqref{reson1}-\eqref{reson3} SM contributions, the insertion of $Q_7,Q_7'$ operators with the $C_7,C_7'$ values given in \eqref{c7bound}, and the contributions from insertion of $Q_{9,10}$ operators with $C_{9,10}^{e,\mu}$ bounded by \eqref{bounds} are summarized in
Table \ref{MSSM-res}. \index{bremsstrahlung}

The maximal branching ratios obtainable in the framework of MSSM with $R$ parity violation are
\begin{align}
\text{Br}(D^0\to e^+e^-\gamma)_{E_{\gamma}\ge 50\; \text{MeV}}^{\not R}&=4.5 \times 10^{-9}, \qquad& \text{Br}(D^0\to \mu^+\mu^-\gamma)_{E_{\gamma}\ge 50\; \text{MeV}}^{\not R}&=50 \times 10^{-9},\\
\text{Br}(D^0\to e^+e^-\gamma)_{E_{\gamma}\ge 100\; \text{MeV}}^{\not R}&=4.5\times 10^{-9}, \qquad &\text{Br}(D^0\to \mu^+\mu^-\gamma)_{E_{\gamma}\ge 100\; \text{MeV}}^{\not R}&=46 \times 10^{-9}.
\end{align}
These are to be compared with the SM predictions \eqref{SM-pred}. Note that the SM predictions are not affected by the cuts on the soft photon energy at the order of $E_{\gamma}\ge 100\; \text{MeV}$, as the bulk of the contribution either comes from the resonances or the low $p^2$ region (while the cut on $E_\gamma$ is the cut on the high $p^2$ region).

\begin{table}[h]
\begin{center}
\begin{tabular}{|c|c|c|} \hline
Contrib. & $\text{Br}(D^0\to e^+e^-\gamma)$ & $\text{Br}(D^0\to \mu^+\mu^-\gamma)$\\ \hline\hline
Nonres. & $12.9 \times 10^{-11}$ & $2.1 \times 10^{-11}$\\ \hline
Reson. &$1.1\times 10^{-9}$ &$1.1\times 10^{-9}$ \\ \hline
$C_7$ &$0.23\times 10^{-9}$ & $0.04\times 10^{-9}$ \\ \hline
$C_9$ &$1.37 \times 10^{-9}$ &$20.5\times 10^{-9}$ \\ \hline
$C_{10}$ & $1.37 \times 10^{-9}$ & $31.3 \times 10^{-9}$\\ \hline\hline
All & $4.52 \times 10^{-9}$ &$50.2 \times 10^{-9}$ \\ \hline
\end{tabular}
\caption[Table of possible contributions to $D^0\to l^+l^-\gamma$ in the context of MSSM with $R$ parity violation]{\footnotesize{ The relative sizes of various possible contributions in the context of MSSM with $R$ parity violation. The photon energy cutoff is taken to be $E_{\gamma}\ge 50$ MeV. Largest possible effects are calculated. The values for nonresonant (Nonres.) and resonant (Reson.) LD contributions are the same as for the SM prediction. The $C_7$ denotes the sum of contributions due to $Q_7$ and $ Q_7'$ insertions, while $C_{9,10}$ denote $Q_{9,10}$ insertions respectively. In the last row the maximal calculated branching
ratios are given.}}\label{MSSM-res}
\end{center}
\end{table}

The enhancement due to possible $R$ parity violating contributions is by more than an order of magnitude in the muon channel compared to the SM prediction. The enhancement also has a distinct signal in the $d\Gamma/dp^2$ decay width distribution. In the SM model the decay $D^0\to l^+l^-\gamma$ either proceeds through $\rho, \omega,\phi$ vector resonances or through nonresonant two-meson exchanges, which are important in the low $p^2$ region. The $R$ parity violating signal on the other hand would arise from the insertion of $Q_{9,10}$ operators and is large in the region of high $p^2$ (small photon energy) region as can be seen from Fig.~\ref{fig-res}. The largest possible effect, however, is below expected experimental sensitivities for the rare charm decays at B-factories \index{B-factories} and CLEO-c, \index{CLEO-c} which are apparently expected to be of the order of $10^{-6}$ \cite{Selen}. \index{rare decays! experimental searches}

\index{beyond the Standard Model|)}\index{R parity@$R$ parity violating terms|)}

\subsection{Summary of $D^0\to l^+l^- \gamma$}\index{calculation! results of|(}
\label{SummaryD0llBarGamma}\index{results}\index{summary}
We have presented a detailed study of $D^0\to e^+ e^- \gamma$ and $D^0\to \mu^+\mu^- \gamma$ decays both in the Standard Model (SM) and in the Minimal Supersymmetric Standard Model (MSSM) with and without $R$ parity violation. As for the SM prediction, the decays are dominated by the inclusion of $Q_{1,2}$ operators, which induce nonperturbative long distance (LD) effects. The penguin operators are suppressed by $V_{cb}^* V_{ub}^*\sim 10^{-4}$ CKM matrix elements and are therefore irrelevant for the processes considered. Nonresonant LD contributions are evaluated by employing the combined heavy quark and chiral symmetries. They are found to be important only in the region of low final lepton pair mass, while their contribution to the integrated decay width is less then 10\%. The decay width is dominated by the decay cascade $D^0\to V\gamma\to l^+l^- \gamma$, where $V=\rho, \omega, \phi$. The Standard Model branching ratio is then predicted to be
\be
\text{Br}(D^0\to l^+l^- \gamma)_{\text{SM}}=(1-3)\times 10^{-9}.
\ee
We also investigated possible enhancements of the decay widths due to new physics contributions. We have found that possible effects coming from gluino-squark exchanges in the context of MSSM with $R$ parity conserved are masked by the LD contributions from the SM. However, if the assumption of $R$ parity conservation is relaxed, the tree level exchange of down squarks can increase the predicted branching ratios by more than an order of magnitude. The largest possible effect comes from the diagrams with photon bremsstrahlung off the leptons in the final state and is IR divergent. \index{infrared divergences} Choosing two different cuts on the photon energy we arrive at \index{bremsstrahlung}
\begin{align}
\text{Br}(D^0\to e^+e^-\gamma)_{E_{\gamma}\ge 50 \text{MeV}}^{\not R}&=4.5 \times 10^{-9}, \qquad& \text{Br}(D^0\to \mu^+\mu^-\gamma)_{E_{\gamma}\ge 50 \text{MeV}}^{\not R}&=50 \times 10^{-9},\\
\text{Br}(D^0\to e^+e^-\gamma)_{E_{\gamma}\ge 100 \text{MeV}}^{\not R}&=4.5 \times 10^{-9}, \qquad &\text{Br}(D^0\to \mu^+\mu^-\gamma)_{E_{\gamma}\ge 100 \text{MeV}}^{\not R}&=46 \times 10^{-9}.
\end{align}
Allowing for the uncertainty in the SM calculation which we discussed
after Eq.\eqref{SM-pred}, we consider that branching ratios in excess of $0.5\times 10^{-8}$
are not accountable by the SM. The effect of the MSSM with R parity violation
in the muon channel is the closest to the experimental sensitivities of $\sim 10^{-6}$
expected at B-factories and CLEO-c \cite{Selen}. \index{CLEO-c} Thus we propose the $D^0\to \mu^+\mu^-\gamma$ 
decay as a possible probe of new physics. \index{calculation! results of|)}
\index{D0llgamma@$D^0\to l^+l^- \gamma$|)}\index{decay!$D^0\to l^+l^- \gamma$|)}\index{rare decays! calculation of|)}\index{rare decays|)}\index{signatures, new physics}\index{supersymmetry|)}

\chapter{Quenching errors in the heavy-to-light transitions}\label{quenching-errors}\index{lattice QCD|(}
In this chapter we will turn to a slightly different subject, namely the application of effective theories in the lattice QCD \cite{Gupta:1997nd}. These have been successfully exploited to tackle the various approximations made in the calculations with the discretized version of QCD \cite{Kronfeld:2002pi}. 

\index{lattice QCD! setup}
In QCD there are several fairly different energy scales, which make the calculations more challenging. There is the energy scale of nonperturbative gluonic effects, $\Lambda$, which for the purpose of this chapter will be be set equal to the chiral perturbation scale $\Lambda=\Lambda_\chi\sim 1$ GeV. The other scales are set by the quark masses. For the light quark masses, we have $m_q\ll \Lambda$, with $m_s\sim 100$ MeV, while the up and down quark masses are about 25 times smaller. On the other hand, for the bottom quark mass we have $m_Q>\Lambda$, while the charm quark mass $m_c\simeq 1.5$ GeV is close to $\Lambda$. The lattice also provides regularization of the theory with both ultraviolet and infrared cutoffs \cite{Gupta:1997nd}. The smallest Compton wavelength realized on the lattice cannot be smaller than the lattice spacing. The largest Compton wavelength realized on the lattice, on the other hand cannot, be larger than the overall size of the lattice. The UV cutoff $\sim \pi/a$ in the momentum space is thus set by the nonzero lattice spacing $a$, while the IR cutoff $1/L$ is provided by the finite size $L$ of the lattice. Ideally one would wish to have as fined grained lattice as possible, so that the UV cutoff would be far above the highest physical quark masses, while the lattice would be larger then the largest Compton wavelengths present in the system, $L\gg 1/m_q$. One would then have the hierarchy
\be
L^{-1}\ll m_q\ll \Lambda\ll m_Q\ll a^{-1}
\ee
In a realistic situation, however, this is not possible. The memory needed for such a calculation is roughly proportional to the number of lattice sites $N=(L/a)^4$ and explodes quickly, if either the size of the lattice is increased and/or the lattice spacing is reduced. In addition, the algorithms slow down, if the number of lattice sites is increased, as well as if the quark masses are reduced. The latter problem arises from the calculation of quark propagators. For it, the inverse of a discretized Dirac operator $(i\sls\! D-m)$ has to be computed. For small quark masses most of the eigenvalues of the Dirac operator are small, making the computation of an inverse numerically extremely demanding. 

In a typical lattice QCD calculation nowadays \cite{Kronfeld:2002pi} thus $N\sim(16-32)^4$ (where different number of lattice points in temporal and spatial directions are also used), with $a^{-1}\sim 1-4 $ GeV (and thus $\pi/a\sim 3-12$ GeV) and $L\sim 1-4$ fm. In typical calculations the ``light'' quarks have masses in the range $0.2-0.4<m_q/m_s<1.2$. For the heavy quarks either $m_Q\sim m_c$ is used, or the static approximation $m_Q\to \infty$ is taken. The values of masses $m_q, m_Q$, the size of lattice $L$ and the lattice spacing $a$ are varied in the calculation. Variation of lattice spacing $a$ is used to numerically extrapolate from nonzero lattice spacing to a continuum $a\to 0$. Arguments from effective field theory approach can be used to describe discretization effects and to improve the extrapolation to continuum.\index{extrapolation! to continuum} This is the so called Symanzik improvement program, that allows for a construction of improved lattice actions \cite{Kronfeld:2002pi,Symanzik:1983dc}. Effective theories, i.e. the chiral perturbation theory, are also used to guide the extrapolation in the quark masses to the physical values. \index{extrapolation! chiral}\index{lattice QCD!approximations in}

Another approximation that is very common in the lattice QCD calculations, but that has not yet been mentioned in the discussion so far, is the so called quenched approximation \cite{Marinari:1981qf}. It is this approximation, that we will focus on in the rest of this chapter. 

\section{Quenched approximation in the lattice QCD}\index{lattice QCD!quenching}\index{quenched approximation|(}
To obtain physical observables in the quantum field theories, one has to be able to compute Green's functions of the form \cite{Gupta:1997nd,Peskin:ev,Weinberg:mt}
\be
\begin{split}
\langle0| T\{ \Psi_{l_1}&(x_1)\dots \Psi_{l_n}(x_n) \bar{\Psi}_{k_1}(y_1)\dots \bar{\Psi}_{k_n}(y_n)\}|0\rangle=\\
&Z^{-1} \int [d\bar{\Psi}] [d\Psi] [dA] \; \exp\!\left({i \int\!\! d^4 x \;\bar{\Psi} (i \;\sls\! D-m) \Psi+ i I_{\text{G}}}\right)\Psi_{l_1}(x_1)\dots \bar{\Psi}_{k_n}(y_n), \label{gen-Green}
\end{split}
\ee
where $\Psi$, $A$ are sets of fermion and gauge fields respectively, with $l_i, k_i$ the Dirac indices, $I_{\text{G}}$ is the purely gluonic action, while 
\be
Z=\int [d\bar{\Psi}] [d\Psi] [dA]\; \exp\!\left({i \int\!\! d^4 x \;\bar{\Psi} (i \;\sls\! D-m) \Psi+ i I_{\text{G}}}\right).
\ee
Note, that we work in the Minkowski metric, while lattice calculations are done in the Euclidean metric. Since we will not introduce lattice QCD actions, nor do any explicit calculations with discretized QCD, this will not be an obstacle. 

Since the action in the path integral \eqref{gen-Green} is quadratic in the fermion fields, these can be integrated out, giving
\be
\begin{split}
\langle \dots \rangle= &Z^{-1} \int [dA] \det(i\;\sls\! D-m) e^{iI_G}\times\\
&\times\left[(i \;\sls\! D-m)^{-1}_{l_1x_1,k_1y_1}\dots (i\;\sls\! D -m)^{-1}_{l_n x_n,k_ny_n}\pm \text{permutations}\right], \label{ferm-int}
\end{split}
\ee
where $l_i, k_i$ are Dirac indices, $x_i, y_i$ are the space-time coordinates, and the sum over all permutations is understood in \eqref{ferm-int}.
The remaining integration over the gluonic fields is done numerically using Monte Carlo methods. As already discussed above, the computation of the Dirac operator inverses is numerically demanding, if the eigenvalues are small, which occurs for small quark masses. Numerically even more demanding is the computation of the Dirac operator determinant. It is thus very tempting to set $\det(i\;\sls\! D-m)=1$ in the calculation of \eqref{ferm-int} and to compensate for the corresponding omission of the sea quarks by adjusting the bare couplings. This is the so called quenched approximation and corresponds to omitting closed fermion loops in the calculation. 

The exact shift in the observables due to the quenched approximation is difficult to estimate without doing the full ``unquenched'' or ``dynamical'' calculations (i.e. with the determinant in \ref{ferm-int} kept in the calculation). For the calculation of the hadronic spectrum the quenched approximation agrees with the experimental data at the level of a few percent \cite{AliKhan:2001tx}. The situation may not be so favorable in other observables, as will be discussed further on. Even though the precise effect of quenching is difficult to predict, there exists an effective theory approach that can guide one toward the sizes of the quenching errors. \index{quenched approximation|)}

\section{Quenched Chiral Perturbation Theory}\index{Q$\chi$PT|(}
An important observation that will lead us to the formulation of the Quenched Chiral Perturbation Theory (Q$\chi$PT) \cite{Bernard:1992mk,Golterman:1994mk}, is that the sea-quark \index{sea-quark} loop effects can be removed by introducing ghost fields. \index{ghost fields} Consider for instance the path integral involving bosonic fields $\tilde {\Psi}$, $\bar{\tilde \Psi}$\footnote{The ghost fields $\tilde {\Psi}$, $\bar{\tilde \Psi}$ have nothing to do with the fermion fields with Goldstone boson degrees of freedom integrated out, that have been introduced in chapter \ref{HQET} (see, e.g., Eq.~\eqref{factor-out}). I apologize to the reader for multiple use of the symbol. Note also, that unlike the fermionic integration variables $\Psi, \bar\Psi$, the ghost fields are not independent \cite{Sharpe:2001fh}. They have to be constrained in order for the functional integrals to converge.}
\be
\int [d\bar{\tilde{\Psi}}][d\tilde{\Psi}]\; \exp\!\left({i \int d^4 x\bar{\tilde{\Psi}} (i\;\sls\! D-m) \tilde{\Psi}}\right)\sim \det (i \;\sls\! D-m)^{-1}. \label{intro-ghost}
\ee
The inverse of the Dirac operator determinant in \eqref{intro-ghost} is exactly what is needed to cancel the determinant in \eqref{ferm-int}. Adding another integration over bosonic fields with the weight function \eqref{intro-ghost} to the Green's function \eqref{gen-Green}, will then result in the cancellation of the Dirac operator determinant in the final result \eqref{ferm-int}. Introduction of the bosonic spin $1/2$ ghost fields \index{ghost fields} (i.e. fields with ``wrong'' statistics) in the QCD Lagrangian thus mimics the quenched approximation. The quenched QCD (QQCD) Lagrangian is then \index{QQCD}
 \begin{equation}
{\cal L}_{\text{QQCD}}={\cal L}_{\text{G}}+\bar{\Psi}\big(i\;\sls\! D-m\big) \Psi +\bar{\tilde \Psi}\big(i\;\sls\! D-m\big) \tilde \Psi, \label{QQCD}
\end{equation}
with ${\cal L}_{\text{G}}$ the purely gluonic part of the Lagrangian. From the Lagrangian \eqref{QQCD} we can anticipate the hadron spectrum of the quenched approximation. In addition to the usual $q\bar{q}$ mesons, we will have bosons consisting of two ghosts $\tilde q \bar{\tilde q}$ as well as fermions consisting of quarks and ghosts, $q\bar{\tilde q}$ or $\tilde q \bar{q}$. 

If the masses $m$ in \eqref{QQCD} are small compared to $\Lambda_\chi$, the quenched QCD \eqref{QQCD} exhibits an approximate chiral symmetry. This has the same Ward identities as the naively expected global symmetry of QQCD \cite{Sharpe:2001fh}
\be
SU(3|3)_L\times SU(3|3)_R \ltimes U(1)_V,\label{group-struct}
\ee
where $\ltimes$ denotes a semi-direct product \footnote{The semi-direct product is defined as follows. Let $A$ and $H$ be subgroups of $G$, where $A$ is normal ($A$ is a normal subgroup if, for all $a\in A$ and all $g\in G$, $g ag^{-1}\in A$). If $A \cap H=\{e\}$, then the semi-direct product $A\ltimes H$ consists of all $ah, a\in A, h\in H$.}. The semi-direct product in \eqref{group-struct} is a consequence of the fact that the generator of $U(1)_V$ group $\bar{I}=\diag(1,1,1,-1,-1, -1)$, and some of the generators of $SU(3|3)_{L,R}$ do not commute. The $SU(3|3)$ is a graded symmetry group that mixes $3$ fermionic and $3$ bosonic degrees of freedom. The fundamental representation of $SU(3|3)$ consists of $6\times 6$ matrices $U$ \index{graded symmetry group}
\be
U=\begin{pmatrix}
A&C\\
D&B
\end{pmatrix},
\ee
with $A$ and $B$ the $3\times 3$ matrices of commuting numbers and, $C,D$ of anticommuting numbers. We define also a cyclic ``supertrace'' $\str U=\tr A-\tr B$, while the ``superdeterminant'' is $\sdet U=\exp(\str \ln U)$, with the property $\sdet (U_1 U_2) =\sdet(U_1) \sdet(U_2)$ \cite{Bernard:1992mk}. For $U\in SU(3|3)$ then $\sdet U=1$ and $U^\dagger U=1$, with the hermitian conjugation defined as usual (complex conjugation of the usual transpose), while the complex conjugation switches the order of anticommuting variables $(\alpha \beta)^*=\beta^*\alpha^*$ \cite{Bernard:1992mk,Weinberg:cr}. \index{superdeterminant}\index{supertrace}

As in the QCD, we assume, that the axial part of the symmetry group \eqref{group-struct} is spontaneously broken \cite{Bernard:1992mk,Sharpe:2001fh}. We then have the following symmetry breaking pattern\index{spontaneous symmetry breaking}
\be
SU(3|3)_L\times SU(3|3)_R \ltimes U(1)_V \to SU(3|3)_V \ltimes U(1)_V. \label{group-break}
\ee
The spontaneous symmetry breaking gives rise to massless Goldstone bosons. From the simple counting of generators of the spontaneously broken symmetry group $SU(3|3)_A$, one would expect $35$ Goldstone bosons. However, the $\eta'$ meson does not decouple as in the usual QCD. This will be discussed in some more detail.

Consider for definiteness the fields
\begin{equation}
\label{eqQ:2}
\Sigma=\exp\left(2i\frac{\Phi}{f}\right), 
\qquad \Phi=
\begin{pmatrix}
\phi&\chi^\dagger\\
\chi&\tilde \phi~
\end{pmatrix}.
\end{equation}
Besides the
ordinary $q\bar{q}$ Goldstone mesons $\pi, K,\eta$, which are represented by the $3\times 3$ matrix $\phi$ in \eqref{eqQ:2}, we have also $\tilde{q} \bar{\tilde{q}}$ mesons $\tilde{\phi}$ as well as $\tilde{q}\bar{q}$ 
pseudoscalar fermions $\chi$. We also define the field $\Phi_0$ that is invariant under the unbroken symmetries in \eqref{group-break} 
\be
\Phi_0=\frac{1}{\sqrt{6}}\str(\Phi)=\frac{1}{\sqrt{2}}(\eta'-\tilde{\eta}').
\ee
 The part of the $\Phi$ matrix that contains $\eta', \tilde \eta'$ fields, is
\be
\Phi_{\eta',\tilde \eta'}=\frac{1}{2 \sqrt{6}}[(\eta'+\tilde\eta')I+(\eta'-\tilde \eta')\bar{I}],
\ee
where $I$ is the $6\times 6$ unit matrix and $\bar{I}=\diag(1,1,1,-1,-1,-1)$ as before. Note, that the unit matrix $I$ is one of the generators of $SU(3|3)$ group, as $\str(I)=0$, while the $\bar{I}$ generates the anomalous $U(1)$. Since the identity operator $I$ is a generator of broken symmetry, the field $ \eta'+\tilde\eta'$ is expected to couple to a Goldstone boson as usual. It takes more effort to show, that also $\Phi_0\sim (\eta'-\tilde \eta')$ couples to a Goldstone boson and has to be kept in the analysis \footnote{This point has been clarified recently in \cite{Sharpe:2001fh} through the use of Ward identities. The important ingredients of this analysis are the facts that $\str I=0$ and $\str(I\bar I)\not =0$.} \cite{Sharpe:2001fh}. We will not show this explicitly, but will give some further arguments why $\eta'$, $\tilde\eta'$ have to be kept in the analysis after the effective Lagrangian is introduced (cf. Eqs.~\eqref{propagator}, \eqref{propagator-expl})

To write down the chiral Lagrangian one then follows the classic analysis of Gasser and Leutwyler with $\eta'$ field kept in the Lagrangian (see section 12. of \cite{Gasser:1984gg}). Since the $\Phi_0$ field is invariant under the non-anomalous part of the graded group, in principle arbitrary functions of $\Phi_0$ can appear in the Lagrangian. Expanding the functions and keeping only the relevant terms for the processes of interest, the Lagrangian up to the order ${\cal O}(p^4)$ is \cite{Bernard:1992mk} \index{alpha0@$\alpha_0$} \index{coupling constants!quenched|(} \index{mu0@$\mu_0$}\index{m0@$m_0$}
\begin{equation}
\label{light}
{\cal L}^{\text{light}}=\frac{f^2}{8}\str(\partial_\mu \Sigma \partial^\mu \Sigma^\dagger)+\frac{f^2 \mu_0}{2}
\str({\cal M}\Sigma+{\cal M}\Sigma^\dagger)
+\alpha_0\partial_\mu\Phi_0\partial^\mu \Phi_0-
m_0^2\Phi_0^2+ {\cal L}_{4},
\end{equation}
with $f=132$ MeV,\index{f, value of@$f$, value of} ${\cal M}=\diag(m_u,m_d,m_s,m_u,m_d,m_s)$. Two new low energy constants $\alpha_0, m_0$ appear in the chiral Lagrangian \eqref{light} \cite{Bernard:1992mk}. The so-called hairpin insertion $m_0^2$ \index{hairpin insertion} is related to the commonly used dimensionless parameter $\delta=m_0^2/(8 \pi^2 f^2 N_f)$, where $N_f=3$. Note that $m_\pi^2\sim 2\mu_0(m_u+m_d)$, $m_K^2\sim 2\mu_0(m_{u,d}+m_s)$, $m_\eta^2\sim\frac{2}{3} \mu_0 (m_u+m_d+4m_s)$, and $m_{\eta'}^2\sim \frac{4}{3}\mu_0 (m_u+m_d+m_s)$, where in the mass of $\eta'$ meson the term $m_0$ has not been included, for the reasons discussed below (cf. Eqs.~\eqref{propagator}, \eqref{propagator-expl}). 

\begin{figure}
\begin{center}
\epsfig{file=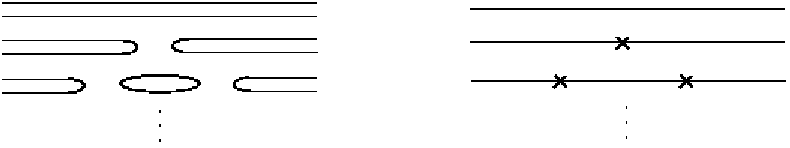, height=2.2cm}
\caption{\footnotesize{The $\eta'$ propagator in the QCD, left, where quark lines and intermediate quark loops are shown. Gluon lines are not shown. On the right, the $\eta'$ propagator in the effective theory is shown, with the cross denoting the $\alpha_0 p^2 -m_0^2$ vertex \eqref{propagator}. Only the upper two diagrams appear in the quenched version of QCD, where quark loops are neglected.}} \label{etaPrprop}\index{m0@$m_0$}
\end{center}
\end{figure}

The presence of $\Phi_0$ terms in \eqref{light} has some unusual consequences for the interactions of $\eta',\tilde \eta'$ bosons. Consider, for simplicity, the case in which all the quark masses are equal. The $\eta', \tilde \eta'$ part of the Lagrangian \eqref{light} is \cite{Golterman:1994mk}
\be
\begin{split}
{\cal L}=&\frac{1}{2} (\partial_\mu \eta')^2-\frac{1}{2}(\partial_\mu \tilde \eta')^2+\frac{1}{2}\alpha_0 (\partial_\mu\eta'-\partial_\mu \tilde \eta')^2\\
&-\frac{1}{2} m_\pi^2 \eta'^2+\frac{1}{2} m_\pi^2 \tilde\eta'^2-\frac{1}{2} m_0^2 (\eta'-\tilde \eta')^2+\dots
\end{split}
\ee
The inverse of a propagator in $(\eta',\tilde\eta')$ basis is then
\be \label{propagator}
(p^2-m_\pi^2)
\begin{pmatrix}
1&0\\
0&-1
\end{pmatrix}
+(\alpha_0 p^2-m_0^2)
\begin{pmatrix}
1&-1\\
-1&1
\end{pmatrix}.
\ee
To diagonalize this propagator, one would have to introduce $p^2$ dependent eigenbasis! Instead, one treats the $(\alpha_0 p^2-m_0^2)$ as an interaction. Note that two subsequent applications of this interaction on one propagator line give a vanishing contribution (i.e. multiplying the first matrix in \eqref{propagator} from the left and from the right with the second matrix in \eqref{propagator} gives zero). This is in agreement with the expectations for the quark flow in $\eta'$ propagator \cite{Kilcup:1989fq,Bernard:1992mk} as depicted on Fig.~\ref{etaPrprop}. In QQCD only the diagrams without quark loops remain (which correspond to only one interaction vertex at the meson level).

The propagator of the $(\eta',\tilde\eta')$ mesons is obtained by inverting \eqref{propagator}
\be\label{propagator-expl}
\frac{1}{p^2-m_\pi^2}
\begin{pmatrix}
1&0\\
0&-1
\end{pmatrix}
-\frac{\alpha_0 p^2-m_0^2}{(p^2-m_\pi^2)^2}
\begin{pmatrix}
1&1\\
1&1
\end{pmatrix}.
\ee
This propagator is very different from what we are used to deal with in the standard $\chi$PT. First of all, the singlet part in \eqref{light} proportional to $m_0^2$, that in unquenched or dynamical $\chi$PT would contribute to the mass of the $\eta'$ meson, now stands in the denominator of the second term in the $\eta',\tilde{\eta}'$ propagator \eqref{propagator-expl}. Taking $m_0\to \infty$, the $\eta'$ meson does not decouple from the other Goldstone bosons, in contrast to what happens in the standard $\chi$PT.\index{m0@$m_0$}

The order ${\cal O}(p^4)$ Lagrangian ${\cal L}_{4}$ contains a plethora of additional terms \cite{Gasser:1984gg}, which we refer to as counterterms. We write down only the terms, which give non-zero contributions to the processes of interest \index{counterterms! quenched}
\begin{equation} 
{\cal L}_{4}=L_4 4 \mu_0\str(\partial_\mu \Sigma \partial^\mu \Sigma^\dagger)\str({\cal M}\Sigma^\dagger+\Sigma 
{\cal M}^\dagger)+
L_5 4 \mu_0\str(\partial_\mu \Sigma^\dagger \partial^\mu \Sigma[{\cal M}\Sigma^\dagger+\Sigma {\cal M}^\dagger])+\dots\label{Lagr4Q}
\end{equation}
Note that the standard $\chi$PT \eqref{chirallagr}, \eqref{eqQ:1} can be obtained from equations \eqref{light}, \eqref{Lagr4Q} by replacing $\str\to \tr$, $\Phi\to\phi$ and setting $\eta'\to 0$.

The heavy mesons are included by combining the heavy-quark symmetry and chiral expansion. The effective Lagrangian in the heavy quark limit and at the next-to-leading order in the chiral expansion is given by \cite{Booth:1994hx,Sharpe:1995qp} \index{g'@$g'$}\index{g@$g$}
\begin{equation}
\begin{split}
\label{heavy}
{\cal L}^{\text{heavy}}=&-\str_a \Tr[\bar{H}_a i v \negcdot D_{ba}H_b]+g \str_a\Tr[\bar{H}_aH_b \gamma_\mu 
{\cal A}_{ba}^\mu \gamma_5]\\
&+g'\str_a\Tr[\bar{H}_aH_a \gamma_\mu \gamma_5]\str({\cal A}^{\mu})+{\cal L}_3^{\text{heavy}},
\end{split}
\end{equation}
where $D_{ba}^\mu H_b=\partial^\mu H_a-H_b{\cal V}_{ba}^\mu$, ${\cal V}^\mu=\frac{1}{2}(\xi^\dagger 
\partial^\mu \xi +\xi \partial^\mu \xi^\dagger)$, ${\cal A}^\mu=\frac{i}{2}(\xi^\dagger \partial^\mu 
\xi -\xi \partial^\mu \xi^\dagger)$ and $\xi\equiv \sqrt{\Sigma}=\exp(i\Phi/f)$. Compared to the usual dynamical heavy meson chiral Lagrangian \eqref{eqQ:3}, a new term appears at the lowest level with the coupling constant $g'$. The higher order terms in the expansion in $v\negcdot p\sim{\cal O}(p) $ and $m_q\sim {\cal O}(p^2)$ are then (up to ${\cal O}(p^3)$
\begin{equation}
\begin{split}
{\cal L}_3^{\text{heavy}}=&2\lambda_1\str_a \Tr[\bar H_aH_b]({\cal M_+})_{ba}+k_1\str_a \Tr[\bar{H}_a i v \negcdot D_{bc}H_b]
({\cal M_+})_{ca}\\
&+k_2\str_a \Tr[\bar{H}_a i v \negcdot D_{ba}H_b]\str_c({\cal M_+})_{cc}+\dots
\end{split}, \label{heavy-counter}
\end{equation}
with ${\cal M_+}=\frac{1}{2}
(\xi^\dagger {\cal M}\xi^\dagger+\xi {\cal M}\xi)$. Dots denote terms that were not written out, since they either do not contribute to the $B\to \pi$ transition or are of higher order in the chiral expansion. The effect of term proportional to $\lambda_1$ is to change the heavy meson propagator. In the case of $s$ quark the shift is $v\negcdot p \to v\negcdot p -\Delta$, where $\Delta=m_{B_S}-m_B$. Note that we do not consider the $1/m_B$ corrections. 
The dynamical heavy meson chiral perturbation theory is obtained by letting $\str\to \tr$, $\Phi\to\phi$, $\eta'\to 0$ and setting $g'\to 0$. 

Next we turn to the heavy-to-light current. At leading and next-to-leading order in the chiral expansion this is (using the heavy quark symmetry) \index{alpha@$\alpha$}\index{kappa12@$\varkappa_{1,2}$} \index{hadronization! currents}
\begin{equation}
\begin{split}
\label{current-Q}
\bar{q}_a\gamma^\mu(1-\gamma_5)Q&\to \frac{i \alpha}{2}\Tr[\gamma^\mu (1-\gamma_5) H_b] \xi_{ba}^\dagger 
[1+V_L'(0)\Phi_0]\nonumber\\
&+\frac{i \alpha}{2}\varkappa_1\Tr[\gamma^\mu (1-\gamma_5) H_c ]\xi_{ba}^\dagger ({\cal M_+})_{cb}\\
&+\frac{i 
\alpha}{2}\varkappa_2\Tr[\gamma^\mu (1-\gamma_5) H_b] \xi_{ba}^\dagger \str({\cal M_+}),
\end{split}
\end{equation}
where ${\cal M_+}=\frac{1}{2}
(\xi^\dagger {\cal M}\xi^\dagger+\xi {\cal M}\xi)$ as before. Only the terms which contribute to the processes of interest up to the next-to-leading 
order in the chiral expansion are displayed. The $V_L'(0)$ is an artifact of the quenched heavy-meson $\chi$PT. The phases of heavy mesons can be chosen such, that $V_L'(0)$ is imaginary, while other coupling constants $\alpha$, $\varkappa_1$, $\varkappa_2$ are real as can be seen by $T$ symmetry considerations.

The values of coupling constants in the dynamical $\chi$PT are extracted from experiment as described in section \ref{coupling}. We use the tree level values of couplings as a first order estimate. Similarly, the coupling constants of quenched chiral Lagrangians \eqref{light}, \eqref{Lagr4Q}, \eqref{heavy}, \eqref{heavy-counter} and the heavy-to-light current \eqref{current-Q} are extracted from the quenched lattice data. From a compilation of quenched determinations of $f_B$ \index{fB@$f_B$} decay constants $f_B=175\pm 20\; {\text{ MeV}}$ \cite{Bernard:2000ki} one arrives at $\alpha_Q=0.40\pm0.05\; \text{GeV}^{3/2}$.\index{alpha@$\alpha$} The $B^*B\pi$ coupling in quenched QCD has been calculated recently \cite{Abada:2002xe} with the value $g_Q=0.67\pm0.09$ (see also exploratory calculation in \cite{deDivitiis:1998kj}). The quenched pion constant is $f_{\pi (Q)}=130\pm 5\; \text{MeV}$ \cite{Blum:2000kn}.

\begin{table} [h]
\begin{center}
\begin{tabular}{|l|c|c|} \hline \index{alpha@$\alpha$}\index{alpha0@$\alpha_0$}\index{g'@$g'$}\index{g@$g$}
Coupl. &Full Th.& Quen. Th.\\ \hline\hline
$\alpha\;[\text{GeV}^{3/2}]$&$0.38\pm0.04 $&$0.40\pm0.05$\\ \hline
 $g$&$0.59\pm0.08 $&$0.67\pm0.09$ \\ \hline
$f \;[\text{MeV}]$&$130.7\pm0.04 $&$130\pm5$ \\ \hline
$m_0\;[\text{GeV}]$&$-$&$0.64\pm0.1$\\ \hline\hline
$L_4\;[\times 10^{-3}]$ &$-0.5\pm0.5 $&$0\pm0.5$ \\ \hline
$L_5\;[\times 10^{-3}]$&$0.6\pm0.5 $&$1\pm0.5$ \\ \hline
 \end{tabular} 
 \caption[Values of coupling constants in Q$\chi$PT]{\footnotesize{Values of coupling constants used in the analysis. Other couplings used in the analysis are $k_2=\varkappa_2=0$, $\alpha_0=0$, while $g'$,$k_1,\varkappa_1,V_L'(0) $ are varied in the ranges $|g'|<g$, $|k_1,\varkappa_1|< 32L_5 \mu_0/f^2$, $|V_L'(0)|<0.5$. }}\index{m0@$m_0$}
\label{tab-koef-Q}\index{kappa12@$\varkappa_{1,2}$}
\end{center}
\end{table}

The value of the quenched log
parameter $\delta=m_0^2 /(24 \pi^2 f_\pi^2)$ is estimated from a number of quenched lattice analyses. The CP-PACS collaboration \cite{CP-PACS} extracted the value of $\delta$ from the spectrum of 
pseudoscalar mesons, and obtain $\delta = 0.10\pm 0.02$, whereas from 
the pseudoscalar meson decay constants they find $\delta = 0.12\pm 0.04$. 
In Ref.~\cite{Bardeen:2000cz} the authors quote 
$\delta = 0.065\pm 0.013$ as their final estimate, while the Ref.~\cite{DeGrand} obtained $\delta = 0.093\pm 0.028$. After combining the above results with the pre-97 world average 
$\delta = 0.1-0.2$~\cite{Sharpe-Latt96} 
we arrive at $\delta = 0.1 \pm 0.04$. 
In terms of $m_0$, by using $f_\pi=131$~MeV, this amounts to
$ m_0= 0.64 \pm 0.1 ~{\rm GeV}$ \index{m0@$m_0$}

No evidence for a nonzero value of the coupling constant $\alpha_0$ \index{alpha0@$\alpha_0$} has been found. Note that what is actually measured in the lattice simulations is $m_0^2-p^2 \alpha_0\sim m_0^2-m_\pi^2\alpha_0$. The claim \cite{Bardeen:2000cz} is that $p^2$ dependence is mild so that the hairpin (cross) vertex on Figure \ref{etaPrprop} is well described by the momentum independent mass insertion $m_0$, therefore we set $\alpha_0=0$ in the numerical analysis. \index{hairpin insertion}

The $g'$ coupling is a $B^*B\eta'$ coupling \eqref{heavy} and there are no lattice results for it. Large $N_c$ arguments suggest that $|g'|< |g|$, so that we vary $g'$ in this range.
The analysis of the counterterms $L_{4,5}$ has been made by the ALPHA collaboration \cite{Heitger:2000ay} in which they get $\alpha_5^Q=8 (4 \pi)^2 L_5^{Q}=0.99\pm0.06$. The $L_4$ is small by $N_c$ arguments and not determined by their analysis. In numerical evaluation we take $L_5^{Q}=(1\pm 0.5)\times 10^{-3}$ and $L_4^{Q}=(0\pm0.5)\times 10^{-3}$. There are no experimental data regarding the sizes of the $k_{1,2}$ and $\varkappa_{1,2}$ counterterms. From large $N_c$ considerations we can conclude that $k_2$ and $\varkappa_2$ are $1/N_c$ suppressed, i.e., the following relations are expected $k_1> k_2$, $\varkappa_1>\varkappa_2$. In the numerical estimates we then set $k_2=\varkappa_2=0$. The approximate size of $k_1$ is determined by observing that $L_5$ term in \eqref{eqQ:1} and $k_1$ term in \eqref{eqQ:3} have similar structure compared to the kinetic term in \eqref{light} and \eqref{heavy} respectively. It is thus reasonable to expect that roughly $|k_1|\sim L_5 4 \mu_0 8/f^2$. Similar reasoning applies for $\varkappa_1$, so that in the numerical evaluation we vary $k_1$ and $\varkappa_1$ in the range $- 32 L_5 \mu_0 /f^2<k_1, \varkappa_1< 32 L_5 \mu_0 /f^2$.\index{coupling constants!quenched|)} \index{counterterms! quenched}\index{Q$\chi$PT|)}
\index{kappa12@$\varkappa_{1,2}$}

\section{Effects of quenching in $B\to \pi, K$ transitions}
In this section we apply the formalism of quenched chiral perturbation theory to the weak $B\to \pi, K$ transitions. Semileptonic decay $B^0\to \pi^- l^+\nu_l$ plays a crucial role in the determination of the Cabibbo-Kobayashi-Maskawa (CKM) matrix element $|V_{ub}|$ from the experiment \cite{Alexander:1996qu}. For this purpose a solid knowledge of the $B\to \pi$ form factors is needed. In principle the form factors can be obtained from lattice simulations, but these are forced to use various approximations at present, as discussed at the beginning of this chapter. We will address the effects of the quenched approximation in this section.

 The quenched version of the HH$\chi$PT has already been used for the heavy meson decay constants, corrections to the heavy meson masses, Isgur-Wise function and bag parameters $B_B$ in \cite{Booth:1994hx,Sharpe:1995qp}. On the other hand, the heavy-to-light weak transition has been considered only in the dynamical HH$\chi$PT in \cite{Falk:1993fr}, while there has been no such study in the quenched version. We study the matrix elements 
$\langle P|\bar q \gamma^{\mu}(1-\gamma_5)b|B\rangle$ and 
$\langle P|\bar q \sigma^{\mu\nu}(1+\gamma_5)b|B\rangle$ with $P=\pi, K$. 
The V-A matrix element is the matrix element of the hadronic current in the $B\to Pl^+\nu_l$ decays, while the tensor matrix element $B\to P$ plays an important role, for instance, in the decay $B\to Kl^+l^-$. This decay might be sensitive to new physics
and has recently been observed experimentally \cite{Abe:2001qh}. 

A general parameterization of the vector current matrix element between two pseudoscalar states is \index{F+0,definition of @$F_{+,0}$, definition of}
\begin{equation}
\langle P(p)|\bar{q}\gamma^\mu (1-\gamma_5)b|B(p_B)\rangle = \Big(p_B^\mu+p^\mu- 
\frac{m_B^2-m_P^2}{q^2}q^\mu\Big) F_+(q^2)+\frac{m_B^2-m_P^2}{q^2}q^\mu F_0(q^2),\label{vect-current}
\end{equation}
where $q^\mu=p_B^\mu-p^\mu$, with $p_B$ the $B$ meson and $p$ the outgoing light pseudoscalar meson momenta. The tensor current matrix element is parametrized as \cite{Isgur:kf}
\begin{equation}
\langle P(p)|\bar{q}\sigma^{\mu\nu} (1+\gamma_5)b|B(v)\rangle=iH[p^\mu v^\nu-p^\nu v^\mu]+S 
\epsilon^{\mu\nu\alpha\beta}v_\alpha p_\beta,
\end{equation}
where $v^\mu$ is the heavy meson velocity defined through $p_B^\mu=m_Q v^\mu$ with $m_Q$ the heavy quark mass. The operator $\bar q \sigma^{\mu\nu} (1+\gamma_5)b$ induces the $B\to Pl^+l^-$ decay, but the form factor $S$ does not contribute to this decay. The form factor $H$, which does contribute, can be related to $F_+$ in the heavy-quark limit. 
The heavy quark symmetry relation $\gamma_0 b=b$ \cite{Isgur:kf} for the heavy meson at rest, $v^\mu=(1,0,0,0)$, (cf. section \ref{heavylightExp}) leads to \footnote{As an intermediate step one can introduce a more convenient parametrization
 \begin{equation*}
\langle P(p)|\bar{q}\gamma^\mu (1-\gamma_5)b|B(v)\rangle =(p^\mu -v \negcdot p \; v^\mu) 
f_p(v\negcdot p)+v^\mu f_v(v\negcdot p).
\end{equation*}
with $v^\mu$ the heavy meson velocity and $v\negcdot p=(m_B^2+m_P^2-q^2)/2m_B$ the pion energy in the $B$ meson rest frame. In the heavy quark limit, with $m_B \to \infty$, then
\begin{equation*}
\label{limit_yes} 
F_+(q^2)=\frac{1}{2}f_p(v\negcdot p)\qquad
F_0(q^2)=\frac{f_v(v\negcdot p)}{m_B},
\end{equation*}
}
\begin{equation}
\langle P(p)|\bar{q}\sigma^{i0} (1+\gamma_5)b|B(v)\rangle=i \langle P(p)|\bar{q}\gamma^{i} 
(1-\gamma_5)b|B(v)\rangle \qquad {\rm and} \qquad H=2 F_+.\label{eq:003}
\end{equation}
Due to the relation \eqref{eq:003}, we do not have to consider the tensor currents separately. The conclusions for $F_+$ apply also for the $H$ form factor.

\begin{figure}
\begin{center}
\epsfig{file=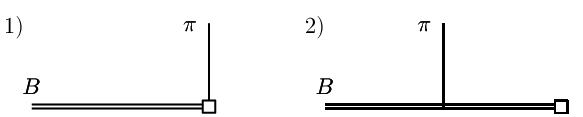, height=1.5cm}
\caption{\footnotesize{The point 1) and pole 2) tree level Feynman diagrams contributing to the $B\to \pi $ transition. The weak current is denoted by the rectangle.}} \label{sl1}
\end{center}
\end{figure}

\begin{figure}
\begin{center}
\epsfig{file=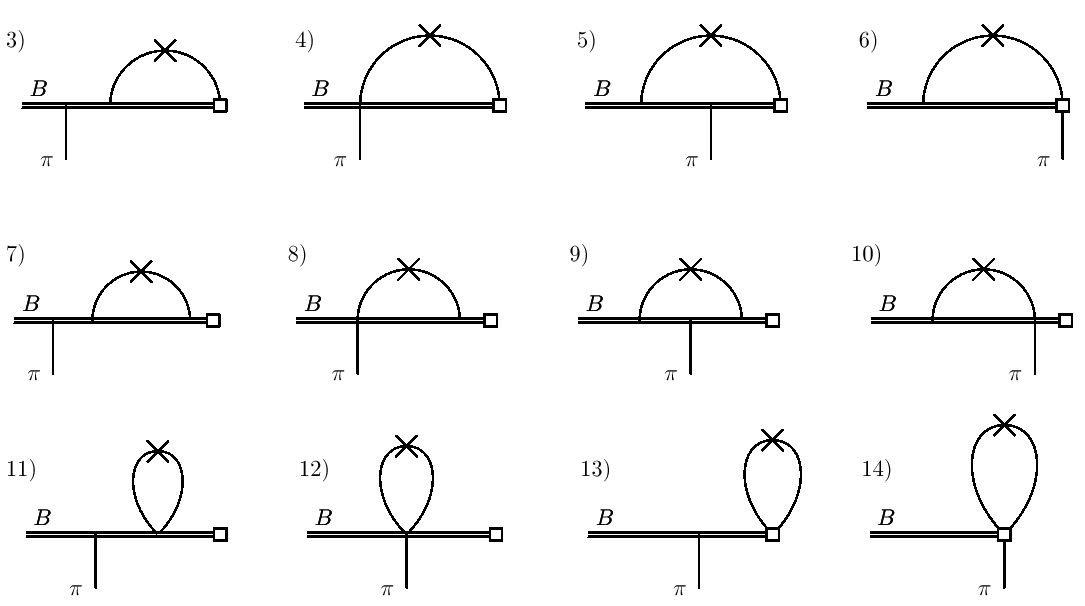, height=5cm}
\caption{\footnotesize{The one loop contributions to the $B\to \pi $ transition. The cross denotes the $m_0$ hairpin vertex \eqref{light}. Each diagram represents two Feynman diagrams in the quenched $\chi$PT, $a$) the diagram without the hairpin vertex (cross) and $b$) the same diagram with a cross. In the full $\chi$PT only diagrams without the cross appear. The amplitudes corresponding to the diagrams are listed in appendix \ref{explicitBpi}. The heavy mesons are denoted by a double line, while the light mesons are denoted by a single line. The weak current insertion is denoted by a rectangle.}} \label{sl2} \index{hairpin insertion}
\end{center}
\end{figure}

\index{quenching effects|(}
The form factors $F_{+,0}$ are calculated to 1-loop using the heavy meson chiral perturbation theory outlined in the previous section. The tree level diagrams are shown in Fig.~\ref{sl1}. In the heavy-quark limit the point diagram gives the leading contributions to $F_0$, while the $F_+$ form factor receives the leading contribution from the pole diagram. The leading order amplitudes are corrected at one-loop through the diagrams shown on Fig.~\eqref{sl2}. Beside the $B\to P$ form factors $F_+$ and $F_0$, we give also the one-loop corrections to the heavy pseudoscalar decay constant $f_{B_j}$ and the light pseudoscalar decay constant $f_{P_{ij}}$. One loop calculation of the decay constants can be found in \cite{Gasser:1984gg,Goity:1992tp,Grinstein:1992qt} for the full (``dynamical'') theory, while in \cite{Booth:1994hx,Sharpe:1995qp,Bernard:1992mk} results for the quenched $\chi$PT have been obtained. The results of the calculation in the quenched $\chi$PT are \index{fB@$f_B$}
\begin{align}
F_{+(Q)}^{B_j\to P_{ij}}&=\frac{1}{2}H_{(Q)}^{B_j\to P_{ij}}=\sqrt{m_B}\frac{\alpha}{2f}\;
\frac{g}{v\negcdot p+\Delta^*_i}\;\Big[1+
\delta F_{+(Q)}^{B_j\to P_{ij}}+\tfrac{1}{2}k_1m_j-4L_5 \frac{4 \mu_0}{f^2}(m_i+m_j)\Big], \label{eq:q1}\\
F_{0(Q)}^{B_j\to P_{ij}}&=\frac{1}{\sqrt{m_B}}\frac{\alpha}{f}\Big[1+\delta F_{0(Q)}^{B_j\to 
P_{ij}}+\big(\tfrac{1}{2}k_1+\varkappa_1\big)m_j-4L_5\frac{ 4 \mu_0}{f^2}(m_i+m_j)\Big],\label{eqQ:6}\\
f_{B_j(Q)}&=\frac{\alpha}{\sqrt{m_B}}\Big[1+\delta f_{B_j(Q)}+\big(\tfrac{1}{2}k_1+
\varkappa_1\big)m_j\Big],\\
f_{P_{ij}(Q)}&=f\Big[1+\delta f_{P{ij}(Q)}+4L_5\frac{ 4\mu_0}{f^2}(m_i+m_j)\Big],
\end{align}
while in the dynamical theory (i.e. in the usual $\chi$PT corresponding to the full QCD) the physical observables are
\begin{align}
\begin{split}
F_{+}^{B_j\to P_{ij}}&=\frac{1}{2}H^{B_j\to P_{ij}}=\sqrt{m_B}\frac{\alpha}{2f}\; \frac{g}
{v\negcdot p+\Delta^*_i}\; \Big[1+\delta 
F_{+}^{B_j\to P_{ij}}\\
~&+\tfrac{1}{2}k_1m_j+\big(\tfrac{1}{2}k_2-8L_4\frac{ 4\mu_0}{f^2}\big)
(m_u+m_d+m_s)-4L_5\frac{ 4\mu_0}{f^2}(m_i+m_j)\Big],\label{eqQ:4}
\end{split}
\\
\begin{split}
F_{0}^{B_j\to P_{ij}}&=\frac{1}{\sqrt{m_B}}\frac{\alpha}{f}\Big[1+\delta F_{0}^{B_j\to P_{ij}}+\big(\tfrac{1}{2}k_1+\varkappa_1\big)m_j\\
&+\big(\tfrac{1}{2}k_2-8L_4\frac{4 \mu_0}{f^2}+\varkappa_2\big)
(m_u+m_d+m_s)-4L_5\frac{4\mu_0}{f^2}(m_i+m_j)\Big],\label{eqQ:5}
\end{split}
\\
f_{B_j}&=\frac{\alpha}{\sqrt{m_B}}\Big[1+\delta f_{B_j}+\big(\tfrac{1}{2}k_1+\varkappa_1\big)m_j+
\big(\tfrac{1}{2}k_2+\varkappa_2\big)(m_u+m_d+m_s)\Big],\\
f_{P_{ij}}&=f\Big[1+\delta f_{P_{ij}}+4L_5\frac{ 4 \mu_0}{f^2}(m_i+m_j)+8L_4\frac{ 4\mu_0}{f^2}(m_u+m_d+m_s),
\Big] \label{eq:q2}
\end{align}
where $m_i$ are the light quark masses \eqref{light}, $\Delta_i^*=m_{B_i^*}-m_{B_j}$, while the meson indices denote quark flavors $B_j\sim b \bar{q}_j$, $P_{ij} \sim q_i \bar{q}_j$. Above we have also introduced the variable $v\negcdot p$, which is equal to the pion energy in the B meson rest frame. It is trivially connected to the $q^2$ variable $v\negcdot p=(m_B^2+m_P^2-q^2)/2m_B$. The $\delta F$, $\delta f$ in \eqref{eq:q1}-\eqref{eq:q2} denote the one-loop chiral corrections which are independent of the counterterms $k_{1,2}$, $\varkappa_{1,2}$, $L_{4,5}$. They depend 
on the couplings $\alpha$, $g$, $f$ and in the quenched theory also on $g^\prime$, $V_L^\prime(0)$, $m_0$ and $\alpha_0$. Note, that the coupling constants $\alpha$, $g$, $f$ need not be the same in the dynamical and quenched versions of the $\chi$PT. We use the values listed in Table \ref{tab-koef-Q}. The explicit expressions for $\delta F$, $\delta f$ are listed in Eqs.~\eqref{eq:q7}-\eqref{eq:End} of appendix \ref{explicitBpi}. Loop diagrams are calculated using dimensional regularization with the renormalization prescription $2/\epsilon-\gamma+\ln(4\pi)+1\to0$ used by Gasser and Leutwyler in \cite{Gasser:1984gg}. The isospin limit $m_u=m_d$ has been used in the calculation, and mass differences between $B$, $B_s$, $B^*$, $B_s^*$ meson states, when appearing in the loop, have been neglected. 

There are several interesting observations regarding expressions \eqref{eq:q1}-\eqref{eq:q2}:
\begin{itemize}
\item The scaling of the form factors with $m_B$ is as required by the heavy quark symmetry, i.e., $F_+\sim \sqrt{m_B}$, $F_0\sim f_B\sim 1/\sqrt{m_B}$. The $1/m_B$ corrections are not included in the analysis. However, for the range of $q^2>18\;\text{GeV}^2$, the corrections can be estimated by $v\negcdot p/m_B \le 0.2$. Errors are expected to be smaller for larger $q^2$, i.e. in the part of phase space where the approach of $\chi$PT is valid.

\item The $F_+$ form factor has a pole at $q^2=m_{B^*}^2$, as it can be seen from the presence of $1/(v\negcdot p+\Delta_i^*)$ term in \eqref{eq:q1}, \eqref{eqQ:4}. No such pole is present in $F_0$. Apart from the pole in $F_+$, all other $q^2$ dependence of $F_+$ and $F_0$ form factors near zero recoil is determined by the 1-loop factors $\delta F_+(q^2) $ and $\delta F_0(q^2)$. There is no $q^2$ dependence coming from the counterterms.

\item The form factors $F_{+(Q)}$, $F_{0(Q)}$ and the decay constant $f_{B(Q)}$ contain quenched logarithms \index{quenched logarithms} of the form $m_0^2 \ln(m_P^2/\mu^2)$. These diverge in the limit $m_P\to 0$ (see Eqs.~\eqref{expans-quenched} below). Divergences of this type are present also in the quark condensate \cite{Bernard:1992mk} and the heavy-meson decay constant $f_B$ \cite{Booth:1994hx,Sharpe:1995qp}, but are not present in $f_P$ for degenerate quarks. 

 
\end{itemize}

\index{expansion! in small $m_P$|(}
To show the last observation explicitly, we expand at fixed $v\negcdot p$ the expressions \eqref{eq:q1}-\eqref{eq:q2} for small pseudoscalar mass $m_P$. For simplicity, the final light pseudoscalar $P$ is assumed to be composed of two mass-degenerate quarks. For small mass expansion in the quenched theory we find
\begin{subequations}\label{expans-quenched} 
\begin{align}
\begin{split}
F_{+(Q)}&=\frac{\sqrt{m_B}~\alpha g}{2 f (v\negcdot p +\Delta^*)}\Big[1- \frac{m_0^2(1+3 g^2)}{96 \pi^2 f^2}\ln\big(\frac{m_P^2}{\mu^2}\big) -\frac{g^2 m_0^2}{12 \pi f^2 v\negcdot p}m_P+\\
&\hskip64mm +c_0+c_2 m_P^2+c_Lm_P^2\ln\big(\frac{m_P^2}{\mu^2}\big) + \cdots\Big],
\end{split}
\\
F_{0(Q)}&= \frac{\alpha }{\sqrt{m_B}f}\Big[1- \frac{m_0^2(1+3 g^2)}{96 \pi^2 f^2} \ln\big(\frac{m_P^2}{\mu^2}\big) + c_0^a+c_2^a m_P^2+c_L^am_P^2\ln\big(\frac{m_P^2}{\mu^2}\big) + \cdots \Big],
\label{importance_of_chiral_log}
\end{align}
\end{subequations}
As before, the $B-B^*$ mass difference of the heavy meson states 
has been neglected in the loops. The coefficients $c_0$, $c_2$ and $c_L$ depend on $v \negcdot p$ and can be obtained by expanding the complete expressions listed in appendix \ref{explicitBpi}. Same leading order terms are obtained also for the expansion in $m_P$ at fixed $q^2$, with the difference from expansion \eqref{expans-quenched} starting at the order ${\cal O}(p^2)$. The expansion for $f_B$ is \index{fB@$f_B$}
\begin{equation}
f_{B(Q)}=\frac{\alpha}{\sqrt{m_B}}\Big[ 1-\frac{m_0^2(1+3 g^2)}{96 \pi^2 f^2}\ln\big(\frac{m_P^2}{\mu^2}\big) +c_0^b+c_2^b m_P^2+c_L^b m_P^2\ln\big(\frac{m_P^2}{\mu^2}\big) + \cdots\Big],\label{expansfBQ}
\end{equation} 
where the coefficients $c_i^b$ do not depend on $v\negcdot p$. The expressions \eqref{expans-quenched}, \eqref{expansfBQ} are to be compared with the unquenched chiral expansions
\begin{subequations}\label{expans-dyn}
\begin{align} 
\begin{split}
F_{+}&=\frac{\sqrt{m_B}~\alpha g}{2 f (v\negcdot p +\Delta^*)}\Big[1 +c_0'+c_2' m_P^2+c_L' m_P^2\ln\big(\frac{m_P^2}{\mu^2}\big) + \cdots\Big],
\end{split}
\\
F_{0}&= \frac{\alpha }{\sqrt{m_B}f}\Big[1+ {c'}_0^a+{c'}_2^a m_P^2+{c'}_L^am_P^2\ln\big(\frac{m_P^2}{\mu^2}\big) + \cdots\Big],
\end{align}
\end{subequations}
where again $c_i'$ depend on $v\cdot p$, while $f_B$ is \footnote{The importance of the chiral logarithms in this expansion for the {\it ratio} of the heavy meson decay constants $f_{B_s}/f_{B}$ can be estimated by comparing it to the ratio of the light pseudoscalar decay constants $f_K/f_\pi$ \cite{Becirevic:2002mh}.}
\begin{equation}
f_{B}=\frac{\alpha}{\sqrt{m_B}}\Big[ 1 +{c'}_0^b+{c'}_2^b m_P^2+{c'}_L^b m_P^2\ln\big(\frac{m_P^2}{\mu^2}\big) + \cdots\Big],\label{expansfB}
\end{equation} 
\index{expansion! in small $m_P$|)}

The difference in the expansions around $m_P\sim 0$ between full and quenched theories is striking. First of all, the chiral limit $m_P\to 0$ does not exist for the quenched case, as then the quenched chiral logs $m_0^2 \ln(m_P^2/\mu^2)$ diverge. Next, in the expansion of $F_{+(Q)}$ a linear term in $m_P$ appears, while in the full theory the expansion begins at $m_P^2$ as usual. The linear term arises from the diagrams 7 and 9 in Fig.~\ref{sl2}. It is absent, however, if $\Delta^*=m_{B^*}-m_B$ is first kept different from zero in the loops and then the limit $m_P\to 0$ is taken. This alternative expansion would then be valid only if $m_{B^*}-m_B>m_P$, which is not the case, e.g., in the $B$ system. In the D meson system on the other hand $m_{D^*}-m_D\sim m_\pi$ for the physical pion in the final state, so that the expansion given in \eqref{expans-quenched}, \eqref{expans-dyn} may not be valid for this limiting case. The expansions \eqref{expans-quenched}, \eqref{expans-dyn} are, however, valid both for the set of the light and heavy quark masses used in the lattice at present, as well as for the $B$ meson system.

\begin{figure}
\begin{center}
\epsfig{file=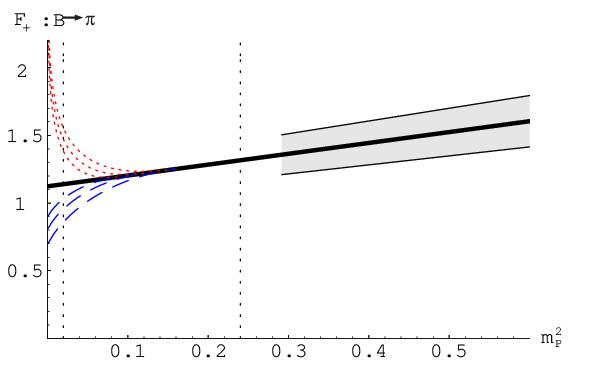,width=6.cm}
\epsfig{file=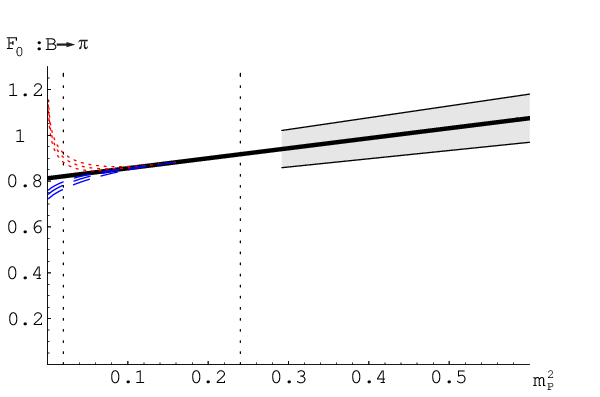,width=6.cm}
\\
\epsfig{file=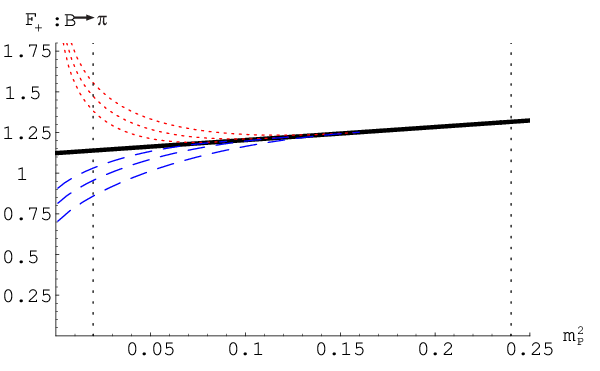,width=6.cm}
\epsfig{file=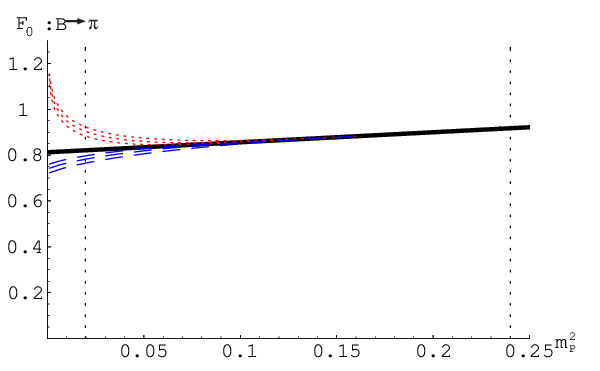,width=6.cm}
\caption{\footnotesize{Illustration of the chiral extrapolations for $F_+$ and $F_0$ at fixed $v\cdot p=0.545$ GeV. On left (right) the $F_+^{B\to\pi}(m^2)$ ($F_0^{B\to\pi}(m^2)$) as a function of final pseudoscalar mass squared, $m_P^2$, are shown. The gray bands denote the lattice data points \cite{Abada:2000ty} together with the error bars. The quenched and dynamical chiral extrapolations at low $m_P^2$ are shown with red dotted and blue dashed lines respectively. Three matching points $m_{\text{m}}=0.3 \;\text{GeV},0.35 \;\text{GeV},0.4 \;\text{GeV}$ are used to match $\chi$PT form factor dependences to the linear extrapolation in $m_P^2$ suggested by the lattice data, shown as solid black line. The two vertical dotted lines denote physical values of $m_\pi^2$ and $m_K^2$ masses. The lower two plots are blow-ups of the upper plots in the region of lower $m_P^2$. }}\label{chiral-extrap}
\end{center}
\end{figure}

Note, that the quenched chiral logarithmic terms do not depend on $v\negcdot p$. The divergent terms in \eqref{importance_of_chiral_log} are positive for $m_P< \mu$. This implies that $F_{+,0(Q)}$, $f_{B(Q)}$ should eventually start increasing with the decreasing $m_P$ (see Figure \ref{chiral-extrap}). A general prediction of our analysis is that at small values of $m_P$, below $m_P\sim 300 \;\text{MeV}$, the $F_{+(Q)}$ form factor increases with smaller $m_P$. The chirally divergent quenched logarithms \index{quenched logarithms} start to dominate the behavior of the form factors at around $200 \text{ MeV}$. The dependence of $F_{+(Q)}$ on $m_P$ is difficult to predict for lager values of $m_P$, as then higher order terms in $\chi$PT become important. 

\index{extrapolation! chiral|(}\index{extrapolation! comparison of quenched and dynamical|(}
We illustrate the difference between the quenched and dynamical chiral extrapolations to the physical pion mass on Figure \ref{chiral-extrap}. Note, that the chiral perturbation theory is valid only for small pseudoscalar masses $m_P\ll 4\pi f$, while the lattice simulations are done at the pion masses $m_P\sim 0.5-0.8$ GeV \cite{Aoki:2001rd,Bowler:1999xn,Abada:2000ty,El-Khadra:2001rv,Shigemitsu:2002wh}. The chiral perturbation theory results thus cannot be directly compared with the lattice data on $B\to \pi$ transition. In order to illustrate the importance of the chiral extrapolations we then proceed as follows. For the dependence of $F_{+,0}$ form factor on $m_P^2$ at high values of the final pseudoscalar mass $m_P\in [0.5,0.8]$ GeV, we use the lattice results. In this range of $m_P$, the lattice data show linear dependence of form factors $F_{+,0}$ on $m_P^2$. For definiteness we use the data of APE \index{APE collaboration} collaboration \cite{Abada:2000ty,Damir:Thanks} at $v\negcdot p=0.545$ GeV. The error bars on the data are indicated by the shaded regions on Figure \ref{chiral-extrap}. 

Linear extrapolation in $m_P^2$ is then used from the region of lattice data points to the value $m_{\text{m}}$ at which the Q$\chi$PT expressions \eqref{eq:q1}-\eqref{eqQ:6} are matched smoothly to the linear dependence of the lattice data. The matching point is varied between $m_{\text{m}}=0.3\;\text{GeV}, 0.35\;\text{GeV}, 0.4\;\text{GeV}$. The form factors at fixed $v\negcdot p$ are thus
\be
F_{+(Q)}(m^2)=\left\{
\begin{aligned}
&F_+^{\text{Latt}}(m^2)&;\qquad &m^2>m_{\text{m}}^2,\\
&\begin{aligned}
&F_{+(Q)}^{\chi\text{PT}}(m^2)-\left(F_{+(Q)}^{\chi\text{PT}}(m_{\text{m}}^2)-F_+^{\text{Latt}}(m_{\text{m}}^2)\right)\\
&-\left(F_{+(Q)}'^{\;\chi\text{PT}}(m_{\text{m}}^2)-F_+'^{\;\text{Latt}}(m_{\text{m}}^{2})\right)\left(m^2-m_{\text{m}}^2\right)
\end{aligned}
&;\qquad& m^2\le m_{\text{m}}^2,
\end{aligned}
\right. \label{F+Glue}
\ee
where $F_{+(Q)}^{\chi\text{PT}}(m^2)$ is given in \eqref{eq:q1}, while $F_+^{\text{Latt}}(m^2)$ denote the linear fit to the lattice data. $F_{+(Q)}'^{\;\chi\text{PT}}$ and $F_+'^{\;\text{Latt}}$ denote derivatives of $F_{+(Q)}^{\;\chi\text{PT}}$ and $F_+^{\;\text{Latt}}$ on $m^2$. Expression \eqref{F+Glue} applies also for the $F_{0(Q)}(m^2)$ form factor with the replacement $F_+\to F_0$.

In exactly the same way as described above also the dynamical $\chi$PT expressions \eqref{eqQ:4}-\eqref{eqQ:5} are matched smoothly to {\it the same quenched lattice data}. Even though the dynamical and quenched lattice data might well not be the same at high $m_P\in [0.5,0.8]$ GeV, this assumption does suffice for our purposes of comparing chiral extrapolations at low pion masses. Note also, that the extrapolation in pion mass for the dynamical $\chi$PT case is done at fixed value of the strange quark mass $m_s$, that is set equal to its physical value.

In Figure \ref{chiral-extrap} the central values of the parameters in Table \ref{tab-koef-Q} have been used. The general features of the analysis are not changed, if the values of the parameters are varied within the ranges given in Table \ref{tab-koef-Q}. Note that the uncertainties due to unknown counterterms and the value of $g'$ coupling constant are small, as these contributions are proportional to $m_P^2$, which is small. The largest variations are obtained by varying values of $m_0^2$ and $g$ as they are multiplying the divergent chiral logarithmic terms \eqref{expans-quenched}. These variations are at the order of $10\%$ at the physical pion mass and, most importantly, do not change the overall $m_P^2$ behaviour of the form factors $F_+$, $F_0$.

There are several things to be learned from Figure \ref{chiral-extrap}. First of all, the behaviour of quenched and dynamical $F_{+,0}(m^2)$ at fixed $v\cdot p$ is strikingly different at low $m_P^2$. The extrapolation of $F_{+,0}$ using quenched $\chi$PT (red dotted line) rises above the linear extrapolation suggested by the lattice data points at higher $m_P^2$ (solid black line). This behaviour is not surprising, and is expected from the presence of quenched chiral logarithms as discussed below Eqs.~\eqref{expans-quenched}-\eqref{expansfB}. On the other hand, the extrapolation of $F_{+,0}$ using dynamical $\chi$PT (blue dashed line) falls below the linear extrapolation. Note, however, that this does not suggest that the values of $F_{+,0}^{B\to\pi}$ obtained by the linear extrapolation from the quenched lattice data lie below the physical values of $F_{+,0}^{B\to\pi}$. We do not know the values of $F_{+,0}^{B\to\pi}$ at high $m_P$ for the dynamical case. These might as well be higher than the quenched lattice data, while we have set them equal. This issue can only be settled by the full dynamical lattice simulation. The other thing to note is, that the chiral extrapolation using dynamical $\chi$PT does not deviate much from the linear extrapolation for the case of the $F_0$ form factor, while the difference might be substantial for the $F_+$ form factor.

\index{extrapolation! chiral|)}

The discussion above suggests, that going to very low quark masses in the quenched lattice calculations of $B\to \pi$ transition is not the most sensible thing to do. Both $F_+$ and $F_0$ form factors blow up at small $m_P$ due to the presence of quenched chiral logarithms. The desirable thing to do is, of course, the dynamical lattice simulation. Chiral perturbation theory can then be used to guide the extrapolation down to the physical pion mass. The $\chi$PT driven extrapolation is important for the case of $F_+$ form factor, while it marginally deviates from the linear extrapolation for $F_0$ form factor (see Fig.~\ref{chiral-extrap}). However, as long as the dynamical lattice calculations do not become available, the linear extrapolations of quenched lattice results seem to be perfectly sensible thing to do. As a very rough estimate of the quenching errors, we might use the difference between Q$\chi$PT extrapolation and the linear one at the mass $m_P=m_\pi$, giving the value of around 30\%. For more detailed quantitative discussion on the quenching errors in $B\to \pi$ transitions, using the same method as outlined in this thesis, we refer the interested reader to the recent work \cite{Becirevic:2002sc}. We stress, however, that the method of (Q)$\chi$PT can merely point toward possible sizes of quenching effects, so that  the issue of quenching errors in $F_{+,0}$ form factors will be completely settled only by a direct comparison of quenched calculations with the future dynamical results. 
\index{extrapolation! comparison of quenched and dynamical|)}\index{lattice QCD|)}\index{quenching effects|)}

\chapter{Concluding remarks} \index{calculation! results of|(}\index{conclusions} \index{errors, estimation}\index{results}
The nonperturbative nature of QCD is a persisting problem of calculations in the hadronic physics. In order to asses these effects in the heavy meson decays, we have used the approach of effective theories. In the processes with momenta exchanges small compared to the chiral scale $\sim 1$ GeV and compared to the scale set by the heavy quark mass, either $m_c\sim 1.5$ GeV or $m_b\sim 4$ GeV, the combined chiral and heavy quark expansion can be used. Keeping the lowest order terms in the chiral and $1/m_Q$ expansion one ends up with a set of interaction terms with unknown couplings. The set of leading order couplings is small enough, so that all of them can be fixed from existing experimental data.

The heavy hadron chiral perturbation theory (HH$\chi$PT) has been applied to $D$ meson decays. The energy of intermediate pseudoscalar mesons in these decays is of order $p\sim m_D/2$,
so that the chiral expansion $p/\Lambda_\chi$ (for $\Lambda_\chi \gtrsim 1$
GeV) is rather close to
unity. Thus, we have extended the possible range
of applicability of the chiral expansion in HH$\chi$PT, compared to the 
previous treatments
like $D^*\to D \pi$, $D^*\to D \gamma$ or $D^*\to D
\gamma \gamma$, in which a heavy meson
appears in the final state. In view of this, the formally next-to-leading terms could be as large as the leading ones. The higher order terms are difficult to evaluate due to the lack of experimental data, but can be estimated by using different renormalization schemes for the chiral loops.

This idea has been tested on the decay mode $D^0\to K^0\bar{K}^0$. A naive application of the widely used factorization approximation leads to a vanishing prediction for this decay mode, in contrast to the experimental situation. In the context of the HH$\chi$PT the factorizable contribution, which is zero, is of ${\cal O}(p)$, while the decay receives nonzero contributions at the one chiral loop order ${\cal O}(p^3)$. The charged pion and kaon exchanges thus introduce the nonfactorizable contributions in the decay. A different nonfactorizable contribution coming from the product of colored currents has been also estimated using gluon condensate. The final results agree with the experimental result. The error due to the neglected higher order terms can be estimated to be at the order of $50\%$.

It is rather difficult to considerably reduce this error with the current approach. However, for the discussion of the flavor changing neutral current (FCNC) rare charm decays, where the formalism has been applied next, the errors of $50\%$ are not disturbingly large. In addition, in the radiative $D$ meson decays $D^0\to \gamma\gamma$, $D^0\to l^+l^-\gamma$ the order ${\cal O}(p^3)$ counterterms do not contribute. No unknown couplings then enter the prediction at the leading order, reducing the error. Also, the chiral loops are finite so that there is no scheme dependence of the predicted decay widths at the leading order.

 The prime interest in the rare heavy meson decays is the search for new physics signatures. Possible effects of the Standard Model extensions can be quite remarkable for some of the inclusive charm decays. For instance, the $R$ parity violating terms can increase the $c\to u l^+l^-$ decay probability by $4$ orders of magnitude over the SM quark level prediction. Experimentally more tractable are, however, the exclusive modes. The perplexity of the exclusive mode calculations is the estimation of the nonperturbative effects. While in the $B$ meson physics it is possible to find exclusive decay modes, that are dominated by the perturbative short distance physics, this is not the case for the $D$ meson rare decays. These are dominated by the nonperturbative, ``long distance'', physics. Nevertheless, as the FCNC charm rare decays are one of the rare probes of the flavor dynamics in the up-quark sector, possible effects of physics beyond the Standard Model can still be large. We have considered rare nonhadronic radiative decays $D^0\to \gamma\gamma$, $D^0\to l^+l^-\gamma$ using HH$\chi$PT. Calculations were done in the context of the Standard Model and its minimal supersymmetric extension with and without conserved $R$ parity. The results are summarized in Table \ref{tab-final}. Note that in $D^0\to l^+l^-\gamma$ calculation the inclusion of intermediate vector mesons was necessary. The results obtained are to be compared with experimental sensitivities of $\sim 10^{-6}$
expected at B-factories and CLEO-c. \index{supersymmetry}

\begin{table} [h]
\begin{center}
\begin{tabular}{|l|c|c|c|} \hline

Decay & SM& MSSM & MSSM $\sls R$ \\ \hline\hline
$D^0\to \gamma\gamma$ &$(1.6\pm 0.8)\times 10^{-8}$&$4.6 \times 10^{-8}$&$4.6\times 10^{-8}$\\\hline
$D^0\to e^+e^-\gamma$ &$(1-3) \times 10^{-9}$&$(1-3) \times 10^{-9}$&$4.5 \times 10^{-9}$\\\hline
$D^0\to \mu^+\mu^-\gamma$ &$(1-3) \times 10^{-9}$&$(1-3) \times 10^{-9}$&$50 \times 10^{-9}$\\\hline
 \end{tabular} 
 \caption[Summary of the results for rare radiative charm decays considered in the thesis]{\footnotesize{The summary of the results for rare radiative charm decays considered in the thesis. Predictions within Standard Model are gathered in the column SM, while the largest possible effects within the Minimal Supersymmetric Standard Model with and without $R$ parity conservation are shown in the last two columns respectively. For $D^0\to l^+l^-\gamma$ decays the cut on the photon energy $E_\gamma >50$ MeV has been used. The $D^0\to \gamma \gamma$ decay amplitude does not receive any additional contributions from possible $R$ parity violating terms.}}\index{summary}
\label{tab-final}
\end{center}
\end{table}

To perform the calculation several technical details had to be resolved as well. To do the calculations in the framework of the HH$\chi$PT, the solutions of one-loop integrals with the heavy-meson propagators had to be found. General results for up to four-point scalar functions are discussed in chapter \ref{scalar-loops} of the thesis. In addition, the existing calculations of renormalization group evolutions in the $B$ and $K$ rare decays have been used to perform the next-to-leading order renormalization group improved evaluation of the Wilson coefficients relevant for the charm decays. 

Finally, a modification of the HH$\chi$PT, the quenched chiral perturbation theory, has been used to discuss the common approximation made in the lattice calculation of $B\to \pi,K$ transitions. The so called quenched approximation consists of omitting the sea-quark loops in the calculations. This can be mimicked by introducing ghost fields in the theory. As a consequence the one chiral loop corrections to the $B\to \pi$ form factors $F_{+,0}$ introduce terms that diverge in the chiral limit $m_\pi\to 0$. This is similar to what has already been found earlier for the heavy meson decay constant $f_B$ and the light quark condensates. The appearance of chirally divergent terms suggests two things, (i) that the dependence of $F_{+,0}$ on $m_P$ is substantially different for the quenched and dynamical (i.e. physical) theories in the region of small final pseudoscalar masses $m_P\sim m_\pi$, and (ii) that making quenched lattice calculations with final pseudoscalar masses close to the physical pion mass is not worth the effort as it may well lead away from the dynamical result.
\index{calculation! results of|)}

\appendix
\renewcommand{\thesection}{\Alph{chapter}.\arabic{section}}
\chapter{One loop scalar and tensor functions, special cases}\label{app-loops}
The general solutions for the scalar functions appearing in the HQET have been found in \cite{Zupan:2002je} and are reviewed in chapter \ref{scalar-loops}. The final expressions for the three-point and four-point functions are quite cumbersome. For this reason we list in this appendix expressions for the three-point and four-point functions for several special sets of parameters. We also list explicit expressions for some of the tensor functions appearing in \eqref{tensor-beg}-\eqref{tensor-end}. At the end we also give a ``glossary'' between notations used in \cite{Eeg:2001un,Fajfer:2001ad} and the present text.

\section{Veltman-Passarino functions, special cases}\index{Veltman-Passarino functions}
In this section we give explicit expressions for some of the Veltman-Passarino functions used in the calculation in chapters \ref{D0K0K0bar}, \ref{rareD}, \ref{quenching-errors}. The one-point scalar function $A_0(m^2)$, and the two-point scalar function $B_0(k^2, m^2, m^2)$ for equal masses in the propagators, are
\begin{align}
A_0(m^2)&=-m^2 \ln\Bigl(\frac{m^2}{\mu^2}\Bigr)+m^2\bar{\Delta},\label{A_0}\\
B_0(k^2, m^2, m^2)&=\bar{\Delta}+1-H\Bigl(\frac{k^2}{m^2}\Bigr)-\ln\Bigl|\frac{m^2}{\mu^2}\Bigr|+i
\pi \Theta\Bigl(-\frac{m^2}{\mu^2}\Bigr)\sign(\mu^2), \label{B_0Veltman}
\end{align}
where $\bar{\Delta}=\frac{2}{\epsilon}-\gamma +\ln(4\pi)+1$. The tensor functions are
\begin{align}
B_1(k^2, m^2, m^2)&=-\frac{1}{2}B_1(k^2, m^2, m^2),\\
\begin{split}
B_{00}(k^2, m^2, m^2)&=-\frac{1}{2}\Bigl(m^2-\frac{k^2}{6}\Bigr)\bar{\Delta}+\frac{1}{2}\Bigl\{\frac{1}{3}\bigl(4
m^2-k^2)\biggl[
1-\frac{1}{2}H\Bigl(\frac{k^2}{m^2}\Bigr)\biggr]-\frac{4}{3}m^2\Bigr.\\
&\Bigl.\qquad\qquad\qquad+\frac{5}{18}k^2+\Bigl(
m^2-\frac{k^2}{6}\Bigr)\Bigl(\ln\Bigl|\frac{m^2}{\mu^2}\Bigr|-i\pi
\Theta\Bigl(-\frac{m^2}{\mu^2}\Bigr)\sign(\mu^2)\Bigr)\Bigr\},
\end{split}\\
\begin{split}
B_{11}(k^2, m^2, m^2)&=-\frac{1}{3}\left[\bar{\Delta}+\frac{7}{6}-2\frac{m^2}{k^2}+2\biggl(\frac{m^2}{k^2}-1\biggr)\biggl(1-\frac{1}{2}H\Bigl(\frac{k^2}{m^2}\Bigr)\biggr)\right.\\
&\left.\qquad\qquad\qquad -\ln\Bigl(\frac{m^2}{\mu^2}\Bigr)+i \pi
\Theta\Bigl(-\frac{m^2}{\mu^2}\Bigr)\sign(\mu^2)\right],
\end{split}\label{B_11Veltman}
\end{align}
and
\begin{equation}
H(a)=\left\{
\begin{aligned}
2&\left(1-\sqrt{4/a-1}\arctan\left(\frac{1}{\sqrt{4/a-1}}\right)\right)\qquad
& 0<a<4,\\
2 &
\left(1-\frac{1}{2}\sqrt{1-4/a}\left[\ln\left|\frac{\sqrt{1-4/a}+1}{\sqrt{1-4/a}-1}\right|-i\pi\Theta(a-4)\right]\right)\qquad
& \text{ otherwise},
\end{aligned}
\right.
\end{equation}
while $m^2$ is assumed to be positive.

Finally, we give explicit expressions for the three-point tensor functions with $k_1^2=k_2^2=0$
\begin{align}
M_2(m,k_1\negcdot k_2)&=C_{00}(0, -2 k_1\negcdot k_2, 0,m^2,m^2),\\
\begin{split}
M_3(m,k_1\negcdot k_2)&=-k_1 \negcdot k_2 C_{11}(0, -2 k_1\negcdot k_2, 0,m^2,m^2)\\
&=-k_1 \negcdot k_2 C_{22}(0, -2 k_1\negcdot k_2, 0,m^2,m^2),
\end{split}
\\
M_4(m,k_1\negcdot k_2)&=- k_1 \negcdot k_2 C_{12}(0, -2 k_1\negcdot k_2, 0,m^2,m^2),
\end{align}
used in the calculation of the $D^0\to\gamma\gamma$ decay (cf. section \ref{D0GammaGamma}). They are
\begin{subequations} \label{Ms}
\begin{align}
\begin{split}
M_2(m, k_1\negcdot k_2)=\frac{1}{2}\bigg[& \frac{1}{2}
\left(\bar{\Delta}-\ln\left(\frac{m^2}{\mu^2}\right)\right)+\frac{1}{a}\bigg(\Li\Big(\frac{2}{1+\sqrt{\quad}}\Big)+\Li\Big(\frac{2}{1-\sqrt{\quad}}\Big)\\
&+1-\sqrt{\quad}\Arcth\bigg(\frac{1}{\sqrt{\quad}}\bigg)\bigg],
\end{split}\\
M_3(m,k_1\negcdot
k_2)=\frac{1}{2}\bigg[&\sqrt{\quad}\Arcth\Big(\frac{1}{\sqrt{\quad}}\Big)-1\bigg),\\
M_4(m,k_1\negcdot
k_2)=\frac{1}{4}+&\frac{1}{2a}\bigg[\Li\Big(\frac{2}{1+\sqrt{\quad}}\Big)+\Li\Big(\frac{2}{1-\sqrt{\quad}}\Big)\bigg],
\end{align}
\end{subequations}
where we have abbreviated $a=2 k_1\negcdot k_2/m^2$ and
$\sqrt{\quad}=\sqrt{1+2
m^2/k_1\negcdot k_2}$, while $\Li(x)$ is a dilogarithm (see chapter \ref{scalar-loops}, Eq. \eqref{eqS:38}).

\section{Heavy-quark scalar and tensor functions, special cases}
In this section we give explicitly the HQET tensor functions that have been used in chapters \ref{D0K0K0bar}, \ref{rareD} for the calculation of the $D^0\to K^0\overline{K^0}$, $D^0\to \gamma\gamma$, and $D^\to l^+l^-\gamma$ decay widths. As a start we list the first few two-point tensor functions
\begin{align}
\bar{B}_1(m,\Delta)=\Delta \bar{B}_0(m,\Delta)& +A_0(m^2),\\
\begin{split}
\bar{B}_{00}(m,\Delta)=-\Delta \Big[(-m^2&+\frac{2}{3}\Delta^2)\ln\left(\frac{m^2}{\mu^2}\right)+\frac{4}{3}(\Delta^2-m^2)F\left(\frac{m}{\Delta}\right)\\
&\qquad\qquad -
\frac{2}{3}\Delta^2(1+\bar{\Delta})+\frac{1}{3}m^2(2+3\bar{\Delta})+\frac{2}{3}m^2-\frac{4}{9}\Delta^2\Big]\label{B_00}
,
\end{split}
\\
\begin{split}
\bar{B}_{11}(m,\Delta)=-\Delta \Big[(2
m^2&-\frac{8}{3}\Delta^2)\ln\left(\frac{m^2}{\mu^2}\right)-\frac{4}{3}(4
\Delta^2-m^2)F\left(\frac{m}{\Delta}\right)\\
&\qquad\qquad+\frac{8}{3}\Delta^2(1+\bar{\Delta})-\frac{2}{3}m^2(1+3\bar{\Delta})-\frac{2}{3}m^2+\frac{4}{9}\Delta^2\Big]\label{B_11}
,
\end{split}
\end{align}
where $A_0(m^2)$ can be found in \eqref{A_0}, $\bar{B}_0(m,\Delta)$ is given in \eqref{eqS:33} together with \eqref{eqS:2}, while $F(m/\Delta)$ is given in \eqref{F_def}.

Next we give the scalar three-point function in two limiting cases, $\bar{C}_0(k,m,m,\Delta)$ with $k^2=0$ and for $\bar{C}_0(-M v,m,m,\Delta+M/2)$ (in the calculation we have $M=m_D$). They are
\begin{align}
\begin{split}
\bar{C}_0(k,m,m,\Delta)|_{k^2=0} &=v\cdot k \Big[ h^2(m,\Delta+v \negcdot
k)-h^2(m,\Delta)\\
&\hskip13mm+\pi
[h(m,\Delta)-h(m,\Delta+v \negcdot k)]\Big],
\end{split}
\\
\begin{split}
\bar{C}_0\left(-M v,m,m,\Delta+\tfrac{M}{2}\right)&=-\frac{2 }{\Delta M} \bigg\{ \Big[\tfrac{\pi}{2}
-
h\left(m,\Delta-\tfrac{M}{2}\right)\Big]\sqrt{m^2 -
\left(\Delta-\tfrac{M}{2}\right)^2- i \delta \;} \\
&\hskip21mm-\Big[\tfrac{\pi}{2} -
h\left(m,\Delta+\tfrac{M}{2}\right)\Big]\sqrt{m^2 - \left(\Delta+\tfrac{M}{2}\right)^2- i
\delta \;} \\
&\hskip21mm-2 h\left(m,\tfrac{M}{2}\right) \sqrt{m^2- \tfrac{M^2}{4} - i \delta \;}\;
\bigg\},
\end{split}
\end{align}
where
\begin{equation}
 h(m,\Delta)=\left\{
\begin{aligned}
\> &\Arctan \left(\frac{\Delta}{\sqrt{m^2-\Delta^2}}\right) ; |m|>
|\Delta|\\
i & \ln \left|
\frac{m}{\Delta-\sqrt{\Delta^2-m^2}}\right|+\sign(\Delta)\frac{\pi}{2};
|m|<|\Delta|
\end{aligned}\right.,
\end{equation}
In the calculation of $D^0\to l^+l^-\gamma$ also a tensor function $\bar{C}_{00}(k,m,m,\Delta)$ with $k^2=0$ appears (cf. Eqs.~\eqref{eta1_2}, \eqref{kp1_2}, \eqref{eta3_2}, \eqref{kp3_2}, \eqref{eta4_4}). It is explicitly
\be
\begin{split}
\bar{C}_{00}(k,m,m,\Delta)|_{k^2=0}=& \frac{m^2}{2 v \negcdot k} C_{0}(k,m,m,\Delta)+\Delta + \frac{v \negcdot k}{2}\\
& - \frac{1}{4 v\negcdot k}\Big[ \Delta \bar{B}_0(m,\Delta)-(\Delta+v\negcdot k)\bar{B}_0(m,\Delta+v
\negcdot k)\Big]. 
\end{split}
\ee
Finally, let us give the expression for the four-point function $\bar{D}_0 (k_1, -k_2, m,m,m,\Delta)$ in the case of $k_1^\mu+k_2^\mu=M v^\mu$, $k_1^2=k_2^2=0$, $v\negcdot
k_1=
v\negcdot k_2=\tfrac{M}{2}$
\be
\begin{split}\label{D0-limit}
 \bar{D}_0 (k_1, -k_2, m,m,m,\Delta)=-\frac{1}{\Delta M^2} \Bigg\{& -2
h^2(m,\tfrac{M}{2}) -2 h^2(m,\Delta) + h^2(m,\Delta-\tfrac{M}{2})
+h^2(m,\Delta+\tfrac{M}{2})\\
&+i \pi \ln \left[\frac{\Delta-\tfrac{M}{2} - i \sqrt{m^2-
(\Delta-\tfrac{M}{2})^2-i \delta \;}}{-\Delta-\tfrac{M}{2} - i
\sqrt{m^2-
(\Delta+\tfrac{M}{2})^2-i \delta \;}}\right]+\\
&+ i \pi \ln \left[\frac{\Delta+ i \sqrt{m^2- \Delta^2-i
\delta \;}}{-\Delta + i
\sqrt{m^2-\Delta^2-i \delta \;}}\right]\Bigg\},
\end{split}
\ee
with $\delta>0$ an infinitesimal positive parameter. This function appears in the calculation of $D^0\to \gamma\gamma$ in section \ref{D0GammaGamma}, where $M=m_D$.

\section{Notational glossary}\label{Notational-glossary}\index{notation|(}
In this section we list relations between notation for the scalar and tensor functions as given in section \ref{Notational-conventions} and the notation commonly used in the 1-loop calculations of the chiral corrections in the heavy meson systems. The relations between one-point and two-point functions as defined in Refs. \cite{Casalbuoni:1996pg,Boyd:1994pa,Stewart:1998ke} and those defined in section \ref{Notational-conventions} are
\begin{align}
I_1(m)&=-A_0(m^2),\\
I_2(m,\Delta)&=-\Delta \bar{B}_0(m,\Delta,)\\
J_1(m,\Delta)&=-\bar{B}_{00}(m,\Delta)/\Delta,\\
J_2(m,\Delta)&=-\bar{B}_{11}(m,\Delta)/\Delta,
\end{align}
the explicit expressions for these functions can be found above, $A_0(m^2)$ in Eq.~\eqref{A_0}, $\bar{B}_0(m,\Delta)$ in \eqref{eqS:33} together with \eqref{eqS:2}, while $\bar{B}_{00}(m,\Delta)$, $\bar{B}_{11}(m,\Delta)$ are given in \eqref{B_00}, \eqref{B_11} respectively. Note that the expressions for $\bar{B}_{00}(m,\Delta)$, $\bar{B}_{11}(m,\Delta)$ differ from the ones given in \cite{Boyd:1994pa} by the last two terms in Eqs.~\eqref{B_00}, \eqref{B_11},
 that are of the order of ${\cal O}(m^2, \Delta^2)$. These
additional finite terms originate from the fact that $\eta^{\mu \nu}$ is
$4-\epsilon$ dimensional metric tensor \cite{Eeg:2001un,Hiorth:2002pp}.

Finally, we give a glossary connecting the functions defined in 
\cite{Eeg:2001un,Fajfer:2001ad} with those defined in section \ref{Notational-conventions}:
\begin{itemize}
\item Veltman-Passarino two-point functions\index{Veltman-Passarino functions}
\begin{align}
N_0(m,k^2)&=-B_0(k^2,m^2,m^2),\\
N_1(m,k^2)&=-B_1(k^2,m^2,m^2),\\
N_2(m,k^2)&=B_{11}(k^2,m^2,m^2),\\
N_3(m,k^2)&=B_{00}(k^2,m^2,m^2),
\end{align}
The explicit expressions for $B_{0,1,00,11}(k^2,m^2,m^2)$ are given in Eqs.~\eqref{B_0Veltman}-\eqref{B_11Veltman}.
\item Veltman-Passarino three-point functions for $k_1^2=k_2^2=0$
\begin{align}
M_2(m,k_1\negcdot k_2)&=C_{00}(0, -2 k_1\negcdot k_2, 0,m^2,m^2),\\
\begin{split}
M_3(m,k_1\negcdot k_2)&=-k_1 \negcdot k_2 C_{11}(0, -2 k_1\negcdot k_2, 0,m^2,m^2)\\
&=-k_1 \negcdot k_2 C_{22}(0, -2 k_1\negcdot k_2, 0,m^2,m^2),
\end{split}
\\
M_4(m,k_1\negcdot k_2)&=- k_1 \negcdot k_2 C_{12}(0, -2 k_1\negcdot k_2, 0,m^2,m^2),
\end{align}
where the explicit expressions are given in \eqref{Ms}.
\item heavy-quark three-point functions
\begin{align}
G_{0,3,4,5,6}(m,\Delta, v\negcdot k)&=- v\negcdot k \;\bar{C}_{0,00,12,22,11}(k,m,m,\Delta)|_{k^2=0},\\
\bar{G}_0(m,\Delta,M)&=-\bar{C}_0(-M v,m,m,\Delta+M/2),
\end{align}
\item heavy-quark four-point function
\be
\bar{M}_0(m,\Delta,M)=-\bar{D}_0(k_1, -k_2,m,m,m,\Delta), \label{M0Bar}
\ee
The relation in \eqref{M0Bar} is valid for the case of $k_1^\mu+k_2^\mu=M v^\mu$, $k_1^2=k_2^2=0$, $v\negcdot
k_1=
v\negcdot k_2=\tfrac{M}{2}$. The explicit expression for this function is given in Eq.~\eqref{D0-limit}.
\end{itemize}

\chapter{Invariant amplitudes for the $D^0\to l^+l^-\gamma$ decay}\label{app-D0llBarGamma}
In this appendix we list explicit expressions for the invariant amplitudes corresponding to the diagrams shown on Figures \ref{fig-1D0llBarGamma}, \ref{SDdiagr}. \index{notation|)}
 
\section{Nonresonant LD invariant amplitudes}\label{app-LDllBarGamma}
In this section we list the analytical results for the diagrams shown on Fig.
\ref{fig-1D0llBarGamma}. They contribute only to the $M_0^{\mu\nu}$ part of the invariant amplitude \eqref{inv_ampl}. Since separate diagrams are not gauge invariant, a general form of an invariant amplitude corresponding to a {\it single} diagram is
\begin{align}
M_0^{i} &=M_0^{i\mu\nu}\epsilon_\mu^*(k)\frac{1}{p^2} \bar{u}(p_1)\gamma_\nu v(p_2),\\
M_0^{i\mu\nu}&=C_{0\eta}^i(p^2)\eta^{\mu\nu}-C_{0kp}^i(p^2)\frac{p^\mu k^\nu}{p\negcdot k}+ D_0^i(p^2) \epsilon^{\mu \nu \alpha \beta} k_\alpha p_\beta.
\end{align}
For a gauge invariant sum of diagrams therefore $\sum_iC_{0\eta}^i=\sum_iC_{0kp}^i$ (cf. \eqref{inv_amp}) has to be true, which represents a very useful numerical check. 

Note that $D_0^i(p^2)$ form factors corresponding to diagrams on Fig.~\ref{fig-1D0llBarGamma} are zero. The analytical expressions for $C^i_{0\eta,kp}(p^2)$ form factors are
\begin{align}
C_{0\eta}^{1\_1}& =i g K\Big\{ V_{us} V^*_{cs} \bar{B}_0(m_K,v\negcdot p+\Delta_s^*)+ V_{ud} V^*_{cd}\bar{B}_0(m_\pi,v\negcdot p+\Delta^*)\Big\},\\
C_{0kp}^{1\_1}&=\frac{k\negcdot p}{m_D^2} C_{0\eta}^{1\_1},\\
C_{0\eta}^{1\_2}& =-2 i g K(V_{us} V^*_{cs} \bar{C}_{00}(-k,m_K,m_K,m_D+\Delta_s^*)+ V_{ud} V^*_{cd}\bar{C}_{00}(-k,m_\pi,m_\pi,m_D+\Delta^*)),\label{eta1_2}\\
\begin{split}
C_{0kp}^{1\_2}& =-2 i g K \frac{k \negcdot p}{m_D^2}\Big\{ V_{us} V^*_{cs} \Big[\bar{C}_{00}(-k,m_K,m_K,m_D+\Delta_s^*) +(m_D-vk)\bar{C}_{12}(-k,m_K,\dots)\Big]\\
&\qquad\qquad\qquad+ V_{ud} V^*_{cd}\Big[ \quad m_K\to m_\pi, \;\Delta_s^*\to \Delta^* \quad\Big]\Big\},
\end{split}\\\label{kp1_2}
C_{0\eta}^{2\_1+2\_2}&=0,
\end{align}
\begin{align}
\begin{split}
C_{0kp}^{2\_1+2\_2}&=- i g K \frac{1}{(v\negcdot k) (v\negcdot p)} \frac{k\negcdot p}{m_D^2}\Big\{ V_{us} V_{cs}^*\Big[ \big(m_K^2-\Delta_s^{*2}\big)\bar{B}_0(m_K,\Delta_s^*)+\big(m_K^2-(m_D+\Delta_s^*)^2\big)\\
&\times \bar{B}_0(m_K,m_D+\Delta_s^*)-\big(m_K^2-(vk+\Delta_s^*)^2)\big)\bar{B}_0(m_K,vk+\Delta_s^*)\\
&-(m_K^2-(vp+\Delta_s^*)^2)\bar{B}_0(m_K,vp+\Delta_s^*)\Big] + V_{ud} V_{cd}^* \Big[ \quad m_K\to m_\pi, \;\Delta_s^*\to \Delta^* \quad \Big]\Big\},
\end{split}
\\
C_{0\eta}^{2\_3}&=0,\\
\begin{split}
C_{0kp}^{2\_3}&= 2 ig K \frac{k \negcdot p}{m_D^2}\Big\{
 V_{us} V_{cs}^* \frac{1}{2(m_D-vk)}\Big[ \bar{B}_1(m_K,vk+\Delta_s^*)-\bar{B}_1(m_K,m_D+\Delta_s^*)\\
&+\bar{B}_1(m_K,\Delta_s^*)-\bar{B}_1(m_K,m_D-vk+\Delta_s^*)\\
&+2\big(m_K^2-\frac{1}{2}k^2-\Delta_s^*(vk+\Delta_s^*)\big)\bar{C}_1(-k,m_K,m_K,vk+\Delta_s^*)\\
&-2\big(m_K^2-\frac{1}{2}k^2-(m_D-vk +\Delta_s^*)(m_D+\Delta_s^*)\big)\bar{C}_1(-k,m_K,m_K,m_D+\Delta_s^*)\Big]
\\
& + V_{ud} V_{cd}^* \frac{1}{2(m_D-vk)}\Big[ \quad m_K\to m_\pi, \;\Delta_s^*\to \Delta^* \quad \Big]\Big\},
\end{split}
\\
C_{0\eta}^{3\_1}&=C_{0\eta}^{1\_1}(\rm{with}\; k \leftrightarrow p),\\
C_{0kp}^{3\_1}&=C_{0kp}^{1\_1}(\rm{with}\; k \leftrightarrow p),\\
C_{0\eta}^{3\_2}&=-2i g K \Big[V_{us} V_{cs}^* \bar{C}_{00}(k,m_K,m_K,\Delta_s^*)+V_{ud} V_{cd}^* \bar{C}_{00}(k,m_\pi,m_\pi,\Delta^*)\Big],\label{eta3_2}\\
\begin{split}
C_{0kp}^{3\_2}&=-2i g K\frac{k\negcdot p}{m_D^2}\Big[V_{us} V_{cs}^*\big(\bar{C}_{00}(k,m_K,m_K,\Delta_s^*)-(m_D-vk)\bar{C}_{12}(k,m_K,m_K,\Delta_s^*)\big)\\
&\qquad\qquad\quad+V_{ud} V_{cd}^* \big(\quad m_K\to m_\pi, \;\Delta_s^*\to \Delta^* \quad\big)\Big],
\end{split}\label{kp3_2}
\\
C_{0\eta}^{4\_1}&=C_{0\eta}^{1\_2}(\rm{with}\; k \leftrightarrow p),\\
C_{0kp}^{4\_1}&=C_{0kp}^{1\_2}(\rm{with}\; k \leftrightarrow p),\\
C_{0\eta}^{4\_2}&=C_{0\eta}^{2\_3}(\rm{with}\; k \leftrightarrow p),\\
C_{0kp}^{4\_2}&=C_{0kp}^{2\_3}(\rm{with}\; k \leftrightarrow p),\\
C_{0\eta}^{4\_3}&=C_{0\eta}^{3\_2}(\rm{with}\; k \leftrightarrow p),\\
C_{0kp}^{4\_3}&=C_{0kp}^{3\_2}(\rm{with}\; k \leftrightarrow p),\\
C_{0\eta}^{4\_4}&=4 i g K\Big[V_{us} V_{cs}^* f_\eta (p,k,m_K,\Delta_s^*)+V_{ud} V_{cd}^* f_\eta(p,k,m_\pi,\Delta^*)\Big],\label{eta4_4}\\
C_{0kp}^{4\_4}&=-4 i g K \frac{k\negcdot p}{m_D^2}\Big[V_{us} V_{cs}^* f_{kp} (p,k,m_K,\Delta_s^*)+V_{ud} V_{cd}^* f_{kp}(p,k,m_\pi,\Delta^*)\Big],\label{kp4_4}\\
C_{0\eta}^{4\_5}&=C_{0\eta}^{4\_4}(\rm{with}\; k \leftrightarrow p),\label{eta4_5}\\
C_{0kp}^{4\_5}&=C_{0kp}^{4\_4}(\rm{with}\; k \leftrightarrow p),\label{kp4_5}
\\
\begin{split}
C_{0\eta}^{4\_6}&=2 i g K\Big\{V_{us} V_{cs}^*\Big[- \bar{B}_0(m_K,m_D+\Delta_s^*)-\bar{C}_0(p+k,m_K,m_K,\Delta_s^*)\times\\
&\times(m_K^2-\Delta_s^{*2}) +m_D B_1(m_D^2,m_K^2,m_K^2)+\Delta_s^*B_0(m_D^2,m_K^2,m_K^2)\Big]
\\
&\qquad + V_{ud} V_{cd}^* \Big[ \quad m_K\to m_\pi, \;\Delta_s^*\to \Delta^* \quad \Big]\Big\},
\end{split}
\\
C_{0kp}^{4\_6}&=0,
\\
\begin{split}
C_{0\eta}^{5\_1+5\_2}&= 2 i K m_D \Big\{ V_{us}V_{cs}^* \Big[\frac{m_K^2}{2}C_0(0,p^2,m_D^2,m_K^2,m_K^2,m_K^2)+\frac{m_D^2}{8 k\negcdot p}B_0(m_D^2,m_K^2,m_K^2)\\
&-\frac{p^2}{8 k\negcdot p}B_0(p^2,m_K^2,m_K^2)+\frac{1}{4}\Big]+ V_{ud} V_{cd}^* \Big[ \quad m_K\to m_\pi, \;\Delta_s^*\to \Delta^* \quad \Big]\Big\},
\end{split}
\end{align}
\begin{align}
\begin{split}
C_{0kp}^{5\_1+5\_2}&= 2 i K {m_D}\Big\{ V_{us}V_{cs}^* \Big[\frac{m_K^2}{2}C_0(0,p^2,m_D^2,m_K^2,m_K^2,m_K^2)+\frac{p^2}{8 k\negcdot p}B_0(m_D^2,m_K^2,m_K^2)\\
&-\frac{p^2}{8 k\negcdot p}B_0(p^2,m_K^2,m_K^2)+\frac{1}{4}\Big]+ V_{ud} V_{cd}^* \Big[ \quad m_K\to m_\pi, \;\Delta_s^*\to \Delta^* \quad \Big]\Big\},
\end{split}
\\
C_{0\eta}^{5\_3}&=-\frac{i K m_D}{2} \Big[ V_{us}V_{cs}^* B_0(m_D^2,m_K^2,m_K^2)+ V_{ud}V_{cd}^* B_0(m_D^2,m_\pi^2,m_\pi^2)\Big],
\\
C_{0kp}^{5\_3}&=0,
\end{align}
where $\Delta_s^*=m_{D_s^*}-m_D$, $\Delta^*=m_{D^*}-m_D$, $K=\sqrt{m_D}{G_F}{a_1 e^3\alpha}/({16 \sqrt{2} \pi^2})$, while in $C_{\eta, kp}^{4\_4}$ we have used the abbreviation
\begin{align}
\begin{split}
f_\eta(p,k,m,\Delta)=&\bar{C}_{00}(k,m,m,vp+\Delta)+(m^2-\Delta^2)\bar{D}_{00}(p,p+k,m,m,m,\Delta)\\
&-vp\; C_{001}(p^2,k^2,(p+k)^2,m^2,m^2,m^2)-m_D C_{002}(p^2,\dots)-\Delta C_{00}(p^2,\dots)
\\
f_{kp}(p,k,m,\Delta)=&m_D \bar{C}_{12}(k,m,m,vp+\Delta)+\bar{C}_{11}(k,m,m,vp+\Delta)\\
&+(m^2-\Delta^2)\Big[\bar{D}_{11}(p,p+k,m,m,m,\Delta)\\
&\qquad+m_D\big(\bar{D}_{12}(p,\dots)+2 \bar{D}_{13}(p,\dots)+\bar{D}_{1}(p,\dots)\big)\\
&\qquad+ m_D^2\big(\bar{D}_{23}(p,\dots)+\bar{D}_{33}(p,\dots)+\bar{D}_{3}(p,\dots)\big)\Big]\\
&-m_D^3\Big[\frac{1}{m_D^2} C_{001}(p^2,k^2,(p+k)^2,m^2,m^2,m^2)+\frac{2}{m_D^2}C_{002}(p^2,\dots)+C_{222}(p^2,\dots)\\
&\qquad\qquad+\frac{vp}{m_D}C_{112}(p^2,\dots)+\left(1+\frac{vp}{m_D}\right) C_{122}(p^2,\dots)+\frac{1}{m_D^2}C_{00}(p^2,\dots)\\
&\qquad\qquad +C_{22}(p^2,\dots)+\frac{vp}{m_D}C_{12}(p^2,\dots) \Big]\\
&-\Delta m_D^2\Big[ C_{22}(p^2,\dots)+C_{12}(p^2,\dots)+C_{2}(p^2,\dots)\Big],
\end{split}\nonumber
\end{align}
with the dots representing the same dependence on the arguments as for the first function in the square brackets.

\section{Nonresonant SD invariant amplitudes} \label{app-C}
In this appendix we list the invariant amplitudes corresponding to the diagrams on Fig.~\ref{SDdiagr}. We use the notation of Eq.~\eqref{inv_ampl}, where we write down only the nonzero form factors
\begin{align}
C_0^{\text{SD}.1}&=i\frac{4}{3} K \frac{V_{ub}V^*_{cb}}{v\negcdot k +\Delta^*}\left(\beta+\frac{1}{m_c}\right) \frac{(k\negcdot p)^2}{m_D^2}\frac{C_7-C_7'}{a_1},
\\
D_0^{\text{SD}.1}
&=\frac{4}{3} K \frac{V_{ub}V^*_{cb}}{v\negcdot k +\Delta^*}\Big(\beta+\frac{1}{m_c}\Big)\frac{v\negcdot p}{m_D}\frac{C_7+C_7'}{a_1},
\\
C_0^{\text{SD}.2}&= C_0^{\text{SD}.1} (\rm{with}\; k \leftrightarrow p),
\\
D_0^{\text{SD}.2}&= D_0^{\text{SD}.1} (\rm{with}\; k \leftrightarrow p),
\\
D_0^{\text{SD}.3}&= -\frac{1}{3}K \frac{ V_{ub}V_{cb}^*}{v\negcdot k+\Delta^*}\Big(\beta+\frac{1}{m_c}\big)\frac{p^2}{m_D}\frac{C_9+C_9'}{a_1},
\end{align}
\begin{align}
D_5^{\text{SD}.4}&= D_0^{\text{SD}.3} (\rm{with}\; C_9^{(')} \to C_{10}^{(')}),\\
M_{\text{BS}}^{\text{SD}.5a}+M_{\text{BS}}^{\text{SD}.5b}&=i \frac{1}{2} K V_{ub}V_{cb}^* \frac{m}{m_D}
\frac{C_{10}-C_{10}'}{a_1},
\end{align}
where $\Delta^*=m_{D^*}-m_D$ and $K=\sqrt{m_D}{G_F}{a_1 e^3\alpha}/({16 \sqrt{2} \pi^2})$ have been used for short, while $m$ is the lepton mass. Note that the ``wrong chirality'' Wilson coefficients $C_{7,9,10}'$ are negligible in the SM.

\chapter{Chiral corrections to $B\to\pi,K$ form factors}\label{explicitBpi}
\section{Quenched theory}
Bellow we list the chiral corrections to the form factors of the $\langle P_{ij}|\bar{q_i}\gamma^{\mu} (1-\gamma_5)b|B_j\rangle$ matrix element, where the flavor content of mesons is $B_j\sim b\bar{q}_j$, $P_{ij}\sim q_i \bar{q}_j$. We work in the isospin limit with $m_{u}=m_d=\hat m\not =m_s$. In the case of $B^-\to \pi^0$, i.e., for the case of $\langle \pi^0|\bar{u}\gamma^{\mu} (1-\gamma_5)b|B_u\rangle$ matrix element, the results 
bellow have to be multiplied by $1/\sqrt{2}$.
In the expressions given below, the abbreviation $M_k^2=4\mu_0 m_k$ is used, where $m_{k=u,d,s}$ are the quark masses. We also use the definitions $T=1$ for $i\not =j$ and $T=0$
 for $i=j$. 
When the expression of the form $[f(M_i^2)-f(M_j^2)]/(M_i^2-M_j^2)$ is to be evaluated for $i=j$ 
(i.e. for $B^-\to \pi^0$), the appropriate derivative in terms of $M^2$ has to be taken. 
 The one-loop chiral corrections for the form factors in the quenched heavy hadron $\chi$PT are (cf. \eqref{eq:q1}, \eqref{eqQ:6})
\begin{align}\label{eq:q7}
\delta F_{+,0(Q)} &=\sum_{I} F_{+,0 (Q)}^{(I)}+\frac{1}{2} \delta Z_B^{\text{loop}}+\frac{1}{2} \delta Z_P^{\text{loop}},
\end{align}
where $\delta F_{+,0}$ are as defined in \eqref{eq:q1}, \eqref{eqQ:6} and do not contain the counterterms. The sum runs over diagrams shown on Figure \ref{sl2}, while $\delta Z_B^{\text{loop}}, \delta Z_P^{\text{loop}}$ are the wave function renormalization factors defined in \eqref{deltaZ} below. In the calculation of one loop contributions we make several approximations in order to simplify the final expressions. We make use of the fact that $v\negcdot p > \Delta^{(*)}$ for the $B\to P$ transition \footnote{The lower limit for $v\negcdot p$ value is the mass of the outgoing pseudoscalar meson $m_P$, while the connection with $q^2$ is $v\negcdot p=(m_B^2+m_P^2-q^2)/2 m_B$.}. Thus we can safely neglect mass differences between $B, B^*, B_s, B_s^*$ states, if they appear in the loop. This limit is physically sensible for most of the diagrams, but it does lead to a spurious singularity at $v\negcdot p\to 0$ \index{double pole} for (7a,b) diagrams. It is thus necessary and appropriate to resum the corresponding diagrams and then subtract the term, that renormalizes the $B^*$ meson mass \cite{Falk:1993fr}. The nonzero corrections $\delta F_+^{(I)}(q^2)$, where $(I)$ denotes the corresponding diagram in 
Fig.~\ref{sl2}, with $a(b)$ denoting diagrams without (with) hairpin insertion, are (in the approximation just described)
\begin{align}\label{eq:q3}
\delta F_+^{(7a)}&=\frac{2 g g'}{16 \pi^2 f^2} 3 
\Big[ J_1(M_i, v\negcdot p)-\frac{1}{vp} \frac{2\pi}{3} M_i^3\Big],
\\
\begin{split}
\delta F_+^{(7b)}&=-\frac{g^2}
{16\pi^2 f^2}\Big[\alpha_0 \Big(J_1(M_i, v\negcdot p)-\frac{1}{vp} \frac{2\pi}{3} M_i^3\Big)\\
&\qquad\qquad\qquad+\Big(\alpha_0
M_i^2-m_0^2\Big)\frac{\partial}{\partial M_i^2}\Big(J_1(M_i, v \negcdot p)-\frac{1}{vp} \frac{2\pi}{3} M_i^3\Big)\Big],
\end{split}
\\
\delta F_+^{(9a)}&=-\frac{ g g^\prime }{16\pi^2f^2}\frac{1}{vp}
\bigg[\frac{2 \pi}{3}\Big(M_i^3+M_j^3\Big)-vp \Big(J_1(M_i,vp)+J_1(M_j,vp)\Big)
\bigg],
\\
\begin{split}
\delta F_+^{(9b)}&=\frac{g^2}{3 }\frac{1}{16\pi^2f^2}
\frac{1}{M_j^2-M_i^2}\frac{1}{vp}
\biggl[\Big(\alpha_0 M_j^2-m_0^2\Big)
\Big[\frac{2\pi}{3} M_j^3-vp J_1(M_j,vp)\Big]\\
&\qquad\qquad\qquad\qquad\qquad\qquad-\Big(\alpha_0 M_i^2-m_0^2\Big)\Big[\frac{2\pi}{3}M_i^3-vp J_1(M_i,
vp)\Big]\biggr],
\end{split}
\\
\begin{split}
\delta F_+^{(12b)}&=T\frac{1}{18}\frac{1}{16\pi^2f^2}\biggl
\{-\frac{2}{M_j^2-M_i^2}\bigg[\Big(\alpha_0 M_j^2-m_0^2\Big)I_1(M_j)-
\Big(\alpha_0 M_i^2-m_0^2\Big)I_1(M_i)\bigg]\\
&+\alpha_0 \Big(I_1(M_i)+I_1(M_j)\Big)+\Big(\alpha_0 M_i^2-m_0^2\Big)
\frac{\partial}{\partial M_i^2}I_1(M_i)+\Big(\alpha_0 M_j^2-m_0^2\Big)
\frac{\partial}{\partial M_j^2}I_1(M_j)\biggr\},
\end{split}
\\
\delta F_+^{(13a)}&=- \frac{i V_L'(0)f}{\sqrt 6}
\frac{1}{16 \pi^2f^2} I_1(M_i),\\
\delta F_+^{(13b)}&=\frac{ 1}{6 }\frac{1}{16 \pi^2 f^2}
\Big[\alpha_0 I_1(M_i)+\Big(\alpha_0 M_i^2-m_0^2\Big)\frac{\partial}
{\partial M_i^2}I_1(M_i)\Big],
\end{align}
The functions $I_1(m)$, $J_1(m,\Delta)$ can be found in appendix \ref{Notational-glossary}. The nonzero corrections $\delta F_0^{(I)}
(q^2)$ are
\begin{align}\label{eq:q4}
\delta F_0^{(4a)}&=-T*\frac{iV_L^\prime(0)f}{2\sqrt{6}}\frac{1}{16\pi^2 f^2}\biggl\{2\Big[I_2(M_j,vp)-I_2(M_i,vp)\Big]+I_1(M_j)-I_1(M_i)\biggr\},
\\
\begin{split}
\delta F_0^{(4b)}&=-T*\frac{1}{6}\frac{1}{16\pi^2 f^2}\biggl\{-\frac{1}{M_j^2-M_i^2}\Big[(\alpha_0
M_j^2-m_0^2)[I_1(M_j)+2 I_2(M_j,v \negcdot p)]\\
&\qquad\qquad\qquad\qquad\qquad\qquad-(\alpha_0 M_i^2-m_0^2)[I_1(M_i)+
2 I_2(M_i,v \negcdot p)]\Big]\\
&\qquad\qquad\qquad+\alpha_0 I_1(M_i)+(\alpha_0 M_i^2-m_0^2)\frac{\partial}
{\partial M_i^2}I_1(M_i)\\
&\qquad\qquad\qquad+2 \Big[\alpha_0 I_2(M_i,vp)+
(\alpha_0 M_i^2-m_0^2)\frac{\partial}{\partial M_i^2}I_2(M_i,vp)\Big]
\biggr\},
\end{split}
\\
\delta F_0^{(14a)}&=-\frac{i V_L'(0)f}{2\sqrt{6}}\frac{1}{16 \pi^2f^2} 
[I_1(M_i)+I_1(M_j)],
\\
\begin{split}
\delta F_0^{(14b)}&=\frac{1}{18 }\frac{1}{16\pi^2f^2}\biggl\{ \frac{1}{M_j^2-M_i^2}
\Big[(\alpha_0 M_j^2-m_0^2)I_1(M_j)-(\alpha_0 M_i^2-m_0^2)I_1(M_i)\Big]\\
&+ \alpha_0 I_1(M_i)+\big(\alpha_0 M_i^2-m_0^2\big)\frac{\partial}
{\partial M_i^2}I_1(M_i)+ \alpha_0 I_1(M_j)+\big(\alpha_0 M_j^2-m_0^2
\big)\frac{\partial}{\partial M_j^2}I_1(M_j)\biggr\},
\end{split}
\end{align}
In the wave function renormalization factors $Z_{B,P}$ we split the one-loop contributions $\delta Z^{\text{loop}}_{B,P}$ and the contributions coming from the counterterms $\delta Z^{\text{\text{c.t.}}}_{B,P}$ 
\begin{equation}
Z_{B,P}=1+\delta Z_{B,P}=1+ \delta Z^{\text{loop}}_{B,P}+\delta Z^{\text{\text{c.t.}}}_{B,P}.\label{deltaZ}
\end{equation}
The loop contributions are
\begin{subequations}
\begin{align}
\delta Z_{B_j}^{\text{loop}}&=\frac{1}{16\pi^2 f^2}\Big[(2 g^2 \alpha_0M_j^2-6 g g'M_j^2-g^2 m_0^2) \ln\Big(\frac{M_j^2}{\mu^2}
\Big)\nonumber\\
&+\alpha_0 g^2M_j^2-m_0^2 g^2 +\Big(-2g^2M_j^2\alpha_0+6 g g' M_j^2 +g^2
m_0^2 \Big)\bar{\Delta}\Big],\\
\delta Z_{P_{ij}}^{\text{loop}}&=\frac{1}{16 \pi^2 f^2}\frac{1}{9} \Big\{ 2 \frac{\ln\big(M_j^2/M_i^2\big)}{M_j^2-M_i^2}\big(\alpha_0 M_j^2 M_i^2-\frac{m_0^2}{2}(M_j^2+M_i^2)\big)\nonumber\\
&\qquad + 2 m_0^2 -\alpha_0 (M_i^2+M_j^2) \Big\} ~.
\end{align}
\end{subequations}
while the counterterms contribute as 
\begin{equation}
\delta Z_{B_j}^{\text{\text{c.t.}}}= k_1 m_j, \hskip2cm \delta Z_{P_{ij}}^{\text{\text{c.t.}}}=-8 L_5\frac{4\mu_0}{f^2}(m_{i} +m_{j}),
\end{equation}
where $m_{i,j}$ are the quark masses as before.
For $\pi, K$ the wave function renormalization factor $Z_{P_{ij}}$ is in the isospin limit $m_u=m_d=\hat m$ 
\begin{subequations}
\begin{align}
Z_\pi&=1 -8 L_5\frac{4\mu_0}{f^2}~2 \hat m,\\
Z_K&=1 +\frac{1}{16 \pi^2 f^2 }\frac{1}{9}\bigg\{ \frac{\ln \big(\frac{2 m_K^2}{m_\pi^2}-1 \big)}
{(m_K^2-m_\pi^2)} \Big(-\alpha_0 m_\pi^4+\alpha_0 2 m_K^2 m_\pi^2 -m_0^2 m_K^2\Big)\nonumber\\
&\qquad + 2 m_0^2 -\alpha_0 2 m_K^2 \Big\} -8 L_5\frac{4 \mu_0}{f^2}(\hat m +m_s)~.
\end{align}
\end{subequations}
We list also the expressions for the heavy 
and light meson decay constants 
\begin{align}
\begin{split}
f_{B_j}&=\frac{\alpha}{\sqrt{m_B}}\Big(1+\frac{1}{16\pi^2 f^2}\Big[\tfrac{1}{6}\alpha_0 I_1(M_j)+
\tfrac{1}{6}(\alpha_0 M_j^2-m_0^2)\frac{\partial}{\partial M_j^2}I_1(M_j)\Big]\\
&\qquad\qquad\qquad\qquad\qquad\qquad\qquad\qquad-\frac{ifV_L^\prime(0)}
{\sqrt{6}16\pi^2 f^2}I_1(M_j)+\varkappa_1m_j+\tfrac{1}{2}\delta Z_{B_j}\Big),
\end{split}
\\
f_{P_{ij}}&=f\Big(1-\frac{1}{16\pi^2 f^2}\frac{2}{9}\bigg\{2 \frac{\ln\big(M_j^2/M_i^2\big)}{M_j^2-M_i^2}\big(\alpha_0 M_j^2 M_i^2-\frac{m_0^2}{2}(M_j^2+M_i^2)\big) \nonumber\\
&\qquad \qquad \qquad\qquad\qquad+2 m_0^2 -\alpha_0 (M_i^2+M_j^2) \Big\}+8L_5\frac{4\mu_0}{f^2}(m_{q_i}+m_{q_j})+\tfrac{1}{2}\delta Z_{P_{ij}}\Big),
\end{align}
where $\delta Z_{B_j},\delta Z_{P_{ij}}$ are defined in \eqref{deltaZ}. For $\pi$ and $K$ this leads in the isospin limit to
\begin{subequations}
\begin{align}
f_K&=f\Big(1-\frac{1}{16\pi^2 f^2}\frac{2}{9}\bigg\{ \frac{\ln \big(\frac{2 m_K^2}{m_\pi^2}-1 \big)}
{(m_K^2-m_\pi^2)} \Big(-\alpha_0 m_\pi^4+\alpha_0 2 m_K^2 m_\pi^2 -m_0^2
m_K^2\Big)\nonumber\\
&\qquad + 2 m_0^2 -\alpha_0 2 m_K^2 \Big\}+8L_5\frac{4\mu_0}{f^2}(m_s+\hat m)+\tfrac{1}{2}\delta Z_K \Big),
\\
f_\pi &=f\Big(1+8L_5\frac{4\mu_0}{f^2}2\hat m+\tfrac{1}{2}\delta Z_\pi\Big).
\end{align}
\end{subequations}
\index{fK, chiral corrections to@$f_K$, chiral corrections to}\index{fpi, chiral corrections to@$f_\pi$, chiral corrections to}

\section{Dynamical theory}

In this subsection the expressions for the form factors in the 
dynamical theory are listed. In the dynamical case the supertrace along $6\times 6$ matrices is replaced by the trace 
along the $3\times 3$ matrices, the $\eta^\prime$ field decouples, the $\phi$ field is traceless, 
$\Phi_0$ is set to zero and there are no diagrams with crosses in Fig.~\ref{sl1}, \ref{sl2}. The non-analytic 
contributions to form factors in this theory have been calculated in \cite{Falk:1993fr}. Our results 
include also the analytic terms and are obtained from the Lagrangians and currents as given in chapter \ref{HQET}. As in the quenched case, we work in the isospin limit $m_u=m_d=\hat m$ and neglect the differences of heavy meson masses in the loops. The chiral corrections $\delta F_{+,0}$ \eqref{eqQ:4}, \eqref{eqQ:5} to the form factors of $B_j \to P_{ij}$ transition $\langle P_{ij}|\bar{q_i}\gamma^\mu (1-\gamma_5)b|B_j\rangle$ are
\begin{equation}
\delta F_{+,0} =\sum_{I} \delta F_{+,0}^{(I)})+\frac{1}{2}\delta Z_{B_j}^{\text{loop}}+\frac{1}{2}\delta Z_{P_{ij}}^{\text{loop}}.
\end{equation}
The sum runs over the diagrams on Figure \ref{sl2}, and $\delta Z_{B_j, P_{ij}}^{\text{loop}}$ are the one loop wave function renormalization factors defined as in \eqref{deltaZ}. One loop chiral corrections to the $F_+$ form factor are 
\begin{align}
\delta F_+^{(7a)}&=\frac{g^2}{16\pi^2f^2}\Big(\sum_{P'} C_{B_j P' P_{ij}}^{(7a)} 3 [ J_1(m_{P'}, vp)-\frac{1}{vp}\frac{2 \pi}{3} m_{P'}^3]\Big),
\\
\delta F_+^{(9a)}&=\frac{g^2}{16\pi^2f^2}\frac{1}{vp}\Big\{ \sum_{P'} C_{B_j P' P_{ij}}^{(9a)}\Big[ \frac{2\pi}{3}m_{P'}^3 -vp J_1(m_{P'}, vp) \Big]\Big\}, \\
\delta F_+^{(12a)}&= \frac{1}{16\pi^2f^2}[\sum_{P'} C_{B_j P' P_{ij}}^{(12a)}I_1(m_{P'})],
\\
\delta F_+^{(13a)}&= \frac{1}{16\pi^2f^2}[\sum_{P'} C_{B_j P' P_{ij}}^{(13a)}I_1(m_{P'})].
\end{align}
The coefficients $C_{B_j P' P_{ij}}$ depend on the final and initial state and are 
\begin{itemize}
\item for $B \to K $ transition \\
\begin{tabular}{l l l l l l l}
$C_{B \pi K}^{(7a)}=0$, & $C_{B K K}^{(7a)}=2$, &$C_{B \eta K}^{(7a)}=\frac{2}{3}$;&\qquad&$
C_{B \pi K}^{(9a)}=0$, &$C_{B K K}^{(9a)}=0$, &$C_{B \eta K}^{(9a)}=\frac{1}{3}$;
\\ 
$C_{B \pi K}^{(12a)}=-\frac{1}{4}$, &$C_{B K K}^{(12a)}=-\frac{1}{2}$, &$C_{B \eta K}^{(12a)}=-\frac{1}{4}$;&\qquad&
$C_{B \pi K}^{(13a)}=0$, &$C_{B K K}^{(13a)}=-1$, &$C_{B \eta K}^{(13a)}=-\frac{1}{3}$;
\end{tabular}

\item for $B\to \pi$ transition \\
\begin{tabular}{l l l l l l l}
 $C_{B \pi \pi}^{(7a)}=\frac{3}{2}$,& $C_{B K \pi}^{(7a)}=1$,& $C_{B \eta \pi}^{(7a)}=\frac{1}{6}$;&\qquad&
 $C_{B \pi \pi}^{(9a)}=\frac{1}{2}$,& $C_{B K \pi}^{(9a)}=0$,& $C_{B \eta \pi}^{(9a)}=-\frac{1}{6}$;
\\ 
$C_{B \pi \pi}^{(12a)}=-\frac{2}{3}$,& $C_{B K \pi}^{(12a)}=-\frac{1}{3}$,&$C_{B \eta \pi}^{(12a)}=0$;&\qquad&
$C_{B \pi \pi}^{(13a)}=-\frac{3}{4}$,& $C_{B K \pi}^{(13a)}=-\frac{1}{2}$,&$C_{B \eta \pi}^{(13a)}=-\frac{1}{12}$; 
\end{tabular}

\item for $B_s \to K$ transition \\
\begin{tabular}{l l l l l l }
$C_{B_s \pi K}^{(7a)}=\frac{3}{2}$,& $C_{B_s K K}^{(7a)}=1$,&$C_{B_s \eta K}^{(7a)}=\frac{1}{6}$; &
$C_{B_s \pi K}^{(9a)}=0$,& $C_{B_s K K}^{(9a)}=0$,& $C_{B_s \eta K}^{(9a)}=\frac{1}{3}$;\\
 $C_{B_s \pi K}^{(12a)}=-\frac{1}{4}$,&$C_{B_s K K}^{(12a)}=-\frac{1}{2}$,& $C_{B_s \eta K}^{(12a)}=-\frac{1}{4}$;&
$C_{B_s \pi K}^{(13a)}=-\frac{3}{4}$,& $C_{B_s K K}^{(13a)}=-\frac{1}{2}$,& $C_{B_s \eta K}^{(13a)}=-\frac{1}{12}$;
\end{tabular}
\end{itemize}
\vskip5mm
\noindent
The nonzero one-loop chiral corrections to the $F_0$ form factor of $B_j \to P_{ij}$ transition are
\begin{align}
\delta F_0^{(4a)}&= \frac{1}{16\pi^2f^2}\Big\{ \sum_{P'} D_{B_jP' P_{ij}}^{(4a)}\Big[
I_2(m_{P'},vp)+\frac{1}{2}I_1(m_{P'})\Big]\Big\},\\
\delta F_0^{(14a)}&=\frac{1}{16\pi^2f^2}[\sum_{P'}D_{B_jP' P_{ij}}^{(14a)}I_1(m_{P'})],
\end{align}
where the coefficients are
\begin{itemize}
\item for $B\to K$\\
$\begin{aligned}
D_{B\pi K}^{(4a)}&=0, D_{B K K}^{(4a)}=2, D_{B \eta K}^{(4a)}=1;& D_{B\pi K}^{(14a)}=-\tfrac{1}{4}, D_{B K K}^{(14a)}=-\tfrac{1}{2}, D_{B \eta K}^{(14a)}=-\tfrac{1}{12};&
\end{aligned}$

\item for $B\to \pi $\\
$\begin{aligned}
 D_{B\pi \pi}^{(4a)}&=2, D_{B K \pi}^{(4a)}=1, D_{B \eta \pi}^{(4a)}=0 ;& D_{B\pi \pi}^{(14a)}=-\tfrac{5}{12}, D_{B K \pi}^{(14a)}=-\tfrac{1}{3}, D_{B \eta \pi}^{(14a)}=-\tfrac{1}{12} ;&
\end{aligned}$

\item for $B_s\to K$\\
$D_{B_s\pi K}^{(4a)}=\tfrac{3}{2}$, $D_{B_s K K}^{(4a)}=1$, $D_{B_s \eta K}^{(4a)}=\tfrac{1}{2}$; $D_{B_s\pi K}^{(14a)}=-\tfrac{1}{4}$, $D_{B_s K K}^{(14a)}=-\tfrac{1}{2}$, $D_{B_s \eta K}^{(14a)}=-\tfrac{1}{12}$;
\end{itemize}
\vskip5mm
\noindent
The wave function renormalization factors $Z$ for $B$ mesons in the dynamical theory are
\begin{align}
Z_{B_{u,d}}&=1-\frac{3g^2}{16\pi^2f^2}\big[\frac{3}{2}
I_1(m_\pi)+I_1(m_K)+\frac{1}{6}I_1(m_\eta)\big]+k_1\hat m+k_2(m_u+m_d+m_s),\\
Z_{B_s}&=1-\frac{3g^2}{16\pi^2f^2}\big[2I_1(m_K)+\frac{2}{3}I_1(m_\eta)\big]+k_1 m_s+k_2(m_u+m_d+m_s),
\end{align}
which are the limiting $\Delta\to 0$ expressions of \eqref{ZDa}, \eqref{ZDSta}. The wave function renormalization factors for the light pseudoscalars $K$, $\pi$ are
\begin{align}
\begin{split}
Z_{K}&=1+\frac{1}{16\pi^2f^2}[I_1(m_K)+\frac{1}{2}I_1(m_\pi)+\frac{1}{2}I_1(m_\eta)]\\
&\qquad\qquad\qquad\qquad
-8L_5\frac{4\mu_0}{f^2}
(\hat m+m_s)-16L_4\frac{4\mu_0}{f^2}(m_u+m_d+m_s),
\end{split}\\
Z_{\pi}&=1+\frac{1}{16\pi^2f^2}[\frac{2}{3}I_1(m_K)+\frac{4}{3}I_1(m_\pi)]-8L_5\frac{4\mu_0}{f^2}~2\hat m-
16L_4\frac{4\mu_0}{f^2}(m_u+m_d+m_s),
\end{align}
Decay constants of $B$ mesons are
\begin{align}
f_{B_s}&=\frac{\alpha}{\sqrt{m_B}}\Big(1-\frac{1}{16\pi^2 f^2}\big[I_1(m_K)+\frac{1}{3}I_1(m_\eta)]+
\varkappa_1m_s+\varkappa_2(m_u+m_d+m_s)+\frac{1}{2}\delta Z_{B_s}\Big),\label{eqQ:7}\\
\begin{split}
f_{B_{u,d}}&=\frac{\alpha}{\sqrt{m_B}}\Big(1-\frac{1}{16\pi^2 f^2}\big[\frac{3}{4}I_1(m_\pi)+\frac{1}
{2}I_1(m_K)+\frac{1}{12}I_1(m_\eta)]\\
&\qquad\qquad\qquad\qquad\qquad\qquad\qquad\qquad
+\varkappa_1\hat m+\varkappa_2(m_u+m_d+m_s)+\frac{1}{2}\delta{Z_{B_{u,d}}}\Big),
\end{split}
\end{align}
while for $\pi, K$ they are
\begin{align}
\begin{split}
f_K&=f\Big(1-\frac{1}{16\pi^2 f^2}[I_1(m_\eta)+2I_1(m_K)+I_1(m_\pi)]+8L_5\frac{4\mu_0}{f^2}(m_s+\hat m)\\
&\qquad\qquad\qquad\qquad\qquad\qquad\qquad\qquad\qquad+
16L_4\frac{4\mu_0}{f^2}(m_u+m_d+m_s)+\frac{1}{2}\delta{Z_K}\Big),
\end{split}
\\
\begin{split}
f_\pi&=f\Big(1-\frac{1}{16\pi^2 f^2}\frac{4}{3}[I_1(m_K)+2I_1(m_\pi)]+8L_5\frac{4\mu_0}{f^2}2\hat m\\
&\qquad\qquad\qquad\qquad\qquad\qquad\qquad\qquad\qquad+
16L_4\frac{4\mu_0}{f^2}(m_u+m_d+m_s)+\frac{1}{2}\delta{Z_\pi}\Big).\label{eq:End}
\end{split}
\end{align}
\index{fK, chiral corrections to@$f_K$, chiral corrections to}\index{fpi, chiral corrections to@$f_\pi$, chiral corrections to}

\index{exclusive decays|see{$D^0\to \gamma\gamma$, $D^0\to K^0\bar{K}^0$, $D^0\to l^+l^-\gamma$}}
\index{formalism, used in thesis|see{{\it approach, used in thesis}}}
\index{form factors! in $B\to \pi$|see{$F_{+,0}$}}
\index{framework, used in thesis|see{{\it approach, used in thesis}}}
\index{FSI|see{{\it final state interaction}s}}
\index{Glashow-Iliopoulos-Maiani|see{{\it GIM suppression}}}
\index{GL, renormalization scheme|see{{\it renormalization scheme, Gasser-Leutwyler}}}
\index{HH$\chi$PT|see{{\it heavy hadron chiral perturbation theory}}}
\index{inclusive decays|see{{\it $c\to u\gamma$, $c\to u l^+l^-$}}}
\index{IR|see{{\it infrared divergences}}}
\index{Lehmann-Symanzik-Zymmermann|see{{\it LSZ}}}
\index{leptonic decay|see{{\it $c\to u l^+l^-$, $D^0 \to l^+l^-\gamma$}}}
\index{loop! corrections|see{{\it contributions, chiral loop}}}
\index{loop! regularization|see{{\it dimensional regularization}}}
\index{mechanism of decay|see{{\it decay}}}
\index{Minimal Supersymmetric Standard Mode|see{{\it MSSM}}}
\index{neglected effects, in thesis|see{{\it difficulties, of approach}}}
\index{Levi-Civita, tensor|see{{\it notation}}}
\index{notation|see{{\it also Notation in the preamble to thesis}}}
\index{operator product expansion|see{{\it OPE}}}
\index{QQCD|see{{\it also quenched approximation}}}
\index{reliability of approach used|see{{\it difficulties, of approach}}}
\index{RG|see{{\it renormalization group}}}
\index{semileptonic operators|see{{\it operators, semileptonic}}}
\index{new physics!see{{\it signatures, new physics}}}
\index{SUSY|see{{\it supersymmetry}}}
\index{trilinear couplings|see{{\it $R$ parity violating terms}}}
\index{vector meson dominance|see{{\it VMD}}}
\index{Wilson expansion|see{{\it OPE}}}
\newpage
\addcontentsline{toc}{chapter}{Index}
\printindex

\chapter*{List of abbreviations}\markboth{LIST OF ABBREVIATIONS}{}\addcontentsline{toc}{chapter}{List of abbreviations}
\begin{tabular}{l |l}
{\bf $\chi$PT}& Chiral Perturbation Theory\\
{\bf $\chi$QM}& Chiral Quark Model\\
{\bf CKM}& Cabibbo-Kobayashi-Maskawa\\
{\bf FCNC}& Flavor Changing Neutral Current\\
{\bf FSI}& Final State Interaction(s)\\
{\bf GIM}& Glashow-Iliopoulos-Maiani \\
{\bf GL}& Gasser-Leutwyler\\
{\bf GUT}& Grand Unified Theory\\
{\bf HH$\chi$PT}& Heavy Hadron Chiral Perturbation Theory\\
{\bf HL$\chi$QM}& Heavy-Light Chiral Quark Model\\
{\bf HQET}& Heavy Quark Effective Theory\\
{\bf IL}& Inami-Lim \\
{\bf IR}& Infrared\\
{\bf LO}& Leading Order\\
{\bf LSZ}& Lehmann-Symanzik-Zymmermann\\
{\bf MSSM}& Minimal Supersymmetric Standard Model\\
{\bf NLO}& Next-to-Leading Order\\
{\bf NNLO }& Next-to-Next-to-Leading Order\\
{\bf OPE}& Operator Product Expansion\\
{\bf $\overline{\text{MS}}$}& Modified Minimal Subtraction \\
{\bf QCD}& Quantum Chromodynamics\\
{\bf Q$\chi$PT}& Quenched Chiral Perturbation Theory\\
{\bf QED}& Quantum Electrodynamics\\
{\bf QQCD}& Quenched Quantum Chromodynamics\\
{\bf RG}& Renormalization Group\\
{\bf RGE}& Renormalization Group Equation(s)/Evolution\\
{\bf SM}& Standard Model\\
{\bf SUSY}& Supersymmetry\\
{\bf UV}& Ultraviolet \\
{\bf VMD}& Vector Meson Dominance
\end{tabular}

\end{document}